\documentclass[a4paper,11pt]{article}
\usepackage{jheppub}
\usepackage[utf8]{inputenc}
\usepackage{url}
\usepackage{mathrsfs}  
\usepackage{pdflscape}
\usepackage{refcount}
\usepackage{booktabs}
\usepackage{todonotes}
\usepackage{enumitem}
\usepackage{adjustbox}
\usepackage{afterpage}

\usepackage{mathtools}

\usepackage{caption}
\usepackage{subcaption}

\usepackage{xstring}
\usepackage{tikz}
\usetikzlibrary{positioning}
\usetikzlibrary{arrows}
\usetikzlibrary{decorations.pathreplacing}
\usetikzlibrary{shapes.geometric}
\usetikzlibrary{calc}
\usetikzlibrary{decorations.pathmorphing}
\usetikzlibrary{shapes.misc}
\usetikzlibrary{snakes}
\usetikzlibrary{decorations.markings} 

\definecolor{goodgreen}{RGB}{55,169,49}

\newcommand{\tQ}{\tilde{Q}}
\newcommand{\tripleslash}{\mathbin{/\mkern-6mu/\mkern-6mu/}}

\usepackage[normalem]{ulem}

\newcommand{\convexpath}[2]{
  [   
  create hullcoords/.code={
    \global\edef\namelist{#1}
    \foreach [count=\counter] \nodename in \namelist {
      \global\edef\numberofnodes{\counter}
      \coordinate (hullcoord\counter) at (\nodename);
    }
    \coordinate (hullcoord0) at (hullcoord\numberofnodes);
    \pgfmathtruncatemacro\lastnumber{\numberofnodes+1}
    \coordinate (hullcoord\lastnumber) at (hullcoord1);
  },
  create hullcoords
  ]
  ($(hullcoord1)!#2!-90:(hullcoord0)$)
  \foreach [
  evaluate=\currentnode as \previousnode using \currentnode-1,
  evaluate=\currentnode as \nextnode using \currentnode+1
  ] \currentnode in {1,...,\numberofnodes} {
    let \p1 = ($(hullcoord\currentnode) - (hullcoord\previousnode)$),
    \n1 = {atan2(\y1,\x1) + 90},
    \p2 = ($(hullcoord\nextnode) - (hullcoord\currentnode)$),
    \n2 = {atan2(\y2,\x2) + 90},
    \n{delta} = {Mod(\n2-\n1,360) - 360}
    in 
    {arc [start angle=\n1, delta angle=\n{delta}, radius=#2]}
    -- ($(hullcoord\nextnode)!#2!-90:(hullcoord\currentnode)$) 
  }
}

\tikzset{flavour/.style={draw=none,minimum size=0.3mm,fill=white, regular polygon,regular polygon sides=4,draw}}
\tikzset{flavor/.style={draw=none,minimum size=0.3mm,fill=white, regular polygon,regular polygon sides=4,draw}}
\tikzset{flavoro/.style={draw=none,minimum size=0.3mm,fill=white, regular polygon,regular polygon sides=4,draw,fill=orange}}
\tikzset{gaugeBig/.style={inner sep=1.3mm,draw=none,fill=white,minimum size=2mm,circle, draw}}
\tikzset{bd/.style={circle, draw=black, inner sep=0pt, fill=black, minimum size=2mm}}
\tikzset{wd/.style={circle, draw=black, inner sep=0pt, fill=white, minimum size=2mm}}
\tikzset{Dynkin/.style={circle, draw=black, inner sep=0pt, fill=white, minimum size=2mm}}
\tikzstyle{ligne}=[draw, very thick] 
\tikzstyle{gridline}=[draw, gray] 
\tikzset{gauge/.style={circle, draw,inner sep=2.5pt}}
\tikzset{gaugeo/.style={circle, draw,inner sep=2.5pt,fill=orange}}
\tikzset{gauger/.style={circle, draw,inner sep=2.5pt,fill=orange}}
\tikzset{gaugeb/.style={circle, draw,inner sep=2.5pt,fill=black}}
\tikzset{gaugeg/.style={circle, draw,inner sep=2.5pt,fill=green}}
\tikzset{hasse/.style={circle, fill,inner sep=1pt}}
\newcommand{\zb}{\bar{z}}
\newcommand{\rect}[5]{
\fill[black!#5] (#1,#2)--(#3,#2)--(#3,#4)--(#1,#4)--(#1,#2);}

\def\ns#1#2{
	\node[circle, draw, red, fill=white] (#2) at (#1){};
	\node[cross out, draw, red] at (#1){};
}

\def\nsm#1#2{
	\node[circle, draw, cyan, fill=white] (#2) at (#1){};
	\node[cross out, draw, cyan] at (#1){};
}

\def\df#1#2{
	\node[cross out, draw,blue] (#2) at (#1){};
}

\newcommand{\torange}[1]{\textcolor{orange}{#1}}

\preprint{\hspace{1cm}}
\title{FI-flows of 3d N=4 Theories}
\author[1,2]{Antoine Bourget,}
\author[3,4]{Simone Giacomelli,}
\author[5]{ Julius F.\ Grimminger,}
\affiliation[1]{Université Paris-Saclay, CNRS, CEA, Institut de physique théorique, 91191, Gif-sur-Yvette, France}
\affiliation[2]{Laboratoire de Physique de l’\'Ecole normale supérieure, ENS, Université PSL, CNRS, Sorbonne
Université, Université de Paris, F-75005 Paris, France}
\affiliation[3]{Dipartimento di Fisica, Universit\`a di Milano-Bicocca, Piazza della Scienza 3, \\ I-20126 Milano, Italy}
\affiliation[4]{INFN, sezione di Milano-Bicocca, Piazza della Scienza 3, \\ I-20126 Milano, Italy}
\affiliation[5]{Theoretical Physics Group, The Blackett Laboratory, Imperial College, London,
Prince Consort Road London, SW7 2AZ, UK}
\emailAdd{antoine.bourget@polytechnique.org}
\emailAdd{simone.giacomelli@unimib.it}
\emailAdd{julius.grimminger17@imperial.ac.uk}

\abstract{We study the 3d $\mathcal{N}=4$ RG-flows triggered by Fayet-Iliopoulos deformations in unitary quiver theories. These deformations can be implemented by a new quiver algorithm which contains at its heart a problem at the intersection of linear algebra and graph theory. When interpreted as magnetic quivers for SQFTs in various dimensions, our results provide a systematic way to explore RG-flows triggered by mass deformations and generalizations thereof. This is illustrated by case studies of SQCD theories and low rank 4d $\mathcal{N}=2$ SCFTs. A delightful by-product of our work is the discovery of an interesting new 3d mirror pair.}

\begin{document}
\maketitle

\pagebreak

\section{Introduction}

\begin{figure}
    \centering
\begin{tikzpicture}
\draw[fill=orange!20] (-1,0)--(6,0)--(8,3)--(1,3)--(-1,0);
\draw[orange,thick] (1,0.5)--(6,2.5);
\draw[orange,thick] (4,0.5)--(3,2.5);
\draw[orange,thick] (1,1.5)--(6,1.5);
\node[circle, fill=orange,inner sep=2.5pt] at (3.5,1.5) {};
\node at (3.5,5) {\begin{tikzpicture}
        \draw (0,0) ellipse (1.5 and .5);
        \draw (-1.5,0)--(0,-2.5)--(1.5,0);
        \node[hasse] at (0,-2.5) {};
\end{tikzpicture}};
\node at (7.5,6) {\begin{tikzpicture}
 \draw (5,0) ellipse (1.5 and .5);
        \draw (3.5,0)--(3.75,-3) (6.25,-2.5)--(6.5,0);
        \draw (3.75,-3) .. controls (4.5,-1) .. (5,-2.5);
        \draw (5,-2.5) .. controls (5.5,-1) .. (6.25,-2.5);
        \node[hasse] at (3.75,-3) {};
        \node[hasse] at (5,-2.5) {};
        \node[hasse] at (6.25,-2.5) {};
\end{tikzpicture}};
\node at (-.5,5) {\begin{tikzpicture}
        \draw (0,0) ellipse (1.5 and .5);
        \draw (-1.5,0)--(-.4,-2) (.4,-2)--(1.5,0);
        \draw (-.4,-2) .. controls (0,-2.5) .. (.4,-2);
\end{tikzpicture}};
\draw[dashed] (-.5,3.2)--(2,2);
\draw[dashed] (3.5,3.2)--(3.5,1.5);
\draw[dashed] (7.5,4.2)--(4.7,2);
\node at (7.5,8.5) {Minimal Deformation};
\node at (3.5,7.0) {No Deformation};
\node at (-.5,7.0) {Generic Deformation};
\node at (1,0.2) {\textcolor{orange}{FI parameter space}};
\end{tikzpicture}
    \caption{Stratification of FI space (orange plane). At the origin, the HB (black) is singular while at a generic point it is resolved. The CB is lifted at a generic point; on the locus of a \emph{minimal} deformation, the CB symmetry drops in rank by one, by definition of minimality.}
    \label{fig:Stratification}
\end{figure}
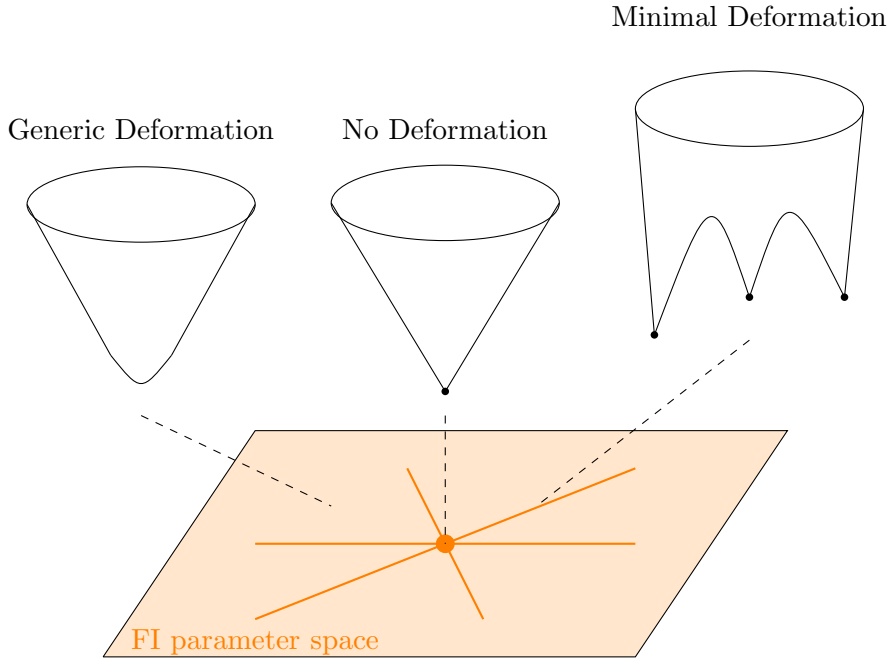

In this paper we address the systematic study of Fayet-Iliopoulos (FI) deformations \cite{Fayet:1974jb} of 3d $\mathcal{N}=4$ unitary quiver gauge theories. In the literature, the moduli spaces of vacua of $3d$ $\mathcal{N}=4$ theories, as well as other quantities, are typically studied in two extreme settings: either all FI terms are set to a generic non-zero value, the Coulomb branch is lifted and the Higgs branch is resolved;\footnote{Alternatively sometimes all mass terms are set to generic values. Masses and FI terms are exchanged by 3d mirror symmetry  \cite{Intriligator:1996ex}.} or all FI terms (and masses) are taken to be zero, Higgs, Coulomb and all mixed branches are present and singular, see Figure \ref{fig:Stratification}. In the following we will consider all options of FI terms such that there is at least one supersymmetric vacuum.

FI parameters of a 3d $\mathcal{N}=4$ theory are a real triplet valued in the Cartan subalgebra of the Coulomb branch symmetry algebra
\begin{equation}
    \zeta\in\mathbb{R}^3\otimes\mathfrak{t}_\mathcal{C}\;.
\end{equation}
Different choices of FI can trigger different RG flows, hence there is a notion of \emph{stratification} in the space of FI deformations \cite{ginzburg2009lectures}. Rather than obtaining such a stratification through some geometric observable on the space of FI deformations, we will use quantum field theory considerations and focus on the notion of \emph{minimal} deformations. We consider a deformation as minimal if the rank of the Coulomb branch global symmetry of the quiver drops by one.\footnote{Typically a FI deformation can trigger an RG flow to multiple SCFTs expressed by various quivers $\mathsf{Q}_i$ with Coulomb branches $\mathcal{C}_i$. In this case is is the rank of the combined global symmetry of all $\mathcal{C}_i$s which should drop by one.}
We conjecture that these deformations precisely parametrize the lowest dimensional loci in the full deformation space, see Figure \ref{fig:Stratification}. Consecutive minimal deformations will eventually lead to completely deforming the theory.  In algebraic geometry we are studying the hyper-K\"ahler quotient
\begin{equation}
    \mathbb{H}^n\underset{\zeta}{\tripleslash}G
\end{equation}
where $G$ is the gauge group and $\zeta\in\mathbb{R}^3\otimes\mathfrak{t}_{\mathcal{Z}(G^\vee)}$ is the level of the moment map. In other words we are studying resolutions / deformations of the Higgs branch \cite{Hitchin:1986ea,Lindstrom:1999pz}. However, it is important to note that the physical deformation parameters present in the quantum field theory need not coincide with the allowed deformation parameters of the Higgs branch as a geometric space. A simple example is a discrete gauge theory with one hypermultiplet charged under the gauge group $\mathbb{Z}_k$: the Higgs branch is $A_{k-1}=\mathbb{C}^2/\mathbb{Z}_k$, which geometrically has $k-1$ deformation parameters, but there are no such parameters in the discrete gauge theory.\footnote{It has been understood for a long time that is it not sufficient to characterize theories by geometry alone, but that one needs to specify allowed deformations. Fundamental instances are the frozen singularities in string theory \cite{Witten:1997bs}, see also \cite{Tachikawa:2015wka}, and this idea is particularly important in the classification of theories from their moduli space geometry \cite{Argyres:2015ffa,Argyres:2015gha,Argyres:2016xmc}.}
We will only use the deformations visible in the Lagrangian theory, i.e.\ the center of the gauge group. For this reason we will not consider orthosymplectic quivers or non-simply laced quivers, and we are not able to probe rank enhancement of topological symmetry in unitary quivers, as happens for moduli spaces of instantons \cite{Cremonesi:2014xha} and other examples \cite{Gledhill:2021cbe}. Furthermore we limit our study to good theories in the sense of \cite{Gaiotto:2008ak}, since FI deformations of bad theories not only deform the Higgs branch but also partially lift it \cite{Yaakov:2013fza,Assel:2017jgo}, see also \cite[Appendix B]{Bourget:2021jwo} for a brane explanation.

This technique of FI deformations is very useful in the context of magnetic quivers, to study mass deformations of theories with 8 supercharges in spacetime dimension higher than 3. FI deformations on magnetic quivers were used successfully to
\begin{itemize}
    \item derive magnetic quivers for 4d $\mathcal{N}=2$ S-fold theories from magnetic quivers for 6d $\mathcal{N}=(1,0)$ SCFTs \cite{Bourget:2020mez},
    \item study mass deformations of 5d $\mathcal{N}=1$ SCFTs, and generate their RG-flow trees \cite{vanBeest:2021xyt},
    \item derive magnetic quivers for 4d $\mathcal{N}=2$ theories on the worldvolume of D3 branes probing 7-branes on orbifolds \cite{Giacomelli:2022drw}.
\end{itemize}

In the following we will implicitly pick a 3d $\mathcal{N}=2$ subalgebra, and work only with complex F-terms / moment maps and complex FI parameters, as opposed to a real triplet. Because of the underlying hyper-K\"ahler structure there is no loss of generality. As discussed in \cite{Bourget:2020mez,vanBeest:2021xyt,Giacomelli:2022drw}, and as will be reviewed and generalized below, one can study the F-term equations of motion in presence of a complex FI term, and use meson propagation to obtain the theory at the endpoint of the flow triggered by the FI deformation. This can be turned into a graphical quiver subtraction algorithm.\footnote{The FI quiver subtraction algorithm computes the type of singularities in the Higgs branch of the theory resolved by an FI term. This is not to be confused with what is commonly referred to as quiver subtraction \cite{Cabrera:2018ann,Bourget:2019aer} which computes symplectic leaves and transverse slices in the Coulomb branch of the theory with mass and FI terms set to zero. The two algorithms are different in a number of ways, as detailed below.} 
The main result of the present paper is a precise rule that identifies the quivers that have to be subtracted, given a minimal FI deformation. We give an executive summary in the next subsection \ref{sec:Summary}.

\subsection{Summary}
\label{sec:Summary}
      
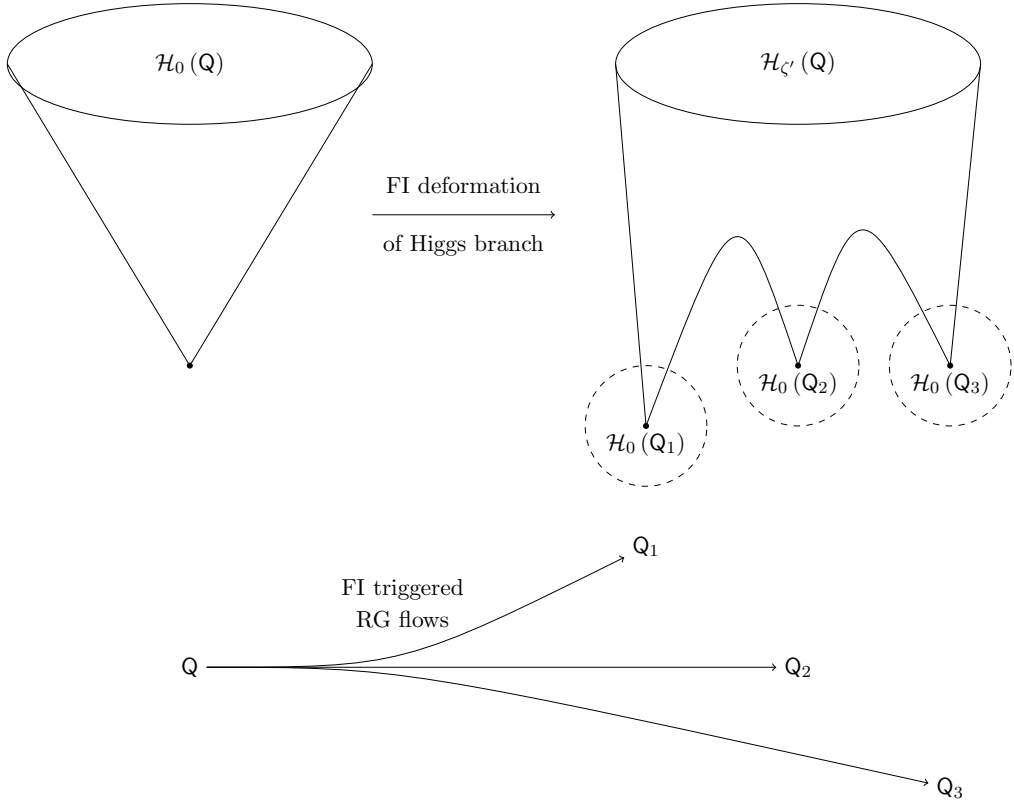
\begin{figure}
    \makebox[\textwidth][c]{\scalebox{0.8}{
    \begin{tikzpicture}
        \draw (0,0) ellipse (3 and 1);
        \draw (-3,0)--(0,-5)--(3,0);
        \node[hasse] at (0,-5) {};

        \node at (0,0) {$\mathcal{H}_{0}\left(\mathsf{Q}\right)$};

        \draw[->] (3,-2.5)--(6,-2.5);
        \node at (4.5,-2) {FI deformation};
        \node at (4.5,-3) {of Higgs branch};
        
        \draw (10,0) ellipse (3 and 1);
        \draw (10-3,0)--(10-2.5,-6) (10+2.5,-5)--(10+3,0);
        \draw (10-2.5,-6) .. controls (10-1,-2) .. (10,-5);
        \draw (10,-5) .. controls (10+1,-2) .. (10+2.5,-5);

        \node[hasse] at (7.5,-6) {};
        \node[hasse] at (10,-5) {};
        \node[hasse] at (12.5,-5) {};

        \node at (10,0) {$\mathcal{H}_{\zeta'}\left(\mathsf{Q}\right)$};

        \draw[dashed] (10-2.5,-6) circle (1);
        \node at (10-2.5,-6.3) {$\mathcal{H}_{0}\left(\mathsf{Q}_1\right)$};
        \draw[dashed] (10,-5) circle (1);
        \node at (10,-5.3) {$\mathcal{H}_{0}\left(\mathsf{Q}_2\right)$};
        \draw[dashed] (10+2.5,-5) circle (1);
        \node at (10+2.5,-5.3) {$\mathcal{H}_{0}\left(\mathsf{Q}_3\right)$};

        \node (a) at (0,-10) {$\mathsf{Q}$};
        \node (b) at (7.5,-8) {$\mathsf{Q}_1$};
        \node (c) at (10,-10) {$\mathsf{Q}_2$};
        \node (d) at (12.5,-12) {$\mathsf{Q}_3$};

        \draw[->] (a) .. controls (3.5,-10) .. (b);
        \draw[->] (a)--(c);
        \draw[->] (a) .. controls (3.5,-10) .. (d);

        \node at (3.5,-8.7) { FI triggered};
        \node at (3.5,-9.2) {RG flows};
        
    \end{tikzpicture}}}
    \caption{Top: Schematic depiction of the conical Higgs branch $\mathcal{H}_0\left(\mathsf{Q}\right)$ of a good unitary quiver $\mathsf{Q}$ with FI term $\zeta=0$ (left), and the deformed Higgs branch $\mathcal{H}_{\zeta'}\left(\mathsf{Q}\right)$ after turning on an FI term $\zeta=\zeta'\neq0$ (right). The singularity of the undeformed Higgs branch splits into several separated singularities, which in turn can be described locally as Higgs branches of good unitary quivers $\mathsf{Q}_i$ which are determined from $\mathsf{Q}$ and $\zeta$.\\
    Bottom: RG flows triggered by $\zeta=\zeta'$. The SCFT $\mathsf{Q}$ may flow to a number of SCFTs $\mathsf{Q}_i$ depending on a precise choice of Higgs branch VEV, i.e.\ choosing one of the singularities in the deformed Higgs branch.\\
    While a split into three singularities is shown, in general any number of separated singularities may be produced by a non-zero FI term. Determining all $\mathsf{Q}_i(\mathsf{Q};\zeta)$ for a choice of $\mathsf{Q}$ and $\zeta$ is the question we address in this paper. }
    \label{fig:FIschematicHiggs}
\end{figure}
    
Let $\mathsf{Q}$ be a good unitary quiver. We work in the framework of complexified gauge transformations, in which the Higgs branch $\mathcal{H}_{\mathsf{Q}}$ is the algebraic variety obtained by imposing F-term constraints and invariance under the complexified gauge group. The FI parameters are deformation parameters for the F-term equations, hence for the Higgs branch $\mathcal{H}_{\mathsf{Q}}$. Under this deformation, the conical singularity generically breaks down into several singularities, see Figure \ref{fig:FIschematicHiggs}. The goal of this paper is to identify the singularities thereby created as Higgs branches of other quivers $\mathsf{Q}_i \equiv \mathsf{Q}_i (\mathsf{Q} ; \zeta)$. 

\begin{figure}
    \centering
\hspace*{-2cm}\scalebox{.8}{\begin{tikzpicture}
\node at (0,0) {
\begin{tikzpicture}[xscale=2,yscale=2]
\node[hasse,black!100] (1) at (0,0) {};
\node[hasse] (11) at (-1,1) {};
\node[hasse,black!100] (12) at (-.2,1) {};
\node[hasse,black!100] (13) at (.3,1) {};
\node[hasse,black!100] (14) at (.8,1) {};
\node[hasse] (21) at (-1.7,2) {};
\node[hasse] (22) at (-1.4,2) {};
\node[hasse] (23) at (-1,2) {};
\node[hasse,black!100] (24) at (-.3,2) {};
\node[hasse,black!100] (25) at (.2,2) {};
\node[hasse] (26) at (.8,2) {};
\node[hasse] (27) at (1.8,2) {};
\node[hasse] (31) at (-1.8,3) {};
\node[hasse] (32) at (-1.2,3) {};
\node[hasse] (33) at (-.6,3) {};
\node[hasse] (34) at (0,3) {};
\node[hasse] (35) at (.8,3) {};
\node[hasse] (36) at (1.2,3) {};
\node[hasse] (41) at (-.9,4) {};
\node[hasse] (42) at (-.2,4) {};
\node[hasse] (43) at (.3,4) {};
\node[hasse] (44) at (.9,4) {};
\node[hasse] (45) at (1.2,4) {};
\node[hasse] (5) at (0,5) {};
\draw[black!100] (1)--(11) (1)--(12) (1)--(13) (1)--(14);
\draw[black!100] (12)--(21) (12)--(24) (13)--(22) (13)--(25) (14)--(23) (14)--(25) (14)--(27) (14)--(26);
\draw (11)--(21) (11)--(22) (11)--(23) ;
\draw[black!100] (24)--(33) (24)--(35) (24)--(34) (25)--(32) (25)--(35) ;
\draw (21)--(31) (21)--(33) (22)--(32) (22)--(33) (23)--(31) (23)--(34) (26)--(34) (26)--(35) (26)--(36) (27)--(34) (27)--(36) (27)--(45);
\draw (31)--(41) (31)--(43) (32)--(42) (32)--(44) (33)--(41) (33)--(45) (34)--(42) (34)--(43) (35)--(45) (36)--(44);
\draw (5)--(41) (5)--(42) (5)--(43) (5)--(44) (5)--(45);
\node at (-2.2,4) {\begin{tabular}{c}
Poset of solutions \\ $\mathcal{S}(\zeta = 0)$
\end{tabular} };
\node at (0,-.3) {$\mathsf{S} = 0$};
\node at (0,5.3) {$\mathsf{S} = \mathsf{Q}$};
\end{tikzpicture} 
};
\node at (10,0) {\begin{tikzpicture}[xscale=2,yscale=2]
\node[hasse,black!15] (1) at (0,0) {};
\node[hasse,orange] (11) at (-1,1) {};
\node[hasse,black!15] (12) at (-.2,1) {};
\node[hasse,black!15] (13) at (.3,1) {};
\node[hasse,black!15] (14) at (.8,1) {};
\node[hasse] (21) at (-1.7,2) {};
\node[hasse] (22) at (-1.4,2) {};
\node[hasse] (23) at (-1,2) {};
\node[hasse,black!15] (24) at (-.3,2) {};
\node[hasse,black!15] (25) at (.2,2) {};
\node[hasse,orange] (26) at (.8,2) {};
\node[hasse,orange] (27) at (1.8,2) {};
\node[hasse] (31) at (-1.8,3) {};
\node[hasse] (32) at (-1.2,3) {};
\node[hasse] (33) at (-.6,3) {};
\node[hasse] (34) at (0,3) {};
\node[hasse] (35) at (.8,3) {};
\node[hasse] (36) at (1.2,3) {};
\node[hasse] (41) at (-.9,4) {};
\node[hasse] (42) at (-.2,4) {};
\node[hasse] (43) at (.3,4) {};
\node[hasse] (44) at (.9,4) {};
\node[hasse] (45) at (1.2,4) {};
\node[hasse] (5) at (0,5) {};
\draw[black!15] (1)--(11) (1)--(12) (1)--(13) (1)--(14);
\draw[black!15] (12)--(21) (12)--(24) (13)--(22) (13)--(25) (14)--(23) (14)--(25) (14)--(27) (14)--(26);
\draw (11)--(21) (11)--(22) (11)--(23) ;
\draw[black!15] (24)--(33) (24)--(35) (24)--(34) (25)--(32) (25)--(35) ;
\draw (21)--(31) (21)--(33) (22)--(32) (22)--(33) (23)--(31) (23)--(34) (26)--(34) (26)--(35) (26)--(36) (27)--(34) (27)--(36) (27)--(45);
\draw (31)--(41) (31)--(43) (32)--(42) (32)--(44) (33)--(41) (33)--(45) (34)--(42) (34)--(43) (35)--(45) (36)--(44);
\draw (5)--(41) (5)--(42) (5)--(43) (5)--(44) (5)--(45);
\node at (0,5.3) {$\mathsf{S} = \mathsf{Q}$};
\node at (0,-.3) {\textcolor{black!15}{$\mathsf{S} = 0$}};
\node at (1.8,1.7) {\textcolor{orange}{$\mathsf{S}_3 (\zeta ')$}};
\node at (.8,1.7) {\textcolor{orange}{$\mathsf{S}_2 (\zeta ')$}};
\node at (-1.0,0.7) {\textcolor{orange}{$\mathsf{S}_1 (\zeta ')$}};
\node at (-2.2,4) {\begin{tabular}{c}
Poset of solutions \\ $\mathcal{S}(\zeta =  \textcolor{orange}{\zeta '})$
\end{tabular} };
\end{tikzpicture} 
};
\draw[->] (5,0)--(7,0);
\end{tikzpicture}}
    \caption{The geometric deformations and RG flows of Figure \ref{fig:FIschematicHiggs} are mirrored in the poset of solutions $\mathcal{S}(\zeta)$ to the FI-Meson problem. When $\zeta = 0$, the poset $\mathcal{S}(\zeta)$ has a unique minimal element $\mathsf{S} = 0$, meaning that there is no deformation. For generic values $\zeta = \zeta ' \neq 0$, the poset $\mathcal{S}(\zeta')$ can have several minima $\mathsf{S}_i (\zeta ')$. These minima are in one-to-one correspondence with the singularities after deformation, using the subtraction \eqref{eq:SubtractionGeneral}. }
    \label{fig:posetsSolutions}
\end{figure}
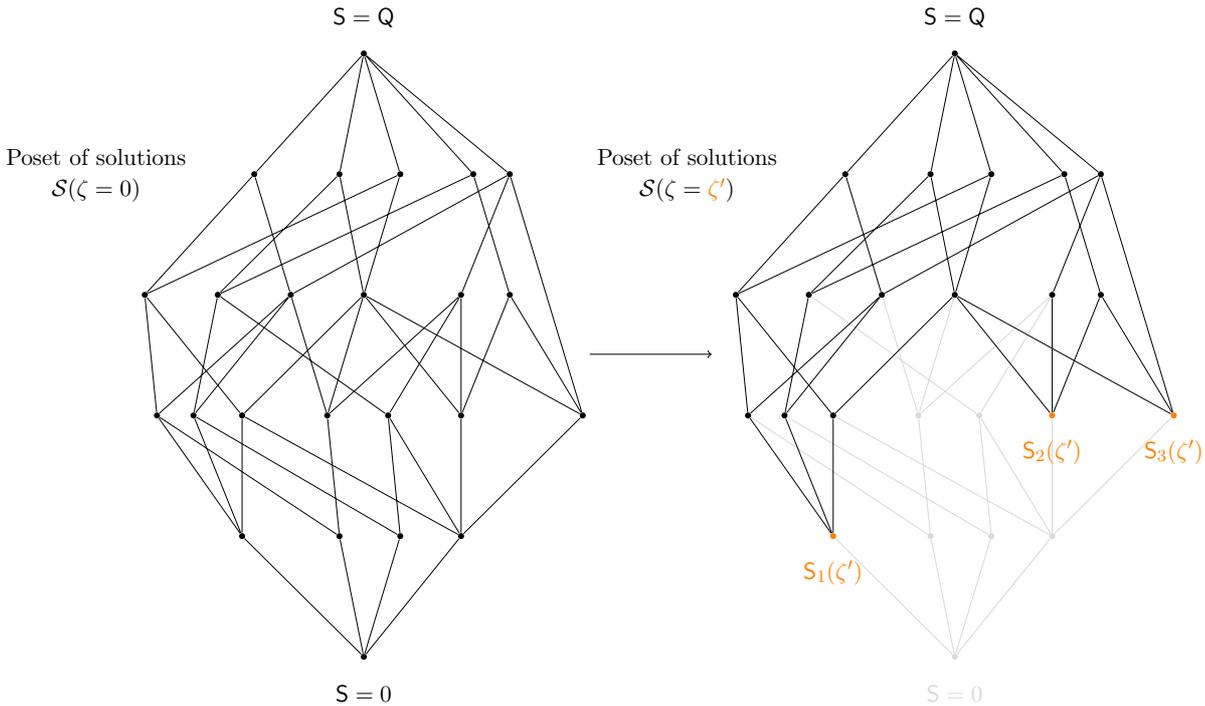

Our first result is a reformulation of this question as a well defined problem, that we call the \emph{FI-Meson problem}, which lies at the intersection between linear algebra and graph theory. This problem takes as an input the quiver $\mathsf{Q}$ and the deformation $\zeta$ and the output is the finite set $\mathcal{S}$ of \emph{solutions}, which are subquivers $\mathsf{S}$ satisfying two conditions:
\begin{enumerate}
    \item[$\alpha$)] Any node in $\mathsf{Q}$ where $\zeta$ takes a non-zero value is present in $\mathsf{S}$ with the same rank
    \item[$\beta$)] The ranks of the other nodes are constrained by \hyperref[theorem1]{Theorem 1}.
\end{enumerate}
The set of solutions is partially ordered, see Figure \ref{fig:posetsSolutions}, and it admits a finite number of local minima that we call the  \emph{extremal solutions} $\mathsf{S}_i (\zeta)$
\begin{equation}
    (\mathsf{Q} , \zeta ) \quad \xrightarrow{\text{FI-Meson}} \quad \mathcal{S} \quad \supset  \quad \mathrm{min} (\mathcal{S}) = \{ \mathsf{S}_i (\zeta) \}\, . 
\end{equation}
Our second result provides a one-to-one correspondence between the extremal solutions $\mathsf{S}_i (\zeta)$ to the FI-Meson problem and the quivers $\mathsf{Q}_i (\mathsf{Q} ; \zeta)$. When the solutions $\mathsf{S}_i (\zeta)$ are known, the quivers $\mathsf{Q}_i (\mathsf{Q} ; \zeta)$ are obtained straightforwardly using a quiver subtraction algorithm, 
\begin{equation}
\label{eq:SubtractionGeneral}
    \mathsf{Q} \quad - \quad \mathsf{S}_i (\zeta) \quad = \quad \mathsf{Q}_i (\mathsf{Q} ; \zeta) \, . 
\end{equation}
This subtraction algorithm has two steps: first one subtracts the ranks of the nodes of $\mathsf{S}_i (\zeta)$ from those of $\mathsf{Q}$, and then one \emph{rebalances} using one or several additional nodes (which can have arbitrarily high rank). 

The main practical difficulty is that the FI-Meson problem remains generically hard to solve. However it can be addressed explicitly for small enough quivers; in addition, when a brane system is available, the analysis can be greatly simplified, as illustrated at length in subsequent sections. In this case the FI deformations are interpreted as brane moves, and the singularities in Figure \ref{fig:FIschematicHiggs} are visible as brane intersections. In general, the solution set $\mathcal{S}$ has several minima. Geometrically, this means that the singularity breaks into several singularities, as illustrated in Figure \ref{fig:FIschematicHiggs} and Figure \ref{fig:posetsSolutions}. Physically, this implies that turning on FI parameters triggers a flow towards \emph{more than one theory}. 

As an application of our methods, we compare the results it produces with what is known from other approaches. One interesting example is SQCD, which although elementary displays a surprisingly rich variety of phenomena. As another example, the graphs of mass deformations of certain families of SCFTs in various dimensions have been worked out from field theory and / or higher dimensional compactification, and they match with our results applied on their magnetic quivers.

\paragraph{Comparison of two quiver subtraction algorithms}

It should be emphasized that the FI subtraction algorithm shares some superficial similarities with the `symplectic leaf quiver subtraction' developed in \cite{Bourget:2019aer,Bourget:2021siw} to uncover the stratification of the Coulomb branch of the quiver (which is a symplectic singularity) into symplectic leaves. This is the geometric incarnation of a generalized Higgs mechanism along Coulomb branch directions. However, \emph{these two algorithms are very different} and should not be confused. As will be amply shown in the present work, 
\begin{itemize}
    \item The identification of the quiver to be subtracted, $\mathsf{S}$, is straightforward in the symplectic leaf algorithm (one simply goes through the list of known minimal symplectic singularities). In the FI algorithm, it is an algorithmically hard problem. 
    \item Typically, in the symplectic leaf algorithm one subtracts quivers of affine type, i.e. good quivers. In the FI algorithm, one typically subtracts bad or ugly quivers. 
    \item The rebalancing process in the FI algorithm is rich, it can involve several non-abelian nodes, while it just involves one abelian node in the symplectic leaf algorithm. 
    \item In the symplectic leaf algorithm, the subtracted quiver $\mathsf{S}$ corresponds to the theory obtained in the IR after Higgsing along the Coulomb branch, while in the FI algorithm, $\mathsf{Q}-\mathsf{S}$ is the IR theory that remains at one of the most singular points in the resolved Higgs branch.
\end{itemize}

\paragraph{FI deforming 3d theories vs mass deforming SCFTs in higher dimensions.}
While studying FI deformations of 3d $\mathcal{N}=4$ theories is interesting in its own right, we are mainly motivated by the possibility to study mass deformation triggered RG flows between SCFTs with 8 supercharges in various dimensions. As depicted in Figure \ref{fig:FIschematicHiggs} a single FI deformation of a 3d $\mathcal{N}=4$ SCFT $\mathsf{Q}$ can trigger different RG flows to distinct 3d $\mathcal{N}=4$ SCFTs $\mathsf{Q}_i$, depending on a choice of vacuum. However for e.g.\ a 5d $\mathcal{N}=1$ SCFTs $\mathcal{T}$ with magnetic quiver $\mathsf{Q}$ any mass deformation triggers a flow to either a single new strongly coupled SCFT $\mathcal{T}'$, or to a weakly coupled theory. Let us consider a mass deformation to a new SCFT. This mass deformation can be mapped to an FI deformation $\zeta'$ of $\mathsf{Q}$. If this FI deformation leads to multiple RG flows $\mathsf{Q}_i$, then the Higgs branch of $\mathcal{T}'$ is the union of the Coulomb branches of $\mathsf{Q}_i$, see Figure \ref{fig:3dVS5dDeform}.

\paragraph{Plan of the paper.}
In Section \ref{sec:3dmirrorer} we present a new 3d mirror which is found by studying FI deformations of the magnetic quiver of SU SQCD. In Section \ref{sec:Basic} we study the simplest examples, $A$- and $D$-type Dynkin quivers, through their brane system and through FI quiver subtraction. These basic examples clarify what happens, when there are several inequivalent solutions to the equations of motion, i.e. several inequivalent FI quiver subtractions. In Section \ref{sec:GeneralMethod} we develop a more general understanding of both solving equations of motion and deriving rules for FI quiver subtraction. In Section \ref{sec:SQCD} we apply what we learned to FI deform magnetic quivers of SQCD theories with unitary and special unitary gauge groups to obtain the mass deformations of SQCD, yielding interesting results about the moduli space of SQCD and the effective low-energy theories at its singular points. We will in particular see that the low-energy theory in general involves matter fields in the determinant representation of a unitary gauge group.\footnote{Many interesting 3d $\mathcal{N}=4$ dualities involving unitary gauge groups and determinant representation have recently been uncovered \cite{Dey:2022ctw,Dey:2022eko}.} In Section \ref{sec:Other} we FI deform magnetic quivers for $4d$ $\mathcal{N}=2$ rank-2 SCFTs and for instanton moduli spaces. We will in particular show that via FI deformations we flow from the magnetic quiver of instanton moduli spaces to that of nilpotent orbits of the same group. In Section \ref{sec:Outlook} we provide some questions we hope will be addressed in the future. In Appendix \ref{app:theorems} we collect various linear algebra results we derived to study the FI-Meson problem. In Appendix \ref{Weylsec} we comment on the action of the Weyl group on FI deformations. In Appendix \ref{app:Mirror} we discuss in more detail our new 3d mirror pair. In Appendix \ref{app:E} we show all minimal FI deformations of $E$-type Dynkin quivers.

\afterpage{
\begin{landscape}
\begin{figure}
    \centering
    \scalebox{0.7}{\begin{tikzpicture}

        \node at (-6,0) {3d $\mathcal{N}=4$ theory $\mathsf{Q}$};

        \draw[blue] (0,0) ellipse (3 and 1);
        \draw[blue] (-3,0)--(0,-5)--(3,0);

        \node at (0,0) {$\mathcal{H}_{0}\left(\mathsf{Q}\right)$};

        \begin{scope}[shift={(-5,-5)},rotate=90]
            \draw[red] (0,0) ellipse (3 and 1);
            \draw[red] (-3,0)--(0,-5)--(3,0); 

            \node at (0,0) {$\mathcal{C}\left(\mathsf{Q}\right)$};
        \end{scope}

        \draw[->] (3,-2.5)--(6,-2.5);
        \node at (4.5,-2) {FI deformation};
        \node at (4.5,-3) {$\zeta=\zeta'$};

        \begin{scope}[shift={(2,0)}]
            \draw[blue] (10,0) ellipse (3 and 1);
            \draw[blue] (10-3,0)--(10-2.5,-5) (10+2.5,-5)--(10+3,0);
            \draw[blue] (10-2.5,-5) .. controls (10,-2) .. (10+2.5,-5);

            \node at (10,0) {$\mathcal{H}_{\zeta'}\left(\mathsf{Q}\right)$};

            \draw[dashed] (10-2.5+1.63,-5+1.63) arc (45:100:2.3);
            \node at (10-1.9,-3.4) {$\mathcal{H}_{0}\left(\mathsf{Q}_1\right)$};
            \draw[dashed] (10+2.5+0.4,-5+2.27) arc (80:135:2.3);
            \node at (10+1.9,-3.4) {$\mathcal{H}_{0}\left(\mathsf{Q}_2\right)$};

            \begin{scope}[shift={(10-5.5,-5)},rotate=90]
                \draw[red] (0,0) ellipse (2 and 1);
                \draw[red] (-2,0)--(0,-3)--(2,0);

                \node at (0,0) {$\mathcal{C}\left(\mathsf{Q}_1\right)$};
            \end{scope}

            \begin{scope}[shift={(16-0.5,-5)},rotate=-90]
                \draw[red] (0,0) ellipse (2 and 1);
                \draw[red] (-2,0)--(0,-3)--(2,0);

                \node at (0,0) {$\mathcal{C}\left(\mathsf{Q}_2\right)$};
            \end{scope}
        \end{scope}

            \draw (-9,-8.5)--(20,-8.5);

        \begin{scope}[shift={(0,-9)}]

            \node at (-6,0) {5d $\mathcal{N}=1$ SCFT $\mathcal{T}$};
        
            \draw[red] (0,-0.5)--(0,-5);

            \node at (0,0) {$\mathcal{C}\left(\mathcal{T}\right)$};

            \begin{scope}[shift={(-5,-5)},rotate=90]
                \draw[blue] (0,0) ellipse (3 and 1);
                \draw[blue] (-3,0)--(0,-5)--(3,0); 

                \node at (0,-1.1) {$\mathcal{H}\left(\mathcal{T}\right)=\mathcal{C}\left(\mathsf{Q}\right)$};
            \end{scope}

            \draw[->] (2,-4)--(7,-4);
            \node at (4.5,-3.5) {mass deformation};
            \node at (4.5,-4.5) {$\mathcal{T}\rightarrow\mathcal{T}'$};

            \begin{scope}[shift={(12,0)}]

                \node at (3,0) {5d $\mathcal{N}=1$ SCFT $\mathcal{T}'$};
                
                \draw[red] (0,-0.5)--(0,-5);

                \node at (0,0) {$\mathcal{C}\left(\mathcal{T}'\right)$};
                
                \begin{scope}[shift={(0,-5)},rotate=145]
                    \draw[blue] (0,3) ellipse (2 and 1);
                    \draw[blue] (-2,3)--(0,0)--(2,3);

                    \node at (0,3) {$\mathcal{C}\left(\mathsf{Q}_1\right)$};
                \end{scope}
                
                \begin{scope}[shift={(0,-5)},rotate=-145]
                    \draw[blue] (0,3) ellipse (2 and 1);
                    \draw[blue] (-2,3)--(0,0)--(2,3); 

                    \node at (0,3) {$\mathcal{C}\left(\mathsf{Q}_2\right)$};
                \end{scope}

                \draw[blue,fill=blue!10] (0,-5)--(-0.1,-8.6)--(0.1,-8.6)--(0,-5);
                \draw (0,-8)--(0,-9);
                \node at (0,-9.3) {$\mathcal{C}\left(\mathsf{Q}_1\right)\cap\mathcal{C}\left(\mathsf{Q}_2\right)$};
                \node at (0,-9.8) {$=\mathcal{C}\left(\mathsf{Q}_{12}\right)$};

                \draw[blue] (3.5,-5)--(4,-5)--(4,-9)--(3.5,-9);
                \node at (6.3,-7) {$\mathcal{H}(\mathcal{T}')=\mathcal{C}\left(\mathsf{Q}_1\right)\cup\mathcal{C}\left(\mathsf{Q}_2\right)$};
                
            \end{scope}
            
        \end{scope}

    \end{tikzpicture}}
    \caption{Top left: Higgs and Coulomb branch of 3d $\mathcal{N}=4$ theory $\mathsf{Q}$ intersecting at the origin of the full moduli space (mixed branches are not drawn). Top right: FI deformed Higgs branch with two most singular points, where Coulomb branches $\mathcal{C}(\mathsf{Q}_1)$ and $\mathcal{C}(\mathsf{Q}_2)$ respectively emanate. Bottom left: Higgs and Coulomb branch of 5d $\mathcal{N}=1$ SCFT $\mathcal{T}$ with magnetic quiver $\mathsf{Q}$. Bottom right: Higgs and Coulomb branch of 5d $\mathcal{N}=1$ SCFT $\mathcal{T}'$ resulting from a flow triggered by a mass deformation of $\mathcal{T}$ which is described by the FI deformation $\zeta'$ of magnetic quiver $\mathsf{Q}$. The Higgs branch of $\mathcal{T}'$ is the union $\mathcal{C}(\mathsf{Q}_1)\cup\mathcal{C}\left(\mathsf{Q}_2\right)$ with non-trivial (but possibly point-like) intersection $\mathcal{C}(\mathsf{Q}_1)\cap\mathcal{C}\left(\mathsf{Q}_2\right)$.}
    \label{fig:3dVS5dDeform}
\end{figure}
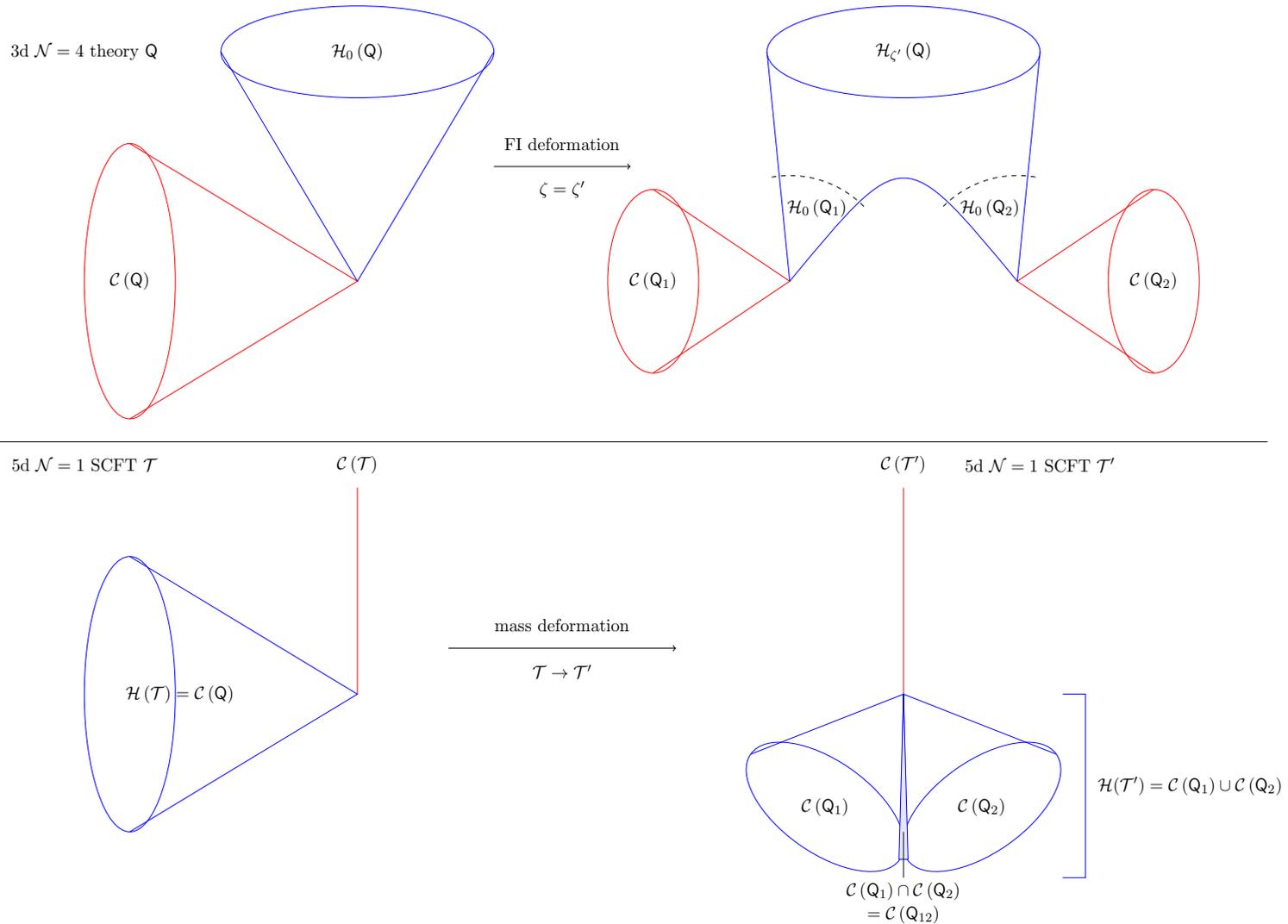
\end{landscape}
}

\subsection{A new 3d mirror pair}\label{sec:3dmirrorer}
As a by-product of our analysis of the FI deformations of SQCD theories, we have found a new 3d mirror pair: 
\begin{subequations}
\begin{equation}\label{mirrornew}
  \raisebox{-.5\height}{  \begin{tikzpicture}
        \node[gauge,label=above:{$1$}] (1) at (1,0) {};
        \node[gauge,label=above:{$2$}] (2) at (2,0) {};
        \node (3) at (3,0) {$\cdots$};
        \node[gauge,label=above:{$M$}] (4) at (4,0) {};
        \node (5) at (5,0) {$\cdots$};
        \node[gauge,label=above:{$M$}] (6) at (6,0) {};
        \node (7) at (7,0) {$\cdots$};
        \node[gauge,label=above:{$2$}] (8) at (8,0) {};
        \node[gauge,label=above:{$1$}] (9) at (9,0) {};
        \draw (1)--(2)--(3)--(4)--(5)--(6)--(7)--(8)--(9);
        \draw [decorate,decoration={brace,amplitude=5pt}] (0.8,0.6)--(9.2,0.6);
        \node at (5,1) {$B-1$};
        
        \node[gauge,label=below:{$1$}] (1d) at (1,-2) {};
        \node[gauge,label=below:{$2$}] (2d) at (2,-2) {};
        \node (3d) at (3,-2) {$\cdots$};
        \node[gauge,label=below:{$N$}] (4d) at (4,-2) {};
        \node (5d) at (5,-2) {$\cdots$};
        \node[gauge,label=below:{$N$}] (6d) at (6,-2) {};
        \node (7d) at (7,-2) {$\cdots$};
        \node[gauge,label=below:{$2$}] (8d) at (8,-2) {};
        \node[gauge,label=below:{$1$}] (9d) at (9,-2) {};
        \draw (1d)--(2d)--(3d)--(4d)--(5d)--(6d)--(7d)--(8d)--(9d);
        \draw [decorate,decoration={brace,amplitude=5pt}] (9.2,-2.6)--(0.8,-2.6);
        \node at (5,-3) {$A-1$};

        \node[gauge,label=left:{$1$}] (4m) at (4,-1) {};
        \node[gauge,label=right:{$1$}] (6m) at (6,-1) {};
        \draw (4d)--(4m)--(4) (6)--(6m)--(6d);
    \end{tikzpicture}}
\end{equation}
is the 3d mirror dual of
\begin{equation}\label{newmirror}
   \raisebox{-.5\height}{ \begin{tikzpicture}
        \node[gauge,label=below:{SU$(N)$}] (0) at (0,0) {};
        \node[gauge,label=below:{U$(1)$}] (1) at (3,0) {};
        \node[gauge,label=below:{SU$(M)$}] (2) at (6,0) {};
        \node (q) at (2.4,0) {$M$};
        \node (n) at (1.5,0.3) {$A$};
        \node (pp) at (3.6,0) {$N$};
        \node (nn) at (4.5,0.3) {$B$};
        \draw[double] (0)--(q)--(1)--(pp)--(2);
        \draw (6.3,-1)--(7.3,0);
        \node at (8,-0.5) {$\mathbb{Z}_{NM}$};
    \end{tikzpicture}}
\end{equation}
with gauge group SU$(N)\times$U$(1)\times$SU$(M)$, $A$ hypermultiplets in the fundamental representation of SU$(N$) of charge $M$ under U$(1)$ and $B$ hypermultiplets in the antifundamental representation of SU$(M)$ of charge $N$ under U$(1)$, gauging the 1-form $\mathbb{Z}_{NM}\subset \mathbb{Z}_{N}\times\mathrm{U}(1)\times\mathbb{Z}_M$. Graphically, the number of hypermultiplets is indicated above the double lines, while the U(1) charges are inserted within the line. 3d mirror symmetry was checked by computing the Higgs and Coulomb branch Hilbert series of both quivers for various choices of $M,N,A,B$. We address this in Appendix \ref{app:Mirror}.
\end{subequations}
Notice that in the special cases $M=1$ \eqref{newmirror} reduces to a $\mathrm{U}(N)$ gauge theory with a generic number of matter hypermultiplets in the fundamental and in the determinant representations, generalizing the findings of \cite{Dey:2021rxw, vanBeest:2021xyt}: 
\begin{subequations}
\begin{equation}
   \raisebox{-.5\height}{ \begin{tikzpicture}
        \node[gauge,label=above:{$1$}] (4) at (4,0) {};
        \node (5) at (5,0) {$\cdots$};
        \node[gauge,label=above:{$1$}] (6) at (6,0) {};
        \draw (4)--(5)--(6);
        \draw [decorate,decoration={brace,amplitude=5pt}] (3.8,0.6)--(6.2,0.6);
        \node at (5,1) {$B-1$};
        
        \node[gauge,label=below:{$1$}] (1d) at (1,-2) {};
        \node[gauge,label=below:{$2$}] (2d) at (2,-2) {};
        \node (3d) at (3,-2) {$\cdots$};
        \node[gauge,label=below:{$N$}] (4d) at (4,-2) {};
        \node (5d) at (5,-2) {$\cdots$};
        \node[gauge,label=below:{$N$}] (6d) at (6,-2) {};
        \node (7d) at (7,-2) {$\cdots$};
        \node[gauge,label=below:{$2$}] (8d) at (8,-2) {};
        \node[gauge,label=below:{$1$}] (9d) at (9,-2) {};
        \draw (1d)--(2d)--(3d)--(4d)--(5d)--(6d)--(7d)--(8d)--(9d);
        \draw [decorate,decoration={brace,amplitude=5pt}] (9.2,-2.6)--(0.8,-2.6);
        \node at (5,-3) {$A-1$};
        \node[gauge,label=left:{$1$}] (4m) at (4,-1) {};
        \node[gauge,label=right:{$1$}] (6m) at (6,-1) {};
        \draw (4d)--(4m)--(4) (6)--(6m)--(6d);
    \end{tikzpicture}}
\end{equation}
is the 3d mirror dual of
\begin{equation} 
  \raisebox{-.5\height}{  \begin{tikzpicture}
        \node[gauge,label=below:{U$(N)$}] (0) at (0,0) {};
        \node[flavour,label=left:{$A$}] (1) at (0,1) {};
        \node[flavour,label=below:{$B$}] (2) at (1,0) {};
        \draw (0)--(1);
        \draw[dotted] (0)--(2);
    \end{tikzpicture}}
\end{equation}
where the line denotes fundamental hypers as usual, and the dotted line denotes the determinant representation of U$(N)$.
\end{subequations}

\section{\texorpdfstring{Warm-up with Branes: affine $A$-\&$D$-type Dynkin Quivers}{}}
\label{sec:Basic}

In this section we consider FI deformations of $A$- and $D$-type Dynkin quivers. These FI deformations are mapped to mass deformations in the respective 3d mirror theories, which are good U(1) and SU(2) gauge theories with fundamental hypermultiplets \cite{Intriligator:1996ex}.
\begin{equation}
    \raisebox{-.5\height}{\begin{tikzpicture}
\node at (0,0) {\begin{tikzpicture}
        \node[gaugeBig, label=right:{$1$}] (1) at (0,0) {};
        \node[flavour,label=right:{$N_f$}] (2) at (0,1) {};
        \draw (1)--(2);
    \end{tikzpicture}};
\node at (5,0) {\begin{tikzpicture}
        \node[gauge,label=below:{$1$}] (1) at (0,0) {};
        \node[gauge,label=below:{$1$}] (2) at (1,0) {};
        \node (3) at (2,0) {$\cdots$};
        \node[gauge,label=below:{$1$}] (4) at (3,0) {};
        \node[gauge,label=below:{$1$}] (5) at (4,0) {};
        \node[gauge,label=above:{$1$}] (6) at (2,1) {};
        \draw (1)--(2)--(3)--(4)--(5)--(6)--(1);\draw[snake=brace]  (4,-.7) -- (0,-.7);
   \node[] at (2,-1.2) {$N_f-1$};
    \end{tikzpicture}};
    \node at (1.5,0) {$\leftrightarrow$};
\end{tikzpicture}}
\label{eq:Atype}
\end{equation}

\begin{equation}
    \raisebox{-.5\height}{\begin{tikzpicture}
\node at (0,0) {\begin{tikzpicture}
        \node[gaugeBig, label=right:{SU(2)}] (1) at (0,0) {};
        \node[flavour,label=right:{PSO($2N_f$)}] (2) at (0,1) {};
        \draw (1)--(2);
    \end{tikzpicture}};
\node at (5,0) {\begin{tikzpicture}
        \node[gauge,label=below:{$1$}] (1) at (0,-0.5) {};
        \node[gauge,label=below:{$1$}] (1u) at (0,0.5) {};
        \node[gauge,label=below:{$2$}] (2) at (1,0) {};
        \node (3) at (2,0) {$\dots$};
        \node[gauge,label=below:{$2$}] (4) at (3,0) {};
        \node[gauge,label=below:{$1$}] (6) at (4,-0.5) {};
        \node[gauge,label=below:{$1$}] (6u) at (4,0.5) {};
        \draw (1)--(2)--(3)--(4)--(6) (1u)--(2) (4)--(6u);\draw[snake=brace]  (3,-.7) -- (1,-.7);
   \node[] at (2,-1.2) {$N_f-3$};
    \end{tikzpicture}};
    \node at (1.5,0) {$\leftrightarrow$};
\end{tikzpicture}}
\label{eq:Dtype}
\end{equation}

\subsection{Type A}

\subsubsection*{Mass deformations of SQED}
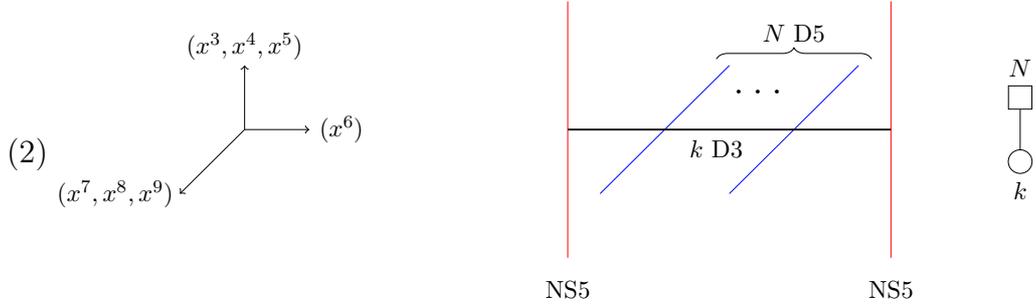
\begin{figure}[t]
\makebox[\textwidth][c]{
\scalebox{.85}{\begin{tikzpicture}
    \node at (0,0) {$\begin{tabular}{c|c|c|c|c|c|c|c|c|c|c}
& $x^0$ & $x^1$ & $x^2$ & $x^3$ & $x^4$ & $x^5$ & $x^6$ & $x^7$ & $x^8$ & $x^9$\\
\hline
{\color{red}NS5} & x & x & x & x & x & x & & & &  \\
\hline
D3 & x & x & x & & & & x & & & \\
\hline
{\color{blue}D5} & x & x & x & & & & & x & x & x
\end{tabular}$};
\node at (0,-5) {$\begin{tikzpicture}
        \begin{scope}[shift={(-2,1)}]
        \draw[->] (-3,1)--(-2,1);
        \node at (-1.5,1) {$(x^6)$};
        \draw[->] (-3,1)--(-3,2);
        \node at (-3,2.3) {$(x^3,x^4,x^5)$};
        \draw[->] (-3,1)--(-4,0);
        \node at (-5,0) {$(x^7,x^8,x^9)$};
        \end{scope}
        \draw[red] (0,0)--(0,4) (5,0)--(5,4);
        \node at (0,-0.5) {NS5};
        \node at (5,-0.5) {NS5};
        \draw[blue] (0.5,1)--(2.5,3) (2.5,1)--(4.5,3);
        \draw[thick] (0,2)--(5,2);
        \node at (2.3,1.7) {$k$ D3};
        \draw [decorate,decoration={brace,amplitude=5pt}] (2.3,3.1)--(4.7,3.1);
        \node at (3.5,3.5) {$N$ D5};
        \node at (3,2.6) {\huge{\dots}};
        \begin{scope}[shift={(7,1.5)}]
        \node[gaugeBig,label=below:{$k$}] (1) at (0,0) {};
        \node[flavour,label=above:{$N$}] (2) at (0,1) {};
        \draw (1)--(2);
        \end{scope}
    \end{tikzpicture}$};
    \node at (-8,0) {\Large(1)};
    \node at (-8,-5) {\Large(2)};
\end{tikzpicture}}}
\caption{(1) The Type IIB set-up: the 'x' mark the spacetime directions spanned by the various branes. The $x^6$ direction can be seen as an $S^1$ direction when we engineer unframed affine quivers later in this section.  (2) Depiction of the brane system for $\mathrm{U}(k)$ with $N$ flavors, with all deformation parameters turned off, at the origin of its moduli space.}
\label{fig:setup}
\end{figure}

The affine $A$-type Dynkin quivers are dual to SQED with charge 1 hypers. We begin this section by studying mass deformations of the latter theory, as it may be more common to the reader. We employ the Hanany-Witten brane system \cite{Hanany:1996ie}. Our conventions are summarized in Figure \ref{fig:setup}. 

\paragraph{Example 1.} Let us begin by studying the example of SQED with two flavors, which is self-mirror, \eqref{eq:Atype} with $N_f=2$. 
This theory has one\footnote{All parameters come as triplets of SU$(2)_C$ or SU$(2)_H$, where the former group acts on the Coulomb branch and the latter group on the Higgs branch. We consider one such triplet as a single parameter.} mass parameter $m$ and one FI parameter $\zeta$.
The moduli space for $m=\zeta=0$ consists of a Coulomb branch $A_1$ and a Higgs branch $a_1$.\footnote{We use the standard notations that $a_N$ denotes the closure of the minimal nilpotent orbit of $\mathfrak{sl}(N+1 , \mathbb{C})$, while $A_N$ is the Kleinian singularity $\mathbb{C}^2 / \mathbb{Z}_{N+1}$. Note that $A_1=a_1$ and we only make the distinction to fit with later examples. } If $m\neq0$ the Higgs branch is lifted and the Coulomb branch is resolved. If $\zeta\neq0$ the Coulomb branch is lifted and the Higgs branch is resolved. If both parameters are turned on then there is no supersymmetric vacuum and we do not consider such a case. We can depict the moduli space of the theory depending on $m$ and $\zeta$ schematically:
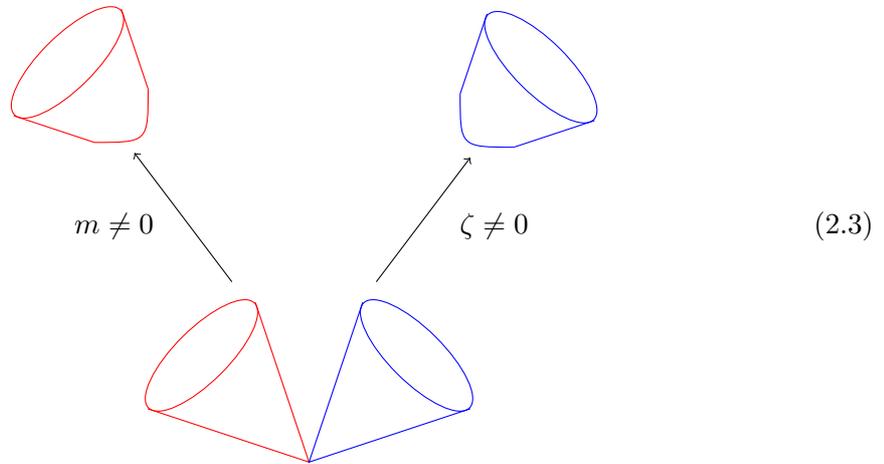
\begin{equation}
    \raisebox{-.5\height}{ \begin{tikzpicture}
        \node (0) at (0,0) {$\vcenter{\hbox{\scalebox{1}{
            \begin{tikzpicture}
                \begin{scope}[rotate=45]
                    \draw[red] (-1,2)--(0,0)--(1,2);
                    \draw[red] (0,2) ellipse (28pt and 10pt);
                \end{scope}
                \begin{scope}[rotate=-45]
                    \draw[blue] (-1,2)--(0,0)--(1,2);
                    \draw[blue] (0,2) ellipse (28pt and 10pt);
                \end{scope}
            \end{tikzpicture}}}}$};
        \node (m) at (-3,4) {$\vcenter{\hbox{\scalebox{1}{\begin{tikzpicture}
                \begin{scope}[rotate=45]
                    \draw[red] (-1,2)--(-0.5,1) (0.5,1)--(1,2);
                    \draw[red] (-0.5,1) .. controls (0,0.5) .. (0.5,1);
                    \draw[red] (0,2) ellipse (28pt and 10pt);
                \end{scope}
        \end{tikzpicture}}}}$};
        \node (f) at (3,4) {$\vcenter{\hbox{\scalebox{1}{\begin{tikzpicture}
                \begin{scope}[rotate=-45]
                    \draw[blue] (-1,2)--(-0.5,1) (0.5,1)--(1,2);
                    \draw[blue] (-0.5,1) .. controls (0,0.5) .. (0.5,1);
                    \draw[blue] (0,2) ellipse (28pt and 10pt);
                \end{scope}
        \end{tikzpicture}}}}$};
        \draw[->] (0)--(m);
        \draw[->] (0)--(f);   
        \node at (-2.5,2) {$m\neq0$};
        \node at (2.5,2) {$\zeta\neq0$};
    \end{tikzpicture}}
\end{equation}
where the {\color{red}Coulomb branch} is depicted in red and the {\color{blue}Higgs branch} is depicted in blue. We can visualize this much less schematically using brane systems and Hasse diagrams:
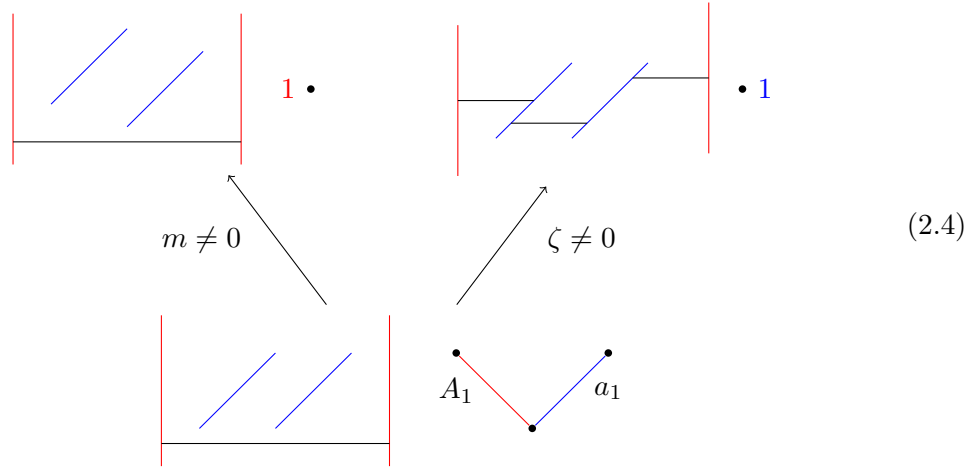
\begin{equation}
    \raisebox{-.5\height}{ \begin{tikzpicture}
        \node (0) at (0,0) {$\vcenter{\hbox{\scalebox{1}{
            \begin{tikzpicture}
                \draw[red] (-1,-1)--(-1,1) (2,-1)--(2,1);
                \draw[blue] (-0.5,-0.5)--(0.5,0.5) (0.5,-0.5)--(1.5,0.5);
                \draw (-1,-0.7)--(2,-0.7);
            \end{tikzpicture}}}}\quad\vcenter{\hbox{\scalebox{1}{
            \begin{tikzpicture}
                \node[hasse] (o) at (0,0) {};
                \node[hasse] (c) at (-1,1) {};
                \node[hasse] (h) at (1,1) {};
                \draw[red] (o)--(c);
                \draw[blue] (o)--(h);
                \node at (-1,0.5) {$A_1$};
                \node at (1,0.5) {$a_1$};
            \end{tikzpicture}}}}$};
            
        \node (m) at (-3,4) {$\vcenter{\hbox{\scalebox{1}{\begin{tikzpicture}
                \draw[red] (-1,-1)--(-1,1) (2,-1)--(2,1);
                \draw[blue] (-0.5,-0.5+0.3)--(0.5,0.5+0.3) (0.5,-0.5)--(1.5,0.5);
                \draw (-1,-0.7)--(2,-0.7);
            \end{tikzpicture}}}}\quad\vcenter{\hbox{\scalebox{1}{\begin{tikzpicture}
                \node[hasse] (o) at (0,0) {};
                \node at (-0.3,0) {\color{red}1};
        \end{tikzpicture}}}}$};
        
        \node (f) at (3,4) {$\vcenter{\hbox{\scalebox{1}{\begin{tikzpicture}
                \draw[red] (-1,-1)--(-1,1) (2+0.3,-1+0.3)--(2+0.3,1+0.3);
                \draw[blue] (-0.5,-0.5)--(0.5,0.5) (0.5,-0.5)--(1.5,0.5);
                \draw (-1,0)--(0,0) (0-0.3,0-0.3)--(1-0.3,0-0.3) (1+0.3,0+0.3)--(2+0.3,0+0.3);
            \end{tikzpicture}}}}\quad\vcenter{\hbox{\scalebox{1}{\begin{tikzpicture}
                \node[hasse] (o) at (0,0) {};
                \node at (0.3,0) {\color{blue}1};
        \end{tikzpicture}}}}$};
        
        \draw[->] (0)--(m);
        \draw[->] (0)--(f);
        
        \node at (-2.5,2) {$m\neq0$};
        \node at (2.5,2) {$\zeta\neq0$};
    \end{tikzpicture}}
\end{equation}
where we denote the dimension of the smooth factor of the Coulomb branch or Higgs branch at the bottom of the Hasse diagram without giving its precise geometry. The notation is taken from \cite{Grimminger:2020dmg}.

\afterpage{
\begin{landscape}
\begin{figure}
    \centering
    \scalebox{0.75}{
    \begin{tikzpicture}
        \node (0) at (0,0) {$\vcenter{\hbox{\scalebox{1}{
            \begin{tikzpicture}
                \draw[red] (-1,-1)--(-1,1) (3,-1)--(3,1);
                \draw[blue] (-0.5,-0.5)--(0.5,0.5) (0,-0.5)--(1,0.5) (0.5,-0.5)--(1.5,0.5) (1,-0.5)--(2,0.5) (1.5,-0.5)--(2.5,0.5);
                \draw (-1,-0.7)--(3,-0.7);
            \end{tikzpicture}}}}\quad\vcenter{\hbox{\scalebox{1}{
            \begin{tikzpicture}
                \node[hasse] (o) at (0,0) {};
                \node[hasse] (c) at (-1,1) {};
                \node[hasse] (h) at (1,1) {};
                \draw[red] (o)--(c);
                \draw[blue] (o)--(h);
                \node at (-1,0.5) {$A_4$};
                \node at (1,0.5) {$a_4$};
            \end{tikzpicture}}}}$};
            
        \node (m1) at (-4,4) {$\vcenter{\hbox{\scalebox{1}{
            \begin{tikzpicture}
                \draw[red] (-1,-1)--(-1,1) (3,-1)--(3,1);
                \draw[blue] (-0.5,-0.5+0.3)--(0.5,0.5+0.3) (0,-0.5)--(1,0.5) (0.5,-0.5)--(1.5,0.5) (1,-0.5)--(2,0.5) (1.5,-0.5)--(2.5,0.5);
                \draw (-1,-0.7)--(3,-0.7);
            \end{tikzpicture}}}}\quad\vcenter{\hbox{\scalebox{1}{
            \begin{tikzpicture}
                \node[hasse] (o) at (0,0) {};
                \node[hasse] (c) at (-1,1) {};
                \node[hasse] (h) at (1,1) {};
                \draw[red] (o)--(c);
                \draw[blue] (o)--(h);
                \node at (-1,0.5) {$A_3$};
                \node at (1,0.5) {$a_3$};
            \end{tikzpicture}}}}$};
            
        \node (m12) at (-8,8) {$\vcenter{\hbox{\scalebox{1}{
            \begin{tikzpicture}
                \draw[red] (-1,-1)--(-1,1) (3,-1)--(3,1);
                \draw[blue] (-0.5,-0.5+0.3+0.3)--(0.5,0.5+0.3+0.3) (0,-0.5+0.3)--(1,0.5+0.3) (0.5,-0.5)--(1.5,0.5) (1,-0.5)--(2,0.5) (1.5,-0.5)--(2.5,0.5);
                \draw (-1,-0.7)--(3,-0.7);
            \end{tikzpicture}}}}\quad\vcenter{\hbox{\scalebox{1}{
            \begin{tikzpicture}
                \node[hasse] (o) at (0,0) {};
                \node[hasse] (c) at (-1,1) {};
                \node[hasse] (h) at (1,1) {};
                \draw[red] (o)--(c);
                \draw[blue] (o)--(h);
                \node at (-1,0.5) {$A_2$};
                \node at (1,0.5) {$a_2$};
            \end{tikzpicture}}}}$};
            
        \node (m123) at (-12,12) {$\vcenter{\hbox{\scalebox{1}{
            \begin{tikzpicture}
                \draw[red] (-1,-1)--(-1,1) (3,-1)--(3,1);
                \draw[blue] (-0.5,-0.5+0.3+0.3+0.3)--(0.5,0.5+0.3+0.3+0.3) (0,-0.5+0.3+0.3)--(1,0.5+0.3+0.3) (0.5,-0.5+0.3)--(1.5,0.5+0.3) (1,-0.5)--(2,0.5) (1.5,-0.5)--(2.5,0.5);
                \draw (-1,-0.7)--(3,-0.7);
            \end{tikzpicture}}}}\quad\vcenter{\hbox{\scalebox{1}{
            \begin{tikzpicture}
                \node[hasse] (o) at (0,0) {};
                \node[hasse] (c) at (-1,1) {};
                \node[hasse] (h) at (1,1) {};
                \draw[red] (o)--(c);
                \draw[blue] (o)--(h);
                \node at (-1,0.5) {$A_1$};
                \node at (1,0.5) {$a_1$};
            \end{tikzpicture}}}}$};
            
        \node (m1234) at (-16,16) {$\vcenter{\hbox{\scalebox{1}{
            \begin{tikzpicture}
                \draw[red] (-1,-1)--(-1,1) (3,-1)--(3,1);
                \draw[blue] (-0.5,-0.5+0.3+0.3+0.3+0.3)--(0.5,0.5+0.3+0.3+0.3+0.3) (0,-0.5+0.3+0.3+0.3)--(1,0.5+0.3+0.3+0.3) (0.5,-0.5+0.3+0.3)--(1.5,0.5+0.3+0.3) (1,-0.5+0.3)--(2,0.5+0.3) (1.5,-0.5)--(2.5,0.5);
                \draw (-1,-0.7)--(3,-0.7);
            \end{tikzpicture}}}}\quad\vcenter{\hbox{\scalebox{1}{
            \begin{tikzpicture}
                \node[hasse] (o) at (0,0) {};
                \node at (-0.3,0) {\color{red}1};
            \end{tikzpicture}}}}$};

        \node (f) at (6,4) {$\vcenter{\hbox{\scalebox{1}{
            \begin{tikzpicture}
                \draw[red] (-1,-1)--(-1,1) (3+0.3,-1+0.3)--(3+0.3,1+0.3);
                \draw[blue] (-0.5,-0.5)--(0.5,0.5) (0,-0.5)--(1,0.5) (0.5,-0.5)--(1.5,0.5) (1,-0.5)--(2,0.5) (1.5,-0.5)--(2.5,0.5);
                \draw (-1,0)--(0,0) (0-0.3,0-0.3)--(0.5-0.3,0-0.3) (0.5,0)--(1,0) (1-0.3,0-0.3)--(1.5-0.3,0-0.3) (1.5,0)--(2,0) (2+0.3,0+0.3)--(3+0.3,0+0.3);
            \end{tikzpicture}}}}\quad\vcenter{\hbox{\scalebox{1}{
            \begin{tikzpicture}
                \node[hasse] (o) at (0,0) {};
                \node at (0.3,0) {\color{blue}4};
            \end{tikzpicture}}}}$};

        \node (fm1) at (2,8) {$\vcenter{\hbox{\scalebox{1}{
            \begin{tikzpicture}
                \draw[red] (-1,-1)--(-1,1) (3+0.3,-1+0.3)--(3+0.3,1+0.3);
                \draw[blue] (-0.5,-0.5+0.3)--(0.5,0.5+0.3) (0,-0.5)--(1,0.5) (0.5,-0.5)--(1.5,0.5) (1,-0.5)--(2,0.5) (1.5,-0.5)--(2.5,0.5);
                \draw (-1,0)--(0.5,0) (0.5-0.3,0-0.3)--(1-0.3,0-0.3) (1,0)--(1.5,0) (1.5-0.3,0-0.3)--(2-0.3,0-0.3) (2+0.3,0+0.3)--(3+0.3,0+0.3);
            \end{tikzpicture}}}}\quad\vcenter{\hbox{\scalebox{1}{
            \begin{tikzpicture}
                \node[hasse] (o) at (0,0) {};
                \node at (0.3,0) {\color{blue}3};
            \end{tikzpicture}}}}$};

        \node (fm12) at (-2,12) {$\vcenter{\hbox{\scalebox{1}{
            \begin{tikzpicture}
                \draw[red] (-1,-1)--(-1,1) (3+0.3,-1+0.3)--(3+0.3,1+0.3);
                \draw[blue] (-0.5,-0.5+0.3+0.3)--(0.5,0.5+0.3+0.3) (0,-0.5+0.3)--(1,0.5+0.3) (0.5,-0.5)--(1.5,0.5) (1,-0.5)--(2,0.5) (1.5,-0.5)--(2.5,0.5);
                \draw (-1,0)--(1,0) (1-0.3,0-0.3)--(1.5-0.3,0-0.3) (1.5,0)--(2,0) (2+0.3,0+0.3)--(3+0.3,0+0.3);
            \end{tikzpicture}}}}\quad\vcenter{\hbox{\scalebox{1}{
            \begin{tikzpicture}
                \node[hasse] (o) at (0,0) {};
                \node at (0.3,0) {\color{blue}2};
            \end{tikzpicture}}}}$};
            
        \node (fm123) at (-6,16) {$\vcenter{\hbox{\scalebox{1}{
            \begin{tikzpicture}
                \draw[red] (-1,-1)--(-1,1) (3+0.3,-1+0.3)--(3+0.3,1+0.3);
                \draw[blue] (-0.5,-0.5+0.3+0.3+0.3)--(0.5,0.5+0.3+0.3+0.3) (0,-0.5+0.3+0.3)--(1,0.5+0.3+0.3) (0.5,-0.5+0.3)--(1.5,0.5+0.3) (1,-0.5)--(2,0.5) (1.5,-0.5)--(2.5,0.5);
                \draw (-1,0)--(1.5,0) (1.5-0.3,-0.3)--(2-0.3,0-0.3) (2+0.3,0+0.3)--(3+0.3,0+0.3);
            \end{tikzpicture}}}}\quad\vcenter{\hbox{\scalebox{1}{
            \begin{tikzpicture}
                \node[hasse] (o) at (0,0) {};
                \node at (0.3,0) {\color{blue}1};
            \end{tikzpicture}}}}$};

        \draw[->,orange] (0)--(m1);
        \draw[->,orange] (m1)--(m12);
        \draw[->,orange] (m12)--(m123);
        \draw[->,orange] (m123)--(m1234);
        
        \node at (-3,2) {$m_1\neq0$};
        \node at (-7,6) {$m_2\neq0$};
        \node at (-11,10) {$m_3\neq0$};
        \node at (-15,14) {$m_4\neq0$};
        
        \draw[->,goodgreen] (0)--(f);
        \draw[->,goodgreen] (m1)--(fm1);
        \draw[->,goodgreen] (m12)--(fm12);
        \draw[->,goodgreen] (m123)--(fm123);
        
        \node at (3.5,1.5) {$\zeta\neq0$};
        \node at (-0.5,5.5) {$\zeta\neq0$};
        \node at (-4.5,9.5) {$\zeta\neq0$};
        \node at (-8.5,13.5) {$\zeta\neq0$};
            
        \draw[->,orange] (f)--(fm1);
        \draw[->,orange] (fm1)--(fm12);
        \draw[->,orange] (fm12)--(fm123);
        
        \node at (-3+8,2+4) {$m_1\neq0$};
        \node at (-7+8,6+4) {$m_2\neq0$};
        \node at (-11+8,10+4) {$m_3\neq0$};
\end{tikzpicture}}
    \caption{Brane diagrams for various phases of SQED with $N_f=5$. There are four mass parameters (turned on sequentially on the orange arrows) and one FI term (green arrow). In each case we draw the Hasse diagram of the full moduli space, with the quaternionic dimension explicitly written when it is smooth. Importantly, not all phases appear on this diagram, see for instance \eqref{eq:m2defBranes}.   }
    \label{fig:1w5web}
\end{figure}
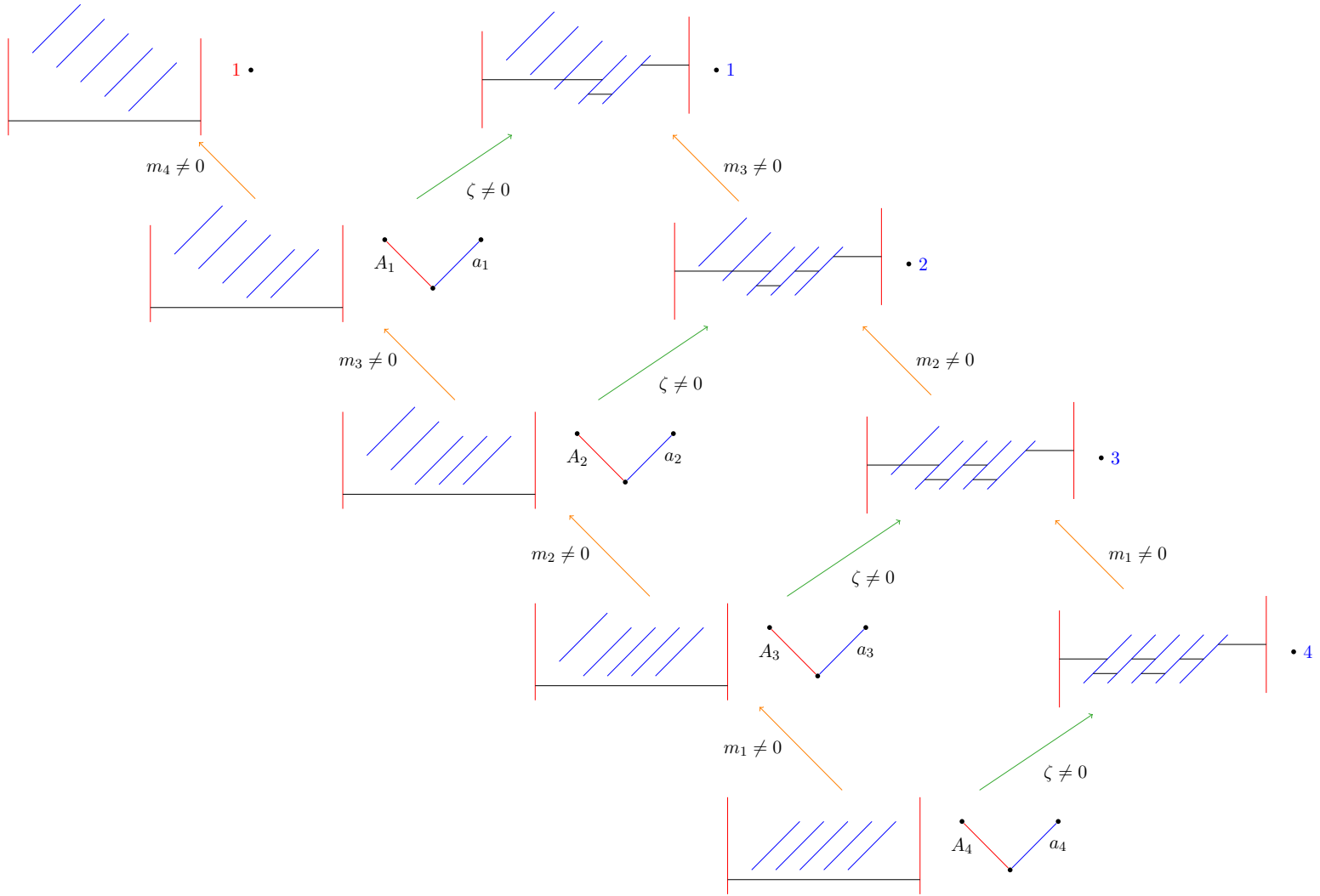
\end{landscape}
}

\paragraph{Example 2.} Let us turn to SQED with 5 flavors, i.e the left hand side of \eqref{eq:Atype} with $N_f=5$. There are four masses $m_1,m_2,m_3,m_4$ and one FI deformation $\zeta$. \footnote{The four masses parametrize the Cartan subalgebra of $\mathfrak{su}(5)$, they correspond to the differences of the mass parameters for the individual hypers. A global shift in all the masses is equivalent to a shift of the scalar in the vector multiplet. } A web of various deformations is depicted in Figure  \ref{fig:1w5web}. There, we see the effect of turning on $m_1 \neq 0$, followed by $m_2 \neq 0$, followed by $m_3 \neq 0$, etc. 
However, there are other mass deformations, for instance 
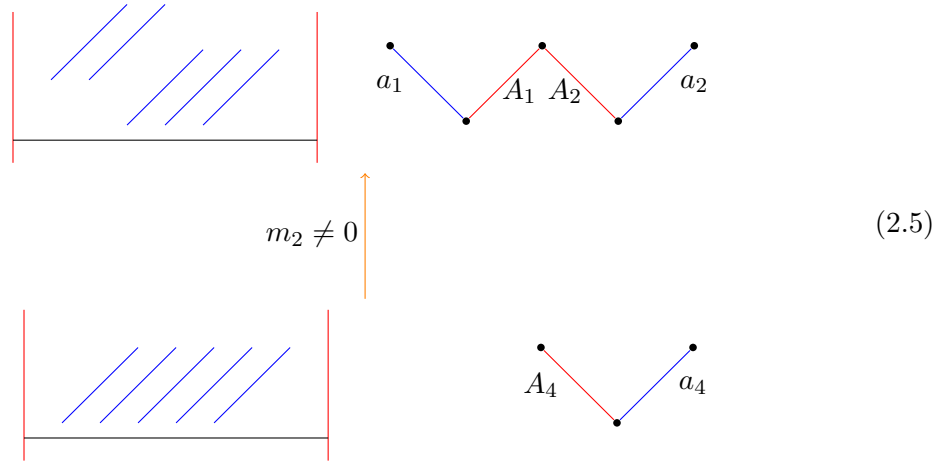
\begin{equation}
\label{eq:m2defBranes}
    \raisebox{-.5\height}{ \begin{tikzpicture}
                        \node (0) at (0,0) {$\vcenter{\hbox{\scalebox{1}{
            \begin{tikzpicture}
                \draw[red] (-1,-1)--(-1,1) (3,-1)--(3,1);
                \draw[blue] (-0.5,-0.5)--(0.5,0.5) (0,-0.5)--(1,0.5) (0.5,-0.5)--(1.5,0.5) (1,-0.5)--(2,0.5) (1.5,-0.5)--(2.5,0.5);
                \draw (-1,-0.7)--(3,-0.7);
            \end{tikzpicture}}}}\qquad \qquad\qquad\vcenter{\hbox{\scalebox{1}{
            \begin{tikzpicture}
                \node[hasse] (o) at (0,0) {};
                \node[hasse] (c) at (-1,1) {};
                \node[hasse] (h) at (1,1) {};
                \draw[red] (o)--(c);
                \draw[blue] (o)--(h);
                \node at (-1,0.5) {$A_4$};
                \node at (1,0.5) {$a_4$};
            \end{tikzpicture}}}}$};
    
                \node (m2) at (0,4) {$\vcenter{\hbox{\scalebox{1}{
            \begin{tikzpicture}
                \draw[red] (-1,-1)--(-1,1) (3,-1)--(3,1);
                \draw[blue] (-0.5,-0.5+0.3+0.3)--(0.5,0.5+0.3+0.3) (0,-0.5+0.3+0.3)--(1,0.5+0.3+0.3) (0.5,-0.5)--(1.5,0.5) (1,-0.5)--(2,0.5) (1.5,-0.5)--(2.5,0.5);
                \draw (-1,-0.7)--(3,-0.7);
            \end{tikzpicture}}}}\quad \vcenter{\hbox{\scalebox{1}{
            \begin{tikzpicture}
                \node (o) at (0,0) {};
                \node (c) at (-1,1) {};
                \node (h) at (1,1) {};
                \node at (0,0.5) {\begin{tikzpicture}
                \node[hasse] (c) at (0,1) {};
                \node[hasse] (o) at (1,0) {};
                \node[hasse] (mm) at (2,1) {};
                \node[hasse] (oo) at (3,0) {};
                \node[hasse] (h) at (4,1) {};
                \draw[blue] (o)--(c) (oo)--(h);
                \draw[red] (o)--(mm)--(oo);
                \node at (0,0.5) {$a_1$};
                \node at (1.7,0.4) {$A_1$};
                \node at (2.3,0.4) {$A_2$};
                \node at (4,0.5) {$a_2$};
            \end{tikzpicture}};
            \end{tikzpicture}}}}$};
            
            \draw[->,orange] (0)--(m2);
        
        \node at (-0.7,2) {$m_2\neq0$};
    \end{tikzpicture}}
\end{equation}
where we set $m_2 \neq 0$ while still keeping the other masses zero. 
The Hasse diagram after deformation deserves some explanations. There are now two singular points in the Coulomb branch (corresponding to the D3 brane hitting each one of the two stacks of D5 branes). As we flow to $m_2\rightarrow\infty$ we have to make a choice, and decide which stack of branes to keep. In the field theory this means the flow can lead to two different theories, depicted on the level of Hasse diagrams as follows: 
\begin{equation}
    \begin{tikzpicture}
\node (0) at (0,0) { \begin{tikzpicture}
                \node[hasse] (c) at (0,1) {};
                \node[hasse] (o) at (1,0) {};
                \node[hasse] (mm) at (2,1) {};
                \draw[blue] (o)--(c);
                \draw[red] (o)--(mm);
                \node at (0,0.5) {$a_4$};
                \node at (1.7,0.4) {$A_4$};
            \end{tikzpicture}};
\node (1) at (5,0) { \begin{tikzpicture}
                \node[hasse] (c) at (0,1) {};
                \node[hasse] (o) at (1,0) {};
                \node[hasse] (mm) at (2,1) {};
                \node[hasse] (oo) at (3,0) {};
                \node[hasse] (h) at (4,1) {};
                \draw[blue] (o)--(c) (oo)--(h);
                \draw[red] (o)--(mm)--(oo);
                \node at (0,0.5) {$a_1$};
                \node at (1.7,0.4) {$A_1$};
                \node at (2.3,0.4) {$A_2$};
                \node at (4,0.5) {$a_2$};
            \end{tikzpicture} };
\node (2) at (12,0) {\begin{tikzpicture}
                \node[hasse] (c) at (0,1) {};
                \node[hasse] (o) at (1,0) {};
                \node[hasse] (mm) at (2,1) {};
                \draw[blue] (o)--(c);
                \draw[red] (o)--(mm);
                \node at (0,0.5) {$a_1$};
                \node at (1.7,0.4) {$A_1$};
            \end{tikzpicture} \hspace*{1em}
            \begin{tikzpicture}
                \node[hasse] (mm) at (2,1) {};
                \node[hasse] (oo) at (3,0) {};
                \node[hasse] (h) at (4,1) {};
                \draw[blue] (oo)--(h);
                \draw[red] (mm)--(oo);
                \node at (2.3,0.4) {$A_2$};
                \node at (4,0.5) {$a_2$};
            \end{tikzpicture} };
            \draw[->] (0) --node[below]{$m_2 \neq 0$} (1); 
            \draw[->] (1) --node[below]{$m_2 \rightarrow \infty$} (2); 
\end{tikzpicture}
\end{equation}
or in terms of quivers:
\begin{equation}
\label{eq:1w5_m2}
    \vcenter{\hbox{\scalebox{1}{\begin{tikzpicture}
        \node[gaugeBig, label=right:{$1$}] (1) at (0,0) {};
        \node[flavour,label=right:{$5$}] (2) at (0,1) {};
        \draw (1)--(2);
    \end{tikzpicture}}}}\quad{}\rightarrow\quad\vcenter{\hbox{\scalebox{1}{\begin{tikzpicture}
        \node[gaugeBig, label=right:{$1$}] (1) at (0,0) {};
        \node[flavour,label=right:{$2$}] (2) at (0,1) {};
        \draw (1)--(2);
    \end{tikzpicture}}}}\quad\vcenter{\hbox{\scalebox{1}{\begin{tikzpicture}
        \node[gaugeBig, label=right:{$1$}] (1) at (0,0) {};
        \node[flavour,label=right:{$3$}] (2) at (0,1) {};
        \draw (1)--(2);
    \end{tikzpicture}}}}\;.
\end{equation}
The two theories on the right hand side of \eqref{eq:1w5_m2} represent the two distinct endpoints of the RG-flow triggered by the mass deformation.

\paragraph{General case. }
The examples above are immediately generalized as follows. For SQED with $N_f$ flavors, there are $2^{N_f -1}$ phases corresponding to whether $m_i$ is zero or not, for $i=1 , \dots  , N_f-1$. The minimal deformations are those where only one mass $m_i$ is non-zero. This mass triggers a flow where two theories can ultimately be reached:   
\begin{equation}
\label{eq:massDeformedSQED}
    \vcenter{\hbox{\scalebox{1}{\begin{tikzpicture}
        \node[gaugeBig, label=right:{$1$}] (1) at (0,0) {};
        \node[flavour,label=right:{$N_f$}] (2) at (0,1) {};
        \draw (1)--(2);
    \end{tikzpicture}}}}\quad{}\rightarrow\quad\vcenter{\hbox{\scalebox{1}{\begin{tikzpicture}
        \node[gaugeBig, label=right:{$1$}] (1) at (0,0) {};
        \node[flavour,label=right:{$i$}] (2) at (0,1) {};
        \draw (1)--(2);
    \end{tikzpicture}}}}\quad\vcenter{\hbox{\scalebox{1}{\begin{tikzpicture}
        \node[gaugeBig, label=right:{$1$}] (1) at (0,0) {};
        \node[flavour,label=right:{$N_f - i$}] (2) at (0,1) {};
        \draw (1)--(2);
    \end{tikzpicture}}}}
\end{equation}
If $i=1$ or $i=N_f-1$, one of the theories is free, and one can safely focus on the interacting one, but even this is an arbitrary choice. Turning on another mass $m_j$ can be analyzed in exactly the same way on the resulting pair of theories. For this reason, we will mostly focus in the following on \emph{minimal} mass deformations. In the next subsection, we give a geometric interpretation of this minimality.

\subsubsection*{FI deformation of magnetic quiver}

The mass deformations for SQED correspond to FI deformations for the mirror quiver, on the right of \eqref{eq:Atype}. When turning on an FI parameter in a unframed unitary quiver -- with gauge nodes U($n_i$) and corresponding FI terms $\zeta_i$ -- one needs to respect the condition \cite[(3.16)]{Bourget:2020mez}
\begin{equation}
\label{eq:FI_cond}
    \sum_i n_i\zeta_i=0\;.
\end{equation}
Using the same labeling of the nodes as before, the minimal mass deformation $m_i \neq 0$ corresponds to $\zeta_i = - \zeta_0 \neq 0$.  

We can realize the unframed mirror quiver in a Hanany-Witten brane setup with the $x^6$ direction as a circular dimension. For concreteness we pick $N_f = 5$. We depict the brane system for our theory in the Higgs phase, suppressing the $(x^3,x^4,x^5)$ directions:
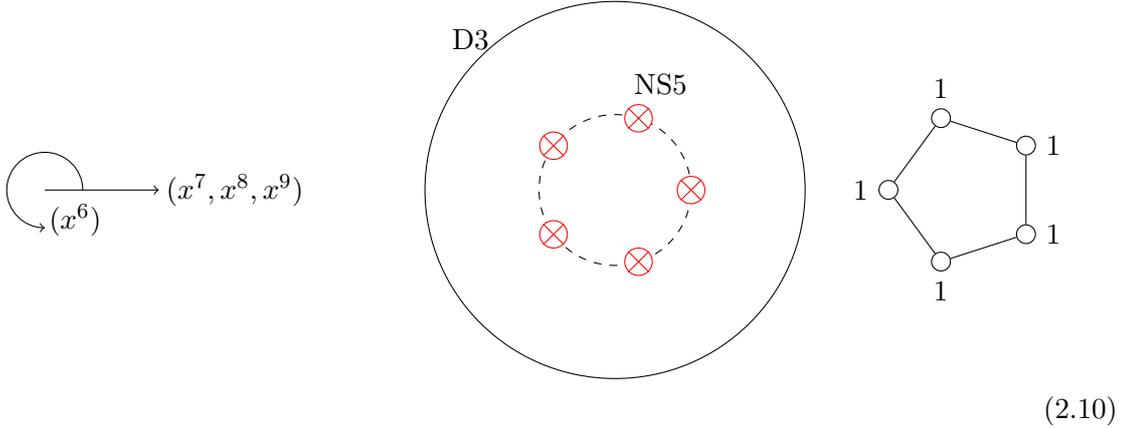
\begin{equation} \label{eq:braneSystemAaffine}
    \begin{tikzpicture}
        \draw[->] (-7,0) arc (0:270:0.5);
        \node at (-7.1,-0.4) {$(x^6)$};
        \draw[->] (-7.5,0)--(-6,0);
        \node at (-5,0) {$(x^7,x^8,x^9)$};   
        \draw[dashed] (0,0) circle (1);
        \draw (0,0) circle (2.5);  
            \ns{1,0};
            \begin{scope}[rotate=360/5]
                \ns{1,0};
            \end{scope}
            \begin{scope}[rotate=2*360/5]
                \ns{1,0};
            \end{scope}
            \begin{scope}[rotate=3*360/5]
                \ns{1,0};
            \end{scope}
            \begin{scope}[rotate=4*360/5]
                \ns{1,0};
            \end{scope}

        \node at (0.6,1.4) {NS5};
        \node at (-1.9,2) {D3};
        \node at (4.5,0) {\begin{tikzpicture}
            \begin{scope}[rotate=1*360/10]
                \node[gauge,label=right:{$1$}] (1) at (1,0) {};
            \end{scope}
            \begin{scope}[rotate=3*360/10]
                \node[gauge,label=above:{$1$}] (2) at (1,0) {};
            \end{scope}
            \begin{scope}[rotate=5*360/10]
                \node[gauge,label=left:{$1$}] (3) at (1,0) {};
            \end{scope}
            \begin{scope}[rotate=7*360/10]
                \node[gauge,label=below:{$1$}] (4) at (1,0) {};
            \end{scope}
            \begin{scope}[rotate=9*360/10]
                \node[gauge,label=right:{$1$}] (5) at (1,0) {};
            \end{scope}
            \draw (1)--(2)--(3)--(4)--(5)--(1);
    \end{tikzpicture}};
    \end{tikzpicture}
\end{equation}
The dashed line denotes the $(x^7,x^8,x^9)$ position the 5 NS5 branes are at. 
Consider the deformation \eqref{eq:m2defBranes}, \eqref{eq:1w5_m2}. 
We depict in orange the two nodes where the {\color{orange}FI} terms $\lambda$ and $-\lambda$ are turned on respectively:
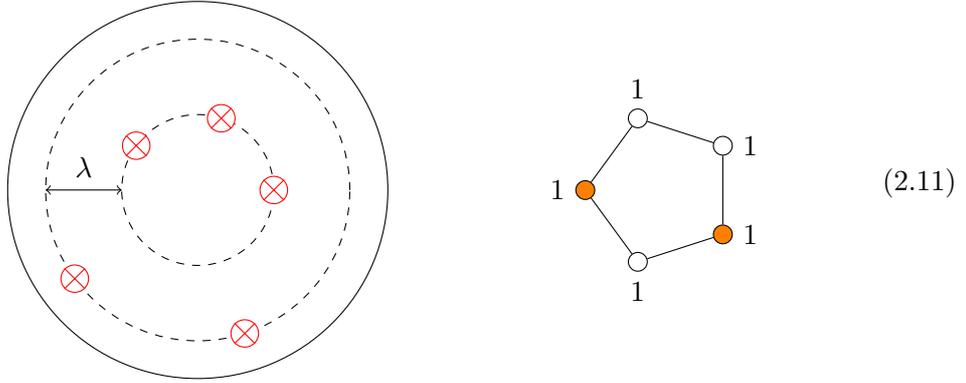
\begin{equation} \label{eq:braneSystemAaffineFI}
  \raisebox{-.5\height}{  \begin{tikzpicture}
        \node at (6,0) {\begin{tikzpicture}
            \begin{scope}[rotate=1*360/10]
                \node[gauge,label=right:{$1$}] (1) at (1,0) {};
            \end{scope}
            \begin{scope}[rotate=3*360/10]
                \node[gauge,label=above:{$1$}] (2) at (1,0) {};
            \end{scope}
            \begin{scope}[rotate=5*360/10]
                \node[gaugeo,label=left:{$1$}] (3) at (1,0) {};
            \end{scope}
            \begin{scope}[rotate=7*360/10]
                \node[gauge,label=below:{$1$}] (4) at (1,0) {};
            \end{scope}
            \begin{scope}[rotate=9*360/10]
                \node[gaugeo,label=right:{$1$}] (5) at (1,0) {};
            \end{scope}
            \draw (1)--(2)--(3)--(4)--(5)--(1);
    \end{tikzpicture}};
    \node at (0,0) {\begin{tikzpicture}
        
        \draw[dashed] (0,0) circle (1);
        \draw[dashed] (0,0) circle (2);
        \draw[<->] (-1,0)--(-2,0);
        \node at (-1.5,0.3) {$\lambda$};
        \draw (0,0) circle (2.5);
        
            \ns{1,0};
            \begin{scope}[rotate=360/5]
                \ns{1,0};
            \end{scope}
            \begin{scope}[rotate=2*360/5]
                \ns{1,0};
            \end{scope}
            \begin{scope}[rotate=3*360/5]
                \ns{2,0};
            \end{scope}
            \begin{scope}[rotate=4*360/5]
                \ns{2,0};
            \end{scope}
    \end{tikzpicture}};
    \end{tikzpicture}}
\end{equation}
We now have 3 NS5 branes at one $(x^7,x^8,x^9)$ position, and 2 NS5 branes at another $(x^7,x^8,x^9)$ position, the two $(x^7,x^8,x^9)$ positions are depicted by two dashed lines. The split of the one dashed line in \eqref{eq:braneSystemAaffine} into two dashed lines in \eqref{eq:braneSystemAaffineFI} reflects that the singular conical Higgs branch of the quiver in \eqref{eq:braneSystemAaffine} is partially resolved creating two singularities. Either of the two singularities locally looks like the Higgs branch of a new quiver.
Taking the limit $\lambda\rightarrow\infty$ we are sending NS5 branes off to infinity. There are two possible scenarios:
\begin{enumerate}
    \item We are left with \begin{equation}
      \raisebox{-.5\height}
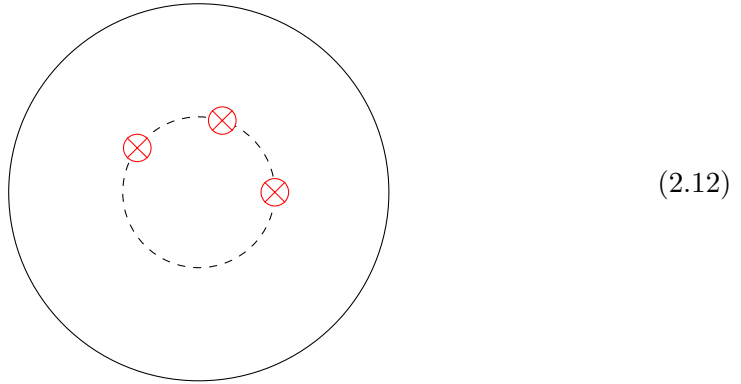
{  \begin{tikzpicture}
        
        \draw[dashed] (0,0) circle (1);
        \draw (0,0) circle (2.5);
        
            \ns{1,0};
            \begin{scope}[rotate=360/5]
                \ns{1,0};
            \end{scope}
            \begin{scope}[rotate=2*360/5]
                \ns{1,0};
            \end{scope}
    \end{tikzpicture}}
    \end{equation}
    which corresponds to the FI quiver subtraction
\begin{equation}
   \raisebox{-.5\height}{ \begin{tikzpicture}
        \node at (0,0) {\begin{tikzpicture}
            \begin{scope}[rotate=1*360/10]
                \node[gauge,label=right:{$1$}] (1) at (1,0) {};
            \end{scope}
            \begin{scope}[rotate=3*360/10]
                \node[gauge,label=above:{$1$}] (2) at (1,0) {};
            \end{scope}
            \begin{scope}[rotate=5*360/10]
                \node[gaugeo,label=left:{$1$}] (3) at (1,0) {};
            \end{scope}
            \begin{scope}[rotate=7*360/10]
                \node[gauge,label=below:{$1$}] (4) at (1,0) {};
            \end{scope}
            \begin{scope}[rotate=9*360/10]
                \node[gaugeo,label=right:{$1$}] (5) at (1,0) {};
            \end{scope}
            \draw (1)--(2)--(3)--(4)--(5)--(1);
    \end{tikzpicture}};
    \node at (4,0) {\begin{tikzpicture}
            \node at (-2,0) {$-$};
            \begin{scope}[rotate=1*360/10]
                \node (1) at (1,0) {};
            \end{scope}
            \begin{scope}[rotate=3*360/10]
                \node (2) at (1,0) {};
            \end{scope}
            \begin{scope}[rotate=5*360/10]
                \node[gaugeo,label=left:{$1$}] (3) at (1,0) {};
            \end{scope}
            \begin{scope}[rotate=7*360/10]
                \node[gauge,label=below:{$1$}] (4) at (1,0) {};
            \end{scope}
            \begin{scope}[rotate=9*360/10]
                \node[gaugeo,label=right:{$1$}] (5) at (1,0) {};
            \end{scope}
            \draw (3)--(4)--(5);
    \end{tikzpicture}};
    \node at (8,0) {\begin{tikzpicture}
            \node at (-2,0) {$=$};
            \begin{scope}[rotate=1*360/10]
                \node[gauge,label=right:{$1$}] (1) at (1,0) {};
            \end{scope}
            \begin{scope}[rotate=3*360/10]
                \node[gauge,label=above:{$1$}] (2) at (1,0) {};
            \end{scope}
            \begin{scope}[rotate=5*360/10]
                \node (3) at (1,0) {};
            \end{scope}
            \begin{scope}[rotate=7*360/10]
                \node (4) at (1,0) {};
            \end{scope}
            \begin{scope}[rotate=9*360/10]
                \node (5) at (1,0) {};
            \end{scope}
            \node[gaugeo,label=below:{$1$}] (rb) at (0,0) {};
            \draw (1)--(2)--(rb)--(1);
    \end{tikzpicture}};
    \end{tikzpicture}}
    \label{eq:sub1}
\end{equation}
    The subtraction algorithm will be explained in full generality in Section \ref{sec:GeneralMethod}; suffice it to say here that the orange node on the right hand side of \eqref{eq:sub1}, which is read from the brane system, will be interpreted later on as a rebalancing node. 
    \item We are left with 
    \begin{equation}
      \raisebox{-.5\height}{  \begin{tikzpicture}
        \draw (0,0) circle (2.5);
        \draw[dashed] (0,0) circle (2);
            \begin{scope}[rotate=3*360/5]
                \ns{2,0};
            \end{scope}
            \begin{scope}[rotate=4*360/5]
                \ns{2,0};
            \end{scope}
    \end{tikzpicture}}
    \end{equation} 
    which corresponds to the FI quiver subtraction
    \begin{equation}
   \raisebox{-.5\height}{ \begin{tikzpicture}
        \node at (0,0) {\begin{tikzpicture}
            \begin{scope}[rotate=1*360/10]
                \node[gauge,label=right:{$1$}] (1) at (1,0) {};
            \end{scope}
            \begin{scope}[rotate=3*360/10]
                \node[gauge,label=above:{$1$}] (2) at (1,0) {};
            \end{scope}
            \begin{scope}[rotate=5*360/10]
                \node[gaugeo,label=left:{$1$}] (3) at (1,0) {};
            \end{scope}
            \begin{scope}[rotate=7*360/10]
                \node[gauge,label=below:{$1$}] (4) at (1,0) {};
            \end{scope}
            \begin{scope}[rotate=9*360/10]
                \node[gaugeo,label=right:{$1$}] (5) at (1,0) {};
            \end{scope}
            \draw (1)--(2)--(3)--(4)--(5)--(1);
    \end{tikzpicture}};
    \node at (4,0) {\begin{tikzpicture}
            \node at (-2,0) {$-$};
            \begin{scope}[rotate=1*360/10]
                \node[gauge,label=right:{$1$}] (1) at (1,0) {};
            \end{scope}
            \begin{scope}[rotate=3*360/10]
                \node[gauge,label=above:{$1$}] (2) at (1,0) {};
            \end{scope}
            \begin{scope}[rotate=5*360/10]
                \node[gaugeo,label=left:{$1$}] (3) at (1,0) {};
            \end{scope}
            \begin{scope}[rotate=7*360/10]
                \node (4) at (1,0) {};
            \end{scope}
            \begin{scope}[rotate=9*360/10]
                \node[gaugeo,label=right:{$1$}] (5) at (1,0) {};
            \end{scope}
            \draw (5)--(1)--(2)--(3);
    \end{tikzpicture}};
    \node at (8,0) {\begin{tikzpicture}
            \node at (-2,0) {$=$};
            \begin{scope}[rotate=1*360/10]
                \node (1) at (1,0) {};
            \end{scope}
            \begin{scope}[rotate=3*360/10]
                \node (2) at (1,0) {};
            \end{scope}
            \begin{scope}[rotate=5*360/10]
                \node (3) at (1,0) {};
            \end{scope}
            \begin{scope}[rotate=7*360/10]
                \node[gauge,label=below:{$1$}] (4) at (1,0) {};
            \end{scope}
            \begin{scope}[rotate=9*360/10]
                \node (5) at (1,0) {};
            \end{scope}
            \node[gaugeo,label=right:{$1$}] (rb) at (0,0) {};
            \draw[double] (4)--(rb);
    \end{tikzpicture}};
    \end{tikzpicture}}
\end{equation}
\end{enumerate}
The two quivers we obtain are the magnetic quivers of the two resulting quivers in \eqref{eq:1w5_m2}. 

\subsubsection*{Higgs branch deformation via bug calculus. }

The conclusions of the previous section can be reproduced by looking at FI deformations of the 3d mirror theory, given on the right hand side of \eqref{eq:Atype}. The nodes are labeled by $i \in \mathbb{Z}_{N_f} = \{0 , 1 , \dots , N_f -1 \}$. Let us denote by $Q_i$ and $\tilde{Q}_i$ the chiral multiplets that form hypermultiplets $(Q_i , \tilde{Q}_i)$ between nodes $i$ and $i+1$. 

It is useful to pick a framed version of the mirror quiver to analyze the FI deformations via `bug calculus' \cite{Lindstrom:1999pz}. 
We can pick without loss of generality the ungauging node to be $i=0$. We then turn on FI terms $\lambda_i \in \mathbb{R}$ on the node $i  = 1 , \dots , N_f -1$. We denote this situation with the diagram 
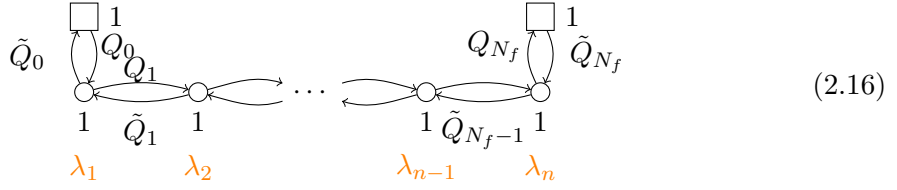
\begin{equation}
   \raisebox{-.5\height}{ \begin{tikzpicture}[xscale=1.5]
            \node[gauge,label=below:{$1$}] (1) at (1,0) {};
            \node[gauge,label=below:{$1$}] (2) at (2,0) {};
            \node (3) at (3,0) {$\cdots$};
            \node[gauge,label=below:{$1$}] (4) at (4,0) {};
            \node[gauge,label=below:{$1$}] (5) at (5,0) {};
            \node[flavour,label=right:{$1$}] (0) at (1,1) {};
            \node[flavour,label=right:{$1$}] (6) at (5,1) {};
            \draw[->] (0) to [out=-70,in=70] (1);  
            \draw[->] (1) to [out=20,in=160] (2);            \draw[->] (2) to [out=20,in=160] (3);          \draw[->] (3) to [out=20,in=160] (4);          \draw[->] (4) to [out=20,in=160] (5);          \draw[->] (5) to [out=110,in=-110] (6);       \draw[->] (6) to [out=-70,in=70] (5);      \draw[->] (5) to [out=200,in=-20] (4);   \draw[->] (4) to [out=200,in=-20] (3);   \draw[->] (3) to [out=200,in=-20] (2);   \draw[->] (2) to [out=200,in=-20] (1); 
            \draw[->] (1) to [out=110,in=-110] (0);
            \node at (1,-1) {\textcolor{orange}{$\lambda_1$}};
            \node at (2,-1) {\textcolor{orange}{$\lambda_2$}};
            \node at (4,-1) {\textcolor{orange}{$\lambda_{n-1}$}};
            \node at (5,-1) {\textcolor{orange}{$\lambda_{n}$}};
            \node at (.5,.5) {$\tilde{Q}_0$};
            \node at (1.5,-.5) {$\tilde{Q}_1$};
            \node at (4.5,-.5) {$\tilde{Q}_{N_f-1}$};
            \node at (5.5,.5) {$\tilde{Q}_{N_f}$};
            \node at (1.3,.6) {$Q_{0}$};
            \node at (1.5,.3) {$Q_{1}$};
            \node at (4.6,.6) {$Q_{N_f}$};
        \end{tikzpicture}}
\end{equation}
The vacuum equations read 
\begin{equation}
    Q_i \tilde{Q}_i -  \tilde{Q}_{i+1} Q_{i+1} = \lambda_{i+1} \, . 
\end{equation}
One can form the usual three invariants $X = Q_{N_f} Q_{N_f-1} \cdots Q_1 Q_0$, $Y = \tilde{Q}_0 \tilde{Q}_1 \cdots \tilde{Q}_{N_f-1} \tilde{Q}_{N_f}$ and $Z=Q_0 \tilde{Q}_0$. 
The equation for the deformed $A_{N_f -1}$ singularity is 
\begin{equation}
    XY = Z (Z-\lambda_1) (Z- (\lambda_1 + \lambda_2)) \cdots (Z- (\lambda_1 + \dots + \lambda_{N_f -1})) \, . 
    \label{eq:KleinADeformed}
\end{equation}
Geometrically, the only thing that really matters is the partition of $N_f$ defined by the factorization on the right-hand side. This partition is unchanged under the action of the Weyl group on the weight space element $\lambda = \lambda_1 \varpi_1 + \dots + \lambda_{N_f-1} \varpi_{N_f-1}$ of $A_{N_f-1}$. Focusing as previously on minimal deformations translates in this language into looking at Weyl orbits of fundamental weights. The space of FI parameters can be seen, in this example, as the space of $A_{N_f-1}$ weights. It admits a natural stratification in terms of facets of the Weyl chambers. The dimension 1 walls, which are generated by fundamental weights, correspond to the  minimal deformations. For the fundamental weight $\varpi_i$, the equation \eqref{eq:KleinADeformed} becomes
\begin{equation}
\label{eq:KleinADeformedMinimal}
    XY = Z^i (Z-1)^{N_f -i} \, . 
\end{equation}
Note that the weights $\varpi_i$ and $\varpi_{N_f -i}$ correspond to the same deformation.

\paragraph{Example. } If we consider $N_f = 3$, the equation is $XY = Z (Z-\lambda_1)(Z-\lambda_1 - \lambda_2)$. We have a two-dimensional space spanned by $(\lambda_1 , \lambda_2)$. There are singularities when the polynomial on the right-hand side has double roots, i.e. when either $\lambda_1 = 0$ or $\lambda_2 = 0$ or $\lambda_1 + \lambda_2 = 0$. Let's represent this graphically: 
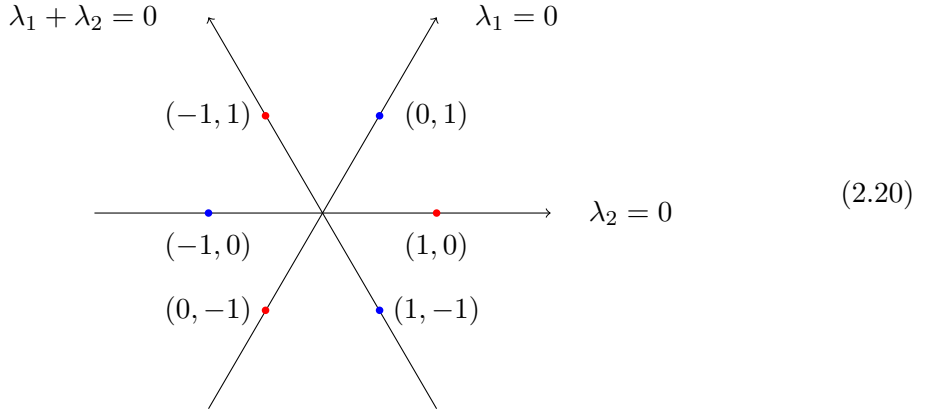
\begin{equation}
\raisebox{-.5\height}{\begin{tikzpicture}[xscale=1.5,yscale=1.5]
\draw[->] (-1,-1.73)--(1,1.73);
\draw[->] (1,-1.73)--(-1,1.73);
\draw[->] (-2,0)--(2,0);
\node at (2.7,0) {$\lambda_2 = 0$};
\node at (1.7,1.73) {$\lambda_1 = 0$};
\node at (-2.1,1.73) {$\lambda_1 + \lambda_2 = 0$};
\node[hasse,red] at (1,0) {};
\node[hasse,blue] at (.5,.86) {};
\node[hasse,red] at (-.5,.86) {};
\node[hasse,blue] at (-1,0) {};
\node[hasse,red] at (-.5,-.86) {};
\node[hasse,blue] at (.5,-.86) {};
\node at (1,-.3) {$(1,0)$};
\node at (1,.86) {$(0,1)$};
\node at (-1,.86) {$(-1,1)$};
\node at (-1,-.3) {$(-1,0)$};
\node at (-1,-.86) {$(0,-1)$};
\node at (1,-.86) {$(1,-1)$};
\end{tikzpicture}}
\end{equation}
The dimension 1 facets of the Weyl chambers are the lines. On these lines the theory is mass deformed as \eqref{eq:massDeformedSQED} with $N_f = 3$ and $i=1$.  This is a precise illustration of the cartoon shown in Figure \ref{fig:Stratification}. 

\paragraph{Summary. }
We have seen several aspects of the mass deformations in SQED. Minimal mass deformations \eqref{eq:massDeformedSQED}  correspond to minimal FI deformations in the mirror, and the latter can be implemented algebraically at the level of the equations of motions \eqref{eq:KleinADeformedMinimal}, or equivalently using quiver subtraction.

\subsubsection{Affine $A$-Type Dynkin Quiver with higher rank nodes}

Let us consider the brane system \eqref{eq:braneSystemAaffine} with two D3 branes instead of one
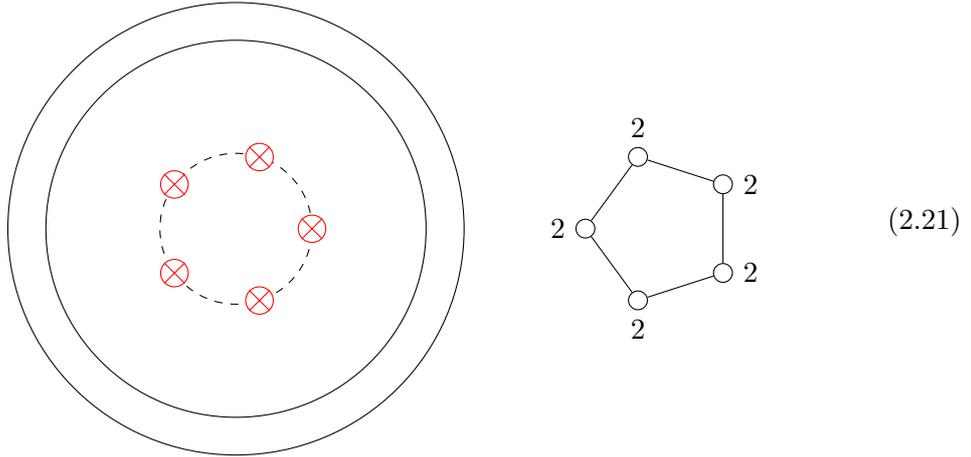
\begin{equation} \label{eq:braneSystemAaffineRank2}
  \raisebox{-.5\height}{  \begin{tikzpicture}
        \draw[dashed] (0,0) circle (1);
        \draw (0,0) circle (2.5);
        \draw (0,0) circle (3);
        
            \ns{1,0};
            \begin{scope}[rotate=360/5]
                \ns{1,0};
            \end{scope}
            \begin{scope}[rotate=2*360/5]
                \ns{1,0};
            \end{scope}
            \begin{scope}[rotate=3*360/5]
                \ns{1,0};
            \end{scope}
            \begin{scope}[rotate=4*360/5]
                \ns{1,0};
            \end{scope}
            
        \node at (5.5,0) {\begin{tikzpicture}
            \begin{scope}[rotate=1*360/10]
                \node[gauge,label=right:{$2$}] (1) at (1,0) {};
            \end{scope}
            \begin{scope}[rotate=3*360/10]
                \node[gauge,label=above:{$2$}] (2) at (1,0) {};
            \end{scope}
            \begin{scope}[rotate=5*360/10]
                \node[gauge,label=left:{$2$}] (3) at (1,0) {};
            \end{scope}
            \begin{scope}[rotate=7*360/10]
                \node[gauge,label=below:{$2$}] (4) at (1,0) {};
            \end{scope}
            \begin{scope}[rotate=9*360/10]
                \node[gauge,label=right:{$2$}] (5) at (1,0) {};
            \end{scope}
            \draw (1)--(2)--(3)--(4)--(5)--(1);
    \end{tikzpicture}};
    \end{tikzpicture}}
\end{equation}
We can now turn on an FI term for the same nodes as in \eqref{eq:braneSystemAaffineFI}. We get
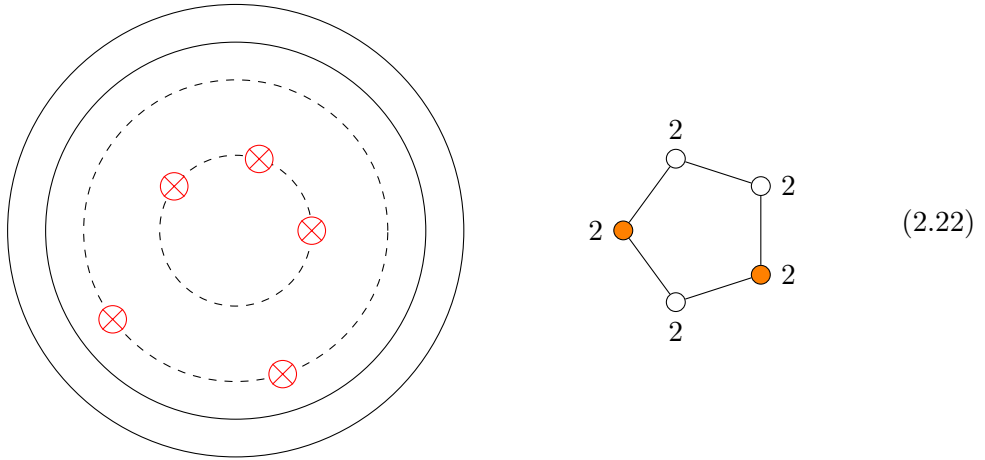
\begin{equation} \label{eq:braneSystemAaffineRank2FI}
   \raisebox{-.5\height}{ \begin{tikzpicture}
        \node at (6,0) {\begin{tikzpicture}
            \begin{scope}[rotate=1*360/10]
                \node[gauge,label=right:{$2$}] (1) at (1,0) {};
            \end{scope}
            \begin{scope}[rotate=3*360/10]
                \node[gauge,label=above:{$2$}] (2) at (1,0) {};
            \end{scope}
            \begin{scope}[rotate=5*360/10]
                \node[gaugeo,label=left:{$2$}] (3) at (1,0) {};
            \end{scope}
            \begin{scope}[rotate=7*360/10]
                \node[gauge,label=below:{$2$}] (4) at (1,0) {};
            \end{scope}
            \begin{scope}[rotate=9*360/10]
                \node[gaugeo,label=right:{$2$}] (5) at (1,0) {};
            \end{scope}
            \draw (1)--(2)--(3)--(4)--(5)--(1);
    \end{tikzpicture}};
    \node at (0,0) {\begin{tikzpicture}
        
        \draw[dashed] (0,0) circle (1);
        \draw[dashed] (0,0) circle (2);
        \draw (0,0) circle (2.5);
        \draw (0,0) circle (3);
        
            \ns{1,0};
            \begin{scope}[rotate=360/5]
                \ns{1,0};
            \end{scope}
            \begin{scope}[rotate=2*360/5]
                \ns{1,0};
            \end{scope}
            \begin{scope}[rotate=3*360/5]
                \ns{2,0};
            \end{scope}
            \begin{scope}[rotate=4*360/5]
                \ns{2,0};
            \end{scope}
    \end{tikzpicture}};
    \end{tikzpicture}}
\end{equation}
The Higgs branch of the quiver in \eqref{eq:braneSystemAaffineRank2FI}\footnote{The Coulomb branch Hilbert series of the quiver in \eqref{eq:braneSystemAaffineRank2} cannot be computed through the monopole formula of \cite{Cremonesi:2013lqa} as it diverges, this doesn't affect our discussion however.} has three most singular points:
\begin{enumerate}
    \item Both D3 branes probe the `inner most' NS5 branes. In this case we have the FI quiver subtraction
    \begin{equation}
   \raisebox{-.5\height}{ \begin{tikzpicture}
        \node at (0,0) {\begin{tikzpicture}
            \begin{scope}[rotate=1*360/10]
                \node[gauge,label=right:{$2$}] (1) at (1,0) {};
            \end{scope}
            \begin{scope}[rotate=3*360/10]
                \node[gauge,label=above:{$2$}] (2) at (1,0) {};
            \end{scope}
            \begin{scope}[rotate=5*360/10]
                \node[gaugeo,label=left:{$2$}] (3) at (1,0) {};
            \end{scope}
            \begin{scope}[rotate=7*360/10]
                \node[gauge,label=below:{$2$}] (4) at (1,0) {};
            \end{scope}
            \begin{scope}[rotate=9*360/10]
                \node[gaugeo,label=right:{$2$}] (5) at (1,0) {};
            \end{scope}
            \draw (1)--(2)--(3)--(4)--(5)--(1);
    \end{tikzpicture}};
    \node at (4,0) {\begin{tikzpicture}
            \node at (-2,0) {$-$};
            \begin{scope}[rotate=1*360/10]
                \node (1) at (1,0) {};
            \end{scope}
            \begin{scope}[rotate=3*360/10]
                \node (2) at (1,0) {};
            \end{scope}
            \begin{scope}[rotate=5*360/10]
                \node[gaugeo,label=left:{$2$}] (3) at (1,0) {};
            \end{scope}
            \begin{scope}[rotate=7*360/10]
                \node[gauge,label=below:{$2$}] (4) at (1,0) {};
            \end{scope}
            \begin{scope}[rotate=9*360/10]
                \node[gaugeo,label=right:{$2$}] (5) at (1,0) {};
            \end{scope}
            \draw (3)--(4)--(5);
    \end{tikzpicture}};
    \node at (8,0) {\begin{tikzpicture}
            \node at (-2,0) {$=$};
            \begin{scope}[rotate=1*360/10]
                \node[gauge,label=right:{$2$}] (1) at (1,0) {};
            \end{scope}
            \begin{scope}[rotate=3*360/10]
                \node[gauge,label=above:{$2$}] (2) at (1,0) {};
            \end{scope}
            \begin{scope}[rotate=5*360/10]
                \node (3) at (1,0) {};
            \end{scope}
            \begin{scope}[rotate=7*360/10]
                \node (4) at (1,0) {};
            \end{scope}
            \begin{scope}[rotate=9*360/10]
                \node (5) at (1,0) {};
            \end{scope}
            \node[gaugeo,label=below:{$2$}] (rb) at (0,0) {};
            \draw (1)--(2)--(rb)--(1);
    \end{tikzpicture}};
    \end{tikzpicture}}
\end{equation}
\item Both D3 branes probe the `outer most' NS5 branes. In this case we have the FI quiver subtraction
\begin{equation}
   \raisebox{-.5\height}{ \begin{tikzpicture}
        \node at (0,0) {\begin{tikzpicture}
            \begin{scope}[rotate=1*360/10]
                \node[gauge,label=right:{$2$}] (1) at (1,0) {};
            \end{scope}
            \begin{scope}[rotate=3*360/10]
                \node[gauge,label=above:{$2$}] (2) at (1,0) {};
            \end{scope}
            \begin{scope}[rotate=5*360/10]
                \node[gaugeo,label=left:{$2$}] (3) at (1,0) {};
            \end{scope}
            \begin{scope}[rotate=7*360/10]
                \node[gauge,label=below:{$2$}] (4) at (1,0) {};
            \end{scope}
            \begin{scope}[rotate=9*360/10]
                \node[gaugeo,label=right:{$2$}] (5) at (1,0) {};
            \end{scope}
            \draw (1)--(2)--(3)--(4)--(5)--(1);
    \end{tikzpicture}};
    \node at (4,0) {\begin{tikzpicture}
            \node at (-2,0) {$-$};
            \begin{scope}[rotate=1*360/10]
                \node[gauge,label=right:{$2$}] (1) at (1,0) {};
            \end{scope}
            \begin{scope}[rotate=3*360/10]
                \node[gauge,label=above:{$2$}] (2) at (1,0) {};
            \end{scope}
            \begin{scope}[rotate=5*360/10]
                \node[gaugeo,label=left:{$2$}] (3) at (1,0) {};
            \end{scope}
            \begin{scope}[rotate=7*360/10]
                \node (4) at (1,0) {};
            \end{scope}
            \begin{scope}[rotate=9*360/10]
                \node[gaugeo,label=right:{$2$}] (5) at (1,0) {};
            \end{scope}
            \draw (5)--(1)--(2)--(3);
    \end{tikzpicture}};
    \node at (8,0) {\begin{tikzpicture}
            \node at (-2,0) {$=$};
            \begin{scope}[rotate=1*360/10]
                \node (1) at (1,0) {};
            \end{scope}
            \begin{scope}[rotate=3*360/10]
                \node (2) at (1,0) {};
            \end{scope}
            \begin{scope}[rotate=5*360/10]
                \node (3) at (1,0) {};
            \end{scope}
            \begin{scope}[rotate=7*360/10]
                \node[gauge,label=below:{$2$}] (4) at (1,0) {};
            \end{scope}
            \begin{scope}[rotate=9*360/10]
                \node (5) at (1,0) {};
            \end{scope}
            \node[gaugeo,label=right:{$2$}] (rb) at (0,0) {};
            \draw[double] (4)--(rb);
    \end{tikzpicture}};
    \end{tikzpicture}}
\end{equation}
\item There is another option, where one D3 brane probes the inner NS5 branes while the other D3 brane probes the outer NS5 branes. This corresponds to an FI quiver subtraction of two quivers\footnote{see e.g.\ \cite[(2.8) and below]{vanBeest:2021xyt} for a previous example of such a double FI quiver subtraction.}
\begin{gather}
\label{eq:doubleQuiverSubtraction}
\mathclap{
    \begin{tikzpicture}
        \node at (0,0) {\begin{tikzpicture}
            \begin{scope}[rotate=1*360/10]
                \node[gauge,label=right:{$2$}] (1) at (1,0) {};
            \end{scope}
            \begin{scope}[rotate=3*360/10]
                \node[gauge,label=above:{$2$}] (2) at (1,0) {};
            \end{scope}
            \begin{scope}[rotate=5*360/10]
                \node[gaugeo,label=left:{$2$}] (3) at (1,0) {};
            \end{scope}
            \begin{scope}[rotate=7*360/10]
                \node[gauge,label=below:{$2$}] (4) at (1,0) {};
            \end{scope}
            \begin{scope}[rotate=9*360/10]
                \node[gaugeo,label=right:{$2$}] (5) at (1,0) {};
            \end{scope}
            \draw (1)--(2)--(3)--(4)--(5)--(1);
    \end{tikzpicture}};
    \node at (4,0) {\begin{tikzpicture}
            \node at (-2,0) {$-$};
            \begin{scope}[rotate=1*360/10]
                \node[gauge,label=right:{$1$}] (1) at (1,0) {};
            \end{scope}
            \begin{scope}[rotate=3*360/10]
                \node[gauge,label=above:{$1$}] (2) at (1,0) {};
            \end{scope}
            \begin{scope}[rotate=5*360/10]
                \node[gaugeo,label=left:{$1$}] (3) at (1,0) {};
            \end{scope}
            \begin{scope}[rotate=7*360/10]
                \node (4) at (1,0) {};
            \end{scope}
            \begin{scope}[rotate=9*360/10]
                \node[gaugeo,label=right:{$1$}] (5) at (1,0) {};
            \end{scope}
            \draw (5)--(1)--(2)--(3);
    \end{tikzpicture}};
    \node at (8,0) {\begin{tikzpicture}
            \node at (-2,0) {$-$};
            \begin{scope}[rotate=1*360/10]
                \node (1) at (1,0) {};
            \end{scope}
            \begin{scope}[rotate=3*360/10]
                \node (2) at (1,0) {};
            \end{scope}
            \begin{scope}[rotate=5*360/10]
                \node[gaugeo,label=left:{$1$}] (3) at (1,0) {};
            \end{scope}
            \begin{scope}[rotate=7*360/10]
                \node[gauge,label=below:{$1$}] (4) at (1,0) {};
            \end{scope}
            \begin{scope}[rotate=9*360/10]
                \node[gaugeo,label=right:{$1$}] (5) at (1,0) {};
            \end{scope}
            \draw (3)--(4)--(5);
    \end{tikzpicture}};
    \node at (12,0) {\begin{tikzpicture}
            \node at (-2,0) {$=$};
            \begin{scope}[rotate=1*360/10]
                \node (1) at (1,0) {};
            \end{scope}
            \begin{scope}[rotate=3*360/10]
                \node (2) at (1,0) {};
            \end{scope}
            \begin{scope}[rotate=5*360/10]
                \node (3) at (1,0) {};
            \end{scope}
            \begin{scope}[rotate=7*360/10]
                \node[gauge,label=below:{$1$}] (4) at (1,0) {};
            \end{scope}
            \begin{scope}[rotate=9*360/10]
                \node (5) at (1,0) {};
            \end{scope}
            \node[gaugeo,label=right:{$1$}] (rb) at (0,0) {};
            \draw[double] (4)--(rb);
    \end{tikzpicture}};
    \node at (14,0) {\begin{tikzpicture}
            \begin{scope}[rotate=1*360/10]
                \node[gauge,label=right:{$1$}] (1) at (1,0) {};
            \end{scope}
            \begin{scope}[rotate=3*360/10]
                \node[gauge,label=above:{$1$}] (2) at (1,0) {};
            \end{scope}
            \begin{scope}[rotate=5*360/10]
                \node (3) at (1,0) {};
            \end{scope}
            \begin{scope}[rotate=7*360/10]
                \node (4) at (1,0) {};
            \end{scope}
            \begin{scope}[rotate=9*360/10]
                \node (5) at (1,0) {};
            \end{scope}
            \node[gaugeo,label=below:{$1$}] (rb) at (0,0) {};
            \draw (1)--(2)--(rb)--(1);
    \end{tikzpicture}};
    \end{tikzpicture}
}\end{gather}
This last case deserves some further remarks. The result of the FI deformation is a disjoint union of two quivers, the moduli space of this theory is the product of the two individual moduli spaces of each quiver. In the limit of $\lambda\rightarrow\infty$ the two sets of NS5 branes are infinitely far apart, yet there is a D3 brane probing each of them. The two D3 branes don't talk to each other since they are infinitely far apart, and hence their worldvolume theory is a disjoint union.
\end{enumerate}

\subsubsection{Affine $A$-Type Dynkin Quiver with higher rank nodes and one flavor}
Let us now consider the brane system \eqref{eq:braneSystemAaffineRank2} with an added D5 brane providing a flavor to one of the nodes.\footnote{The Coulomb branch of the resulting quiver is a moduli space of instantons and its Hilbert series can be computed straight forwardly using the monopole formula. Other FI deformations of this quiver are discussed in Section \ref{sec:Other}.}
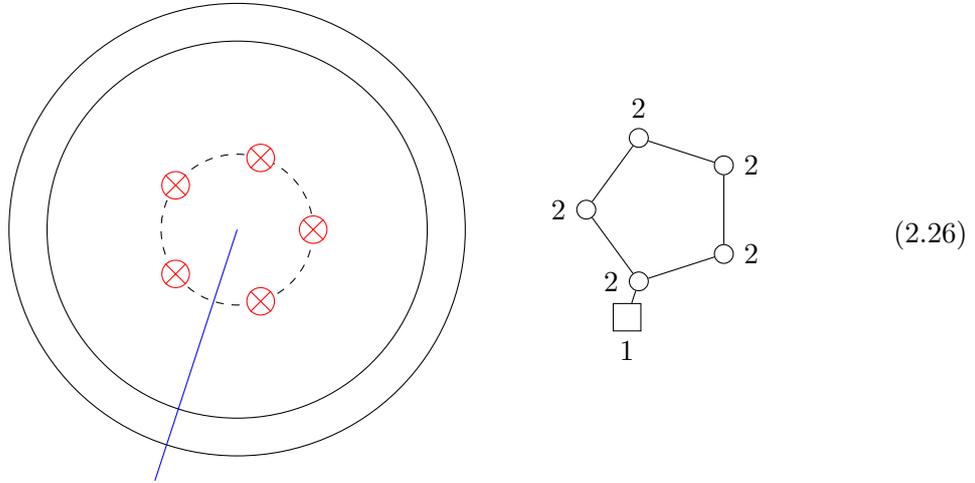
\begin{equation} \label{eq:braneSystemAaffineRank2Flav}
    \raisebox{-.5\height}{\begin{tikzpicture}
        
        \draw[dashed] (0,0) circle (1);
        \draw (0,0) circle (2.5);
        \draw (0,0) circle (3);
        \begin{scope}[rotate=7*360/10]
            \draw[blue] (0,0)--(3.5,0);
        \end{scope}

            \ns{1,0};
            \begin{scope}[rotate=360/5]
                \ns{1,0};
            \end{scope}
            \begin{scope}[rotate=2*360/5]
                \ns{1,0};
            \end{scope}
            \begin{scope}[rotate=3*360/5]
                \ns{1,0};
            \end{scope}
            \begin{scope}[rotate=4*360/5]
                \ns{1,0};
            \end{scope}
            
        \node at (5.5,0) {\begin{tikzpicture}
            \begin{scope}[rotate=1*360/10]
                \node[gauge,label=right:{$2$}] (1) at (1,0) {};
            \end{scope}
            \begin{scope}[rotate=3*360/10]
                \node[gauge,label=above:{$2$}] (2) at (1,0) {};
            \end{scope}
            \begin{scope}[rotate=5*360/10]
                \node[gauge,label=left:{$2$}] (3) at (1,0) {};
            \end{scope}
            \begin{scope}[rotate=7*360/10]
                \node[gauge,label=left:{$2$}] (4) at (1,0) {};
            \end{scope}
            \begin{scope}[rotate=7*360/10]
                \node[flavour,label=below:{$1$}] (4f) at (1.5,0) {};
            \end{scope}
            \begin{scope}[rotate=9*360/10]
                \node[gauge,label=right:{$2$}] (5) at (1,0) {};
            \end{scope}
            \draw (1)--(2)--(3)--(4)--(5)--(1) (4)--(4f);
    \end{tikzpicture}};
    \end{tikzpicture}}
\end{equation}
Again we turn on FI terms on two nodes just as before
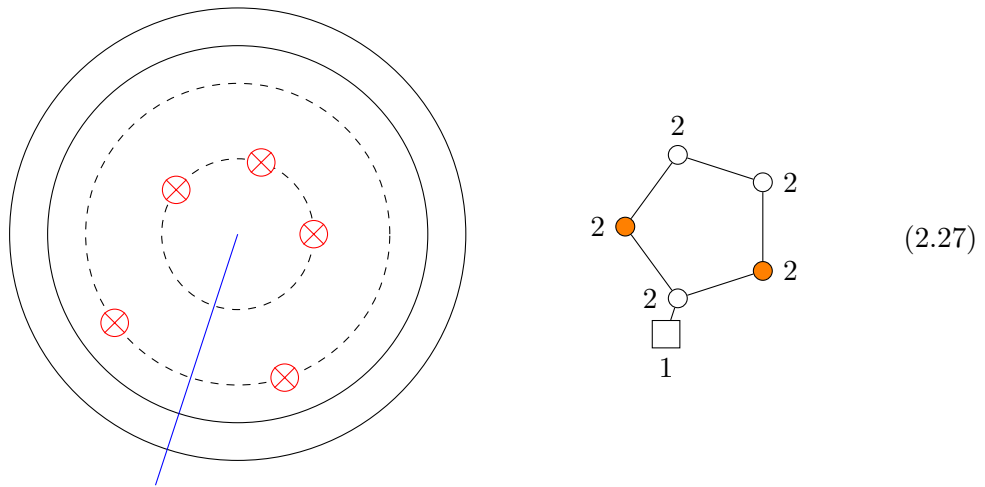
\begin{equation} \label{eq:braneSystemAaffineRank2FlavFI}
 \raisebox{-.5\height}{   \begin{tikzpicture}
        \node at (6,0) {\begin{tikzpicture}
            \begin{scope}[rotate=1*360/10]
                \node[gauge,label=right:{$2$}] (1) at (1,0) {};
            \end{scope}
            \begin{scope}[rotate=3*360/10]
                \node[gauge,label=above:{$2$}] (2) at (1,0) {};
            \end{scope}
            \begin{scope}[rotate=5*360/10]
                \node[gaugeo,label=left:{$2$}] (3) at (1,0) {};
            \end{scope}
            \begin{scope}[rotate=7*360/10]
                \node[gauge,label=left:{$2$}] (4) at (1,0) {};
            \end{scope}
            \begin{scope}[rotate=7*360/10]
                \node[flavour,label=below:{$1$}] (4f) at (1.5,0) {};
            \end{scope}
            \begin{scope}[rotate=9*360/10]
                \node[gaugeo,label=right:{$2$}] (5) at (1,0) {};
            \end{scope}
            \draw (1)--(2)--(3)--(4)--(5)--(1) (4)--(4f);
    \end{tikzpicture}};
    \node at (0,0) {\begin{tikzpicture}
        
        \draw[dashed] (0,0) circle (1);
        \draw[dashed] (0,0) circle (2);
        \draw (0,0) circle (2.5);
        \draw (0,0) circle (3);
        \begin{scope}[rotate=7*360/10]
            \draw[blue] (0,0)--(3.5,0);
        \end{scope}
        
            \ns{1,0};
            \begin{scope}[rotate=360/5]
                \ns{1,0};
            \end{scope}
            \begin{scope}[rotate=2*360/5]
                \ns{1,0};
            \end{scope}
            \begin{scope}[rotate=3*360/5]
                \ns{2,0};
            \end{scope}
            \begin{scope}[rotate=4*360/5]
                \ns{2,0};
            \end{scope}
    \end{tikzpicture}};
    \end{tikzpicture}}
\end{equation}
Again there are three most singular points in the resolved Higgs branch:
\begin{enumerate}
    \item Both D3 branes probe the `inner most' NS5 branes. In this case we have the FI quiver subtraction
    \begin{equation}
   \raisebox{-.5\height}{ \begin{tikzpicture}
        \node at (0,0) {\begin{tikzpicture}
            \begin{scope}[rotate=1*360/10]
                \node[gauge,label=right:{$2$}] (1) at (1,0) {};
            \end{scope}
            \begin{scope}[rotate=3*360/10]
                \node[gauge,label=above:{$2$}] (2) at (1,0) {};
            \end{scope}
            \begin{scope}[rotate=5*360/10]
                \node[gaugeo,label=left:{$2$}] (3) at (1,0) {};
            \end{scope}
            \begin{scope}[rotate=7*360/10]
                \node[gauge,label=left:{$2$}] (4) at (1,0) {};
                \node[flavour] at (1.5,0) {};
                \node[gauge,label=below:{$1$}] (4f) at (1.5,0) {};
            \end{scope}
            \begin{scope}[rotate=9*360/10]
                \node[gaugeo,label=right:{$2$}] (5) at (1,0) {};
            \end{scope}
            \draw (1)--(2)--(3)--(4)--(5)--(1) (4)--(4f);
    \end{tikzpicture}};
    \node at (4,0) {\begin{tikzpicture}
            \node at (-2,0) {$-$};
            \begin{scope}[rotate=1*360/10]
                \node (1) at (1,0) {};
            \end{scope}
            \begin{scope}[rotate=3*360/10]
                \node (2) at (1,0) {};
            \end{scope}
            \begin{scope}[rotate=5*360/10]
                \node[gaugeo,label=left:{$2$}] (3) at (1,0) {};
            \end{scope}
            \begin{scope}[rotate=7*360/10]
                \node[gauge,label=below:{$2$}] (4) at (1,0) {};
            \end{scope}
            \begin{scope}[rotate=9*360/10]
                \node[gaugeo,label=right:{$2$}] (5) at (1,0) {};
            \end{scope}
            \draw (3)--(4)--(5);
    \end{tikzpicture}};
    \node at (8,0) {\begin{tikzpicture}
            \node at (-2,0) {$=$};
            \begin{scope}[rotate=1*360/10]
                \node[gauge,label=right:{$2$}] (1) at (1,0) {};
            \end{scope}
            \begin{scope}[rotate=3*360/10]
                \node[gauge,label=above:{$2$}] (2) at (1,0) {};
            \end{scope}
            \begin{scope}[rotate=5*360/10]
                \node (3) at (1,0) {};
            \end{scope}
            \begin{scope}[rotate=7*360/10]
                \node (4) at (1,0) {};
            \end{scope}
            \begin{scope}[rotate=9*360/10]
                \node (5) at (1,0) {};
            \end{scope}
            \node[gaugeo,label=below:{$2$}] (rb) at (0,0) {};
                \node[flavour] at (-0.5,-0.5) {};
            \node[gauge,label=left:{$1$}] (f) at (-0.5,-0.5) {};
            \draw (1)--(2)--(rb)--(1) (rb)--(f);
    \end{tikzpicture}};
    \end{tikzpicture}}
\end{equation}

\item Both D3 branes probe the `outer most' NS5 branes. In this case we have the FI quiver subtraction
\begin{equation}
   \raisebox{-.5\height}{ \begin{tikzpicture}
        \node at (0,0) {\begin{tikzpicture}
            \begin{scope}[rotate=1*360/10]
                \node[gauge,label=right:{$2$}] (1) at (1,0) {};
            \end{scope}
            \begin{scope}[rotate=3*360/10]
                \node[gauge,label=above:{$2$}] (2) at (1,0) {};
            \end{scope}
            \begin{scope}[rotate=5*360/10]
                \node[gaugeo,label=left:{$2$}] (3) at (1,0) {};
            \end{scope}
            \begin{scope}[rotate=7*360/10]
                \node[gauge,label=left:{$2$}] (4) at (1,0) {};
                \node[flavour] at (1.5,0) {};
                \node[gauge,label=below:{$1$}] (4f) at (1.5,0) {};
            \end{scope}
            \begin{scope}[rotate=9*360/10]
                \node[gaugeo,label=right:{$2$}] (5) at (1,0) {};
            \end{scope}
            \draw (1)--(2)--(3)--(4)--(5)--(1) (4)--(4f);
    \end{tikzpicture}};
    \node at (4,0) {\begin{tikzpicture}
            \node at (-2,0) {$-$};
            \begin{scope}[rotate=1*360/10]
                \node[gauge,label=right:{$2$}] (1) at (1,0) {};
            \end{scope}
            \begin{scope}[rotate=3*360/10]
                \node[gauge,label=above:{$2$}] (2) at (1,0) {};
            \end{scope}
            \begin{scope}[rotate=5*360/10]
                \node[gaugeo,label=left:{$2$}] (3) at (1,0) {};
            \end{scope}
            \begin{scope}[rotate=7*360/10]
                \node (4) at (1,0) {};
            \end{scope}
            \begin{scope}[rotate=9*360/10]
                \node[gaugeo,label=right:{$2$}] (5) at (1,0) {};
            \end{scope}
            \draw (5)--(1)--(2)--(3);
    \end{tikzpicture}};
    \node at (8,0) {\begin{tikzpicture}
            \node at (-2,0) {$=$};
            \begin{scope}[rotate=1*360/10]
                \node (1) at (1,0) {};
            \end{scope}
            \begin{scope}[rotate=3*360/10]
                \node (2) at (1,0) {};
            \end{scope}
            \begin{scope}[rotate=5*360/10]
                \node (3) at (1,0) {};
            \end{scope}
            \begin{scope}[rotate=7*360/10]
                \node[gauge,label=left:{$2$}] (4) at (1,0) {};
                \node[flavour] at (1.5,0) {};
                \node[gauge,label=below:{$1$}] (4f) at (1.5,0) {};
            \end{scope}
            \begin{scope}[rotate=9*360/10]
                \node (5) at (1,0) {};
            \end{scope}
            \node[gaugeo,label=right:{$2$}] (rb) at (0,0) {};
            \draw[double] (4)--(rb);
            \draw (4)--(4f);
    \end{tikzpicture}};
    \end{tikzpicture}}
\end{equation}

\item One D3 brane probes the inner NS5 branes while the other D3 brane probes the outer NS5 branes
\begin{equation}
    \begin{tikzpicture}
        \node at (0,0) {\begin{tikzpicture}
            \begin{scope}[rotate=1*360/10]
                \node[gauge,label=right:{$2$}] (1) at (1,0) {};
            \end{scope}
            \begin{scope}[rotate=3*360/10]
                \node[gauge,label=above:{$2$}] (2) at (1,0) {};
            \end{scope}
            \begin{scope}[rotate=5*360/10]
                \node[gaugeo,label=left:{$2$}] (3) at (1,0) {};
            \end{scope}
            \begin{scope}[rotate=7*360/10]
                \node[gauge,label=left:{$2$}] (4) at (1,0) {};
                \node[flavour] at (1.5,0) {};
                \node[gauge,label=below:{$1$}] (4f) at (1.5,0) {};
            \end{scope}
            \begin{scope}[rotate=9*360/10]
                \node[gaugeo,label=right:{$2$}] (5) at (1,0) {};
            \end{scope}
            \draw (1)--(2)--(3)--(4)--(5)--(1) (4)--(4f);
    \end{tikzpicture}};
    \node at (4,0) {\begin{tikzpicture}
            \node at (-2,0) {$-$};
            \begin{scope}[rotate=1*360/10]
                \node[gauge,label=right:{$1$}] (1) at (1,0) {};
            \end{scope}
            \begin{scope}[rotate=3*360/10]
                \node[gauge,label=above:{$1$}] (2) at (1,0) {};
            \end{scope}
            \begin{scope}[rotate=5*360/10]
                \node[gaugeo,label=left:{$1$}] (3) at (1,0) {};
            \end{scope}
            \begin{scope}[rotate=7*360/10]
                \node (4) at (1,0) {};
            \end{scope}
            \begin{scope}[rotate=9*360/10]
                \node[gaugeo,label=right:{$1$}] (5) at (1,0) {};
            \end{scope}
            \draw (5)--(1)--(2)--(3);
    \end{tikzpicture}};
    \node at (8,0) {\begin{tikzpicture}
            \node at (-2,0) {$-$};
            \begin{scope}[rotate=1*360/10]
                \node (1) at (1,0) {};
            \end{scope}
            \begin{scope}[rotate=3*360/10]
                \node (2) at (1,0) {};
            \end{scope}
            \begin{scope}[rotate=5*360/10]
                \node[gaugeo,label=left:{$1$}] (3) at (1,0) {};
            \end{scope}
            \begin{scope}[rotate=7*360/10]
                \node[gauge,label=below:{$1$}] (4) at (1,0) {};
            \end{scope}
            \begin{scope}[rotate=9*360/10]
                \node[gaugeo,label=right:{$1$}] (5) at (1,0) {};
            \end{scope}
            \draw (3)--(4)--(5);
    \end{tikzpicture}};
    \node at (12,0) {\begin{tikzpicture}
            \node at (-2,0) {$=$};
            \begin{scope}[rotate=1*360/10]
                \node (1) at (1,0) {};
            \end{scope}
            \begin{scope}[rotate=3*360/10]
                \node (2) at (1,0) {};
            \end{scope}
            \begin{scope}[rotate=5*360/10]
                \node (3) at (1,0) {};
            \end{scope}
            \begin{scope}[rotate=7*360/10]
                \node[gauge,label=below:{$1$}] (4) at (1,0) {};
            \end{scope}
            \begin{scope}[rotate=9*360/10]
                \node (5) at (1,0) {};
            \end{scope}
            \node[gaugeo,label=right:{$1$}] (rb) at (0,0) {};
            \draw[double] (4)--(rb);

            \begin{scope}[shift={(2,0)}]
            \begin{scope}[rotate=1*360/10]
                \node[gauge,label=right:{$1$}] (1n) at (1,0) {};
            \end{scope}
            \begin{scope}[rotate=3*360/10]
                \node[gauge,label=above:{$1$}] (2n) at (1,0) {};
            \end{scope}
            \begin{scope}[rotate=5*360/10]
                \node (3n) at (1,0) {};
            \end{scope}
            \begin{scope}[rotate=7*360/10]
                \node (4n) at (1,0) {};
                \node[flavour] at (1.5,0) {};
                \node[gauge,label=below:{$1$}] (4fn) at (1.5,0) {};
            \end{scope}
            \begin{scope}[rotate=9*360/10]
                \node (5n) at (1,0) {};
            \end{scope}
            \end{scope}
            \node[gaugeo,label=below:{$1$}] (rbn) at (2,0) {};
            \draw (1n)--(2n)--(rbn)--(1n);
            \draw (rbn)--(4fn)--(4);
    \end{tikzpicture}};
    \end{tikzpicture}
\end{equation}
The resulting unframed quiver is connected, it can be disconnected however e.g.\ by framing the U$(1)$ node which corresponds to the (flavour) D5 brane in the brane system.
\end{enumerate}

\subsection{Type D}
\label{sec:Dtype}

We now turn to the the SU(2) gauge theory with fundamental matter \eqref{eq:Dtype}. 

\subsubsection*{Branes}

The right hand side of \eqref{eq:Dtype} is realized in a brane system on a circular dimension with two {\color{cyan}ON${}^-$} planes 
as follows (drawn here for $N_f = 7$) \cite{Hanany:1999sj, Hanany:2001iy,Cremonesi:2014xha}: 
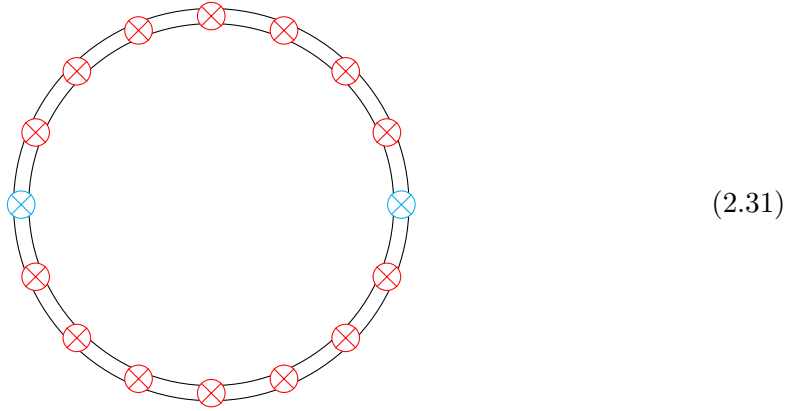
\begin{equation}
    \vcenter{\hbox{\scalebox{1}{\begin{tikzpicture}
        \draw (0,0) circle (2.4);
        \draw (0,0) circle (2.6);
        
        
            \nsm{-2.5,0};
            \nsm{2.5,0};
        
        
            \begin{scope}[rotate=22.5]
                \ns{2.5,0};
            \end{scope}
            \begin{scope}[rotate=45]
                \ns{2.5,0};
            \end{scope}
            \begin{scope}[rotate=67.5]
                \ns{2.5,0};
            \end{scope}
            \begin{scope}[rotate=90]
                \ns{2.5,0};
            \end{scope}
            \begin{scope}[rotate=112.5]
                \ns{2.5,0};
            \end{scope}
            \begin{scope}[rotate=135]
                \ns{2.5,0};
            \end{scope}
            \begin{scope}[rotate=157.5]
                \ns{2.5,0};
            \end{scope}
        
        
            \begin{scope}[rotate=-22.5]
                \ns{2.5,0};
            \end{scope}
            \begin{scope}[rotate=-45]
                \ns{2.5,0};
            \end{scope}
            \begin{scope}[rotate=-67.5]
                \ns{2.5,0};
            \end{scope}
            \begin{scope}[rotate=-90]
                \ns{2.5,0};
            \end{scope}
            \begin{scope}[rotate=-112.5]
                \ns{2.5,0};
            \end{scope}
            \begin{scope}[rotate=-135]
                \ns{2.5,0};
            \end{scope}
            \begin{scope}[rotate=-157.5]
                \ns{2.5,0};
            \end{scope}
    \end{tikzpicture}}}}
    \label{eq:drawingDtypeBranes}
\end{equation}
The orientifold planes act as $(x^6 , x^7 , x^8 , x^9) \rightarrow (-x^6 , -x^7 , -x^8 , -x^9)$, which corresponds in this picture to a reflection through the horizontal line containing the two orientifolds and a reflection with respect to the circle they belong to. 

Consider for instance the FI deformations 
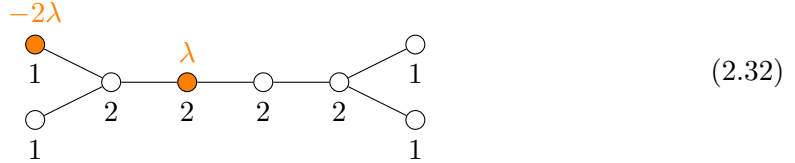
\begin{equation}
   \raisebox{-.5\height}{ 
\begin{tikzpicture}
        \node[gauge,label=below:{$1$}] (1) at (0,-0.5) {};
        \node[gaugeo,label=below:{$1$},label=above:{\textcolor{orange}{$- 2 \lambda$}}] (1u) at (0,0.5) {};
        \node[gauge,label=below:{$2$}] (2) at (1,0) {};
        \node[gaugeo,label=below:{$2$},label=above:{\textcolor{orange}{$\lambda$}}] (3) at (2,0) {};
        \node[gauge,label=below:{$2$}] (4) at (3,0) {};
        \node[gauge,label=below:{$2$}] (5) at (4,0) {};
        \node[gauge,label=below:{$1$}] (6) at (5,-0.5) {};
        \node[gauge,label=below:{$1$}] (6u) at (5,0.5) {};
        \draw (1)--(2)--(3)--(4)--(5)--(6) (1u)--(2) (5)--(6u);
    \end{tikzpicture}}
\end{equation}
The non abelian nodes in this quiver correspond to the four D3 brane intervals that stretch between the five central NS5 branes in \eqref{eq:drawingDtypeBranes}, while the abelian nodes correspond to the two D3 brane intervals next to the orientifold planes. 
Therefore the FI deformation is accomplished in the brane system by lifting three NS5 branes adjacent to the orientifold on the left. The D3 branes, whose motion is restricted by the orientifold action, give rise to two distinct phases: 
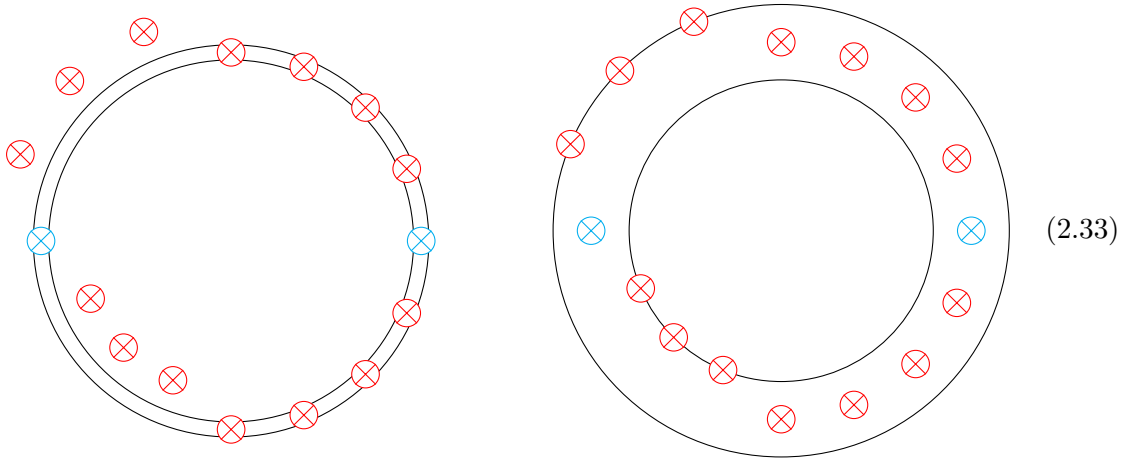
\begin{equation}
    \vcenter{\hbox{\scalebox{1}{\begin{tikzpicture}
        \draw (0,0) circle (2.4);
        \draw (0,0) circle (2.6);
        
        
            \nsm{-2.5,0};
            \nsm{2.5,0};
        
        
            \begin{scope}[rotate=22.5]
                \ns{2.5,0};
            \end{scope}
            \begin{scope}[rotate=45]
                \ns{2.5,0};
            \end{scope}
            \begin{scope}[rotate=67.5]
                \ns{2.5,0};
            \end{scope}
            \begin{scope}[rotate=90]
                \ns{2.5,0};
            \end{scope}
            \begin{scope}[rotate=112.5]
                \ns{3,0};
            \end{scope}
            \begin{scope}[rotate=135]
                \ns{3,0};
            \end{scope}
            \begin{scope}[rotate=157.5]
                \ns{3,0};
            \end{scope}
        
        
            \begin{scope}[rotate=-22.5]
                \ns{2.5,0};
            \end{scope}
            \begin{scope}[rotate=-45]
                \ns{2.5,0};
            \end{scope}
            \begin{scope}[rotate=-67.5]
                \ns{2.5,0};
            \end{scope}
            \begin{scope}[rotate=-90]
                \ns{2.5,0};
            \end{scope}
            \begin{scope}[rotate=-112.5]
                \ns{2,0};
            \end{scope}
            \begin{scope}[rotate=-135]
                \ns{2,0};
            \end{scope}
            \begin{scope}[rotate=-157.5]
                \ns{2,0};
            \end{scope}
    \end{tikzpicture}}}}
    \qquad\qquad
    \vcenter{\hbox{\scalebox{1}{\begin{tikzpicture}
        \draw (0,0) circle (3);
        \draw (0,0) circle (2);
        
        
            \nsm{-2.5,0};
            \nsm{2.5,0};
        
        
            \begin{scope}[rotate=22.5]
                \ns{2.5,0};
            \end{scope}
            \begin{scope}[rotate=45]
                \ns{2.5,0};
            \end{scope}
            \begin{scope}[rotate=67.5]
                \ns{2.5,0};
            \end{scope}
            \begin{scope}[rotate=90]
                \ns{2.5,0};
            \end{scope}
            \begin{scope}[rotate=112.5]
                \ns{3,0};
            \end{scope}
            \begin{scope}[rotate=135]
                \ns{3,0};
            \end{scope}
            \begin{scope}[rotate=157.5]
                \ns{3,0};
            \end{scope}
        
        
            \begin{scope}[rotate=-22.5]
                \ns{2.5,0};
            \end{scope}
            \begin{scope}[rotate=-45]
                \ns{2.5,0};
            \end{scope}
            \begin{scope}[rotate=-67.5]
                \ns{2.5,0};
            \end{scope}
            \begin{scope}[rotate=-90]
                \ns{2.5,0};
            \end{scope}
            \begin{scope}[rotate=-112.5]
                \ns{2,0};
            \end{scope}
            \begin{scope}[rotate=-135]
                \ns{2,0};
            \end{scope}
            \begin{scope}[rotate=-157.5]
                \ns{2,0};
            \end{scope}
    \end{tikzpicture}}}}
    \label{eq:drawingTwoPhasesDtypeBranes}
\end{equation}
In one phase (left), the world-volume of the D3 branes contains the same branes as in \eqref{eq:drawingDtypeBranes} but with $N_f = 4$. In the other phase (right), the D3 branes have left the orientifold fixed locus, and the local physics on their world-volumes is that of \eqref{eq:braneSystemAaffine} with $N_f = 3$. This constitutes the brane proof that the singularity $D_7$ splits under the FI deformation to two singularities $D_4$ and $A_2$. 

The previous result can be recovered directly via quiver subtraction as follows: 
\begin{equation}\label{D7higgs}
    \begin{tikzpicture}
        \node at (0,0) {$\begin{tikzpicture}
            \node[gauge,label=below:{$1$}] (1) at (0,-0.5) {};
            \node[gaugeo,label=below:{$1$}] (1u) at (0,0.5) {};
            \node[gauge,label=below:{$2$}] (2) at (1,0) {};
            \node[gaugeo,label=below:{$2$}] (3) at (2,0) {};
            \node[gauge,label=below:{$2$}] (4) at (3,0) {};
            \node[gauge,label=below:{$2$}] (5) at (4,0) {};
            \node[gauge,label=below:{$1$}] (6) at (5,-0.5) {};
            \node[gauge,label=below:{$1$}] (6u) at (5,0.5) {};
            \draw (1)--(2)--(3)--(4)--(5)--(6) (1u)--(2) (5)--(6u);
        \end{tikzpicture}$};
        \node at (-4,-3) {$\begin{tikzpicture}
            \node at (-0.5,0) {$-$};
            \node[gauge,label=below:{$1$}] (1) at (0,-0.5) {};
            \node[gaugeo,label=below:{$1$}] (1u) at (0,0.5) {};
            \node[gauge,label=below:{$2$}] (2) at (1,0) {};
            \node[gaugeo,label=below:{$2$}] (3) at (2,0) {};
            \node[gauge,label=below:{$1$}] (4) at (3,0) {};
            \node (5) at (4,0) {};
            \node (6) at (5,-0.5) {};
            \node (6u) at (5,0.5) {};
            \draw (1)--(2)--(3)--(4) (1u)--(2);
        \end{tikzpicture}$};
        \node at (-4,-6) {$\begin{tikzpicture}
            \node (1) at (0,-0.5) {};
            \node (1u) at (0,0.5) {};
            \node (2) at (1,0) {};
            \node (3) at (2,0) {};
            \node[gauge,label=below:{$1$}] (4) at (3,0) {};
            \node[gauge,label=below:{$2$}] (5) at (4,0) {};
            \node[gauge,label=below:{$1$}] (6) at (5,-0.5) {};
            \node[gauge,label=below:{$1$}] (6u) at (5,0.5) {};
            \node[gaugeo,label=left:{$1$}] (r) at (4,0.5) {};
            \draw (4)--(5)--(6) (5)--(6u) (r)--(5);
        \end{tikzpicture}$};
        \node at (4,-3) {$\begin{tikzpicture}
            \node at (-0.5,0) {$-$};
            \node (1) at (0,-0.5) {};
            \node[gaugeo,label=below:{$1$}] (1u) at (0,0.5) {};
            \node[gauge,label=below:{$1$}] (2) at (1,0) {};
            \node[gaugeo,label=below:{$2$}] (3) at (2,0) {};
            \node[gauge,label=below:{$2$}] (4) at (3,0) {};
            \node[gauge,label=below:{$2$}] (5) at (4,0) {};
            \node[gauge,label=below:{$1$}] (6) at (5,-0.5) {};
            \node[gauge,label=below:{$1$}] (6u) at (5,0.5) {};
            \draw (2)--(3)--(4)--(5)--(6) (1u)--(2) (5)--(6u);
        \end{tikzpicture}$};
        \node at (4,-6) {$\begin{tikzpicture}
            \node[gauge,label=below:{$1$}] (1) at (0,-0.5) {};
            \node (1u) at (0,0.5) {};
            \node[gauge,label=below:{$1$}] (2) at (1,0) {};
            \node (3) at (2,0) {};
            \node (4) at (3,0) {};
            \node (5) at (4,0) {};
            \node (6) at (5,-0.5) {};
            \node (6u) at (5,0.5) {};
            \node[gaugeo,label=left:{$1$}] (r) at (0.5,0.5) {};
            \draw (1)--(2)--(r)--(1);
        \end{tikzpicture}$};
        \draw[->] (-2,-1)--(-2,-4.5);
        \draw[->] (0,-1)--(1,-4.5);
    \end{tikzpicture}
\end{equation}
The quivers that need to be subtracted can be read directly from the brane systems \eqref{eq:drawingTwoPhasesDtypeBranes}. Note that on the left we subtract a \emph{finite} Dynkin diagram of type $D_5$, while on the right we subtract another quiver that has the shape of the $D_7$ Dynkin diagram but different ranks. Both quivers are \emph{ugly}, in the terminology of \cite{Gaiotto:2008ak}, and their 3d $\mathcal{N}=4$ Coulomb branches are freely generated.

As is hopefully clear from these examples, the FI deformations can be implemented easily from a quiver subtraction algorithm, without referring to a brane system. However, the brane system was essential in identifying which quiver to subtract. In general situation, with an arbitrary quiver, brane systems are in general not available. Therefore, we need an \emph{independent way of identifying which quivers to subtract}. Note in particular that we should be able to reproduce the subtraction of disconnected quivers as in \eqref{eq:doubleQuiverSubtraction}. The whole problem of FI deformations reduces to this question, which we now address in full generality in the next section.

\section{FI-flows via FI-subtractions}
\label{sec:GeneralMethod}

When the Higgs branch of a given theory can be described by a magnetic quiver, assumed to be unitary and simply laced, its mass deformations correspond to FI deformations for the magnetic quiver. It is therefore of interest to \emph{identify and characterize the minimal FI deformations of a unitary quiver}. 

In this section, we formalize this question and show that it is in general a difficult algebraic problem. We show that it can always be treated as a quiver subtraction procedure, generalizing what we saw in the previous section, but identifying which quivers to subtract is the central difficulty.  In particular, it should be stressed that it is not enough to subtract finite Dynkin quivers or restrict to subtracting quivers with a freely generated Coulomb branch. 

\subsection{Rephrasing into Linear Algebra}
\label{sec:hardProblem}

Consider a quiver with unitary gauge nodes and only bifundamental matter.\footnote{Throughout this paper, we use quivers without framing, it being understood that a diagonal $\mathrm{U}(1)$ has to be ungauged. In some of the examples we used framed versions for convenience. } We assume that every gauge node $\mathrm{U}(k)$ has at least $2k$ fundamentals. Formally, the quiver is defined by a set of vertices $V$, a list\footnote{The list can have repetitions. Note that the edges are oriented. } $E$ of edges $e \in V^2$ and a list of ranks $k_v \in \mathbb{Z}_{>0}$ for each $v \in V$. We turn on FI terms $\zeta = (\zeta_v)_{v \in V} \in \mathbb{C}^V$ at the vertices, which have to satisfy 
\begin{equation}
\label{eq:constraintsFI}
    \sum\limits_{v \in V}  \zeta_v  k_v = 0 \, . 
\end{equation}

In terms of fields, we associate to each edge $e = (e_1 , e_2)$, with $e_1 , e_2 \in V$, two chiral multiplets $Q_e$ and $\tilde{Q}_e$ transforming respectively in the representations $\mathbf{e_1} \times \overline{\mathbf{e_2}}$ and $\mathbf{e_2} \times \overline{\mathbf{e_1}}$. We see $Q_e$ as a $e_1 \times e_2$ matrix and $\tilde{Q}_e$ as a $e_2 \times e_1$ matrix. 
To each edge $e = (e_1 , e_2)$ one can associate two mesons: $M_e = Q_e \tilde{Q}_e$, transforming in the adjoint representation of $\mathrm{U}(k_{e_1})$, and $\tilde{M}_e = \tilde{Q}_e  Q_e$ transforming in the adjoint representation of $\mathrm{U}(k_{e_2})$. 
The F-term equations read 
\begin{equation}
\label{eq:Fterms}
    \sum\limits_{e | e_1 = v} M_e -\sum\limits_{e | e_2 = v} \tilde{M}_e  = \zeta_v \mathbf{1}_{k_v} \, .  
\end{equation}
The Higgs branch of the magnetic quiver is parametrized by the gauge invariant combinations of the $Q_e$ and $\tilde{Q}_e$ subject to \eqref{eq:Fterms}. 

The problem can now be stated as follows. Given a quiver, and a list of FI parameters $(\zeta_v)_{v \in V}$, what is the set of mesons $(M_e , \tilde{M}_e)$ satisfying \eqref{eq:Fterms}, up to gauge transformations?  There are two aspects to this question, which we attack in two steps. First, we need to characterize the possible pairs $(M_e , \tilde{M}_e)$ of matrices associated to a given edge, and second, we need to find an arrangement of these matrices in such a way that the sum \eqref{eq:Fterms} at each vertex is satisfied. Both steps involve simple linear algebra, but in combination they yield a difficult problem. We look at these in turn. 
\begin{itemize}
    \item \textbf{Edges: }
Let us look closer at the relation between $M_e$ and $\tilde{M}_e$ for a given edge $e \in E$. This boils down to the following linear algebra problem: what is the relation between two matrices of the form $Q \tilde{Q}$ and $\tilde{Q} Q$? The answer to this question is provided by the \hyperref[theorem1]{Theorem 1} stated and proved in Appendix \ref{app:theorems}. The bottom line is that the Jordan blocks of $M_e$ and $\tilde{M}_e$ with non zero eigenvalues are the same, while the Jordan blocks with 0 eigenvalue have sizes that can differ by at most one. Two matrices that satisfy this condition can always be written as $Q \tilde{Q}$ and $\tilde{Q} Q$ for some rectangular matrices $Q$ and $\tilde{Q}$ of appropriate shape. 
\item \textbf{Vertices: }
The vertices $v$ are distinguished by the number of edges reaching them, and the value of $\zeta_v$. If only two edges reach $v$, then \eqref{eq:Fterms} is straightforward to solve. Assume for simplicity that the orientations are such that the equation reads $M_e - \tilde{M}_{e'} = \zeta_v \mathbf{1}_{k_v}$. Then  $M_e  = \tilde{M}_{e'} +  \zeta_v \mathbf{1}_{k_v}$, and this simply shifts the eigenvalue of all the Jordan blocks of $\tilde{M}_{e'}$ by $\zeta_v$. At bifurcations however the problem is more intricate, even if $\zeta_v = 0$. Consider for instance the equation $M_e = \tilde{M}_{e'} + \tilde{M}_{e''}$. The constraints from the edges characterize the types of Jordan blocks for these three matrices, it is difficult to characterize the blocks of a sum from those of its summands. 
\end{itemize}

Given a set of meson pairs $(M_e , \tilde{M}_e)$ satisfying \eqref{eq:Fterms}, one constructs a subquiver $\mathsf{S}$ of $\mathsf{Q}$ as follows. The sets of vertices and nodes are the same as those of $\mathsf{Q}$, and the ranks are 
\begin{equation}
    k'_v = \max \left( \max\limits_{e | v \in e} r_e , k_v (1-\delta_{\zeta_v}) \right) 
\end{equation}
where $\delta_{\zeta_v}$ is a delta function whose value is zero whenever the FI parameter at $v$ vanishes and is one otherwise and
\begin{equation}
    r_e = \max ( \mathrm{rank} (M_e) , \mathrm{rank} (\tilde{M}_e) ) \, . 
\end{equation}
The quivers $\mathsf{S}$ that can be obtained this way are called \emph{solutions} of the FI-Meson problem. 

The set $\mathcal{S}$ of all solutions is partially ordered is the obvious way: for two solutions $\mathsf{S}^1$ and $\mathsf{S}^2$ (with nodes of ranks $k^1_v$ and $k^2_v$) we say that $\mathsf{S}^1 \preceq \mathsf{S}^2$ if for all $v \in V$, $k^1_v \leq k^2_v$. With this partial order the solution set $\mathcal{S}$, which is finite, has a number of minimal elements that we call the \emph{extremal solutions}: 
\begin{equation}
    \mathcal{S}^{\mathrm{extr}} = \min \mathcal{S} = \{  \mathsf{S}_i | i=1 , \dots , n \} \, . 
\end{equation}

\begin{figure} 
\begin{center}

\vspace*{1cm}

\begin{tikzpicture}
\node at (0,.5) {$(a)$};
            \node[gauge,label=below:{$1$}] (1) at (1,0) {};
            \node[gauge,label=below:{$2$}] (2) at (2,0) {};
            \node (3) at (3,0) {$\cdots$};
            \node[gauge,label=below:{$2$}] (4) at (4,0) {};
            \node[gauge,label=below:{$2$}] (5) at (5,0) {};
            \node[gaugeo,label=below:{$2$}] (6) at (6,0) {};
            \node[gauge,label=below:{$2$}] (7) at (7,0) {};
            \node[gauge,label=below:{$2$}] (8) at (8,0) {};
            \node (9) at (9,0) {$\cdots$};
            \node[gauge,label=below:{$2$}] (10) at (10,0) {};
            \node[gauge,label=above:{$1$}] (11) at (11,0) {};
            \node[gauge,label=right:{$1$}] (10u) at (10,1) {};
            \node[gaugeo,label=left:{$1$}] (2u) at (2,1) {};
            \draw (1)--(2)--(3)--(4)--(5)--(6)--(7)--(8)--(9)--(10)--(11) (10u)--(10) (2)--(2u);
            \node (a) at (1,.7) {\scalebox{.8}{$(2)$}};
            \node (b) at (3.5,1) {\scalebox{.8}{$  \left(\begin{array}{cc}
           1&1\\1&1
            \end{array}\right)$}};
            \node (c) at (1.7,-1.5) {\scalebox{.8}{$  \left(\begin{array}{cc}
           0&0\\1&0
            \end{array}\right)$}};
            \node (d) at (4,-1.5) {\scalebox{.8}{$  \left(\begin{array}{cc}
           1&1\\0&1
            \end{array}\right)$}};
            \node (e) at (6.3,1) {\scalebox{.8}{$  \left(\begin{array}{cc}
           0&1\\0&0
            \end{array}\right)$}};
            \draw[->,dotted] (a)--(2,.7);
            \draw[->,dotted] (b)--(2,.3);
            \draw[->,dotted] (c)--(1.7,0);
            \draw[->,dotted] (d)--(2.3,0);
            \draw[->,dotted] (d)--(5.7,0);
            \draw[->,dotted] (e)--(6.3,0);
        \end{tikzpicture}     
        
        \vspace{1cm}
        
\begin{tikzpicture}
\node at (0,.5) {$(b)$};
            \node[gauge,label=below:{$1$}] (1) at (1,0) {};
            \node[gauge,label=below:{$2$}] (2) at (2,0) {};
            \node (3) at (3,0) {$\cdots$};
            \node[gauge,label=below:{$2$}] (4) at (4,0) {};
            \node[gauge,label=below:{$2$}] (5) at (5,0) {};
            \node[gaugeo,label=below:{$2$}] (6) at (6,0) {};
            \node[gauge,label=below:{$2$}] (7) at (7,0) {};
            \node[gauge,label=below:{$2$}] (8) at (8,0) {};
            \node (9) at (9,0) {$\cdots$};
            \node[gauge,label=below:{$2$}] (10) at (10,0) {};
            \node[gauge,label=above:{$1$}] (11) at (11,0) {};
            \node[gauge,label=right:{$1$}] (10u) at (10,1) {};
            \node[gaugeo,label=left:{$1$}] (2u) at (2,1) {};
            \draw (1)--(2)--(3)--(4)--(5)--(6)--(7)--(8)--(9)--(10)--(11) (10u)--(10) (2)--(2u);
            \node (a) at (1,.7) {\scalebox{.8}{$(2)$}};
            \node (b) at (5,1.5) {\scalebox{.8}{$  \left(\begin{array}{cc}
           2&0\\0&0
            \end{array}\right)$}};
            \node (c) at (6.3,-1.5) {\scalebox{.8}{$  \left(\begin{array}{cc}
           1&0\\0&-1
            \end{array}\right)$}};
            \node (d) at (7,1.5) {\scalebox{.8}{$  \left(\begin{array}{cc}
           0&1\\1&0
            \end{array}\right)$}};
            \node (e) at (12,1.5) {\scalebox{.8}{$  \left(\begin{array}{cc}
           0&1\\0&0
            \end{array}\right)$}};
            \node (f) at (10.3,-1.5) {\scalebox{.8}{$  \left(\begin{array}{cc}
           0&0\\1&0
            \end{array}\right)$}};
            \draw[->,dotted] (a)--(2,.7);
            \draw[->,dotted] (b)--(2,0.3);
            \draw[->,dotted] (b)--(2.3,0);
            \draw[->,dotted] (b)--(5.7,0);
            \draw[->,dotted] (c)--(6.3,0);
            \draw[->,dotted] (d)--(6.7,0);
            \draw[->,dotted] (d)--(9.7,0);
            \draw[->,dotted] (e)--(10,.3);
            \draw[->,dotted] (f)--(10.3,0);
        \end{tikzpicture}
       
        \vspace{1cm}
        
        \hspace*{-1cm}\scalebox{.9}{\begin{tikzpicture}
        \node (a) at (0,0) {
        $\begin{tikzpicture}
            \node[gauge,label=below:{$1$}] (1) at (1,0) {};
            \node[gauge,label=below:{$2$}] (2) at (2,0) {};
            \node (3) at (3,0) {$ \cdots $};
            \node[gauge,label=below:{$2$}] (4) at (4,0) {};
            \node[gauge,label=below:{$2$}] (5) at (5,0) {};
            \node[gaugeo,label=below:{$2$}] (6) at (6,0) {};
            \node[gauge,label=below:{$2$}] (7) at (7,0) {};
            \node[gauge,label=below:{$2$}] (8) at (8,0) {};
            \node (9) at (9,0) {$ \cdots $};
            \node[gauge,label=below:{$2$}] (10) at (10,0) {};
            \node[gauge,label=below:{$1$}] (11) at (11,0) {};
            \node[gauge,label=right:{$1$}] (10u) at (10,1) {};
            \node[gaugeo,label=left:{$1$}] (2u) at (2,1) {};
            \draw (1)--(2)--(3)--(4)--(5)--(6)--(7)--(8)--(9)--(10)--(11) (10u)--(10) (2)--(2u);
        \end{tikzpicture}$
        };
		\node at (-4,-3) {$(a)$};
        \node (b) at (-4,-3.5) {
        $\begin{tikzpicture}
            \node at (0.5,.5) {$-$};
            \node[gauge,label=below:{$1$}] (1) at (1,0) {};
            \node[gauge,label=below:{$2$}] (2) at (2,0) {};
            \node[gauge,label=below:{$2$}] (3) at (3,0) {};
            \node (4) at (4,0) {$ \cdots $};
            \node[gauge,label=below:{$2$}] (5) at (5,0) {};
            \node[gaugeo,label=below:{$2$}] (6) at (6,0) {};
            \node[gauge,label=below:{$1$}] (7) at (7,0) {};
            \node[gaugeo,label=left:{$1$}] (2u) at (2,1) {};
            \draw (1)--(2)--(3)--(4)--(5)--(6)--(7) (2)--(2u);
        \end{tikzpicture}$
        };
        \node (c) at (-4,-7) {
        $\begin{tikzpicture}
            \node[gauge,label=below:{$1$}] (7) at (7,0) {};
            \node[gauge,label=below:{$2$}] (8) at (8,0) {};
            \node (9) at (9,0) {$ \cdots $};
            \node[gauge,label=below:{$2$}] (10) at (10,0) {};
            \node[gauge,label=below:{$1$}] (11) at (11,0) {};
            \node[gauge,label=right:{$1$}] (10u) at (10,1) {};
            \node[gaugeo,label=left:{$1$}] (8u) at (8,1) {};
            \draw (7)--(8)--(9)--(10)--(11) (10u)--(10) (8)--(8u);
        \end{tikzpicture}$
        };
		\node at (6,-3) {$(b)$};
        \node (d) at (6,-3.5) {
        $\begin{tikzpicture}
            \node at (1.5,.5) {$-$};
            \node[gauge,label=below:{$1$}] (2) at (2,0) {};
            \node (3) at (3,0) {$ \cdots $};
            \node[gauge,label=below:{$1$}] (4) at (4,0) {};
            \node[gauge,label=below:{$1$}] (5) at (5,0) {};
            \node[gaugeo,label=below:{$2$}] (6) at (6,0) {};
            \node[gauge,label=below:{$2$}] (7) at (7,0) {};
            \node[gauge,label=below:{$2$}] (8) at (8,0) {};
            \node (9) at (9,0) {$ \cdots $};
            \node[gauge,label=below:{$2$}] (10) at (10,0) {};
            \node[gauge,label=below:{$1$}] (11) at (11,0) {};
            \node[gauge,label=right:{$1$}] (10u) at (10,1) {};
            \node[gaugeo,label=left:{$1$}] (2u) at (2,1) {};
            \draw (2)--(3)--(4)--(5)--(6)--(7)--(8)--(9)--(10)--(11) (10u)--(10) (2)--(2u);
        \end{tikzpicture}$
        };        
        \node (e) at (6,-7) {
        $\begin{tikzpicture}
            \node[gauge,label=below:{$1$}] (7) at (7,0) {};
            \node[gauge,label=below:{$1$}] (8) at (8,0) {};
            \node (9) at (9,0) {$ \cdots $};
            \node[gauge,label=below:{$1$}] (10) at (10,0) {};
            \node[gauge,label=below:{$1$}] (11) at (11,0) {};
            \node[gaugeo,label=right:{$1$}] (10u) at (9,1) {};
            \draw (7)--(8)--(9)--(10)--(11)--(10u)--(7);
        \end{tikzpicture}$
        };
        \draw[->] (a) .. controls (0,-4.5) .. (c);
        \draw[->] (a) .. controls (0,-4.5) .. (e);
    \end{tikzpicture}}
\end{center}
\caption{Two solutions for D type and corresponding subtractions. The subtractions realize the splitting $D_n \rightarrow A_{k-1} D_{n-k}$. The orange node in the middle is the $k$th starting from the left. This reproduces the subtraction \eqref{D7higgs} obtained from the brane system.  }
\label{fig:solutionsDtype}
\end{figure}
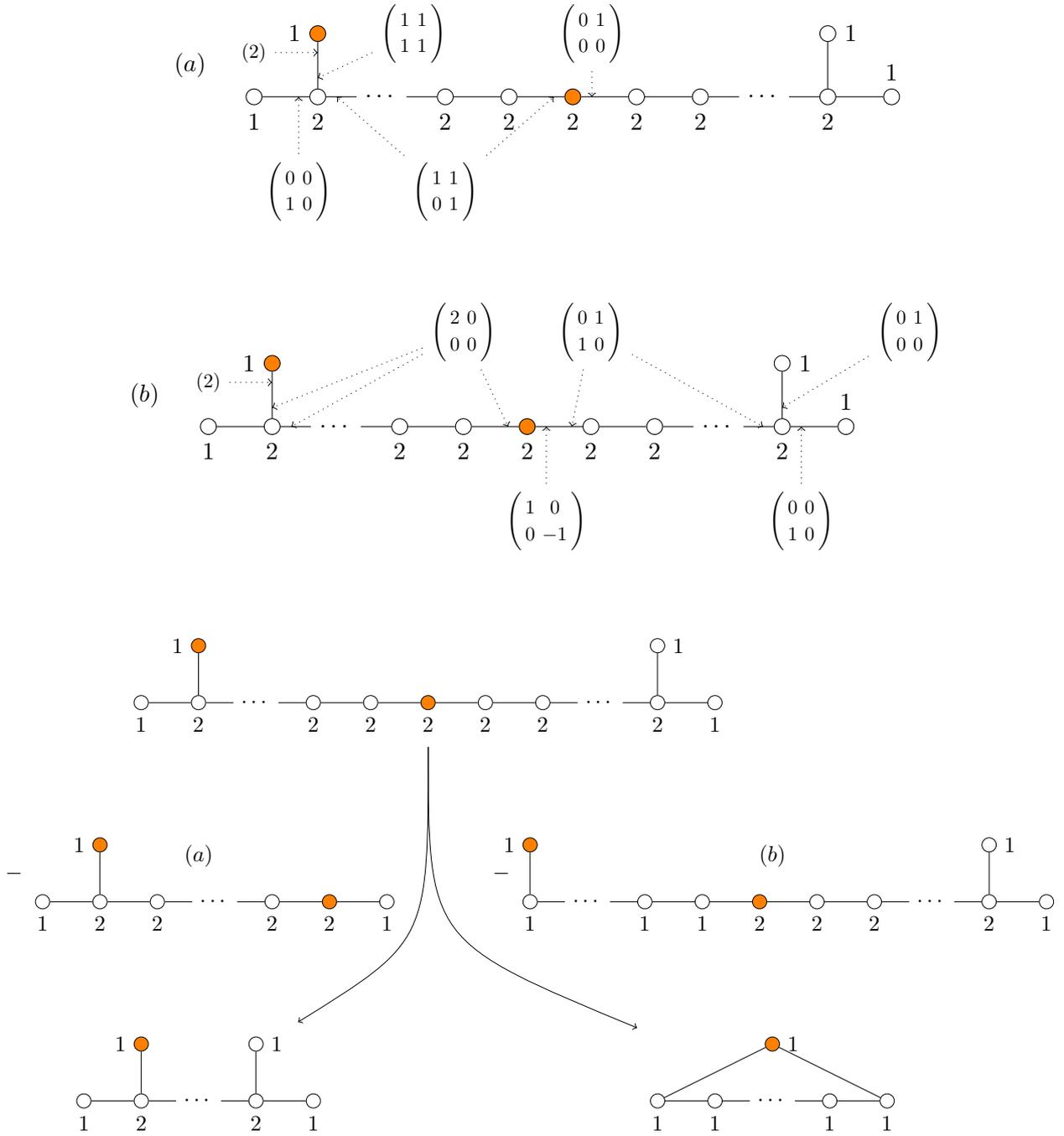

\paragraph{Notations. } In the following, it is convenient to show solutions of \eqref{eq:Fterms} in a graphical way, without having to explicitly name the edges and vertices. We do this using the following notation. For each edge, we show the two matrices corresponding to $M_e$ and $\tilde{M}_e$ at the ends of the edge $e$, next to the vertex for which these matrices transform in the adjoint representation: 
\begin{equation}
\label{eq:notationMesons}
      \raisebox{-.5\height}{    \begin{tikzpicture}
            \node[gauge,label=below:{$k_1$}] (2) at (3,0) {};
            \node[gauge,label=below:{$k_2$}] (3) at (6,0) {};
            \node (3l) at (5,-1) {$\tilde{M}$}; 
            \node (2r) at (4,-1) {$M$};
            \draw[->,dotted] (3l)--(5.5,0);
            \draw[->,dotted] (2r)--(3.5,0);
            \draw (2)--(3);
        \end{tikzpicture} }
\end{equation}
This notation means that $M$ is an $k_1 \times k_1$ matrix, $\tilde{M}$ is an $k_2 \times k_2$ matrix, and they are related as given by \hyperref[theorem1]{Theorem 1}. For instance, 
\begin{equation}
   \raisebox{-.5\height}{ \begin{tikzpicture}
            \node[gauge,label=below:{$2$}] (2) at (3,0) {};
            \node[gauge,label=below:{$3$}] (3) at (6,0) {};
            \node (3l) at (6,-2) {\scalebox{.8}{$  \left(\begin{array}{ccc}
            2&-1&-1\\1&0&-1\\1&-1&0
            \end{array}\right)$}}; 
            \node (2r) at (3,-2) {\scalebox{.8}{$  \left(\begin{array}{cc}
           1&0\\0&1
            \end{array}\right)$}};
            \draw[->,dotted] (3l)--(5.5,0);
            \draw[->,dotted] (2r)--(3.5,0);
            \draw (2)--(3);
        \end{tikzpicture} }
\end{equation}
is admissible, as the Jordan block decomposition of the $3 \times 3$ matrix is $J_2 (1) \oplus J_1 (0)$. Here we use the following standard notation for $k \times k$ Jordan blocks: 
\begin{equation}
    J_k (\lambda) = \begin{pmatrix}
  \, \, \lambda \, \, & 1 & & &  \\
         & \, \, \lambda \, \, & 1 & & \\ 
         & &  \, \, \lambda \, \, & \ddots & \\ 
         & & & \ddots & 1 \\
         & & & &  \, \, \lambda \, \, 
\end{pmatrix}\,   . 
\end{equation}
Vertices $v$ where $\zeta_v = 0$ are painted in white, and the sum of adjacent matrices has to sum to 0 (with appropriate signs). Vertices $v$ with  $\zeta_v \neq 0$ are painted in orange, the value of $\zeta_v$ is indicated in orange next to the vertex, and the adjacent matrices have to sum to $\zeta_v$ times the identity matrix (again with appropriate signs). When only two nodes are painted, there is no need to indicate the value of $\zeta_v$ as it is fixed up to an overall constant. Thus, for a given quiver the solution in terms of mesons can be drawn by putting all the (non-zero) meson matrices as in \eqref{eq:notationMesons}. From there the solution quivers $\mathsf{S}$ can be read directly by looking at the ranks of the meson matrices on edges adjacent to each vertex. For instance, the D-type Klein singularities studied in Section \ref{sec:Dtype} are shown in Figure \ref{fig:solutionsDtype}. One can check that the solutions presented there are extremal. 

\paragraph{Subtraction algorithm. } 
Let us first assume that the solution is irreducible, i.e. there is no basis in which all the matrices $M_e$ and $\tilde{M}_e$ can be written into block diagonal form in a non trivial way, with each block being a solution of the equations. In this case, there is only a $\mathrm{U} (1)$ group, acting on the $Q_e$ and $\tilde{Q}_e$, that leaves the set of all these matrices invariant. This $\mathrm{U} (1)$ is what replaces the group 
\begin{equation}
    \prod\limits_{v \in V} \mathrm{U} ( k'_v ) \, . 
\end{equation}
This is illustrated in Figure \ref{fig:solutionsDtype}. 
More generally, if the solution can be decomposed into the direct sum of $K$ copies of a given solution, the group is broken to $\mathrm{U} (K)$. A new quiver is obtained, with nodes $\mathrm{U}(k_v - k'_v)$ and an additional node $\mathrm{U}(K)$ and edges used to rebalance. If the solution can be decomposed into $K_1$ times a solution, plus $K_2$ times another one, etc, then one adds nodes $\mathrm{U}(K_1)$, $\mathrm{U}(K_2)$, etc. This is what is at play in the "double subtraction" in \eqref{eq:doubleQuiverSubtraction}. Several other examples of this are given in Section \ref{sec:SQCDUnitary}.

\subsection{Example Solutions and Complexity}

In general, it is difficult to find explicit matrix solutions to \eqref{eq:Fterms}. The reason is that while the Jordan structure of the matrices at each junction is known, the basis in which the canonical Jordan form is reached depends on each leg. To make the problem manifest, we look at two examples among the possible deformations of $E$-type Klein singularities. It should be noted that these are still very simple spaces, but the complexity of the linear algebra problem already requires computer assistance. 

\paragraph{Example 1. }
For an illustration of the complexity of the problem, one can have a look at the quiver 
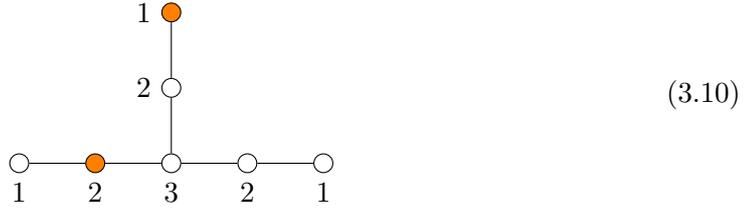
\begin{equation}
  \raisebox{-.5\height}{  \begin{tikzpicture}
            \node[gauge,label=below:{$1$}] (1) at (1,0) {};
            \node[gaugeo,label=below:{$2$}] (2) at (2,0) {};
            \node[gauge,label=below:{$3$}] (3) at (3,0) {};
            \node[gauge,label=below:{$2$}] (4) at (4,0) {};
            \node[gauge,label=below:{$1$}] (5) at (5,0) {};
            \node[gauge,label=left:{$2$}] (3u) at (3,1) {};
            \node[gaugeo,label=left:{$1$}] (3uu) at (3,2) {};
            \draw (1)--(2)--(3)--(4)--(5) (3)--(3u)--(3uu);
        \end{tikzpicture}}
        \label{eq:FIdefE6Example}
\end{equation} 
We consider the affine $E_6$ quiver, and turn on FI terms as indicated by the orange numbers. For simplicity, label the three legs by $1,2,3$, oriented from the central node, leg 1 being the one where an FI term is turned on for a $\mathrm{U}(1)$ node, leg 2 having an FI term for a $\mathrm{U}(2)$, and leg 3 having no FI term. In any given solution, leg $i$ gives a $3 \times 3$ matrix $M_i$, and these have to satisfy $M_1 + M_2 + M_3 = 0$. Let's try to solve this equation. Consider first the Jordan types of the $M_i$, which are constrained by \hyperref[theorem1]{Theorem 1} in Appendix \ref{app:theorems}: 
\begin{itemize}
    \item We begin with $M_1$: since the Jordan type of the $1 \times 1$ matrix at the end of the leg is $J_1 (2)$, the type of $M_1$ is either $J_1 (2) \oplus J_2 (0)$ or $J_1 (2)  \oplus J_1 (0)\oplus J_1 (0)$. 
    \item For $M_2$, the type of the $2 \times 2$ matrix before the FI node is either $J_2 (0)$ or $J_1 (0)\oplus J_1 (0)$. Therefore the type of $M_2$ is either $J_2 (1) \oplus J_1 (0)$ or $J_1 (1)\oplus J_1 (1)\oplus J_1 (0)$. 
    \item Finally, $M_3$ is nilpotent, its Jordan type is either $J_3 (0)$, $J_2 (0) \oplus J_1 (0)$ or $J_1 (0)\oplus J_1 (0)\oplus J_1 (0)$. 
\end{itemize}

\begin{table}[]
    \centering
    \hspace*{-.5cm}\begin{tabular}{|c|ccc|} \hline 
Matrices & $M_1$ & $M_2$ & $M_3$ \\ \hline 
Possibilities & \begin{tabular}{c}
$J_1 (2) \oplus J_2 (0)$ \\ $J_1 (2) \oplus J_1 (0)  \oplus J_1 (0)$ 
\end{tabular} & 
\begin{tabular}{c}
$J_2 (-1) \oplus J_1 (0)$ \\ $J_1 (-1) \oplus J_1 (-1) \oplus J_1 (0)$ 
\end{tabular} & \begin{tabular}{c}
$J_3 (0)$ \\ $J_2 (0) \oplus J_1 (0)$ \\ $J_1 (0) \oplus J_1 (0) \oplus J_1 (0)$ 
\end{tabular} \\ \hline 
Solution 1 & \begin{tabular}{c}
$J_1 (2) \oplus J_1 (0)  \oplus J_1 (0)$ \\ $  \left(\begin{array}{ccc}
           1&1&0\\1&1&0\\0&0&0
            \end{array}\right)$
\end{tabular}   &  \begin{tabular}{c}
$J_2 (-1) \oplus J_1 (0)$   \\ $  \left(\begin{array}{ccc}
           -1&-1&0\\0&-1&0\\0&0&0
            \end{array}\right)$
\end{tabular}  &  \begin{tabular}{c}
  $J_2 (0) \oplus J_1 (0)$   \\ $  \left(\begin{array}{ccc}
           0&0&0\\-1&0&0\\0&0&0
            \end{array}\right)$
\end{tabular}\\ \hline 
Solution 2 &  \begin{tabular}{c}
 $J_1 (2) \oplus J_2 (0)$ \\ $  \left(\begin{array}{ccc}
           2&-1&-1\\0&0&-1\\0&0&0
            \end{array}\right)$
\end{tabular} & \begin{tabular}{c}
$J_1 (-1) \oplus J_1 (-1) \oplus J_1 (0)$ \\ $\left(\begin{array}{ccc}
            -2&1&1\\-1&0&1\\-1&1&0
            \end{array}\right)$
\end{tabular}   & \begin{tabular}{c}
$J_3 (0)$   \\ $  \left(\begin{array}{ccc}
            0&0&0\\1&0&0\\1&-1&0
            \end{array}\right)$
\end{tabular} \\ \hline 
\end{tabular}
    \caption{Solutions for the FI deformations \eqref{eq:FIdefE6Example}. The columns are labelled by the $3 \times 3$ meson matrices, one for each leg in the quiver. The possibilities are the possible Jordan types compatible with the FI constraints. Then we show the two extremal solutions in terms of Jordan forms, and with a choice of representative. }
    \label{tab:E6solution}
\end{table}

\begin{figure}
\begin{center}
\begin{tikzpicture}
            \node[gauge,label=below:{$1$}] (1) at (0,0) {};
            \node[gaugeo,label=below:{$2$},label=above:{\torange{$1$}}] (2) at (3,0) {};
            \node[gauge,label=below:{$3$}] (3) at (6,0) {};
            \node[gauge,label=below:{$2$}] (4) at (9,0) {};
            \node[gauge,label=below:{$1$}] (5) at (12,0) {};
            \node[gauge,label=right:{$2$}] (6) at (6,3) {};
            \node[gaugeo,label=right:{$1$},label=left:{\torange{$-2$}}] (7) at (6,6) {};
            \draw (1)--(2)--(3)--(4)--(5) (3)--(6)--(7);
            \node (3l) at (4,-2) {\scalebox{.8}{$  \left(\begin{array}{ccc}
            -1&-1&0\\0&-1&0\\0&0&0
            \end{array}\right)$}}; 
            \node (3r) at (8,-2) {\scalebox{.8}{$  \left(\begin{array}{ccc}
            0&0&0\\-1&0&0\\0&0&0
            \end{array}\right)$}};
            \node (3u) at (8,+2) {\scalebox{.8}{$  \left(\begin{array}{ccc}
           1&1&0\\1&1&0\\0&0&0
            \end{array}\right)$}};
            \node (2r) at (2,+2) {\scalebox{.8}{$  \left(\begin{array}{cc}
           -1&-1\\0&-1
            \end{array}\right)$}};
            \node (6u) at (3,+4) {\scalebox{.8}{$  \left(\begin{array}{cc}
           1&1\\1&1
            \end{array}\right)$}};
            \node (7u) at (3,+6) {\scalebox{.8}{$ (2)$}};
            \node (4r) at (11,-2) {\scalebox{.8}{$  \left(\begin{array}{cc}
           0&0\\0&0
            \end{array}\right)$}};
            \node (2l) at (1,-2) {\scalebox{.8}{$ \left(\begin{array}{cc}
           0&-1\\0&0
            \end{array}\right)$}};
            \draw[->,dotted] (3l)--(5.5,0);
            \draw[->,dotted] (3r)--(6.5,0);
            \draw[->,dotted] (3u)--(6,.5);
            \draw[->,dotted] (2r)--(3.5,0);
            \draw[->,dotted] (2l)--(2.5,0);
            \draw[->,dotted] (4r)--(9.5,0);
            \draw[->,dotted] (4r)--(8.5,0);
            \draw[->,dotted] (6u)--(6,3.5);
            \draw[->,dotted] (6u)--(6,2.5);
            \draw[->,dotted] (7u)--(6,5.5);
        \end{tikzpicture} 

\vspace*{2cm}

\begin{tikzpicture}
            \node[gauge,label=below:{$1$}] (1) at (0,0) {};
            \node[gaugeo,label=below:{$2$},label=above:{\textcolor{orange}{$1$}}] (2) at (3,0) {};
            \node[gauge,label=below:{$3$}] (3) at (6,0) {};
            \node[gauge,label=below:{$2$}] (4) at (9,0) {};
            \node[gauge,label=below:{$1$}] (5) at (12,0) {};
            \node[gauge,label=right:{$2$}] (6) at (6,3) {};
            \node[gaugeo,label=right:{$1$},label=left:{\textcolor{orange}{$-2$}}] (7) at (6,6) {};
            \draw (1)--(2)--(3)--(4)--(5) (3)--(6)--(7);
            \node (3l) at (4,-2) {\scalebox{.8}{$  \left(\begin{array}{ccc}
            -2&1&1\\-1&0&1\\-1&1&0
            \end{array}\right)$}}; 
            \node (3r) at (8,-2) {\scalebox{.8}{$  \left(\begin{array}{ccc}
            0&0&0\\1&0&0\\1&-1&0
            \end{array}\right)$}};
            \node (3u) at (8,+2) {\scalebox{.8}{$  \left(\begin{array}{ccc}
           2&-1&-1\\0&0&-1\\0&0&0
            \end{array}\right)$}};
            \node (2r) at (2,+2) {\scalebox{.8}{$  \left(\begin{array}{cc}
           -1&0\\0&-1
            \end{array}\right)$}};
            \node (6u) at (3,+4) {\scalebox{.8}{$  \left(\begin{array}{cc}
           2&0\\0&0
            \end{array}\right)$}};
            \node (7u) at (3,+6) {\scalebox{.8}{$ (2)$}};
            \node (4r) at (11,-2) {\scalebox{.8}{$  \left(\begin{array}{cc}
           0&0\\1&0
            \end{array}\right)$}};
            \node (2l) at (1,-2) {\scalebox{.8}{$ \left(\begin{array}{cc}
           0&0\\0&0
            \end{array}\right)$}};
            \draw[->,dotted] (3l)--(5.5,0);
            \draw[->,dotted] (3r)--(6.5,0);
            \draw[->,dotted] (3u)--(6,.5);
            \draw[->,dotted] (2r)--(3.5,0);
            \draw[->,dotted] (2l)--(2.5,0);
            \draw[->,dotted] (4r)--(9.5,0);
            \draw[->,dotted] (4r)--(8.5,0);
            \draw[->,dotted] (6u)--(6,3.5);
            \draw[->,dotted] (6u)--(6,2.5);
            \draw[->,dotted] (7u)--(6,5.5);
        \end{tikzpicture} 
\end{center}
\caption{The two minimal deformations corresponding to the given FI in orange. The top quiver displays Solution 1 in Table \ref{tab:E6solution}, and the bottom quiver displays Solution 2.  }
\label{fig:solutionE6}
\end{figure}

We now are left with solving the equation $M_1 + M_2 + M_3 = 0$ with these Jordan form constraints. Using brute force computation, we find exactly two extremal solutions up to conjugation. This is summarized in Table \ref{tab:E6solution}. 
Once these solutions are known, it is easy to complete all the meson matrices, as depicted in Figure \ref{fig:solutionE6}. This means that two quiver subtractions are possible: they are shown in Figure \ref{fig:solutionE6}.
Geometrically, this means that the $E_6$ singularity is deformed to a space with two singularities, of types $A_1$ and $A_4$. 

\paragraph{Example 2. }
We now turn to the quiver 
\begin{equation}\label{eq:FIdefE8Example}
   \raisebox{-.5\height}{ \begin{tikzpicture}
            \node[gaugeo,label=below:{$1$}] (1) at (1,0) {};
            \node[gauge,label=below:{$2$}] (2) at (2,0) {};
            \node[gauge,label=below:{$3$}] (3) at (3,0) {};
            \node[gauge,label=below:{$4$}] (4) at (4,0) {};
            \node[gauge,label=below:{$5$}] (5) at (5,0) {};
            \node[gauge,label=below:{$6$}] (6) at (6,0) {};
            \node[gauge,label=below:{$4$}] (7) at (7,0) {};
            \node[gaugeo,label=below:{$2$}] (8) at (8,0) {};
            \node[gauge,label=left:{$3$}] (6u) at (6,1) {};
            \draw (1)--(2)--(3)--(4)--(5)--(6)--(7)--(8) (6)--(6u);
        \end{tikzpicture}}
\end{equation}
One can perform exactly the same analysis as for the previous example. Given the lengths of the legs and using the theorem, the possibilities are given respectively by the 7 partitions of 5, the 3 partitions of 4 with no entry $>2$, and the 4 partitions of 6 with no entry $>2$.  This is summarized in Table \ref{tab:E8Solution}. We find a unique extremal solution up to conjugation, and it leads to the subtraction shown in the Appendix, see \eqref{eq:E8D7}. 

\begin{table}[]
    \centering
\hspace*{-1cm}\begin{tabular}{|c|ccc|} \hline 
Matrices & $M_1$ & $M_2$ & $M_3$ \\ \hline 
Possibilities & \begin{tabular}{c}
$J_1 (2) \oplus 5 J_1 (0)$ \\ $J_1 (2) \oplus  J_2 (0) \oplus 3 J_1 (0) $ \\ $J_1 (2) \oplus 2 J_2 (0) \oplus  J_1 (0) $ \\ $J_1 (2) \oplus J_3 (0) \oplus  2 J_1 (0) $ \\ $J_1 (2) \oplus J_3 (0) \oplus  J_2 (0) $ \\ $J_1 (2) \oplus J_4 (0) \oplus   J_1 (0) $ \\ $J_1 (2) \oplus J_5 (0)  $ 
\end{tabular} & 
\begin{tabular}{c}
$J_1 (-1) \oplus J_1 (-1) \oplus 4 J_1 (0)$\\ $J_1 (-1) \oplus J_1 (-1) \oplus J_2 (0) \oplus 2 J_1 (0)$ \\ $J_1 (-1) \oplus J_1 (-1) \oplus 2 J_2 (0)$
\end{tabular} & \begin{tabular}{c}
$6 J_1 (0) $ \\ $J_2 (0) \oplus 4 J_1 (0)$ \\ $2 J_2 (0) \oplus 2 J_1 (0)$ \\  $3 J_2 (0)$
\end{tabular} \\ \hline 
Solution & \begin{tabular}{c}
 $J_1 (2) \oplus J_3 (0) \oplus  2 J_1 (0) $  \\ $ \left(
\begin{array}{cccc}
 1 & -1 & 0 & -1 \\
 -1 & 3 & 1 & 1 \\
 1 & -3 & -1 & -1 \\
 1 & -2 & -1 & -1 \\
\end{array}
\right)$
\end{tabular}   &  \begin{tabular}{c}
$J_1 (-1) \oplus J_1 (-1) \oplus J_2 (0) \oplus 2 J_1 (0)$   \\ $ \left(
\begin{array}{cccc}
 -1 & 0 & 0 & 0 \\
 0 & -1 & 0 & 0 \\
 0 & 0 & 0 & -1 \\
 0 & 0 & 0 & 0 \\
\end{array}
\right)$
\end{tabular}  &  \begin{tabular}{c}
$2 J_2 (0) \oplus 2 J_1 (0)$  \\ $ \left(
\begin{array}{cccc}
 0 & 1 & 0 & 1 \\
 1 & -2 & -1 & -1 \\
 -1 & 3 & 1 & 2 \\
 -1 & 2 & 1 & 1 \\
\end{array}
\right)$
\end{tabular}\\ \hline 
\end{tabular}
    \caption{Solutions for the FI deformations \eqref{eq:FIdefE8Example}. Note that a trivial block $2 J_1 (0)$ appears in the solution everywhere, and we show only the non trivial $4 \times 4$ block in the explicit solution.  }
    \label{tab:E8Solution}
\end{table}

\begin{figure}
    \centering
\begin{tikzpicture}[xscale=2,yscale=2]
\node[hasse,black!15] (1) at (0,0) {};
\node[hasse,orange] (11) at (-1,1) {};
\node[hasse,black!15] (12) at (-.2,1) {};
\node[hasse,black!15] (13) at (.3,1) {};
\node[hasse,black!15] (14) at (.8,1) {};
\node[hasse] (21) at (-1.7,2) {};
\node[hasse,red] (22) at (-1.4,2) {};
\node[hasse] (23) at (-1,2) {};
\node[hasse,black!15] (24) at (-.3,2) {};
\node[hasse,black!15] (25) at (.2,2) {};
\node[hasse,orange] (26) at (.8,2) {};
\node[hasse,orange] (27) at (1.8,2) {};
\node[hasse] (31) at (-1.8,3) {};
\node[hasse] (32) at (-1.2,3) {};
\node[hasse] (33) at (-.6,3) {};
\node[hasse] (34) at (0,3) {};
\node[hasse,red] (35) at (.8,3) {};
\node[hasse] (36) at (1.2,3) {};
\node[hasse] (41) at (-.9,4) {};
\node[hasse] (42) at (-.2,4) {};
\node[hasse] (43) at (.3,4) {};
\node[hasse] (44) at (.9,4) {};
\node[hasse] (45) at (1.2,4) {};
\node[hasse] (5) at (0,5) {};
\draw[black!15] (1)--(11) (1)--(12) (1)--(13) (1)--(14);
\draw[black!15] (12)--(21) (12)--(24) (13)--(22) (13)--(25) (14)--(23) (14)--(25) (14)--(27) (14)--(26);
\draw (11)--(21) (11)--(22) (11)--(23) ;
\draw[black!15] (24)--(33) (24)--(35) (24)--(34) (25)--(32) (25)--(35) ;
\draw (21)--(31) (21)--(33) (22)--(32) (22)--(33) (23)--(31) (23)--(34) (26)--(34) (26)--(35) (26)--(36) (27)--(34) (27)--(36) (27)--(45);
\draw (31)--(41) (31)--(43) (32)--(42) (32)--(44) (33)--(41) (33)--(45) (34)--(42) (34)--(43) (35)--(45) (36)--(44);
\draw (5)--(41) (5)--(42) (5)--(43) (5)--(44) (5)--(45);
\node at (0,5.3) {$\mathsf{S} = \mathsf{Q}$};
\node at (0,-.3) {\textcolor{black!15}{$\mathsf{S} = 0$}};
\node (f1) at (-2.5,1.7) {\textcolor{red}{$\mathsf{S}^{\textrm{free}}_1 (\zeta ')$}};
\node (f2) at (2.1,3.4) {\textcolor{red}{$\mathsf{S}^{\textrm{free}}_2 (\zeta ')$}};
\node (f3) at (2.5,2) {\textcolor{red}{$\mathsf{S}^{\textrm{free}}_3 (\zeta ')$}};
\node at (1.8,1.7) {\textcolor{orange}{$\mathsf{S}_3 (\zeta ')$}};
\node at (.8,1.7) {\textcolor{orange}{$\mathsf{S}_2 (\zeta ')$}};
\node at (-1.0,0.7) {\textcolor{orange}{$\mathsf{S}_1 (\zeta ')$}};
\node at (-2.2,4) {\begin{tabular}{c}
Poset of solutions \\ $\mathcal{S}(\zeta =  \textcolor{orange}{\zeta '})$
\end{tabular} };
\draw[->,dotted] (f1)--(22);
\draw[->,dotted] (f2)--(35);
\draw[->,dotted] (f3)--(27);
\end{tikzpicture} 
    \caption{Given an FI deformation $\zeta '$ one can look for free quiver solutions, shown here in red. They can happen to be extremal (as is the case here for $\mathsf{S}^{\textrm{free}}_3 (\zeta ') =\mathsf{S}_3 (\zeta ')$. If they are not, they can be used as seeds to restrict the space of potential extremal solutions to explore to the solutions that lie below the free solutions.}
    \label{fig:freeSeeds}
\end{figure}
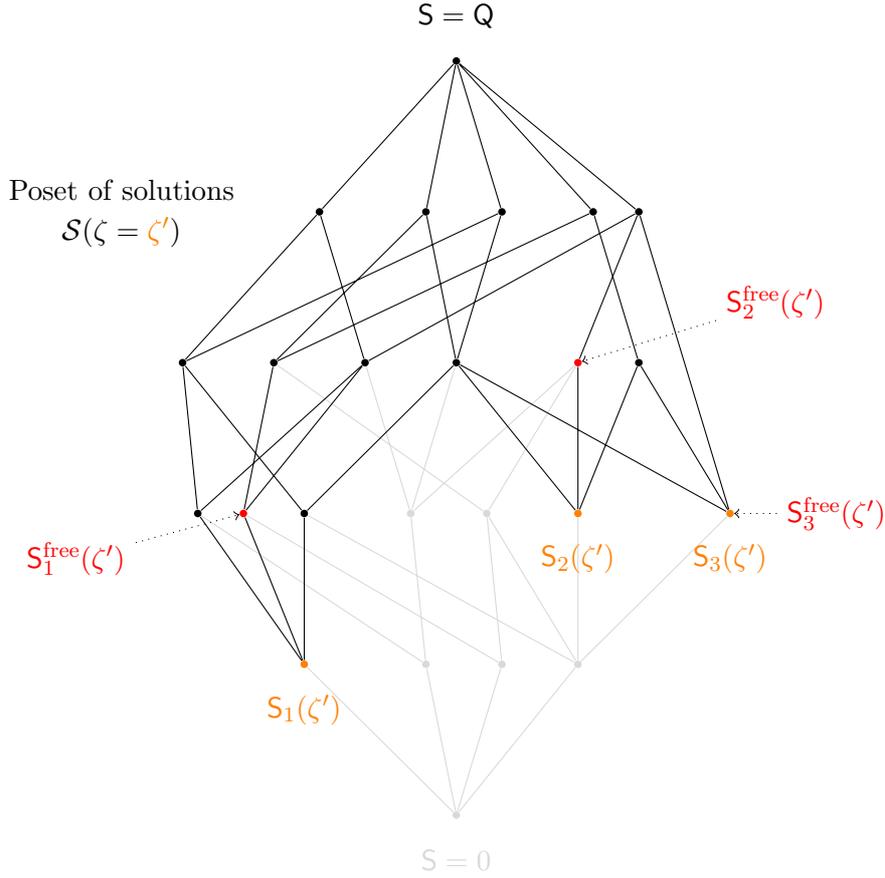

 \subsection{How to find extremal solutions?}

The previous subsection shows that finding extremal solutions for a given quiver $\mathsf{Q}$ and FI parameters $\zeta$ can be difficult in general. Here we list a few methods that have helped us identifying extremal solutions for specific examples. 
\begin{itemize}
    \item The first method is brute force computation, exploring all possible cases. We first use \hyperref[theorem1]{Theorem 1} in Appendix \ref{app:theorems} to restrict the possible Jordan forms of various meson matrices, and solve the graph constraints using generic basis transformations on the intersecting nodes. This boils down to studying an algebraic variety embedded in affine space of dimension $\sim \sum_v k_v^2$. This quickly becomes intractable as the ranks increase. 
    \item A handy set of extremal solutions is provided by the Klein singularities $\mathbb{C}^2 / \Gamma_{ADE}$. Indeed, there is a well known resolution of the singularities involving a set of $r$ compact $\mathbb{P}^1$ divisors intersecting transversely \cite{slodowy1980four}, where $r$ is the rank of the ADE algebra. A minimal FI deformation corresponds to a partial resolution where one of these compact divisors is blown up, while the $r-1$ other compact divisors can be kept at zero size. This guarantees that there is exactly one minimal FI deformation for each simple root $\alpha_i$ of the algebra, i.e. each node in the finite ADE Dynkin diagram. In our language, an FI term is turned on at that node, and at the affine node, in order to guarantee the vanishing \eqref{eq:constraintsFI}. Then we have exactly one extremal solution to the FI-Meson problem for each connected component of the finite Dynkin diagram with node $\alpha_i$ removed. It turns out they correspond to \emph{free quivers}. All these cases are shown explicitly in Appendix \ref{app:E}. 
    \item Building up on the previous point, we have observed that given a quiver $\mathsf{Q}$ with prescribed FI deformation $\zeta$, one can usually easily find solutions $\mathsf{S}_i^{\textrm{free}}$ which are \emph{free quivers} (meaning that their 3d Coulomb branch is spanned by free hypermultiplets). In the case of the Klein singularities, these are extremal. \emph{This is not the case in general}, as many examples in Sections \ref{sec:SQCD} and \ref{sec:Other} demonstrate. However, it appears heuristically useful to use the free solutions $\mathsf{S}_i^{\textrm{free}}$ as starting points in the poset of solutions $\mathcal{S}(\zeta)$, and to find extremal solutions $\mathsf{S}_i^{\textrm{extremal}} \preceq \mathsf{S}_i^{\textrm{free}}$, see Figure \ref{fig:freeSeeds}. 
    \item Given an extremal solution $\mathsf{S}$ for a quiver $\mathsf{Q}$, if we replace in $\mathsf{Q}$ a node $\mathrm{U}(k)$ by a chain of nodes $\mathrm{U}(k)$, thus forming a new quiver $\tilde{\mathsf{Q}}$, then an extremal solution $\tilde{\mathsf{S}}$ is obtained by performing a similar replacement of a node by a chain in $\mathsf{S}$. 
\end{itemize}

\clearpage

\section{SQCD} 
\label{sec:SQCD}

In this section, we consider $\mathrm{U}(k)$ and $\mathrm{SU}(k)$ SQCD with $N_f \geq 4$ flavors. The relevant mirror pairs are well-known \cite{Hanany:1996ie, Gaiotto:2008ak}: 
\begin{equation}
   \raisebox{-.5\height}{ \begin{tikzpicture}
\node at (0,0) {\begin{tikzpicture}
            \node[gauge,label=below:{\tiny $\mathrm{U}(k)$}] (0) at (0,0) {};
            \node[flavor,label=above:{\tiny $N_f$}] (1) at (0,1) {};
            \draw (0)--(1);
\end{tikzpicture}};
\node at (7,0) {\begin{tikzpicture}
            \node[gauge,label=below:{\tiny $1$}] (0) at (0,-4) {};
            \node[] (1) at (1,-4) {$\dots$};
            \node[gauge,label=below:{\tiny $k$}] (2) at (2,-4) {};
            \node[] (4) at (4,-4) {$\dots$};
            \node[gauge,label=below:{\tiny $k$}] (3) at (3,-4) {};
            \node[gauge,label=below:{\tiny $k$}] (5) at (5,-4) {};
            \node[gauge,label=below:{\tiny $k$}] (6) at (6,-4) {};
            \node[] (7) at (7,-4) {$\dots$};
            \node[gauge,label=above:{\tiny $1$}] (8) at (4,-3) {};
            \node[gauge,label=below:{\tiny $1$}] (9) at (8,-4) {};
            \draw (0)--(1)--(2)--(3)--(4)--(5)--(6)--(7)--(9) (2)--(8)--(6);
            \draw[snake=brace]  (8.1,-4.7) -- (-.1,-4.7);
   \node[] at (4,-5.1) {\scriptsize $N_f-1$};
\end{tikzpicture}};
\node at (1.5,0) {$\leftrightarrow$};
\end{tikzpicture}}
\end{equation}

\begin{equation}
    \raisebox{-.5\height}{\begin{tikzpicture}
\node at (0,0) {\begin{tikzpicture}
            \node[gauge,label=below:{\tiny $\mathrm{SU}(k)$}] (0) at (0,0) {};
            \node[flavor,label=above:{\tiny $N_f$}] (1) at (0,1) {};
            \draw (0)--(1);
\end{tikzpicture}};
\node at (7,0) {\begin{tikzpicture}
            \node[gauge,label=below:{\tiny $1$}] (0) at (0,-4) {};
            \node[] (1) at (1,-4) {$\dots$};
            \node[gauge,label=below:{\tiny $k$}] (2) at (2,-4) {};
            \node[] (4) at (4,-4) {$\dots$};
            \node[gauge,label=below:{\tiny $k$}] (3) at (3,-4) {};
            \node[gauge,label=below:{\tiny $k$}] (5) at (5,-4) {};
            \node[gauge,label=below:{\tiny $k$}] (6) at (6,-4) {};
            \node[] (7) at (7,-4) {$\dots$};
            \node[gauge,label=above:{\tiny $1$}] (8) at (2,-3) {};
            \node[gauge,label=above:{\tiny $1$}] (8b) at (6,-3) {};
            \node[gauge,label=below:{\tiny $1$}] (9) at (8,-4) {};
            \draw (0)--(1)--(2)--(3)--(4)--(5)--(6)--(7)--(9) (2)--(8) (8b)--(6);
            \draw[snake=brace]  (8.1,-4.7) -- (-.1,-4.7);
   \node[] at (4,-5.1) {\scriptsize $N_f-1$};
\end{tikzpicture}};
\node at (1.5,0) {$\leftrightarrow$};
\end{tikzpicture}}
\label{eq:SQCD-SU}
\end{equation}

This is a good testing ground for our methods as the mass deformations of the SQCD theories are easily understood, and we furthermore have a straightforward brane system to engineer the theories. On the other side, the FI-flows on the magnetic quivers shown above display a vast array of phenomena. Rather than attempting a cumbersome fully general discussion in terms of $N_f$ and $k$, we focus on specific examples and put emphasis on interesting phenomena: subtraction of bad and ugly quivers, rebalancing with multiple nodes, and Weyl group symmetry.

\subsection{Unitary gauge group} 
\label{sec:SQCDUnitary}

\afterpage{
\begin{landscape}
\begin{figure}
    \centering
\scalebox{.9}{
\begin{tabular}{|c|c|c|c|}
 \hline 
Frame $F$ & Brane system & Solution $\mathsf{S}_F$ & Quiver  $\mathsf{Q}_F = \mathsf{Q} - \mathsf{S}_F$ \\ \hline 
I & \begin{tikzpicture}
 \draw[red] (2.5,-2)--(2.5,0) (3,-1)--(3,1) (4.5,-1)--(4.5,1) (5,-1)--(5,1) (5.5,-2)--(5.5,0) (6,-2)--(6,0) (6.5,-1)--(6.5,1) (7,-1)--(7,1) (8.5,-1)--(8.5,1) (9,-2)--(9,0);
   \draw[blue]  (3.3,-1.5)--(4.3,0.5) (7.2,-1.5)--(8.2,0.5);
   \draw (4.5,0.6)--(7,0.6) (3,0.75)--(8.5,0.75) (4,0)--(4.5,0) (7.9,0)--(7,0) (2.5,-1.2)--(3.4,-1.2) (7.4,-1.2)--(9,-1.2); 
\end{tikzpicture}     &  \begin{tikzpicture}
            \node[gauge,fill=orange,label=below:{\tiny $1$},label=above:{\textcolor{orange}{$\lambda$}}] (1) at (1,-4) {};
            \node[] at (2,-3) {\tiny $\left(\begin{array}{cc}1 &0\\ 0 & 0\end{array}\right)$};
            \node[gauge,label=below:{\tiny $2$}] (2) at (2,-4) {};
            \node[gauge,label=below:{\tiny $2$}] (3) at (3,-4) {};
            \node[gauge,fill=orange,label=below:{\tiny $2$},label=above:{\textcolor{orange}{$-\lambda$}}] (4) at (4,-4) {};
            \node[gauge,label=below:{\tiny $2$}] (5) at (5,-4) {};
            \node[] at (5,-5) {\tiny $\left(\begin{array}{cc}1 &0\\ 0 & 1\end{array}\right)$};
            \node[gauge,fill=orange,label=below:{\tiny $2$},label=above:{\textcolor{orange}{$\lambda$}}] (6) at (6,-4) {};
            \node[gauge,label=below:{\tiny $2$}] (7) at (7,-4) {};
            \node[gauge,label=below:{\tiny $2$}] (8) at (8,-4) {};
            \node[] at (8,-3) {\tiny $\left(\begin{array}{cc}1 &0\\ 0 & 0\end{array}\right)$};
            \node[gauge,fill=orange,label=below:{\tiny $1$},label=above:{\textcolor{orange}{$-\lambda$}}] (9) at (9,-4) {};
            \node[gauge,label=above:{\tiny $1$}] (10) at (5,-3) {};
            \draw (1)--(2)--(3)--(4)--(5)--(6)--(7)--(8)--(9) (2)--(10)--(8);
            \draw[->] (2,-3.3)--(1.8,-3.8);
            \draw[->] (2.2,-3.3)--(2.3,-3.7);
            \draw[->] (8,-3.3)--(8.2,-3.8);
            \draw[->] (7.8,-3.3)--(7.7,-3.7);
            \draw[->] (4.8,-4.7)--(4.7,-4.2);
            \draw[->] (5.2,-4.7)--(5.3,-4.2);
\end{tikzpicture} & 
\begin{tikzpicture}
        \node[gauge,label=above:{$1$}] (1) at (0,0) {};
        \node[gauge,label=above:{$2$}] (2) at (1,0) {};
        \node[gauge,label=above:{$2$}] (3) at (2,0) {};
        \node[gauge,label=above:{$2$}] (4) at (3,0) {};
        \node[gauge,label=above:{$1$}] (5) at (4,0) {};
        \node[gauge,label=below:{$1$}] (6) at (2,-1) {};
        \draw (1)--(2)--(3)--(4)--(5) (2)--(6)--(4);
    \end{tikzpicture}      \\ \hline 
II & \begin{tikzpicture}
 \draw[red] (2.5,-2)--(2.5,0) (3,-1)--(3,1) (4.5,-1)--(4.5,1) (5,-1)--(5,1) (5.5,-2)--(5.5,0) (6,-2)--(6,0) (6.5,-1)--(6.5,1) (7,-1)--(7,1) (8.5,-1)--(8.5,1) (9,-2)--(9,0);
   \draw[blue]  (3.3,-1.5)--(4.3,0.5) (7.2,-1.5)--(8.2,0.5);
   \draw (3,0.75)--(8.5,0.75) (2.5,-1.7)--(9,-1.7); 
\end{tikzpicture}    & \begin{tikzpicture}
            \node[gauge,fill=orange,label=below:{\tiny $1$},label=above:{\textcolor{orange}{$\lambda$}}] (1) at (1,-4) {};
            \node[] at (2,-5) {\tiny $\left(\begin{array}{cc}1 &0\\ 0 & 0\end{array}\right)$};
            \node[gauge,label=below:{\tiny $2$}] (2) at (2,-4) {};
            \node[gauge,label=below:{\tiny $2$}] (3) at (3,-4) {};
            \node[gauge,fill=orange,label=below:{\tiny $2$},label=above:{\textcolor{orange}{$-\lambda$}}] (4) at (4,-4) {};
            \node[gauge,label=below:{\tiny $2$}] (5) at (5,-4) {};
            \node[] at (5,-5) {\tiny $\left(\begin{array}{cc}0 &0\\ 0 & 1\end{array}\right)$};
            \node[gauge,fill=orange,label=below:{\tiny $2$},label=above:{\textcolor{orange}{$\lambda$}}] (6) at (6,-4) {};
            \node[gauge,label=below:{\tiny $2$}] (7) at (7,-4) {};
            \node[gauge,label=below:{\tiny $2$}] (8) at (8,-4) {};
            \node[] at (8,-5) {\tiny $\left(\begin{array}{cc}-1 &0\\ 0 & 0\end{array}\right)$};
            \node[gauge,fill=orange,label=below:{\tiny $1$},label=above:\textcolor{orange}{{$-\lambda$}}] (9) at (9,-4) {};
            \node[gauge,label=above:{\tiny $1$}] (10) at (5,-3) {};
            \draw (1)--(2)--(3)--(4)--(5)--(6)--(7)--(8)--(9) (2)--(10)--(8);
            \draw[->] (2.2,-4.7)--(2.3,-4.2);
            \draw[->] (1.8,-4.7)--(1.7,-4.2);
            \draw[->] (8.2,-4.7)--(8.3,-4.2);
            \draw[->] (7.8,-4.7)--(7.7,-4.2);
            \draw[->] (4.8,-4.7)--(4.7,-4.2);
            \draw[->] (5.2,-4.7)--(5.3,-4.2);
\end{tikzpicture} &  \begin{tikzpicture}
        \node[gauge,label=above:{$1$}] (1) at (0,0) {};
        \node[gauge,label=above:{$1$}] (2) at (1,0) {};
        \node[gauge,label=above:{$1$}] (3) at (2,0) {};
        \node[gauge,label=above:{$1$}] (4) at (3,0) {};
        \node[gauge,label=above:{$1$}] (5) at (4,0) {};
        \node[gauge,label=below:{$1$}] (7) at (1,-2) {};
        \node[gauge,label=below:{$1$}] (8) at (2,-2) {};
        \node[gauge,label=below:{$1$}] (9) at (3,-2) {};
        \node[gauge,label=below:{$1$}] (6) at (2,-1) {};
        \draw (1)--(2)--(3)--(4)--(5) (1)--(6)--(5) (6)--(7)--(8)--(9)--(6);
    \end{tikzpicture}   \\ \hline
III & \begin{tikzpicture}
 \draw[red] (2.5,-2)--(2.5,0) (3,-1)--(3,1) (4.5,-1)--(4.5,1) (5,-1)--(5,1) (5.5,-2)--(5.5,0) (6,-2)--(6,0) (6.5,-1)--(6.5,1) (7,-1)--(7,1) (8.5,-1)--(8.5,1) (9,-2)--(9,0);
   \draw[blue]  (3.3,-1.5)--(4.3,0.5) (7.2,-1.5)--(8.2,0.5);
   \draw (5.5,-1.6)--(6,-1.6) (2.5,-1.75)--(9,-1.75) (4,0)--(3,0) (7.9,0)--(8.5,0) (5.5,-1.2)--(3.4,-1.2) (7.4,-1.2)--(6,-1.2); 
\end{tikzpicture}     & \begin{tikzpicture}
            \node[gauge,fill=orange,label=below:{\tiny $1$},label=above:{\textcolor{orange}{$\lambda$}}] (1) at (1,-4) {};
            \node[] at (1.5,-5) {\tiny $\left(\begin{array}{cc}1 &0\\ 0 & 0\end{array}\right)$};
            \node[] at (2.5,-5) {\tiny $\left(\begin{array}{cc}1 &0\\ 0 & 1\end{array}\right)$};
            \node[] at (2,-3) {\tiny $\left(\begin{array}{cc}0 &0\\ 0 & -1\end{array}\right)$};
            \node[gauge,label=below:{\tiny $2$}] (2) at (2,-4) {};
            \node[gauge,label=below:{\tiny $2$}] (3) at (3,-4) {};
            \node[gauge,fill=orange,label=below:{\tiny $2$},label=above:{\textcolor{orange}{$-\lambda$}}] (4) at (4,-4) {};
            \node[gauge,label=below:{\tiny $2$}] (5) at (5,-4) {};
            \node[] at (5,-5) {\tiny $\left(\begin{array}{cc}0 &0\\ 0 & 0\end{array}\right)$};
            \node[gauge,fill=orange,label=below:{\tiny $2$},label=above:{\textcolor{orange}{$\lambda$}}] (6) at (6,-4) {};
            \node[gauge,label=below:{\tiny $2$}] (7) at (7,-4) {};
            \node[gauge,label=below:{\tiny $2$}] (8) at (8,-4) {};
            \node[] at (7.4,-5) {\tiny $\left(\begin{array}{cc}-1 &0\\ 0 & -1\end{array}\right)$};
             \node[] at (8.6,-5) {\tiny $\left(\begin{array}{cc}-1 &0\\ 0 & 0\end{array}\right)$};
             \node[] at (8,-3) {\tiny $\left(\begin{array}{cc}0 &0\\ 0 & -1\end{array}\right)$};
            \node[gauge,fill=orange,label=below:{\tiny $1$},label=above:{\textcolor{orange}{$-\lambda$}}] (9) at (9,-4) {};
            \node[gauge,label=above:{\tiny $1$}] (10) at (5,-3) {};
            \draw (1)--(2)--(3)--(4)--(5)--(6)--(7)--(8)--(9) (2)--(10)--(8);
            \draw[->] (2.5,-4.7)--(2.5,-4.2);
            \draw[->] (1.5,-4.7)--(1.5,-4.2);
            \draw[->] (2.2,-3.3)--(2.3,-3.7);
            \draw[->] (7.8,-3.3)--(7.7,-3.7);
            \draw[->] (8.6,-4.7)--(8.6,-4.2);
            \draw[->] (7.4,-4.7)--(7.4,-4.2);
            \draw[->] (4.8,-4.7)--(4.7,-4.2);
            \draw[->] (5.2,-4.7)--(5.3,-4.2);
\end{tikzpicture} & \begin{tikzpicture}
        \node[gauge,label=below:{$1$}] (1) at (0,0) {};
        \node[gauge,label=below:{$2$}] (2) at (1,0) {};
        \node[gauge,label=below:{$1$}] (3) at (2,0) {};
        \node[gauge,label=right:{$1$}] (4u) at (1,1) {};
        \draw (1)--(2)--(3);
        \draw[transform canvas={xshift=-1pt}] (2)--(4u);
        \draw[transform canvas={xshift=1pt}] (2)--(4u);
    \end{tikzpicture}  \\ \hline 
\end{tabular}}   
 \caption{This table summarizes the FI flows for the deformation \eqref{u2vev}. First we use brane systems and identify three phases, that we call frames $F \in \{ I , II , III\}$. Correspondingly, there are three extremal solutions $\mathsf{S}_F$, and three resulting quivers $\mathsf{Q}_F$. }
    \label{fig:threeFramesSQCDU}
\end{figure}
\end{landscape}
}

Let us focus on $\mathrm{U}(2)$ SQCD with $N_f = 10$ flavors and study its mass deformations. The brane system describing the magnetic theory is 
\begin{equation}
\raisebox{-.5\height}{\begin{tikzpicture}
 \draw[red] (2.5,-1)--(2.5,1) (3,-1)--(3,1) (4.5,-1)--(4.5,1) (5,-1)--(5,1) (5.5,-1)--(5.5,1) (6,-1)--(6,1) (6.5,-1)--(6.5,1) (7,-1)--(7,1) (8.5,-1)--(8.5,1) (9,-1)--(9,1);
\draw[blue]  (3.3,-0.1)--(4.3,0.9) (7.2,-0.1)--(8.2,0.9);
\draw (2.5,-0.5)--(9,-0.5) (3,-0.35)--(8.5,-0.35); 
\end{tikzpicture}    }
\end{equation} 
Now we turn on a mass deformation in the electric theory, which in the brane system is implemented by moving a subset of NS5 branes. Let us consider for example moving the first, 5th, 6th and 10th branes in the above figure. Clearly this requires rearranging the D3 branes in order to preserve supersymmetry. We have three possibilities and for each one of them we provide the corresponding brane system in Figure \ref{fig:threeFramesSQCDU}.

The mass deformation is equivalent to the following FI deformation in the magnetic quiver $\mathsf{Q}$: 
\begin{equation}\label{u2vev}
\raisebox{-.5\height}{\begin{tikzpicture}
            \node[gauge,fill=orange,label=below:{\tiny $1$},label=above:{\textcolor{orange}{$\lambda$}}] (1) at (1,-4) {};
            \node[gauge,label=below:{\tiny $2$}] (2) at (2,-4) {};
            \node[gauge,label=below:{\tiny $2$}] (3) at (3,-4) {};
            \node[gauge,fill=orange,label=below:{\tiny $2$},label=above:{\textcolor{orange}{$-\lambda$}}] (4) at (4,-4) {};
            \node[gauge,label=below:{\tiny $2$}] (5) at (5,-4) {};
            \node[gauge,fill=orange,label=below:{\tiny $2$},label=above:{\textcolor{orange}{$\lambda$}}] (6) at (6,-4) {};
            \node[gauge,label=below:{\tiny $2$}] (7) at (7,-4) {};
            \node[gauge,label=below:{\tiny $2$}] (8) at (8,-4) {};
            \node[gauge,fill=orange,label=below:{\tiny $1$},label=above:{\textcolor{orange}{$-\lambda$}}] (9) at (9,-4) {};
            \node[gauge,label=above:{\tiny $1$}] (10) at (5,-3) {};
            \draw (1)--(2)--(3)--(4)--(5)--(6)--(7)--(8)--(9) (2)--(10)--(8);
\end{tikzpicture}}
\end{equation} 
In this context the three frames arise because there are three extremal solutions to the FI-Meson problem. The first frame is realized by propagating the vev from the leftmost node in \eqref{u2vev} to the rightmost going through the abelian node on top. 
The second frame is obtained by propagating the vev along the $A_9$ quiver, without involving the abelian node on top. 
The third frame instead arises by leaving the central $\mathrm{U}(2)$ node unbroken. 
These extremal solutions $\mathsf{S}_F$, for each frame $F = I, II, III$, are drawn in Figure \ref{fig:threeFramesSQCDU}, along with the resulting quivers $\mathsf{Q}_F = \mathsf{Q} - \mathsf{S}_F$.

\afterpage{
\begin{landscape}
\begin{figure}
    \centering
\begin{tabular}{|c|c|c|} \hline 
Frame &  Weight $[1,0,0,-1,0,1,0,0,-1]$  & Weight $[0,0,0,1,0,0,0,0,0]$  \\  \hline  
I & 
\begin{tikzpicture}
\node at (1,-2.5) {};
            \node[gaugeo,label=below:{\tiny $1$}] (1) at (1,-4) {};
            \node[gauge,label=below:{\tiny $1$}] (2) at (2,-4) {};
            \node[gaugeo,label=below:{\tiny $2$}] (4) at (4,-4) {};
            \node[gauge,label=below:{\tiny $2$}] (5) at (5,-4) {};
            \node[gaugeo,label=below:{\tiny $2$}] (6) at (6,-4) {};
            \node[gauge,label=below:{\tiny $1$}] (8) at (8,-4) {};
            \node[gaugeo,label=below:{\tiny $1$}] (9) at (9,-4) {};
            \node[gauge,label=below:{\tiny $1$}] (10) at (5,-3) {};
            \draw (1)--(2) (4)--(5)--(6) (8)--(9) (2)--(10)--(8);
\end{tikzpicture} &

\begin{tikzpicture}
            \node[gauge,label=below:{\tiny $1$}] (1) at (1,-4) {};
            \node[gauge,label=below:{\tiny $2$}] (2) at (2,-4) {};
            \node[gauge,label=below:{\tiny $2$}] (3) at (3,-4) {};
            \node[gauge,fill=orange,label=below:{\tiny $2$}] (4) at (4,-4) {};
            \node[gauge,label=below:{\tiny $1$}] (5) at (5,-4) {};
            \node[gauge,fill=orange,label=below:{\tiny $1$}] (10) at (5,-3) {};
            \draw (1)--(2)--(3)--(4)--(5) (2)--(10);
\end{tikzpicture} \\ \hline  
II & 
\begin{tikzpicture}
            \node[gaugeo,label=below:{\tiny $1$}] (1) at (1,-4) {};
            \node[gauge,label=below:{\tiny $1$}] (2) at (2,-4) {};
            \node[gauge,label=below:{\tiny $1$}] (3) at (3,-4) {};
            \node[gaugeo,label=below:{\tiny $1$}] (4) at (4,-4) {};
            \node[gaugeo,label=below:{\tiny $1$}] (6) at (6,-4) {};
            \node[gaugeo,label=below:{\tiny $1$}] (4b) at (4,-3) {};
            \node[gauge,label=below:{\tiny $1$}] (5b) at (5,-3) {};
            \node[gaugeo,label=below:{\tiny $1$}] (6b) at (6,-3) {};
            \node[gauge,label=below:{\tiny $1$}] (7) at (7,-4) {};
            \node[gauge,label=below:{\tiny $1$}] (8) at (8,-4) {};
            \node[gaugeo,label=below:{\tiny $1$}] (9) at (9,-4) {};
            \draw (1)--(2)--(3)--(4) (6)--(7)--(8)--(9) (4b)--(5b)--(6b);
\end{tikzpicture}
  & \begin{tikzpicture}
\node at (1,-2.5) {};
            \node[gauge,label=below:{\tiny $1$}] (2) at (2,-4) {};
            \node[gauge,label=below:{\tiny $1$}] (3) at (3,-4) {};
            \node[gauge,fill=orange,label=below:{\tiny $1$}] (4) at (4,-3.7) {};
            \node[gauge,fill=orange,label=below:{\tiny $1$}] (4b) at (4,-4.5) {};
            \node[gauge,label=below:{\tiny $1$}] (5) at (5,-4) {};
            \node[gauge,label=below:{\tiny $1$}] (6) at (6,-4) {};
            \node[gauge,label=below:{\tiny $1$}] (7) at (7,-4) {};
            \node[gauge,label=below:{\tiny $1$}] (8) at (8,-4) {};
            \node[gauge,fill=orange,label=below:{\tiny $1$}] (10) at (5,-3) {};
            \draw (2)--(3)--(4) (4b)--(5)--(6)--(7)--(8) (2)--(10)--(8);
\end{tikzpicture} \\ \hline 
III & \begin{tikzpicture}
\node at (1,-2.5) {};
            \node[gaugeo,label=below:{\tiny $1$}] (1b) at (1,-5) {};
            \node[gauge,label=below:{\tiny $1$}] (2b) at (2,-5) {};
            \node[gauge,label=below:{\tiny $1$}] (3b) at (3,-5) {};
            \node[gaugeo,label=below:{\tiny $1$}] (4b) at (4,-5) {};
            \node[gauge,label=below:{\tiny $1$}] (2) at (2,-4) {};
            \node[gauge,label=below:{\tiny $1$}] (3) at (3,-4) {};
            \node[gaugeo,label=below:{\tiny $1$}] (4) at (4,-4) {};
            \node[gaugeo,label=below:{\tiny $1$}] (6) at (6,-4) {};
            \node[gauge,label=below:{\tiny $1$}] (7) at (7,-4) {};
            \node[gauge,label=below:{\tiny $1$}] (8) at (8,-4) {};
            \node[gaugeo,label=below:{\tiny $1$}] (6b) at (6,-5) {};
            \node[gauge,label=below:{\tiny $1$}] (7b) at (7,-5) {};
            \node[gauge,label=below:{\tiny $1$}] (8b) at (8,-5) {};
            \node[gaugeo,label=below:{\tiny $1$}] (9b) at (9,-5) {};
            \node[gauge,label=below:{\tiny $1$}] (10) at (5,-3) {};
            \draw (1b)--(2b)--(3b)--(4b) (2)--(3)--(4) (6)--(7)--(8) (6b)--(7b)--(8b)--(9b) (2)--(10)--(8);
\end{tikzpicture} & 
\begin{tikzpicture}
\node at (6,-5.0) {};
            \node[gauge,label=below:{\tiny $1$}] (3) at (3,-4) {};
            \node[gauge,fill=orange,label=below:{\tiny $2$}] (4) at (4,-4) {};
            \node[gauge,label=below:{\tiny $2$}] (5) at (5,-4) {};
            \node[gauge,label=below:{\tiny $2$}] (6) at (6,-4) {};
            \node[gauge,label=below:{\tiny $2$}] (7) at (7,-4) {};
            \node[gauge,label=below:{\tiny $2$}] (8) at (8,-4) {};
            \node[gauge,label=below:{\tiny $1$}] (9) at (9,-4) {};
            \node[gauge,fill=orange,label=below:{\tiny $1$}] (10) at (5,-3) {};
            \draw (3)--(4)--(5)--(6)--(7)--(8)--(9) (10)--(8);
\end{tikzpicture}  \\  \hline  
\end{tabular}
    \caption{Quivers to be subtracted in each frame, for two different weights in the same Weyl orbit. For the weight $[1,0,0,-1,0,1,0,0,-1]$, there are three quivers to subtract (counting multiplicities), so the rebalancing nodes form a rank 3 group, while for the other weight, it forms a rank 1 group. Therefore the sum of ranks of quivers in the $[1,0,0,-1,0,1,0,0,-1]$ column is two more than the corresponding quiver in the $[0,0,0,1,0,0,0,0,0]$  column. }
    \label{fig:quiversSubtractedTwoWeights}
\end{figure}
\end{landscape}
}

Interestingly, it can be checked that the same results can be obtained by simplifying the FI deformation taking into account the action of the Weyl group of the global symmetry, as explained in Appendix \ref{Weylsec}. Here the global symmetry is $A_{N_f - 1} = A_9$, which is generated in the magnetic quiver by the 9 balanced nodes in the chain. The FI deformation is labeled by a weight of $A_9$, that we write in the basis of fundamental weights $\varpi = [1,0,0,-1,0,0,1,0,0,-1]$. We observe that $\varpi$ is in the Weyl orbit of the fundamental weight $\varpi_4 := [0,0,0,1,0,0,0,0,0]$, and therefore we can equivalently consider the deformation 
\begin{equation} 
\raisebox{-.5\height}{\begin{tikzpicture}
            \node[gauge,label=below:{\tiny $1$}] (1) at (1,-4) {};
            \node[gauge,label=below:{\tiny $2$}] (2) at (2,-4) {};
            \node[gauge,label=below:{\tiny $2$}] (3) at (3,-4) {};
            \node[gauge,fill=orange,label=below:{\tiny $2$}] (4) at (4,-4) {};
            \node[gauge,label=below:{\tiny $2$}] (5) at (5,-4) {};
            \node[gauge,label=below:{\tiny $2$}] (6) at (6,-4) {};
            \node[gauge,label=below:{\tiny $2$}] (7) at (7,-4) {};
            \node[gauge,label=below:{\tiny $2$}] (8) at (8,-4) {};
            \node[gauge,label=below:{\tiny $1$}] (9) at (9,-4) {};
            \node[gauge,fill=orange,label=below:{\tiny $1$}] (10) at (5,-3) {};
            \draw (1)--(2)--(3)--(4)--(5)--(6)--(7)--(8)--(9) (2)--(10)--(8);
\end{tikzpicture}}
\end{equation} 
The quivers that need to be subtracted are depicted in the rightmost column of Figure \ref{fig:quiversSubtractedTwoWeights}. For Frames I and III, the solutions to the F-term equations are clear (they are similar to those shown in Figure \ref{fig:solutionsDtype} for the D-type singularities). Frame II is more interesting, and an extremal solution to the FI-Meson problem is easily worked out to be 
\begin{equation}
    \raisebox{-.5\height}{\begin{tikzpicture}
            \node[gauge,label=below:{\tiny $1$}] (1) at (1,-4) {};
            \node[] at (3.5,-5) {\tiny $\left(\begin{array}{cc}-1 &0\\ 0 & 0\end{array}\right)$};
            \node[] at (4.5,-5) {\tiny $\left(\begin{array}{cc}0 &0\\ 0 & 1\end{array}\right)$};
            \node[] at (2,-3) {\tiny $\left(\begin{array}{cc}-1 &0\\ 0 & 0\end{array}\right)$};
            \node[] at (4,-2.5) {\tiny $\left( -1 \right)$};
            \node[] at (6,-2.5) {\tiny $\left( 1 \right)$};
            \node[gauge,label=below:{\tiny $2$}] (2) at (2,-4) {};
            \node[gauge,label=below:{\tiny $2$}] (3) at (3,-4) {};
            \node[gauge,fill=orange,label=below:{\tiny $2$}] (4) at (4,-4) {};
            \node[gauge,label=below:{\tiny $2$}] (5) at (5,-4) {};
            \node[gauge,label=below:{\tiny $2$}] (6) at (6,-4) {};
            \node[gauge,label=below:{\tiny $2$}] (7) at (7,-4) {};
            \node[gauge,label=below:{\tiny $2$}] (8) at (8,-4) {};
             \node[] at (8,-3) {\tiny $\left(\begin{array}{cc}0 &0\\ 0 & 1\end{array}\right)$};
            \node[gauge,label=below:{\tiny $1$}] (9) at (9,-4) {};
        \node[gauge,fill=orange,label=above:{\tiny $1$}] (10) at (5,-3) {};
            \draw (1)--(2)--(3)--(4)--(5)--(6)--(7)--(8)--(9) (2)--(10)--(8);
            \draw[->] (3.5,-4.7)--(3.5,-4.2);
            \draw[->] (4.5,-4.7)--(4.5,-4.2);
            \draw[->] (2.2,-3.3)--(2.3,-3.7);
            \draw[->] (4,-2.8)--(4.2,-3);
            \draw[->] (6,-2.8)--(5.8,-3);
            \draw[->] (7.8,-3.3)--(7.7,-3.7);
\end{tikzpicture}}
\label{eq:solutionBad}
\end{equation}
This solution instructs us to subtract a "bad" quiver: 
\begin{equation}
  \raisebox{-.5\height}{  \begin{tikzpicture}
            \node[gauge,label=below:{\tiny $1$}] (2) at (2,-4) {};
            \node[gauge,label=below:{\tiny $1$}] (3) at (3,-4) {};
            \node[gauge,fill=orange,label=below:{\tiny $2$}] (4) at (4,-4) {};
            \node[gauge,label=below:{\tiny $1$}] (5) at (5,-4) {};
            \node[gauge,label=below:{\tiny $1$}] (6) at (6,-4) {};
            \node[gauge,label=below:{\tiny $1$}] (7) at (7,-4) {};
            \node[gauge,label=below:{\tiny $1$}] (8) at (8,-4) {};
\node[gauge,fill=orange,label=above:{\tiny $1$}] (10) at (5,-3) {};
            \draw (2)--(3)--(4)--(5)--(6)--(7)--(8) (2)--(10)--(8);
\end{tikzpicture}}
\end{equation}
Note however that the matrices in \eqref{eq:solutionBad} provide a rationale why this is, and show that an alternative interpretation is as shown in Figure \ref{fig:quiversSubtractedTwoWeights} (Frame II, weight $[0,0,0,1,0,0,0,0,0]$).

In the more general case of $\mathrm{U}(N_c)$ SQCD we find for every minimal FI deformation $N_c+1$ frames, depending on how we distribute the $D3$ branes between the two stacks of five branes. Let us illustrate this at the level of the brane system for $\mathrm{U}(3)$ SQCD, where we take the number of flavors to be 12 for definiteness. The relevant brane diagram in Type IIB is 
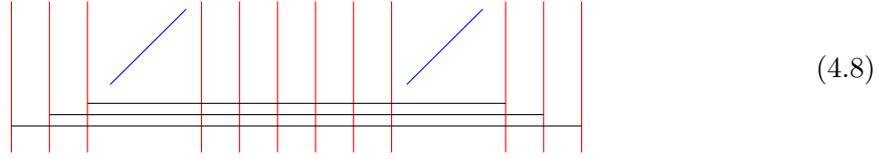
\begin{equation}
\raisebox{-.5\height}{\begin{tikzpicture}
 \draw[red] (2,-1)--(2,1) (2.5,-1)--(2.5,1) (3,-1)--(3,1) (4.5,-1)--(4.5,1) (5,-1)--(5,1) (5.5,-1)--(5.5,1) (6,-1)--(6,1) (6.5,-1)--(6.5,1) (7,-1)--(7,1) (8.5,-1)--(8.5,1) (9,-1)--(9,1) (9.5,-1)--(9.5,1);
   \draw[blue]  (3.3,-0.1)--(4.3,0.9) (7.2,-0.1)--(8.2,0.9);
   \draw (2,-0.65)--(9.5,-0.65) (2.5,-0.5)--(9,-0.5) (3,-0.35)--(8.5,-0.35); 
\end{tikzpicture}    }
\end{equation} 
If we now displace the $D5$ branes $1,2,6,7,11,12$ along the direction wrapped by the $NS5$ branes we find the following four frames:
\begin{itemize}
\item {\bf Frame I} 
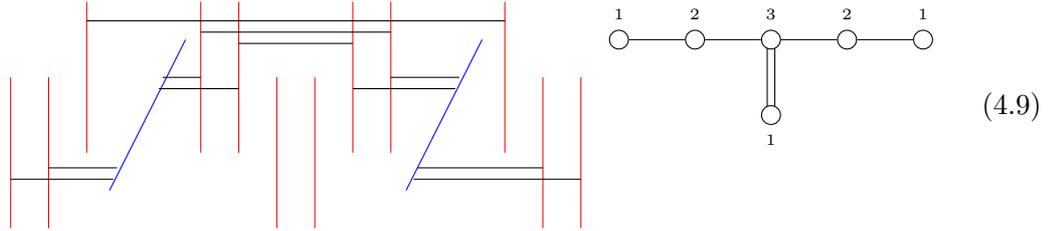
\begin{equation}
\raisebox{-.5\height}{
\begin{tikzpicture}
 \draw[red] (2,-2)--(2,0) (2.5,-2)--(2.5,0) (3,-1)--(3,1) (4.5,-1)--(4.5,1) (5,-1)--(5,1) (5.5,-2)--(5.5,0) (6,-2)--(6,0) (6.5,-1)--(6.5,1) (7,-1)--(7,1) (8.5,-1)--(8.5,1) (9,-2)--(9,0) (9.5,-2)--(9.5,0);
   \draw[blue]  (3.3,-1.5)--(4.3,0.5) (7.2,-1.5)--(8.2,0.5);
   \draw (5,0.45)--(6.5,0.45) (4.5,0.6)--(7,0.6) (3,0.75)--(8.5,0.75) (4,0)--(4.5,0) (7.9,0)--(7,0) (3.95,-0.15)--(5,-0.15) (7.85,-0.15)--(6.5,-0.15) (2.5,-1.2)--(3.4,-1.2) (7.35,-1.2)--(9,-1.2) (2,-1.35)--(3.35,-1.35) (7.3,-1.35)--(9.5,-1.35); 
          \draw (11.95,0.43)--(11.95,-0.43) (12.05,0.43)--(12.05,-0.43);
  \node[gauge,label=above:{\tiny $1$}] (1) at (10,0.5) {};
        \node[gauge,label=above:{\tiny $2$},fill=white] (2) at (11,0.5) {};
        \node[gauge,label=above:{\tiny $3$},fill=white] (3) at (12,0.5) {};
        \node[gauge,label=above:{\tiny $2$}] (4) at (13,0.5) {};
        \node[gauge,label=above:{\tiny $1$}] (5) at (14,0.5) {};
        \node[gauge,label=below:{\tiny $1$},fill=white] (6) at (12,-0.5) {};
          \draw (1)--(2)--(3)--(4)--(5);
\end{tikzpicture}    }
\end{equation} 
\item {\bf Frame II} 
\begin{equation}
\raisebox{-.5\height}
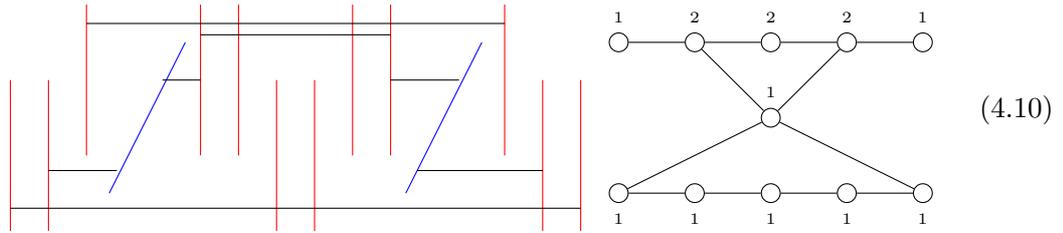
{\begin{tikzpicture}
 \draw[red] (2,-2)--(2,0) (2.5,-2)--(2.5,0) (3,-1)--(3,1) (4.5,-1)--(4.5,1) (5,-1)--(5,1) (5.5,-2)--(5.5,0) (6,-2)--(6,0) (6.5,-1)--(6.5,1) (7,-1)--(7,1) (8.5,-1)--(8.5,1) (9,-2)--(9,0) (9.5,-2)--(9.5,0);
   \draw[blue]  (3.3,-1.5)--(4.3,0.5) (7.2,-1.5)--(8.2,0.5);
   \draw (3,0.75)--(8.5,0.75) (4.5,0.6)--(7,0.6) (2,-1.7)--(9.5,-1.7) (4,0)--(4.5,0) (7.9,0)--(7,0) (2.5,-1.2)--(3.4,-1.2) (7.35,-1.2)--(9,-1.2); 
  \node[gauge,label=above:{\tiny $1$}] (1) at (10,0.5) {};
        \node[gauge,label=above:{\tiny $2$}] (2) at (11,0.5) {};
        \node[gauge,label=above:{\tiny $2$}] (3) at (12,0.5) {};
        \node[gauge,label=above:{\tiny $2$}] (4) at (13,0.5) {};
        \node[gauge,label=above:{\tiny $1$}] (5) at (14,0.5) {};
        \node[gauge,label=above:{\tiny $1$}] (6) at (12,-0.5) {};
        \node[gauge,label=below:{\tiny $1$}] (7) at (10,-1.5) {};
        \node[gauge,label=below:{\tiny $1$}] (8) at (11,-1.5) {};
        \node[gauge,label=below:{\tiny $1$}] (9) at (12,-1.5) {};
        \node[gauge,label=below:{\tiny $1$}] (10) at (13,-1.5) {};
        \node[gauge,label=below:{\tiny $1$}] (11) at (14,-1.5) {};
        \draw (1)--(2)--(3)--(4)--(5) (2)--(6)--(4) (6)--(7)--(8)--(9)--(10)--(11)--(6);
\end{tikzpicture} }   
\end{equation} 
\item {\bf Frame III}
\begin{equation}
\raisebox{-.5\height}
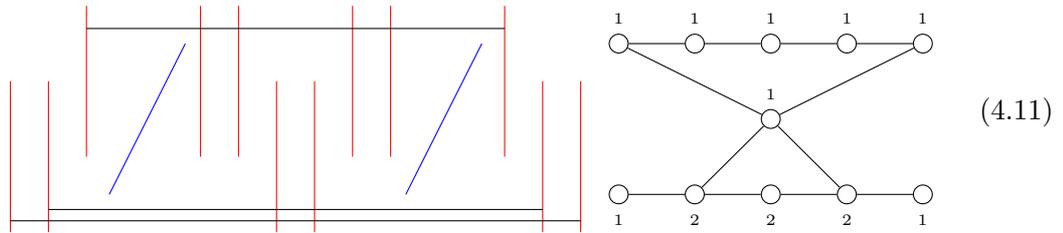
{\begin{tikzpicture}
 \draw[red] (2,-2)--(2,0) (2.5,-2)--(2.5,0) (3,-1)--(3,1) (4.5,-1)--(4.5,1) (5,-1)--(5,1) (5.5,-2)--(5.5,0) (6,-2)--(6,0) (6.5,-1)--(6.5,1) (7,-1)--(7,1) (8.5,-1)--(8.5,1) (9,-2)--(9,0) (9.5,-2)--(9.5,0);
   \draw[blue]  (3.3,-1.5)--(4.3,0.5) (7.2,-1.5)--(8.2,0.5);
   \draw (2,-1.85)--(9.5,-1.85) (2.5,-1.7)--(9,-1.7) (3,0.7)--(8.5,0.7);  
   \node[gauge,label=above:{\tiny $1$}] (1) at (10,0.5) {};
        \node[gauge,label=above:{\tiny $1$}] (2) at (11,0.5) {};
        \node[gauge,label=above:{\tiny $1$}] (3) at (12,0.5) {};
        \node[gauge,label=above:{\tiny $1$}] (4) at (13,0.5) {};
        \node[gauge,label=above:{\tiny $1$}] (5) at (14,0.5) {};
        \node[gauge,label=above:{\tiny $1$}] (6) at (12,-0.5) {};
        \node[gauge,label=below:{\tiny $1$}] (7) at (10,-1.5) {};
        \node[gauge,label=below:{\tiny $2$}] (8) at (11,-1.5) {};
        \node[gauge,label=below:{\tiny $2$}] (9) at (12,-1.5) {};
        \node[gauge,label=below:{\tiny $2$}] (10) at (13,-1.5) {};
        \node[gauge,label=below:{\tiny $1$}] (11) at (14,-1.5) {};
        \draw (6)--(1)--(2)--(3)--(4)--(5)--(6) (10)--(6)--(8) (7)--(8)--(9)--(10)--(11);     
\end{tikzpicture}    }
\end{equation} 
\item {\bf Frame IV} 
\begin{equation}
\raisebox{-.5\height}
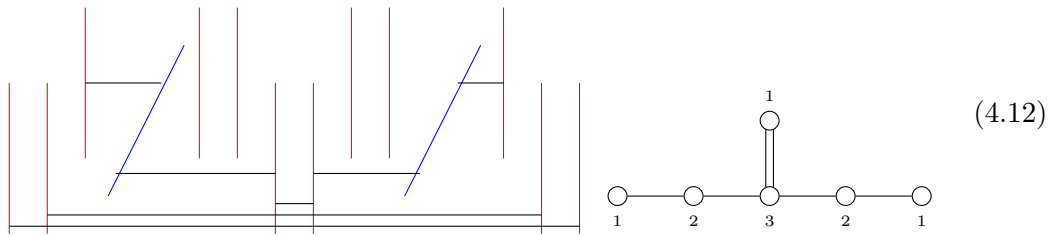
{
\begin{tikzpicture}
 \draw[red] (2,-2)--(2,0) (2.5,-2)--(2.5,0) (3,-1)--(3,1) (4.5,-1)--(4.5,1) (5,-1)--(5,1) (5.5,-2)--(5.5,0) (6,-2)--(6,0) (6.5,-1)--(6.5,1) (7,-1)--(7,1) (8.5,-1)--(8.5,1) (9,-2)--(9,0) (9.5,-2)--(9.5,0);
   \draw[blue]  (3.3,-1.5)--(4.3,0.5) (7.2,-1.5)--(8.2,0.5);
   \draw (5.5,-1.6)--(6,-1.6) (2.5,-1.75)--(9,-1.75) (2,-1.9)--(9.5,-1.9) (4,0)--(3,0) (7.9,0)--(8.5,0) (5.5,-1.2)--(3.4,-1.2) (7.4,-1.2)--(6,-1.2); 
           \draw (11.95,-1.43)--(11.95,-0.57) (12.05,-1.43)--(12.05,-0.57);
        \node[gauge,label=below:{\tiny $1$}] (1) at (10,-1.5) {};
        \node[gauge,label=below:{\tiny $2$}] (2) at (11,-1.5) {};
        \node[gauge,label=below:{\tiny $3$},fill=white] (3) at (12,-1.5) {};
        \node[gauge,label=below:{\tiny $2$}] (4) at (13,-1.5) {};
        \node[gauge,label=below:{\tiny $1$}] (5) at (14,-1.5) {};
        \node[gauge,label=above:{\tiny $1$},fill=white] (6) at (12,-0.5) {};
        \draw (1)--(2)--(3)--(4)--(5);
\end{tikzpicture}   }
\end{equation} 
\end{itemize}
One easily checks that these results can be recovered via quiver subtraction, as we have done for $\mathrm{U}(2)$ SQCD in table \ref{fig:quiversSubtractedTwoWeights}.

\subsection{Special Unitary Gauge Group}
\label{sec:SpecialUnitaryGaugeGroup}

\subsubsection*{A new mirror pair}

Let us now discuss the special unitary case \eqref{eq:SQCD-SU}.
One particularly interesting subset of mass deformations is described by the following FI deformation of the mirror dual:
\begin{equation}\label{massdefSU1}
\raisebox{-.5\height}{\begin{tikzpicture}
            \node[gauge,label=below:{$1$}] (7) at (9.5,0) {};
            \node (8) at (10.25,0) {$\cdots$};
            \node[gauge,label=below:{$k$}, fill=orange] (9) at (11,0) {};
            \node (10) at (11.75,0) {$\cdots$};
            \node[gauge,label=below:{$N+k$}] (11) at (12.5,0) {};
            \node (12) at (13.25,0) {$\cdots$};
            \node[gauge,label=below:{$N+k$}] (13) at (14,0) {};
            \node (14) at (14.75,0) {$\cdots$};
            \node[gauge,label=below:{$k$},fill=orange] (15) at (15.5,0) {};
            \node (16) at (16.25,0) {$\cdots$};
            \node[gauge,label=below:{$1$}] (17) at (17,0) {};
            \node[gauge,label=above:{$1$}] (11a) at (12.5,1) {};
            \node[gauge,label=above:{$1$}] (13a) at (14,1) {};
            \draw (7)--(8)--(9)--(10)--(11)--(12)--(13)--(14)--(15)--(16)--(17) (11)--(11a) (13)--(13a);
            \draw[snake=brace]  (17.1,-0.75) -- (9.4,-0.75);
   \node[] at (13.25,-1.1) {\scriptsize $N_f-1$};
   \draw[->](17.5,0) -- (18.5,0); 
            \node[gauge,label=below:{$1$}] (a) at (19,0) {};
            \node (b) at (19.75,0) {$\cdots$};
            \node[gauge,label=below:{$N$}] (c) at (20.5,0) {};
            \node (d) at (21.25,0) {$\cdots$};
            \node[gauge,label=below:{$N$}] (e) at (22,0) {};
            \node (f) at (22.75,0) {$\cdots$};
            \node[gauge,label=below:{$1$}] (g) at (23.5,0) {};
            \node[gauge,label=left:{$1$}] (cc) at (20.5,1) {};
            \node[gauge,label=right:{$1$}] (ee) at (22,1) {};
            \node[gauge,label=above:{$k$}] (h) at (21.25,2) {};
            \node[gauge,label=above:{$1$}] (a1) at (19.75,2) {};
            \node (a2) at (20.5,2) {$\cdots$};
            \node[gauge,label=above:{$1$}] (a4) at (22.75,2) {};
            \node (a3) at (22,2) {$\cdots$};
            \draw (a)--(b)--(c)--(d)--(e)--(f)--(g) (c)--(cc)--(h)--(ee)--(e) (a1)--(a2)--(h)--(a3)--(a4);
            \draw[snake=brace]  (23.6,-0.75) -- (18.9,-0.75);
   \node[] at (21.25,-1.1) {\scriptsize $N_f-2k-1$};
        \end{tikzpicture}}
        \end{equation} 
In this case, the FI-Meson problem is trivial, the unique extremal solution $\mathsf{S}$ is the direct sum of $k$ quivers $A_{N_f - 2k}$. The subtraction algorithm instructs us to subtract these quivers (or equivalently subtract a linear quiver of length $N_f +1 - 2k$ with $\mathrm{U}(k)$ nodes) and rebalance with a $\mathrm{U}(k)$ node. 
The result is the quiver on the right, which is just a special case of \eqref{mirrornew} with $B=2M=2k$ and can be interpreted as the mirror dual of the following theory (see \eqref{newmirror}): 
\begin{equation}\label{mirrorSU}
    \vcenter{\hbox{\scalebox{1}{\begin{tikzpicture}
        \node[gauge,label=below:{SU$(N)$}] (0) at (0,0) {};
        \node[gauge,label=below:{U$(1)$}] (1) at (3,0) {};
        \node[gauge,label=below:{SU$(k)$}] (2) at (6,0) {};
        \node (q) at (2.5,0) {$k$};
        \node (n) at (1.5,0.3) {$N_f-2k$};
        \node (pp) at (3.5,0) {$N$};
        \node (nn) at (4.5,0.3) {$2k$};
        \draw[double] (0)--(q)--(1)--(pp)--(2);
        \draw (6.2,-1)--(7.2,0);
        \node at (7.5,-0.5) {$\mathbb{Z}_{kN}$};
    \end{tikzpicture}}}}
\end{equation}
with gauge group SU$(N)\times$U$(1)\times$SU$(k)$, $N_f-2k$ hypermultiplets in the fundamental representation of SU$(N$) of charge $k$ under U$(1)$ and $2k$ hypermultiplets in the antifundamental representation of SU$(k)$ of charge $N$ under U$(1)$, and gauged 1-form $\mathbb{Z}_{kN}$ symmetry indicated by the quotient. Graphically, the number of hypermultiplets is indicated above the double lines, while the $\mathrm{U}(1)$ charges are inserted within the line. 
As supporting evidence for our claim, we observe that the moduli space dimensions and Hilbert series of the candidate dual theories match. We actually find that the matching of Hilbert series extends to the more general family \eqref{mirrornew} and \eqref{newmirror}. We defer a detailed discussion of the Hilbert series to Appendix \ref{app:Mirror}.

In the special case $k=1$ the theory \eqref{mirrorSU} is equivalent to a $\mathrm{U}(N)$ gauge theory with $N_f-2$ fundamentals and two hypermultiplets in the determinant representation. If we instead set $N=1$ \eqref{mirrorSU} reduces to a $\mathrm{U}(k)$ gauge theory with $2k$ fundamentals and $N_f-2k$ hypermultiplets in the determinant representation, whose mirror dual according to \eqref{massdefSU1} is 
\begin{equation} 
\raisebox{-.5\height}{\begin{tikzpicture}
            \node (f) at (0,1) {$\cdots$};
            \node[gauge,label=above:{$1$}] (cc) at (-1,1) {};
            \node[gauge,label=above:{$1$}] (ee) at (1,1) {};
            \node[gauge,label=below:{$k$}] (h) at (0,0) {};
            \node[gauge,label=below:{$1$}] (a1) at (-2,0) {};
            \node (a2) at (-1,0) {$\cdots$};
            \node[gauge,label=below:{$1$}] (a4) at (2,0) {};
            \node (a3) at (1,0) {$\cdots$};
            \draw (f)--(cc)--(h)--(ee)--(f) (a1)--(a2)--(h)--(a3)--(a4);
            \draw[snake=brace]  (-1.1,1.75) -- (1.1,1.75);
   \node[] at (0,2.1) {\scriptsize $N_f-2k+1$};
\end{tikzpicture}}
\end{equation} 
Based on the above results it is natural to conjecture that the mirror dual of $\mathrm{U}(N)$ SQCD with $N_f\geq 2N$ flavors and $n$ hypermultiplets in the determinant representation is given by the quiver 
\begin{equation} 
 \raisebox{-.5\height}{\begin{tikzpicture}
\node at (1,0) { \begin{tikzpicture}
        \node[gauge,label=below:{$\textrm{U}(N)$}] (3) at (2,0) {};
        \node[flavor,label=left:{$N_f$}] (3u) at (1.5,1) {};
        \node[flavor,label=right:{$n$}] (4u) at (2.5,1) {};
        \draw (3)--(3u);
        \draw[dotted] (3)--(4u);
    \end{tikzpicture} };
\node at (8,0) { \begin{tikzpicture}
            \node (f) at (0,1) {$\cdots$};
            \node[gauge,label=above:{$1$}] (cc) at (-1,1) {};
            \node[gauge,label=above:{$1$}] (ee) at (1,1) {};
            \node[gauge,label=below:{$N$}] (h) at (-1,0) {}; 
            \node (l) at (0,0) {$\cdots$};
            \node[gauge,label=below:{$N$}] (k) at (1,0) {};
            \node[gauge,label=below:{$1$}] (a1) at (-3,0) {};
            \node (a2) at (-2,0) {$\cdots$};
            \node[gauge,label=below:{$1$}] (a4) at (3,0) {};
            \node (a3) at (2,0) {$\cdots$};
            \draw (h)--(cc)--(f)--(ee)--(k) (a1)--(a2)--(h)--(l)--(k)--(a3)--(a4);
            \draw[snake=brace]  (-1.1,1.75) -- (1.1,1.75);
   \node[] at (0,2.1) {\scriptsize $n+1$};
            \draw[snake=brace]  (3.1,-0.75) -- (-3.1,-0.75);
   \node[] at (0,-1.1) {\scriptsize $N_f-1$};
\end{tikzpicture}};
\draw[<->] (3,0)--(4,0);
    \end{tikzpicture} }
\end{equation} 
This claim is clearly consistent with the special cases we have considered above and with the case $N_f=2N+1$, $n=1$ already studied in \cite{Dey:2021rxw, vanBeest:2021xyt}. One can also easily check that the Coulomb and Higgs branch dimensions match our expectations.

\subsubsection*{An exhaustive case study}

In this subsection we carry out the exhaustive study of deformations for the case \eqref{eq:SQCD-SU} with $k=3$ and $N_f = 7$. More precisely, we consider all the FI-deformations which involve exactly two nodes in the magnetic quiver.\footnote{For these deformations we don't write explicitly the value of the FI terms on the quivers, as it is fixed uniquely up to an irrelevant overall magnitude by the ratio of the ranks of the two gauge nodes. It is enough to paint two nodes in orange to fully specify the FI deformation. } In general, for the SQCD theory $\mathrm{SU}(k)$ with and odd number $N_f>2k$ of flavors, and using charge conjugation symmetries to eliminate redundant cases, there are precisely $(\frac{N_f +1}{2})^2$ cases to consider. Here, with $N_f = 7$ this gives 16 deformations; in each case the number of extremal solution appears to be between one and three.  The results are gathered in Tables \ref{tab:SU3with7FlavorsPart1}, \ref{tab:SU3with7FlavorsPart2}, and \ref{tab:SU3with7FlavorsPart3}, where we have picked a numbering of the 16 minimal deformations, from $\# 1$ to $\# 16$. 

Many extremal solutions correspond to cases that have already appeared in earlier examples: those involving two nodes of the same rank were discussed in Section \ref{sec:SpecialUnitaryGaugeGroup}, and those involving a $\mathrm{U}(1)$ node and a $\mathrm{U}(2)$ node were given in Figure \ref{fig:solutionsDtype}. We are left with deformations involving one $\mathrm{U}(3)$ node and either a $\mathrm{U}(1)$ or a $\mathrm{U}(2)$ node, which are more interesting and which we discuss now in detail. For case $\# 3$, there are three solutions, but two of them are related by a graph isomorphism, so it is enough to give the two non isomorphic solutions, which are given explicitly as follows: 
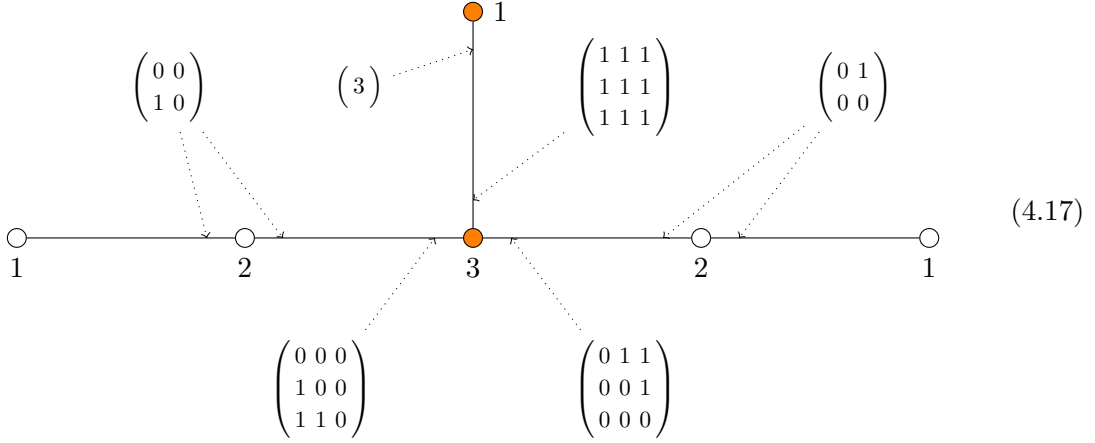
\begin{equation}
   \raisebox{-.5\height}{ \begin{tikzpicture}
            \node[gauge,label=below:{$1$}] (1) at (0,0) {};
            \node[gauge,label=below:{$2$}] (2) at (3,0) {};
            \node[gaugeo,label=below:{$3$}] (3) at (6,0) {};
            \node[gauge,label=below:{$2$}] (4) at (9,0) {};
            \node[gauge,label=below:{$1$}] (5) at (12,0) {};
            \node[gaugeo,label=right:{$1$}] (6) at (6,3) {};
            \draw (1)--(2)--(3)--(4)--(5) (3)--(6);
            \node (3l) at (4,-2) {\scalebox{.8}{$  \left(\begin{array}{ccc}
            0&0&0\\1&0&0\\1&1&0
            \end{array}\right)$}}; 
            \node (3r) at (8,-2) {\scalebox{.8}{$  \left(\begin{array}{ccc}
            0&1&1\\0&0&1\\0&0&0
            \end{array}\right)$}};
            \node (3u) at (8,+2) {\scalebox{.8}{$  \left(\begin{array}{ccc}
           1&1&1\\1&1&1\\1&1&1
            \end{array}\right)$}};
            \node (2r) at (2,+2) {\scalebox{.8}{$  \left(\begin{array}{cc}
           0&0\\1&0
            \end{array}\right)$}};
            \node (6u) at (4.5,2) {\scalebox{.8}{$  \left(\begin{array}{c}
          3
            \end{array}\right)$}};
            \node (4r) at (11,+2) {\scalebox{.8}{$  \left(\begin{array}{cc}
           0&1\\0&0
            \end{array}\right)$}};
            \draw[->,dotted] (3l)--(5.5,0);
            \draw[->,dotted] (3r)--(6.5,0);
            \draw[->,dotted] (3u)--(6,.5);
            \draw[->,dotted] (2r)--(3.5,0);
            \draw[->,dotted] (2r)--(2.5,0);
            \draw[->,dotted] (4r)--(9.5,0);
            \draw[->,dotted] (4r)--(8.5,0);
            \draw[->,dotted] (6u)--(6,2.5);
        \end{tikzpicture} }
        \label{eq:solutionCase3a}
\end{equation}
and 
\begin{equation}
  \raisebox{-.5\height}
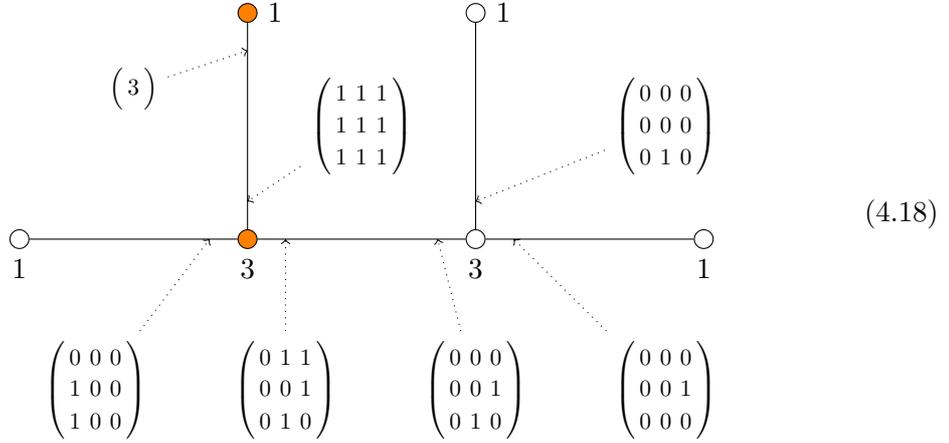
{  \begin{tikzpicture}
            \node[gauge,label=below:{$1$}] (2) at (3,0) {};
            \node[gaugeo,label=below:{$3$}] (3) at (6,0) {};
            \node[gauge,label=below:{$3$}] (4) at (9,0) {};
            \node[gauge,label=below:{$1$}] (5) at (12,0) {};
            \node[gaugeo,label=right:{$1$}] (6) at (6,3) {};
            \node[gauge,label=right:{$1$}] (7) at (9,3) {};
            \draw (2)--(3)--(4)--(5) (3)--(6) (4)--(7);
            \node (3l) at (4,-2) {\scalebox{.8}{$  \left(\begin{array}{ccc}
            0&0&0\\1&0&0\\1&0&0
            \end{array}\right)$}}; 
            \node (3r) at (6.5,-2) {\scalebox{.8}{$  \left(\begin{array}{ccc}
            0&1&1\\0&0&1\\0&1&0
            \end{array}\right)$}};
            \node (4l) at (9,-2) {\scalebox{.8}{$  \left(\begin{array}{ccc}
            0&0&0\\0&0&1\\0&1&0
            \end{array}\right)$}};
            \node (3u) at (7.5,+1.5) {\scalebox{.8}{$  \left(\begin{array}{ccc}
           1&1&1\\1&1&1\\1&1&1
            \end{array}\right)$}};
            \node (6u) at (4.5,2) {\scalebox{.8}{$  \left(\begin{array}{c}
          3
            \end{array}\right)$}};
            \node (4r) at (11.5,-2) {\scalebox{.8}{$  \left(\begin{array}{ccc}
            0&0&0\\0&0&1\\0&0&0
            \end{array}\right)$}};
            \node (4u) at (11.5,+1.5) {\scalebox{.8}{$  \left(\begin{array}{ccc}
            0&0&0\\0&0&0\\0&1&0
            \end{array}\right)$}};
            \draw[->,dotted] (3l)--(5.5,0);
            \draw[->,dotted] (3r)--(6.5,0);
            \draw[->,dotted] (3u)--(6,.5);
            \draw[->,dotted] (4l)--(8.5,0);
            \draw[->,dotted] (4r)--(9.5,0);
            \draw[->,dotted] (4u)--(9,.5);
            \draw[->,dotted] (6u)--(6,2.5);
        \end{tikzpicture} }
        \label{eq:solutionCase3c}
\end{equation}
For case $\# 4$, the first two solutions are deduced straightforwardly from \eqref{eq:solutionCase3a} and \eqref{eq:solutionCase3c}. The third solution appears in the deformations of the Klein $E_6$ surface, and for this reason is given in Figure \ref{fig:solutionE6} in Appendix \ref{app:E}. Next, we move on to case $\# 9$, where the $E_6$ solution is used again (with a trivial extension in the third solution). The same applies to the first solution in case $\# 10$. For the second solution, one can use a slight modification of \eqref{eq:solutionCase3c}. The most difficult case is $\# 13$, where both nodes with FI terms are non-abelian. The first extremal solution is given explicitly as follows: 
\begin{equation}
  \raisebox{-.5\height}{
  }
        \label{eq:exEa}
\end{equation}
The other solutions are obtained by making use of the pair of matrices \eqref{eq:matrixToDecompose}. Case number $\# 14$ is similar, with the only exception being the last entry in case $\# 14$, for which we find the following solution 
\begin{equation}
    \raisebox{-.5\height}{    %
 }
\end{equation}
We note that we can only conjecture these solutions are extremal, and we acknowledge it would be helpful to have a systematic way to check extremality.

\begin{table}[p]
    \centering  
\scalebox{.6}{   
%
    
}
    \caption{Analysis for SU(3) with 7 flavors (1/3). }
    \label{tab:SU3with7FlavorsPart1}
\end{table}

\begin{table}[p]
    \centering  
\scalebox{.6}{   
%
    
}
    \caption{Analysis for SU(3) with 7 flavors (2/3).}
    \label{tab:SU3with7FlavorsPart2}
\end{table}

\begin{table}[p]
    \centering  
\scalebox{.6}{   
%
   \\  \hline 
\end{tabular}    
}
    \caption{Analysis for SU(3) with 7 flavors (3/3). The first column specifies the FI deformation, the second column shows the extremal solutions and the third column gives the result of the subtraction. The last column gives a 3d mirror. }
    \label{tab:SU3with7FlavorsPart3}
\end{table}

\clearpage 

\section{SCFTs and Non-Lagrangian Theories}
\label{sec:Other} 

In this section we will explore FI deformations of magnetic quivers for higher dimensional theories. In particular we will discuss mass deformations for rank two $\mathcal{N}=2$ SCFTs in four dimensions, which are quite well-understood, and then show how FI deformations relate multi instanton moduli spaces to non-minimal nilpotent orbits for the same group $G$.

\subsection{4d SCFTs} 

Four dimensional SCFTs form a set of theories of prime importance, and display a large varieties of behaviors: the theories can be Lagrangian or not, they can have Coulomb branch operators of fractional scaling dimension (Argyres-Douglas theories \cite{Argyres:1995jj}), some possess a conformal manifold while others are isolated, etc. The low rank theories have been studied in \cite{Argyres:2015ffa,Argyres:2016xua,Martone:2021ixp,Argyres:2022lah}. Their Higgs branches can be studied using magnetic quivers \cite{Benini:2010uu, Xie:2012hs, DelZotto:2014kka,Cremonesi:2014xha,Cabrera:2018jxt,Cabrera:2019izd,Cabrera:2019dob, Bourget:2019rtl, Bourget:2020asf, Bourget:2020gzi, Beratto:2020wmn, Closset:2020scj, Bourget:2020mez, vanBeest:2020kou, Giacomelli:2020gee, VanBeest:2020kxw, Closset:2020afy, Akhond:2021knl, Martone:2021ixp, Arias-Tamargo:2021ppf, Bourget:2021xex, Carta:2021dyx, Xie:2021ewm, Nawata:2021nse, Closset:2021lwy, Bhardwaj:2021mzl, Kang:2022zsl,Nawata:2023rdx}. For theories of ranks 1 and 2, a systematic exploration of the magnetic quivers was given in \cite{Bourget:2020asf,Bourget:2021csg}. This shows that a full description of 4d SCFTs requires at least non simply laced quivers, with orthogonal and symplectic gauge groups in addition to unitary ones. Here, in line with the rest of this paper, we will focus for simplicity on 4d theories with a simply-laced unitary magnetic quiver. For concreteness, we concentrate on rank 2 theories, where the mass deformations have been analyzed in detail in \cite{Martone:2021drm}, and is reproduced with magnetic quivers in \cite[Fig. 3]{Bourget:2021csg}. In this section, we show how these results are confirmed by our FI-flows. 

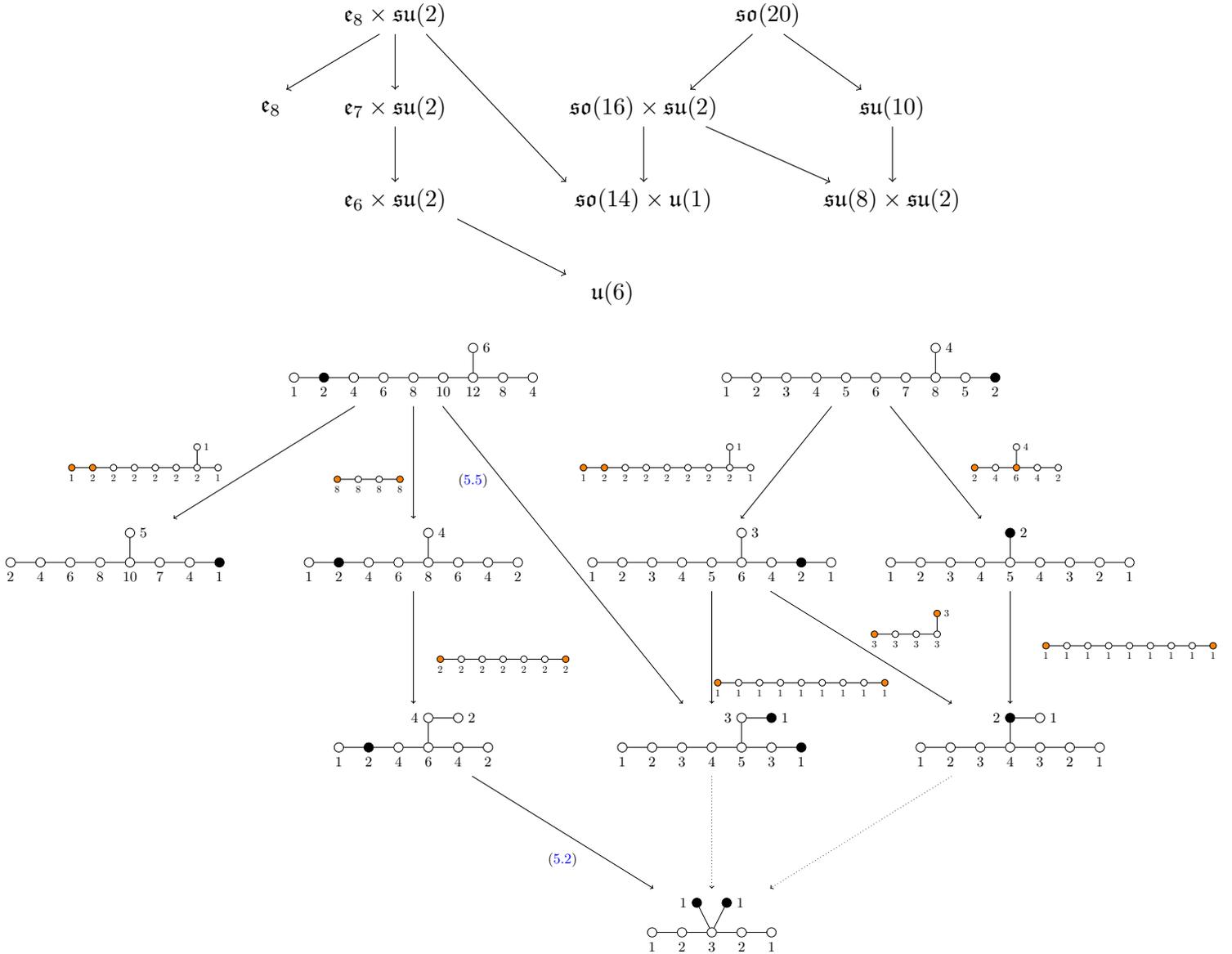
\begin{figure}
    \centering
   
\newcommand{\quiv}[1]{%
    \IfEqCase{#1}{%
{1}{\raisebox{-.5\height}{\begin{tikzpicture}[x=.8cm,y=.8cm,decoration={markings,mark=at position 0.5 with {\arrow{>},arrowhead=1cm}}]
\node at (1,.5) {};
\node (g1) at (1,0) [gauge,label=below:{1}] {};
\node (g2) at (2,0) [gaugeb,label=below:{2}] {};
\node (g3) at (3,0) [gauge,label=below:{4}] {};
\node (g4) at (4,0) [gauge,label=below:{6}] {};
\node (g5) at (5,0) [gauge,label=below:{8}] {};
\node (g6) at (6,0) [gauge,label=below:{10}] {};
\node (g7) at (7,0) [gauge,label=below:{12}] {};
\node (g8) at (8,0) [gauge,label=below:{8}] {};
\node (g9) at (9,0) [gauge,label=below:{4}] {};
\node (g10) at (7,1) [gauge,label=right:{6}] {};
\draw (g1)--(g2)--(g3)--(g4)--(g5)--(g6)--(g7)--(g8)--(g9) (g7)--(g10);
\end{tikzpicture}}}%
{1001}{\raisebox{-.5\height}{\begin{tikzpicture}[x=.8cm,y=.8cm,decoration={markings,mark=at position 0.5 with {\arrow{>},arrowhead=1cm}}]
\node at (1,.5) {};
\node (g1) at (1,0) [gaugeo,label=below:{1}] {};
\node (g2) at (2,0) [gaugeo,label=below:{2}] {};
\node (g3) at (3,0) [gauge,label=below:{2}] {};
\node (g4) at (4,0) [gauge,label=below:{2}] {};
\node (g5) at (5,0) [gauge,label=below:{2}] {};
\node (g6) at (6,0) [gauge,label=below:{2}] {};
\node (g7) at (7,0) [gauge,label=below:{2}] {};
\node (g8) at (8,0) [gauge,label=below:{1}] {};
\node (g10) at (7,1) [gauge,label=right:{1}] {};
\draw (g1)--(g2)--(g3)--(g4)--(g5)--(g6)--(g7)--(g8) (g7)--(g10);
\end{tikzpicture}}}%
{1002}{\raisebox{-.5\height}{\begin{tikzpicture}[x=.8cm,y=.8cm,decoration={markings,mark=at position 0.5 with {\arrow{>},arrowhead=1cm}}]
\node at (1,.5) {};
\node (g1) at (1,0) [gaugeo,label=below:{8}] {};
\node (g2) at (2,0) [gauge,label=below:{8}] {};
\node (g3) at (3,0) [gauge,label=below:{8}] {};
\node (g4) at (4,0) [gaugeo,label=below:{8}] {};
\draw (g1)--(g2)--(g3)--(g4);
\end{tikzpicture}}}%
{2}{\raisebox{-.5\height}{\begin{tikzpicture}[x=.8cm,y=.8cm,decoration={markings,mark=at position 0.5 with {\arrow{>},arrowhead=1cm}}]
\node at (1,.5) {};
\node (g1) at (1,0) [gauge,label=below:{1}] {};
\node (g2) at (2,0) [gauge,label=below:{2}] {};
\node (g3) at (3,0) [gauge,label=below:{3}] {};
\node (g4) at (4,0) [gauge,label=below:{4}] {};
\node (g5) at (5,0) [gauge,label=below:{5}] {};
\node (g6) at (6,0) [gauge,label=below:{6}] {};
\node (g7) at (7,0) [gauge,label=below:{7}] {};
\node (g8) at (8,0) [gauge,label=below:{8}] {};
\node (g9) at (9,0) [gauge,label=below:{5}] {};
\node (g10) at (10,0) [gaugeb,label=below:{2}] {};
\node (g11) at (8,1) [gauge,label=right:{4}] {};
\draw (g1)--(g2)--(g3)--(g4)--(g5)--(g6)--(g7)--(g8)--(g9)--(g10) (g8)--(g11);
\end{tikzpicture}}}%
{2001}{\raisebox{-.5\height}{\begin{tikzpicture}[x=.8cm,y=.8cm,decoration={markings,mark=at position 0.5 with {\arrow{>},arrowhead=1cm}}]
\node at (1,.5) {};
\node (g1) at (1,0) [gaugeo,label=below:{1}] {};
\node (g2) at (2,0) [gaugeo,label=below:{2}] {};
\node (g3) at (3,0) [gauge,label=below:{2}] {};
\node (g4) at (4,0) [gauge,label=below:{2}] {};
\node (g5) at (5,0) [gauge,label=below:{2}] {};
\node (g6) at (6,0) [gauge,label=below:{2}] {};
\node (g7) at (7,0) [gauge,label=below:{2}] {};
\node (g8) at (8,0) [gauge,label=below:{2}] {};
\node (g9) at (9,0) [gauge,label=below:{1}] {};
\node (g11) at (8,1) [gauge,label=right:{1}] {};
\draw (g1)--(g2)--(g3)--(g4)--(g5)--(g6)--(g7)--(g8)--(g9) (g8)--(g11);
\end{tikzpicture}}}%
{2002}{\raisebox{-.5\height}{\begin{tikzpicture}[x=.8cm,y=.8cm,decoration={markings,mark=at position 0.5 with {\arrow{>},arrowhead=1cm}}]
\node at (1,.5) {};
\node (g6) at (6,0) [gaugeo,label=below:{2}] {};
\node (g7) at (7,0) [gauge,label=below:{4}] {};
\node (g8) at (8,0) [gaugeo,label=below:{6}] {};
\node (g9) at (9,0) [gauge,label=below:{4}] {};
\node (g10) at (10,0) [gauge,label=below:{2}] {};
\node (g11) at (8,1) [gauge,label=right:{4}] {};
\draw (g6)--(g7)--(g8)--(g9)--(g10) (g8)--(g11);
\end{tikzpicture}}}%
{3}{\raisebox{-.5\height}{\begin{tikzpicture}[x=.8cm,y=.8cm,decoration={markings,mark=at position 0.5 with {\arrow{>},arrowhead=1cm}}]
\node at (1,.5) {};
\node (g1) at (1,0) [gauge,label=below:{2}] {};
\node (g2) at (2,0) [gauge,label=below:{4}] {};
\node (g3) at (3,0) [gauge,label=below:{6}] {};
\node (g4) at (4,0) [gauge,label=below:{8}] {};
\node (g5) at (5,0) [gauge,label=below:{10}] {};
\node (g6) at (6,0) [gauge,label=below:{7}] {};
\node (g7) at (7,0) [gauge,label=below:{4}] {};
\node (g8) at (8,0) [gaugeb,label=below:{1}] {};
\node (g9) at (5,1) [gauge,label=right:{5}] {};
\draw (g1)--(g2)--(g3)--(g4)--(g5)--(g6)--(g7)--(g8) (g9)--(g5);
\end{tikzpicture}}}%
{4}{\raisebox{-.5\height}{\begin{tikzpicture}[x=.8cm,y=.8cm,decoration={markings,mark=at position 0.5 with {\arrow{>},arrowhead=1cm}}]
\node at (1,.5) {};
\node (g1) at (1,0) [gauge,label=below:{1}] {};
\node (g2) at (2,0) [gaugeb,label=below:{2}] {};
\node (g3) at (3,0) [gauge,label=below:{4}] {};
\node (g4) at (4,0) [gauge,label=below:{6}] {};
\node (g5) at (5,0) [gauge,label=below:{8}] {};
\node (g6) at (6,0) [gauge,label=below:{6}] {};
\node (g7) at (7,0) [gauge,label=below:{4}] {};
\node (g8) at (8,0) [gauge,label=below:{2}] {};
\node (g9) at (5,1) [gauge,label=right:{4}] {};
\draw (g1)--(g2)--(g3)--(g4)--(g5)--(g6)--(g7)--(g8) (g9)--(g5);
\end{tikzpicture}}}%
{4002}{\raisebox{-.5\height}{\begin{tikzpicture}[x=.8cm,y=.8cm,decoration={markings,mark=at position 0.5 with {\arrow{>},arrowhead=1cm}}]
\node (g2) at (2,0) [gaugeo,label=below:{2}] {};
\node (g3) at (3,0) [gauge,label=below:{2}] {};
\node (g4) at (4,0) [gauge,label=below:{2}] {};
\node (g5) at (5,0) [gauge,label=below:{2}] {};
\node (g6) at (6,0) [gauge,label=below:{2}] {};
\node (g7) at (7,0) [gauge,label=below:{2}] {};
\node (g8) at (8,0) [gaugeo,label=below:{2}] {};
\draw (g2)--(g3)--(g4)--(g5)--(g6)--(g7)--(g8);
\end{tikzpicture}}}%
{5}{\raisebox{-.5\height}{\begin{tikzpicture}[x=.8cm,y=.8cm,decoration={markings,mark=at position 0.5 with {\arrow{>},arrowhead=1cm}}]
\node at (1,.5) {};
\node (g1) at (1,0) [gauge,label=below:{1}] {};
\node (g2) at (2,0) [gauge,label=below:{2}] {};
\node (g3) at (3,0) [gauge,label=below:{3}] {};
\node (g4) at (4,0) [gauge,label=below:{4}] {};
\node (g5) at (5,0) [gauge,label=below:{5}] {};
\node (g6) at (6,0) [gauge,label=below:{6}] {};
\node (g7) at (7,0) [gauge,label=below:{4}] {};
\node (g8) at (8,0) [gaugeb,label=below:{2}] {};
\node (g9) at (9,0) [gauge,label=below:{1}] {};
\node (g10) at (6,1) [gauge,label=right:{3}] {};
\draw (g1)--(g2)--(g3)--(g4)--(g5)--(g6)--(g7)--(g8)--(g9) (g6)--(g10);
\end{tikzpicture}}}%
{5001}{\raisebox{-.5\height}{\begin{tikzpicture}[x=.8cm,y=.8cm,decoration={markings,mark=at position 0.5 with {\arrow{>},arrowhead=1cm}}]
\node at (1,.5) {};
\node (g1) at (1,0) [gaugeo,label=below:{1}] {};
\node (g2) at (2,0) [gauge,label=below:{1}] {};
\node (g3) at (3,0) [gauge,label=below:{1}] {};
\node (g4) at (4,0) [gauge,label=below:{1}] {};
\node (g5) at (5,0) [gauge,label=below:{1}] {};
\node (g6) at (6,0) [gauge,label=below:{1}] {};
\node (g7) at (7,0) [gauge,label=below:{1}] {};
\node (g8) at (8,0) [gauge,label=below:{1}] {};
\node (g9) at (9,0) [gaugeo,label=below:{1}] {};
\draw (g1)--(g2)--(g3)--(g4)--(g5)--(g6)--(g7)--(g8)--(g9);
\end{tikzpicture}}}%
{5002}{\raisebox{-.5\height}{\begin{tikzpicture}[x=.8cm,y=.8cm,decoration={markings,mark=at position 0.5 with {\arrow{>},arrowhead=1cm}}]
\node at (1,.5) {};
\node (g3) at (3,0) [gaugeo,label=below:{3}] {};
\node (g4) at (4,0) [gauge,label=below:{3}] {};
\node (g5) at (5,0) [gauge,label=below:{3}] {};
\node (g6) at (6,0) [gauge,label=below:{3}] {};
\node (g10) at (6,1) [gaugeo,label=right:{3}] {};
\draw (g3)--(g4)--(g5)--(g6)--(g10);
\end{tikzpicture}}}%
{6}{\raisebox{-.5\height}{\begin{tikzpicture}[x=.8cm,y=.8cm,decoration={markings,mark=at position 0.5 with {\arrow{>},arrowhead=1cm}}]
\node at (1,.5) {};
\node (g1) at (1,0) [gauge,label=below:{1}] {};
\node (g2) at (2,0) [gauge,label=below:{2}] {};
\node (g3) at (3,0) [gauge,label=below:{3}] {};
\node (g4) at (4,0) [gauge,label=below:{4}] {};
\node (g5) at (5,0) [gauge,label=below:{5}] {};
\node (g6) at (6,0) [gauge,label=below:{4}] {};
\node (g7) at (7,0) [gauge,label=below:{3}] {};
\node (g8) at (8,0) [gauge,label=below:{2}] {};
\node (g9) at (9,0) [gauge,label=below:{1}] {};
\node (g10) at (5,1) [gaugeb,label=right:{2}] {};
\draw (g1)--(g2)--(g3)--(g4)--(g5)--(g6)--(g7)--(g8)--(g9) (g5)--(g10);
\end{tikzpicture}}}%
{6001}{\raisebox{-.5\height}{\begin{tikzpicture}[x=.8cm,y=.8cm,decoration={markings,mark=at position 0.5 with {\arrow{>},arrowhead=1cm}}]
\node at (1,.5) {};
\node (g1) at (1,0) [gaugeo,label=below:{1}] {};
\node (g2) at (2,0) [gauge,label=below:{1}] {};
\node (g3) at (3,0) [gauge,label=below:{1}] {};
\node (g4) at (4,0) [gauge,label=below:{1}] {};
\node (g5) at (5,0) [gauge,label=below:{1}] {};
\node (g6) at (6,0) [gauge,label=below:{1}] {};
\node (g7) at (7,0) [gauge,label=below:{1}] {};
\node (g8) at (8,0) [gauge,label=below:{1}] {};
\node (g9) at (9,0) [gaugeo,label=below:{1}] {};
\draw (g1)--(g2)--(g3)--(g4)--(g5)--(g6)--(g7)--(g8)--(g9);
\end{tikzpicture}}}%
{7}{\raisebox{-.5\height}{\begin{tikzpicture}[x=.8cm,y=.8cm,decoration={markings,mark=at position 0.5 with {\arrow{>},arrowhead=1cm}}]
\node at (1,.5) {};
\node (g1) at (1,0) [gauge,label=below:{1}] {};
\node (g2) at (2,0) [gaugeb,label=below:{2}] {};
\node (g3) at (3,0) [gauge,label=below:{4}] {};
\node (g4) at (4,0) [gauge,label=below:{6}] {};
\node (g5) at (5,0) [gauge,label=below:{4}] {};
\node (g6) at (6,0) [gauge,label=below:{2}] {};
\node (g8) at (4,1) [gauge,label=left:{4}] {};
\node (g9) at (5,1) [gauge,label=right:{2}] {};
\draw (g1)--(g2)--(g3)--(g4)--(g5)--(g6) (g4)--(g8)--(g9);
\end{tikzpicture}}}%
{8}{\raisebox{-.5\height}{\begin{tikzpicture}[x=.8cm,y=.8cm,decoration={markings,mark=at position 0.5 with {\arrow{>},arrowhead=1cm}}]
\node at (1,.5) {};
\node (g1) at (1,0) [gauge,label=below:{1}] {};
\node (g2) at (2,0) [gauge,label=below:{2}] {};
\node (g3) at (3,0) [gauge,label=below:{3}] {};
\node (g4) at (4,0) [gauge,label=below:{4}] {};
\node (g5) at (5,0) [gauge,label=below:{5}] {};
\node (g6) at (6,0) [gauge,label=below:{3}] {};
\node (g7) at (7,0) [gaugeb,label=below:{1}] {};
\node (g8) at (5,1) [gauge,label=left:{3}] {};
\node (g9) at (6,1) [gaugeb,label=right:{1}] {};
\draw (g1)--(g2)--(g3)--(g4)--(g5)--(g6)--(g7) (g5)--(g8)--(g9);
\end{tikzpicture}}}%
{9}{\raisebox{-.5\height}{\begin{tikzpicture}[x=.8cm,y=.8cm,decoration={markings,mark=at position 0.5 with {\arrow{>},arrowhead=1cm}}]
\node at (1,.5) {};
\node (g1) at (1,0) [gauge,label=below:{1}] {};
\node (g2) at (2,0) [gauge,label=below:{2}] {};
\node (g3) at (3,0) [gauge,label=below:{3}] {};
\node (g4) at (4,0) [gauge,label=below:{4}] {};
\node (g5) at (5,0) [gauge,label=below:{3}] {};
\node (g6) at (6,0) [gauge,label=below:{2}] {};
\node (g7) at (7,0) [gauge,label=below:{1}] {};
\node (g8) at (4,1) [gaugeb,label=left:{2}] {};
\node (g9) at (5,1) [gauge,label=right:{1}] {};
\draw (g1)--(g2)--(g3)--(g4)--(g5)--(g6)--(g7) (g4)--(g8)--(g9);
\end{tikzpicture}}}%
{10}{\raisebox{-.5\height}{\begin{tikzpicture}[x=.8cm,y=.8cm,decoration={markings,mark=at position 0.5 with {\arrow{>},arrowhead=1cm}}]
\node at (1,.5) {};
\node (g1) at (1,0) [gauge,label=below:{1}] {};
\node (g2) at (2,0) [gauge,label=below:{2}] {};
\node (g3) at (3,0) [gauge,label=below:{3}] {};
\node (g4) at (4,0) [gauge,label=below:{4}] {};
\node (g5) at (5,0) [gauge,label=below:{2}] {};
\node (g6) at (4.5,1) [gauge,label=right:{2}] {};
\node (g7) at (3.5,1) [gaugeb,label=left:{1}] {};
\draw (g1)--(g2)--(g3)--(g4)--(g5) (g4)--(g6) (g4)--(g7);
\end{tikzpicture}}}%
{11}{\raisebox{-.5\height}{\begin{tikzpicture}[x=.8cm,y=.8cm,decoration={markings,mark=at position 0.5 with {\arrow{>},arrowhead=1cm}}]
\node at (1,.5) {};
\node (g1) at (1,0) [gauge,label=below:{1}] {};
\node (g2) at (2,0) [gaugeb,label=below:{2}] {};
\node (g3) at (3,0) [gauge,label=below:{4}] {};
\node (g4) at (4,0) [gauge,label=below:{2}] {};
\node (g6) at (3.5,1) [gauge,label=right:{2}] {};
\node (g7) at (2.5,1) [gauge,label=left:{2}] {};
\draw (g1)--(g2)--(g3)--(g4) (g3)--(g6) (g3)--(g7);
\end{tikzpicture}}}%
{12}{\raisebox{-.5\height}{\begin{tikzpicture}[x=.8cm,y=.8cm,decoration={markings,mark=at position 0.5 with {\arrow{>},arrowhead=1cm}}]
\node at (1,.5) {};
\node (g1) at (1,0) [gauge,label=below:{1}] {};
\node (g2) at (2,0) [gauge,label=below:{2}] {};
\node (g3) at (3,0) [gauge,label=below:{3}] {};
\node (g4) at (4,0) [gauge,label=below:{2}] {};
\node (g5) at (5,0) [gauge,label=below:{1}] {};
\node (g6) at (3.5,1) [gaugeb,label=right:{1}] {};
\node (g7) at (2.5,1) [gaugeb,label=left:{1}] {};
\draw (g1)--(g2)--(g3)--(g4)--(g5) (g3)--(g6) (g3)--(g7);
\end{tikzpicture}}}%
}}

\hspace*{-2.5cm}\begin{tikzpicture}
\node at (0,8) {
\begin{tikzpicture}
\node[] (1) at (0,0) {$\mathfrak{e}_8\times \mathfrak{su}(2)$}; 
\node[] (2) at (6,0) {$\mathfrak{so}(20)$}; 
\node[] (3) at (-2,-1.5) {$\mathfrak{e}_8$};
\node[] (4) at (0,-1.5) {$\mathfrak{e}_7\times \mathfrak{su}(2)$}; 
\node[] (5) at (0,-3) {$\mathfrak{e}_6\times \mathfrak{su}(2)$};
\node[] (6) at (4,-1.5) {$\mathfrak{so}(16)\times \mathfrak{su}(2)$}; 
\node[] (7) at (4,-3) {$\mathfrak{so}(14)\times \mathfrak{u}(1)$};
\node[] (8) at (3.5,-4.5) {$\mathfrak{u}(6)$};
\node[] (9) at (8,-1.5) {$\mathfrak{su}(10)$}; 
\node[] (10) at (8,-3) {$\mathfrak{su}(8)\times \mathfrak{su}(2)$};
\draw[->] (-0.25,-0.3)--(-1.75,-1.2);
\draw[->] (0,-0.3)--(0,-1.2);
\draw[->] (0,-1.8)--(0,-2.7);
\draw[->] (1,-3.3)--(2.75,-4.2);
\draw[->] (0.5,-0.3)--(2.75,-2.7);
\draw[->] (5.75,-0.3)--(4.75,-1.2);
\draw[->] (6.25,-0.3)--(7.5,-1.2); 
\draw[->] (4,-1.8)--(4,-2.7);
\draw[->] (8,-1.8)--(8,-2.7);
\draw[->] (5,-1.8)--(7,-2.7);
\end{tikzpicture} };
\node at (0,0) {\scalebox{.6}{
\begin{tikzpicture}[xscale=8,yscale=5]
\node (t1) at (1,8) {\quiv{1}};
\node at (.1,7.5) {\scalebox{.7}{\quiv{1001}}};
\node at (.85,7.4) {\scalebox{.7}{\quiv{1002}}};
\node (t2) at (2.5,8) {\quiv{2}};
\node at (1.85,7.5) {\scalebox{.7}{\quiv{2001}}};
\node at (2.85,7.5) {\scalebox{.7}{\quiv{2002}}};
\node (t3) at (0,7) {\quiv{3}};
\node (t4) at (1,7) {\quiv{4}};
\node at (1.3,6.4) {\scalebox{.7}{\quiv{4002}}};
\node (t5) at (2,7) {\quiv{5}};
\node at (2.3,6.3) {\scalebox{.7}{\quiv{5001}}};
\node at (2.6,6.6) {\scalebox{.7}{\quiv{5002}}};
\node (t6) at (3,7) {\quiv{6}};
\node at (3.4,6.5) {\scalebox{.7}{\quiv{6001}}};
\node (t7) at (1,6) {\quiv{7}};
\node (t8) at (2,6) {\quiv{8}};
\node (t9) at (3,6) {\quiv{9}};
\node (t12) at (2,5) {\quiv{12}};
\draw[->] (t1)--(t3);
\draw[->] (t1)--(t4);
\draw[->] (t1)--(t8);
\draw[->] (t2)--(t5);
\draw[->] (t2)--(t6);
\draw[->] (t4)--(t7);
\draw[->] (t5)--(t8);
\draw[->] (t5)--(t9);
\draw[->] (t6)--(t9);
\draw[->] (t7)--(t12);
\draw[->,dotted] (t8)--(t12);
\draw[->,dotted] (t9)--(t12);
\node (s) at (1.5,5.35) {\eqref{eq:trans2}};
\node (s) at (1.2,7.4) {\eqref{eq:trans1}};
\end{tikzpicture}
} };
\end{tikzpicture}
    \caption{Top: Part of the graph of mass deformation for 4d $\mathcal{N}=2$ SCFTs of rank 2. We stop at the point where the theories become Lagrangian (here the SU(3) SQCD theory). Bottom: Corresponding graph of magnetic quivers, along with the solutions to be subtracted to implement the mass deformations. Most subtractions are straightforward. The non-trivial ones are discussed in the text. }
    \label{fig:rank2Tree}
\end{figure}

The top of the mass deformation tree is given by two models: the rank two E-string theory and the $\mathfrak{so}(20)$ model. 
The corresponding magnetic quivers are drawn on top of Figure \ref{fig:rank2Tree}. 
We will now identify the FI deformations which implement the RG flows to their descendant theories. 

Let us start from the rank two E-string theory. It was already shown in \cite{vanBeest:2021xyt} that we can flow to the rank two $\mathfrak{e}_7$ theory and then from there to the $\mathfrak{e}_6$ theory reproducing the known pattern for rank one models. 
A novelty with respect to the rank one case is the existence of the flow from the rank two $\mathfrak{e}_6$ theory to $\mathfrak{su}(3)$ SQCD with six flavors 
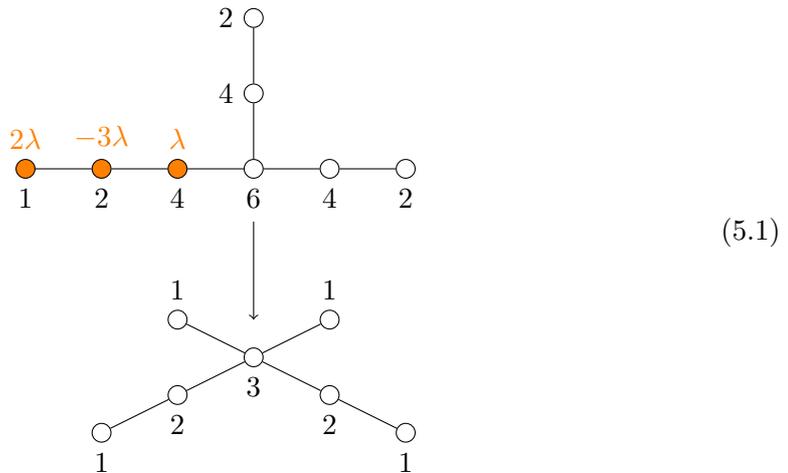
\begin{equation}
   \raisebox{-.5\height}{ \begin{tikzpicture}
    \label{eq:flowE6SU3}
            \node[gauge, fill=orange,label=below:{$1$},label=above:{\textcolor{orange}{$2\lambda$}}] (2) at (2,0) {};
            \node[gauge, fill=orange,label=below:{$2$},label=above:{\textcolor{orange}{$-3\lambda$}}] (3) at (3,0) {};
            \node[gauge,fill=orange,label=below:{$4$},label=above:{\textcolor{orange}{$\lambda$}}] (4) at (4,0) {};
            \node[gauge,label=below:{$6$}] (5) at (5,0) {};
            \node[gauge,label=below:{$4$}] (6) at (6,0) {};
             \node[gauge,label=below:{$2$}] (7) at (7,0) {};
            \node[gauge,label=left:{$4$}] (8) at (5,1) {};
            \node[gauge,label=left:{$2$}] (9) at (5,2) {};
            \draw (2)--(3)--(4)--(5)--(6)--(7) (5)--(8)--(9);
            \draw[->] (5,-0.7) to (5,-2); 
            \node[gauge,label=below:{$1$}] (c) at (3,-3.5) {};
            \node[gauge,label=below:{$2$}] (d) at (4,-3) {};
            \node[gauge,label=below:{$3$}] (e) at (5,-2.5) {};
            \node[gauge,label=above:{$1$}] (f) at (6,-2) {};
             \node[gauge,label=above:{$1$}] (g) at (4,-2) {};
            \node[gauge,label=below:{$2$}] (h) at (6,-3) {};
            \node[gauge,label=below:{$1$}] (i) at (7,-3.5) {};
             \draw (c)--(d)--(e)--(h)--(i) (e)--(f) (e)--(g);
        \end{tikzpicture}}
\end{equation} 
An explicit solution for this case is 
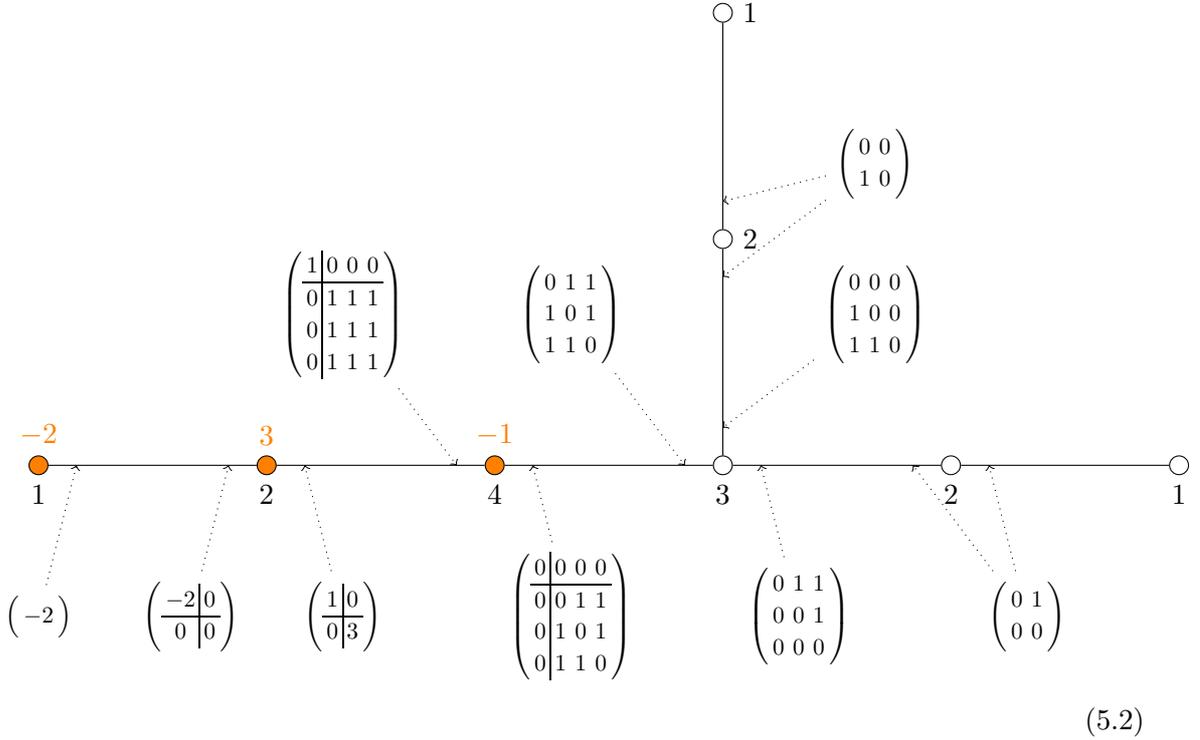
\begin{equation}
\label{eq:trans2}
            \begin{tikzpicture}
            \node[gaugeo,label=below:{$1$},label=above:{\textcolor{orange}{$-2$}}] (1) at (0,0) {};
            \node[gaugeo,label=below:{$2$},label=above:{\textcolor{orange}{$3$}}] (2) at (3,0) {};
            \node[gaugeo,label=below:{$4$},label=above:{\textcolor{orange}{$-1$}}] (3) at (6,0) {};
            \node[gauge,label=below:{$3$}] (4) at (9,0) {};
            \node[gauge,label=below:{$2$}] (5) at (12,0) {};
            \node[gauge,label=below:{$1$}] (6) at (15,0) {};
            \node[gauge,label=right:{$2$}] (7) at (9,3) {};
            \node[gauge,label=right:{$1$}] (8) at (9,6) {};
            \draw (1)--(2)--(3)--(4)--(5)--(6) (4)--(7)--(8);            \node (1r) at (0,-2) {\scalebox{.8}{$  \left(\begin{array}{c}
          - 2
            \end{array}\right)$}}; 
\node (2l) at (2,-2) {\scalebox{.8}{$  \left(\begin{array}{c|c}
           -2 & 0 \\ \hline  0 & 0 
            \end{array}\right)$}}; 
\node (2r) at (4,-2) {\scalebox{.8}{$  \left(\begin{array}{c|c}
           1 & 0 \\  \hline 0 & 3 
            \end{array}\right)$}}; 
\node (3l) at (4,+2) {\scalebox{.8}{$  \left(\begin{array}{c| ccc}
           1 & 0 & 0 & 0 \\ \hline 0 & 1 & 1 & 1 \\0 & 1 & 1 & 1 \\0 & 1 & 1 & 1 
            \end{array}\right)$}}; 
\node (3r) at (7,-2) {\scalebox{.8}{$  \left(\begin{array}{c|ccc}
           0 & 0 & 0 & 0 \\ \hline 0 & 0 & 1 & 1 \\0 & 1 & 0 & 1 \\0 & 1 & 1 & 0 
            \end{array}\right)$}}; 
\node (4l) at (7,+2) {\scalebox{.8}{$  \left(\begin{array}{ccc}
        0 & 1 & 1 \\1 & 0 & 1 \\ 1 & 1 & 0 
            \end{array}\right)$}}; 
\node (4r) at (10,-2) {\scalebox{.8}{$  \left(\begin{array}{ccc}
        0 & 1 & 1 \\0 & 0 & 1 \\ 0 & 0 & 0 
            \end{array}\right)$}}; 
\node (5l) at (13,-2) {\scalebox{.8}{$  \left(\begin{array}{cc}
        0 & 1 \\0 & 0 
            \end{array}\right)$}}; 
\node (4u) at (11,2) {\scalebox{.8}{$  \left(\begin{array}{ccc}
        0 & 0 & 0 \\1 & 0 & 0 \\ 1 & 1 & 0 
            \end{array}\right)$}}; 
\node (6d) at (11,4) {\scalebox{.8}{$  \left(\begin{array}{cc}
        0 & 0 \\1 & 0 
            \end{array}\right)$}}; 
            \draw[->,dotted] (1r)--(0.5,0);
            \draw[->,dotted] (2l)--(2.5,0);
            \draw[->,dotted] (2r)--(3.5,0);
            \draw[->,dotted] (3l)--(5.5,0);
            \draw[->,dotted] (3r)--(6.5,0);
            \draw[->,dotted] (4l)--(8.5,0);
            \draw[->,dotted] (4r)--(9.5,0);
            \draw[->,dotted] (5l)--(11.5,0);
            \draw[->,dotted] (5l)--(12.5,0);
            \draw[->,dotted] (4u)--(9,.5);
            \draw[->,dotted] (6d)--(9,2.5);
            \draw[->,dotted] (6d)--(9,3.5);
        \end{tikzpicture} 
\end{equation}
The solution is not irreducible, as indicated by the separation in the matrices. This means the subtraction involves two quivers, or equivalently, the rebalancing is done with two $\mathrm{U}(1)$ nodes. These are the two abelian nodes on top of the resulting quiver in \eqref{eq:flowE6SU3}. 

Apart from these flows, we also have the mass deformation to the $\mathfrak{e}_8$ model discussed in \cite{Giacomelli:2017ckh, Martone:2021ixp}: 
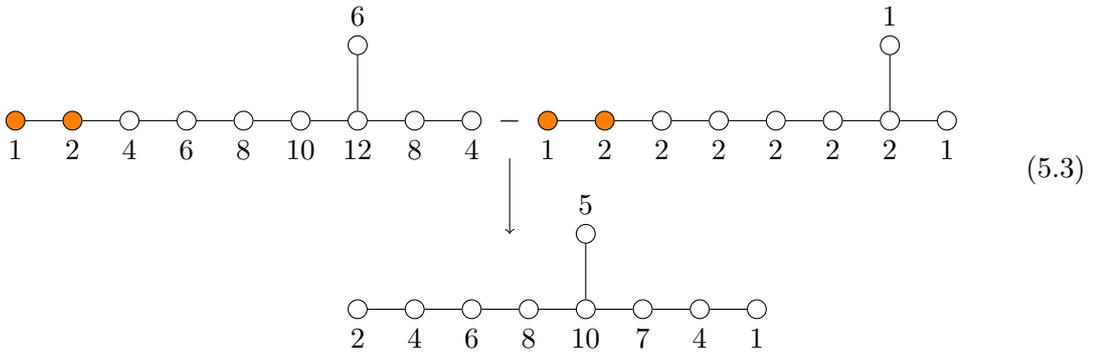
\begin{equation}\label{massdef1}
\raisebox{-.5\height}{\begin{tikzpicture}
            \node[gauge,label=below:{$1$}, fill=orange] (1) at (-6,0) {};
            \node[gauge,label=below:{$2$}, fill=orange] (2) at (-5.25,0) {};
            \node[gauge,label=below:{$4$}] (3) at (-4.5,0) {};
            \node[gauge,label=below:{$6$}] (4) at (-3.75,0) {};
            \node[gauge,label=below:{$8$}] (5) at (-3,0) {};
            \node[gauge,label=below:{$10$}] (6) at (-2.25,0) {};
            \node[gauge,label=below:{$12$}] (7) at (-1.5,0) {};
            \node[gauge,label=below:{$8$}] (8) at (-0.75,0) {};
            \node[gauge,label=below:{$4$}] (9) at (0,0) {};
            \node[gauge,label=above:{$6$}] (10) at (-1.5,1) {};
            \draw (1)--(2)--(3)--(4)--(5)--(6)--(7)--(8)--(9) (7)--(10);
            \node[] at (0.5,0) {$-$}; 
            \node[gaugeo,label=below:{$1$}] (a) at (1,0) {};
            \node[gaugeo,label=below:{$2$}] (b) at (1.75,0) {};
            \node[gauge,label=below:{$2$}] (c) at (2.5,0) {};
            \node[gauge,label=below:{$2$}] (d) at (3.25,0) {};
            \node[gauge,label=below:{$2$}] (e) at (4,0) {};
            \node[gauge,label=below:{$2$}] (f) at (4.75,0) {};
            \node[gauge,label=below:{$2$}] (g) at (5.5,0) {};
            \node[gauge,label=below:{$1$}] (h) at (6.25,0) {};
            \node[gauge,label=above:{$1$}] (i) at (5.5,1) {};
            \draw (a)--(b)--(c)--(d)--(e)--(f)--(g)--(h) (g)--(i); 
            \draw[->] (0.5,-0.5)--(0.5,-1.5); 
            \node[gauge,label=below:{$2$}] (A) at (-1.5,-2.5) {};
            \node[gauge,label=below:{$4$}] (B) at (-0.75,-2.5) {};
            \node[gauge,label=below:{$6$}] (C) at (0,-2.5) {};
            \node[gauge,label=below:{$8$}] (D) at (0.75,-2.5) {};
            \node[gauge,label=below:{$10$}] (E) at (1.5,-2.5) {};
            \node[gauge,label=below:{$7$}] (F) at (2.25,-2.5) {};
            \node[gauge,label=below:{$4$}] (G) at (3,-2.5) {};
            \node[gauge,label=below:{$1$}] (H) at (3.75,-2.5) {};
            \node[gauge,label=above:{$5$}] (I) at (1.5,-1.5) {};
            \draw (A)--(B)--(C)--(D)--(E)--(F)--(G)--(H) (E)--(I);
\end{tikzpicture}}
\end{equation} 
Finally, we can also reproduce the flow to the $\mathfrak{so}(14)\times \mathfrak{u}(1)$ SCFT: 
\begin{equation}
    \raisebox{-.5\height}{\begin{tikzpicture}
            \node[gauge,fill=orange,label=below:{$1$},label=above:{\textcolor{orange}{$2\lambda$}}] (1) at (1,0) {};
            \node[gauge,label=below:{$2$}] (2) at (2,0) {};
            \node[gauge,label=below:{$4$}] (3) at (3,0) {};
            \node[gauge,fill=orange,label=below:{$6$},label=above:{\textcolor{orange}{$-2\lambda$}}] (4) at (4,0) {};
            \node[gauge,label=below:{$8$}] (5) at (5,0) {};
            \node[gauge,fill=orange,label=below:{$10$},label=above:{\textcolor{orange}{$\lambda$}}] (6) at (6,0) {};
             \node[gauge,label=below:{$12$}] (7) at (7,0) {};
            \node[gauge,label=below:{$8$}] (8) at (8,0) {};
            \node[gauge,label=below:{$4$}] (9) at (9,0) {};
            \node[gauge,label=above:{$6$}] (10) at (7,1) {};
            \draw (1)--(2)--(3)--(4)--(5)--(6)--(7)--(8)--(9) (7)--(10);
            \draw[->] (5,-1) to (5,-3); 
            \node[gauge,label=below:{$1$}] (a) at (1,-4) {};
            \node[gauge,label=below:{$2$}] (b) at (2,-4) {};
            \node[gauge,label=below:{$3$}] (c) at (3,-4) {};
            \node[gauge,label=below:{$4$}] (d) at (4,-4) {};
            \node[gauge,label=below:{$5$}] (e) at (5,-4) {};
            \node[gauge,label=above:{$3$}] (f) at (6,-3.5) {};
             \node[gauge,label=above:{$1$}] (g) at (7,-3) {};
            \node[gauge,label=below:{$3$}] (h) at (6,-4.5) {};
            \node[gauge,label=below:{$1$}] (i) at (7,-5) {};
             \draw (a)--(b)--(c)--(d)--(e)--(f)--(g) (e)--(h)--(i);
        \end{tikzpicture}}
\end{equation} 
Notice that also in this case the deformation is implemented by subtracting two quivers: 
\begin{equation}
\label{eq:trans1}
    \raisebox{-.5\height}{\begin{tikzpicture}
            \node[gauge,fill=orange,label=below:{$1$},label=above:{\textcolor{orange}{$2\lambda$}}] (1) at (1,0) {};
            \node[gauge,label=below:{$2$}] (2) at (2,0) {};
            \node[gauge,label=below:{$4$}] (3) at (3,0) {};
            \node[gauge,fill=orange,label=below:{$6$},label=above:{\textcolor{orange}{$-2\lambda$}}] (4) at (4,0) {};
            \node[gauge,label=below:{$8$}] (5) at (5,0) {};
            \node[gauge,fill=orange,label=below:{$10$},label=above:{\textcolor{orange}{$\lambda$}}] (6) at (6,0) {};
             \node[gauge,label=below:{$12$}] (7) at (7,0) {};
            \node[gauge,label=below:{$8$}] (8) at (8,0) {};
            \node[gauge,label=below:{$4$}] (9) at (9,0) {};
            \node[gauge,label=above:{$6$}] (10) at (7,1) {};
            \draw (1)--(2)--(3)--(4)--(5)--(6)--(7)--(8)--(9) (7)--(10);
            \node[] at (10,0) {$-$};
            \node[gaugeo,label=below:{$1$}] (a) at (1,-2) {};
            \node[gauge,label=below:{$1$}] (b) at (2,-2) {};
            \node[gauge,label=below:{$1$}] (c) at (3,-2) {};
            \node[gaugeo,label=below:{$1$}] (d) at (4,-2) {};
            \node[gaugeo,label=below:{$5$}] (e) at (4,-3) {};
            \node[gauge,label=below:{$5$}] (f) at (5,-3) {};
             \node[gaugeo,label=below:{$10$}] (g) at (6,-3) {};
            \node[gauge,label=below:{$10$}] (h) at (7,-3) {};
            \node[gauge,label=below:{$5$}] (i) at (8,-3) {};
            \node[gauge,label=above:{$5$}] (l) at (7,-2) {};
             \draw (a)--(b)--(c)--(d) (e)--(f)--(g)--(h)--(i) (h)--(l);
        \end{tikzpicture}}
\end{equation} 
and the rebalancing is done with a $\mathrm{U}(1)$ node and also a $\mathrm{U}(5)$ which ends up being the central node of the magnetic quiver for the $\mathfrak{so}(14)\times \mathfrak{u}(1)$ SCFT.

Let us now consider the mass deformations of the $\mathfrak{so}(20)$ model. These involve only subtractions of finite quivers of types A and D. 
Overall, we recover the known tree of mass deformations of 4d $\mathcal{N}=2$ rank two SCFTs which admit a simply-laced magnetic quiver \cite{Martone:2021drm}\footnote{Here we have omitted the RG flows involving two or more mass parameters, focusing on the minimal FI deformations.}, shown in Figure \ref{fig:rank2Tree}.

As a final remark, we notice that studying mass deformations of 4d $\mathcal{N}=2$ Lagrangian SCFTs via their magnetic quivers is subtle, since minimal mass deformations can look non minimal at the level of FI deformations of the corresponding magnetic quivers and for this reason we refrain from discussing them in detail. 
In order to illustrate this point, let us focus on the $\mathrm{SU}(3)$ theory with six flavors we have mentioned before. It is known that this model can be mass deformed both to the $D_2(\mathrm{SU}(5))$ SCFT (see \cite{Cecotti:2013lda, Martone:2021drm}) and to $\mathrm{U}(2)$ SQCD with six flavors\footnote{More precisely, the flow to $\mathrm{U}(2)$ SQCD with six flavors involves both giving a common mass to all the flavors and also moving on the Coulomb branch of the theory.}. Moreover, both models can be mass deformed to $\mathrm{U}(2)$ SQCD with five flavors. The main point now is that upon dimensional reduction to 3d the $D_2(\mathrm{SU}(5))$ theory becomes equivalent to $\mathrm{U}(2)$ with five flavors \cite{Closset:2020afy, Giacomelli:2020ryy} and therefore the flow from $\mathrm{SU}(3)$ SQCD with six flavors to $D_2(\mathrm{SU}(5))$, which is a minimal mass deformation in 4d, in 3d is equivalent to a sequence of two deformations: to $\mathrm{U}(2)$ SQCD with six flavors first and then to the $\mathrm{U}(2)$ theory with five flavors and this is precisely what we see at the level of FI deformations of the magnetic quiver.
The above discussion can be summarized as follows:
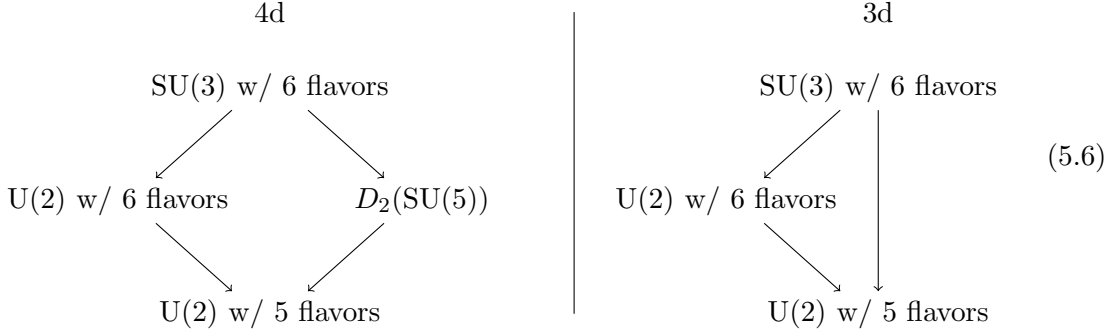
\begin{equation}
\raisebox{-.5\height}{\begin{tikzpicture}
\node[] at (-4,0) {4d}; 
\node[] at (4,0) {3d}; 
\draw (0,0)--(0,-4); 
\node[] at (-4,-1) {$\mathrm{SU}(3)$ w/ 6 flavors};
\node[] at (-2,-2.5) {$D_2(\mathrm{SU}(5))$};
\node[] at (-6,-2.5) {$\mathrm{U}(2)$ w/ 6 flavors};
\node[] at (-4,-4) {$\mathrm{U}(2)$ w/ 5 flavors};
\draw[->] (-4.5,-1.3)--(-5.5,-2.2);
\draw[->] (-3.5,-1.3)--(-2.5,-2.2);
\draw[->] (-5.5,-2.8)--(-4.5,-3.7);
\draw[->] (-2.5,-2.8)--(-3.5,-3.7);

\node[] at (4,-1) {$\mathrm{SU}(3)$ w/ 6 flavors};
\node[] at (2,-2.5) {$\mathrm{U}(2)$ w/ 6 flavors};
\node[] at (4,-4) {$\mathrm{U}(2)$ w/ 5 flavors};
\draw[->] (3.5,-1.3)--(2.5,-2.2);
\draw[->] (4,-1.3)--(4,-3.7);
\draw[->] (2.5,-2.8)--(3.5,-3.7);
\end{tikzpicture}}
\end{equation}
The corresponding FI deformations leading from $\mathrm{SU}(3)$ SQCD to the $\mathrm{U}(2)$ theory with five flavors are
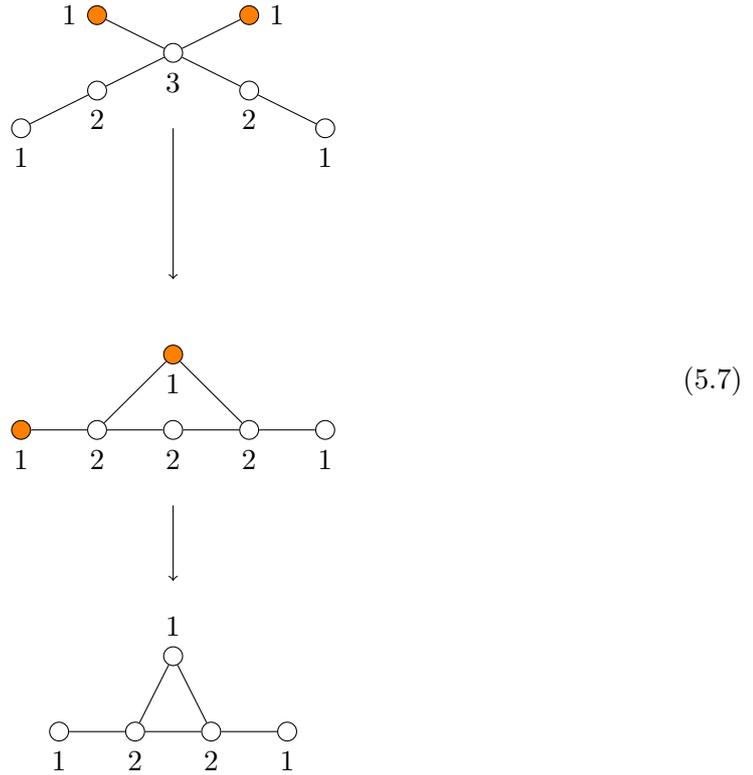
\begin{equation}
   \raisebox{-.5\height}{ \begin{tikzpicture}
            \node[gauge,label=below:{$1$}] (c) at (3,-1) {};
            \node[gauge,label=below:{$2$}] (d) at (4,-0.5) {};
            \node[gauge,label=below:{$3$}] (e) at (5,0) {};
            \node[gauge,label=right:{$1$}, fill=orange] (f) at (6,0.5) {};
             \node[gauge,label=left:{$1$}, fill=orange] (g) at (4,0.5) {};
            \node[gauge,label=below:{$2$}] (h) at (6,-0.5) {};
            \node[gauge,label=below:{$1$}] (i) at (7,-1) {};
             \draw (c)--(d)--(e)--(h)--(i) (e)--(f) (e)--(g);
            \draw[->] (5,-1) to (5,-3); 
            \node[gauge,label=below:{$1$},fill=orange] (c) at (3,-5) {};
            \node[gauge,label=below:{$2$}] (d) at (4,-5) {};
            \node[gauge,label=below:{$2$}] (e) at (5,-5) {};
            \node[gauge,label=below:{$1$},  fill=orange] (f) at (5,-4) {};
            \node[gauge,label=below:{$2$}] (h) at (6,-5) {};
            \node[gauge,label=below:{$1$}] (i) at (7,-5) {};
             \draw (c)--(d)--(e)--(h)--(i) (d)--(f)--(h); 
             \draw[->] (5,-6) to (5,-7);
            \node[gauge,label=below:{$1$}] (c) at (3.5,-9) {};
            \node[gauge,label=below:{$2$}] (d) at (4.5,-9) {};
            \node[gauge,label=above:{$1$}] (f) at (5,-8) {};
            \node[gauge,label=below:{$2$}] (h) at (5.5,-9) {};
            \node[gauge,label=below:{$1$}] (i) at (6.5,-9) {};
             \draw (c)--(d)--(h)--(i) (d)--(f)--(h); 
        \end{tikzpicture}}
\end{equation} 
The phenomenon we have just described is rather common and arises also for rank one SCFTs: The $\mathrm{SU}(2)$ theory with four flavors can be mass deformed either to $D_2(SU(3))$ or to SQED with four flavors (see e.g. \cite{Argyres:2015gha}) and these two theories can both flow to SQED with three flavors. Upon compactification to 3d, it is known that the $D_2(SU(3))$ SCFT reduces to SQED with three flavors \cite{Benvenuti:2017kud, Benvenuti:2017bpg}.

\subsection{Instanton Moduli Spaces}

In this section we will see that via FI deformations the magnetic quivers describing $G$ instanton moduli spaces reduce to those describing nilpotent orbits of $G$. This statement can be understood by considering the cases $G=A,D$: The moduli space of $k$ instantons of $\mathrm{SU}(N)$ is known to coincide with the Higgs branch of $\mathrm{U}(k)$ SQCD with $N$ flavors and one adjoint hypermultiplet. This theory has $\mathrm{SU}(2)\times \mathrm{SU}(N)$ symmetry and the $\mathrm{SU}(2)$ clearly acts on the adjoint field only. By turning on a mass term for $\mathrm{SU}(2)$ we make the adjoint massive and the theory reduces to $\mathrm{U}(k)$ SQCD with $N$ flavors, whose Higgs branch describes the nilpotent orbit of $\mathrm{SU}(N)$ labeled by the partition $[N-k,k]$. 

At the level of magnetic quivers this is implemented as follows: As is well known, the mirror dual of the adjoint SQCD theory is given by the quiver \cite{Intriligator:1996ex} 
 \begin{equation}\label{Acase}
 \raisebox{-.5\height}{\begin{tikzpicture}
            \node[gauge,label=below:{$k$}] (7) at (7,0) {};
            \node[gauge,label=below:{$k$}] (8) at (8,0) {};
            \node (9) at (9,0) {$\cdots$};
            \node[gauge,label=below:{$k$}] (10) at (10,0) {};
            \node[gauge,label=below:{$k$}] (11) at (11,0) {};
            \node[gauge,label=below:{$k$}, fill=orange] (10u) at (9,1) {};
             \node[gauge,label=above:{$1$}, fill=orange] (10uu) at (9,2) {};
            \draw (7)--(8)--(9)--(10)--(11)--(10u)--(7) (10u)--(10uu);
            \draw[snake=brace]  (11.1,-0.75) -- (6.9,-0.75);
   \node[] at (9,-1) {\scriptsize $N-1$};
        \end{tikzpicture}}
        \end{equation}
where we have colored in orange the nodes at which we turn on the FI deformation describing the $\mathrm{SU}(2)$ mass term. The equations of motion can be solved by setting the vev of the $\mathrm{U}(1)\times \mathrm{U}(k)$ bifundamentals $Q$ and $\widetilde{Q}$ to 
\begin{equation} \widetilde{Q}^T=Q=\sqrt{\lambda}(1,1,\dots,1).\end{equation} 
As a result, the vev of the meson in the $\mathrm{U}(k)$ adjoint becomes a $k\times k$ matrix whose entries are all equal to $\lambda$. It is then consistent to assign nilpotent vev to the adjoint mesons built out of $\mathrm{U}(k)\times \mathrm{U}(k)$ bifundamentals on the left and on the right of the top $\mathrm{U}(k)$ gauge node in \eqref{Acase}: one meson will have all entries above the main diagonal equal to $\lambda$ whereas the other will have non-vanishing entries below the main diagonal only. These vevs will propagate along the quiver, partially higgsing the various $\mathrm{U}(k)$ groups. This results in the following RG flow 
\begin{equation}\label{massdefA}
\raisebox{-.5\height}{\begin{tikzpicture}
            \node[gauge,label=below:{$k$}] (7) at (7,0) {};
            \node[gauge,label=below:{$k$}] (8) at (8,0) {};
            \node (9) at (9,0) {$\cdots$};
            \node[gauge,label=below:{$k$}] (10) at (10,0) {};
            \node[gauge,label=below:{$k$}] (11) at (11,0) {};
            \node[gauge,label=below:{$k$}, fill=orange] (10u) at (9,1) {};
             \node[gauge,label=above:{$1$}, fill=orange] (10uu) at (9,2) {};
            \draw (7)--(8)--(9)--(10)--(11)--(10u)--(7) (10u)--(10uu);
            \draw[snake=brace]  (11.1,-0.75) -- (6.9,-0.75);
   \node[] at (9,-1) {\scriptsize $N-1$};
   \draw[->](11,1) -- (12,1); 
             \node[gauge,label=below:{\tiny $1$}] (1) at (12,0) {};
            \node[gauge,label=below:{\tiny $2$}] (2) at (13,0) {};
            \node[] (3) at (14,0) {$\dots$};
            \node[gauge,label=below:{\tiny $k$}] (4) at (15,0) {};
             \node[] (5) at (16,0) {$\dots$};
            \node[gauge,label=below:{\tiny $k$}] (6) at (17,0) {}; 
             \node[] (a) at (18,0) {$\dots$};
            \node[gauge,label=below:{\tiny $2$}] (b) at (19,0) {};
            \node[gauge,label=below:{\tiny $1$}] (c) at (20,0) {};
            \node[gauge,label=above:{\tiny $1$}] (d) at (16,1) {};
            \draw (1)--(2)--(3)--(4)--(5)--(6)--(a)--(b)--(c) (4)--(d)--(6);
            \draw[snake=brace]  (20.1,-0.75) -- (11.9,-0.75);
   \node[] at (16,-1) {\scriptsize $N-1$};
        \end{tikzpicture}}
        \end{equation}
where on the right of \eqref{massdefA} we recognize the magnetic quiver of $\mathrm{U}(k)$ SQCD with $N$ flavors and no adjoints, as expected. 

We can similarly analyze the $k$-instanton moduli space of $\mathrm{SO}(2N)$, which coincides with the Higgs branch of $\mathrm{USp}(2k)$ SQCD with $N$ hypermultiplets in the fundamental representation and one in the rank-2 antisymmetric. As in the previous case, the global symmetry of the theory is $\mathrm{SO}(2N)\times \mathrm{SU}(2)$ where the $\mathrm{SU}(2)$ acts on the antisymmetric only. If we turn on a mass term for the $\mathrm{SU}(2)$ we therefore remove the hyper in the antisymmetric from the infrared spectrum and the theory reduces to the $\mathrm{USp}(2k)$ theory with $N$ fundamentals. 

Let us reproduce this result from the mirror dual side using FI deformations. We focus for simplicity on the case $k=2$. The mirror dual of $\mathrm{USp}(4)$ SQCD with $N>3$ flavors and an antisymmetric hypermultiplet is given by the quiver (see \cite{Intriligator:1996ex} and also \cite{deBoer:1996mp, Kapustin:1998fa} for the corresponding brane construction) 
\begin{equation}\label{USp4}
\raisebox{-.5\height}{\begin{tikzpicture}
            \node[gauge,label=below:{$1$}] (1) at (-2,-1) {};
            \node[gauge,label=below:{$2$}] (2) at (-1,-0.5) {};
            \node (4) at (1,0) {$\cdots$};
            \node[gauge,label=below:{$4$}] (3) at (0,0) {};
            \node[gauge,label=below:{$4$}] (5) at (2,0) {};
            \node[gauge,label=below:{$2$}] (6) at (3,-0.5) {};
            \node[gauge,label=above:{$2$}] (7) at (3,0.5) {};
            \node[gauge,label=above:{$2$}] (8) at (-1,0.5) {};
            \draw (1)--(2)--(3)--(4)--(5)--(6) (7)--(5) (3)--(8);
            \draw[snake=brace]  (-0.1,0.3) -- (2.1,0.3);
   \node[] at (1,0.6) {\scriptsize $N-3$}; 
\end{tikzpicture}}\end{equation} 
and, as in the previous case, turning on a $\mathrm{SU}(2)$ mass term is equivalent to activating a FI deformation at the abelian node and at the $\mathrm{U}(2)$ next to it. As we have seen, this RG flow is described by a D-type quiver subtraction:
\begin{equation}\label{massdefD}\raisebox{-.5\height}{\begin{tikzpicture}
            \node[gauge,label=below:{$1$}, fill=orange] (1) at (-2,-1) {};
            \node[gauge,label=below:{$2$}, fill=orange] (2) at (-1,-0.5) {};
            \node (4) at (1,0) {$\cdots$};
            \node[gauge,label=below:{$4$}] (3) at (0,0) {};
            \node[gauge,label=below:{$4$}] (5) at (2,0) {};
            \node[gauge,label=below:{$2$}] (6) at (3,-0.5) {};
            \node[gauge,label=above:{$2$}] (7) at (3,0.5) {};
            \node[gauge,label=above:{$2$}] (8) at (-1,0.5) {};
            \draw (1)--(2)--(3)--(4)--(5)--(6) (7)--(5) (3)--(8);
            \draw[snake=brace]  (-0.1,0.3) -- (2.1,0.3);
   \node[] at (1,0.6) {\scriptsize $N-3$}; 
   \node[] at (3.7,0) {$-$}; 
   \node[gauge,label=below:{$1$}] (1) at (4,-1) {};
            \node[gauge,label=below:{$2$}] (2) at (5,-0.5) {};
            \node[gauge,label=above:{$1$}] (4) at (5,0.5) {};
            \node[gauge,label=below:{$2$}] (3) at (6,0) {};
            \node[gauge,label=below:{$1$}] (5) at (7,0) {};
            \draw (1)--(2)--(3)--(5) (3)--(4);
            \node[] at (8,0) {$=$};
            \node[gauge,label=above:{$1$}] (a) at (0,-2.5) {};
            \node[gauge,label=below:{$2$}] (b) at (1,-3) {};
            \node[gauge,label=below:{$3$}] (c) at (2,-3) {};
            \node[gauge,label=below:{$4$}] (d) at (3,-3) {};
            \node[gauge,label=above:{$1$}] (g) at (3,-2) {};
            \node (e) at (4,-3) {$\cdots$};
            \node[gauge,label=below:{$4$}] (f) at (5,-3) {};
            \node[gauge,label=above:{$2$}] (h) at (6,-2.5) {};
            \node[gauge,label=below:{$2$}] (l) at (6,-3.5) {};
            \draw[snake=brace]  (5.1,-3.7) -- (2.9,-3.7);
   \node[] at (4,-4) {\scriptsize $N-5$}; 
            \draw (a)--(b)--(c)--(d)--(e)--(f)--(l) (f)--(h) (d)--(g);
\end{tikzpicture}}\end{equation} 
where the bottom quiver in \eqref{massdefD} is the mirror dual of $\mathrm{USp}(4)$ SQCD with $N$ fundamentals as expected (see \cite{Hanany:1999sj}). 

In the exceptional case we have to rely on the magnetic quiver analysis only, since we do not have a Lagrangian description of the theory whose Higgs branch is the instanton moduli space of $E_{6,7,8}$. We consider again the two instanton case and turn on FI deformations which correspond to a mass term for the $\mathrm{SU}(2)$ global symmetry. Also in this case the deformation is implemented via quiver subtraction and we end up with the magnetic quivers associated with the next-to-minimal nilpotent orbits of $E_{6,7,8}$ \cite{Hanany:2017ooe}. 
\begin{equation}\label{massdefE}\begin{tikzpicture}
            \node[gauge,label=below:{$1$}, fill=orange] (1) at (-6,0) {};
            \node[gauge,label=below:{$2$}, fill=orange] (2) at (-5.25,0) {};
            \node[gauge,label=below:{$4$}] (3) at (-4.5,0) {};
            \node[gauge,label=below:{$6$}] (4) at (-3.75,0) {};
            \node[gauge,label=below:{$8$}] (5) at (-3,0) {};
            \node[gauge,label=below:{$10$}] (6) at (-2.25,0) {};
            \node[gauge,label=below:{$12$}] (7) at (-1.5,0) {};
            \node[gauge,label=below:{$8$}] (8) at (-0.75,0) {};
            \node[gauge,label=below:{$4$}] (9) at (0,0) {};
            \node[gauge,label=above:{$6$}] (10) at (-1.5,1) {};
            \draw (1)--(2)--(3)--(4)--(5)--(6)--(7)--(8)--(9) (7)--(10);
            \draw[->] (0,0.5)--(1,0.5); 
            \node[gauge,label=below:{$2$}] (a) at (1,0) {};
            \node[gauge,label=below:{$4$}] (b) at (1.75,0) {};
            \node[gauge,label=below:{$6$}] (c) at (2.5,0) {};
            \node[gauge,label=below:{$8$}] (d) at (3.25,0) {};
            \node[gauge,label=below:{$10$}] (e) at (4,0) {};
            \node[gauge,label=below:{$7$}] (f) at (4.75,0) {};
            \node[gauge,label=below:{$4$}] (g) at (5.5,0) {};
            \node[gauge,label=below:{$1$}] (h) at (6.25,0) {};
            \node[gauge,label=above:{$5$}] (i) at (4,1) {};
            \draw (a)--(b)--(c)--(d)--(e)--(f)--(g)--(h) (e)--(i);
            
            \node[gauge,label=below:{$1$},fill=orange] (A1) at (-7,-3) {};
            \node[gauge,label=below:{$2$},fill=orange] (A2) at (-6,-3) {};
            \node[gauge,label=below:{$4$}] (A3) at (-5,-3) {};
            \node[gauge,label=below:{$6$}] (A4) at (-4,-3) {};
            \node[gauge,label=below:{$8$}] (A5) at (-3,-3) {};
            \node[gauge,label=below:{$6$}] (A6) at (-2,-3) {};
            \node[gauge,label=below:{$4$}] (A7) at (-1,-3) {};
            \node[gauge,label=below:{$2$}] (A8) at (0,-3) {};
            \node[gauge,label=above:{$4$}] (A9) at (-3,-2) {};
            \draw (A1)--(A2)--(A3)--(A4)--(A5)--(A6)--(A7)--(A8) (A5)--(A9);
            \draw[->] (0,-2.5)--(1,-2.5); 
            \node[gauge,label=below:{$2$}] (Aa) at (1,-3) {};
            \node[gauge,label=below:{$4$}] (Ab) at (2,-3) {};
            \node[gauge,label=below:{$6$}] (Ac) at (3,-3) {};
            \node[gauge,label=below:{$5$}] (Ad) at (4,-3) {};
            \node[gauge,label=below:{$4$}] (Ae) at (5,-3) {};
            \node[gauge,label=below:{$2$}] (Af) at (6,-3) {};
            \node[gauge,label=above:{$1$}] (Ag) at (5,-2) {};
            \node[gauge,label=above:{$3$}] (Ah) at (3,-2) {};
            \draw (Aa)--(Ab)--(Ac)--(Ad)--(Ae)--(Af) (Ag)--(Ae) (Ac)--(Ah);
            
             \node[gauge,label=below:{$1$},  fill=orange] (B1) at (-5,-7) {};
            \node[gauge,label=below:{$2$}, fill=orange] (B2) at (-4,-7) {};
            \node[gauge,label=below:{$4$}] (B3) at (-3,-7) {};
            \node[gauge,label=above:{$2$}] (B8) at (-2,-5) {};
            \node[gauge,label=below:{$6$}] (B4) at (-2,-7) {};
            \node[gauge,label=below:{$4$}] (B5) at (-1,-7) {};
            \node[gauge,label=below:{$2$}] (B6) at (0,-7) {};
            \node[gauge,label=left:{$4$}] (B7) at (-2,-6) {};
            \draw (B1)--(B2)--(B3)--(B4)--(B5)--(B6) (B8)--(B7)--(B4);
            \draw[->] (0,-6.5)--(1,-6.5); 
            \node[gauge,label=below:{$2$}] (Ba) at (1,-7) {};
            \node[gauge,label=below:{$4$}] (Bb) at (2,-7) {};
            \node[gauge,label=above:{$3$}] (Bc) at (3,-6.5) {};
            \node[gauge,label=below:{$3$}] (Bd) at (3,-7.5) {};
            \node[gauge,label=above:{$2$}] (Be) at (4,-6.5) {};
            \node[gauge,label=below:{$2$}] (Bf) at (4,-7.5) {};
            \node[gauge,label=below:{$1$}] (Bg) at (5,-7) {};
            \draw (Ba)--(Bb)--(Bc)--(Be)--(Bg) (Bb)--(Bd)--(Bf)--(Bg);
\end{tikzpicture}\end{equation}

\section{Outlook}
\label{sec:Outlook}

In this paper we have seen that we can efficiently study mass deformations of field theories with eight supercharges by considering FI deformations of their magnetic quivers. Remarkably this holds true even when the theory is strongly coupled, providing us with an efficient tool to study RG flows which would be otherwise intractable. As the many examples we have discussed in the paper clearly show, the method is very general and allows us to understand deformations for a wide variety of moduli spaces.

Although we have not found an algorithmic method to solve the algebraic problem in general, the reformulation in terms of quiver subtractions allows us to greatly simplify the analysis and in several cases identify the answer rather easily. Moreover, for a large class of quivers the problem can be addressed by exploiting brane technology, making it much simpler to identify the infrared quiver.

Apart from finding a systematic way of solving the FI-meson problem, there are several interesting directions for future investigations. One is the study of FI deformations for non simply-laced quivers, which have no known Lagrangian description. Sometimes these can be understood by first unfolding the quiver, deforming it and then refolding, although this method is not applicable in general. Studying this problem in detail would definitely shed light on the physics of non simply-laced quivers. 
Another point we did not address in this paper is the role of adjoint matter (or even more general representations) in the quiver when considering FI deformations. This is known to arise e.g. in magnetic quivers for SCFTs in dimensions 4-6. 
A more challenging problem is to generalize the technique to orthosymplectic quivers, where the FI deformation involves monopole operators and therefore does not have a simple Lagrangian description. We hope to come back to some of these problems in the future. 

Finally, we would like to remark that in this work we have primarily focused on understanding the RG flow triggered by a given FI deformation. This however is one side of the problem: In order to study systematically RG flows for SCFTs with eight supercharges with this method, one also needs to understand the correspondence between relevant deformations of the SCFT and FI deformations of the corresponding magnetic quiver. More precisely, it would be important to identify the map between the generalized Coulomb branch (which includes the space parameterized by mass parameters) of the SCFT and the space of FI deformations. In this work we have exploited as a guiding principle to identify the relevant FI deformation the global symmetry preserved along the RG flow. This significantly constrains the set of candidate FI deformations and in many cases allows us to identify it unambiguously, up to the action of the Weyl group of the Coulomb branch global symmetry of the quiver. In general however, this is not enough to fully specify the map we are after. In \cite{vanBeest:2021xyt} a partial answer was provided in the context of relevant deformations of 5d SCFTs, by noticing that the generalized toric polygon describing the 5d theory provides constraints on the choice of FI deformation. It would be important to fill in this gap and provide a general rule to identify the FI deformation describing the RG flow of interest.

\section*{Acknowledgments}
We thank Andrew Dancer, Amihay Hanany and Dan Waldram for discussions.
AB is supported by the ERC Consolidator Grant 772408-Stringlandscape, and by the LabEx ENS-ICFP: ANR-10-LABX-0010/ANR-10-IDEX-0001-02 PSL*.
The work of SG is supported by the INFN grant “Per attivit\`a di formazione per sostenere progetti di ricerca” (GRANT 73/STRONGQFT).
JFG is supported by STFC grant ST/T000791/1.

\appendix

\section{The Meson Propagation Theorem}
\label{app:theorems}

In this Appendix, we prove the central theorem which underlies the FI-Meson problem. It gives a necessary and sufficient condition for two matrices to be expressible as $AB$ and $BA$ respectively. This plays a central role in Section \ref{sec:GeneralMethod}. 

Let $M$ and $\tilde{M}$ be two square matrices with complex coefficients, of respective sizes $m$ and $n$. We write 
\begin{equation}
\label{eq:relation}
M \leftrightsquigarrow \tilde{M}
\end{equation}
if there exists two matrices $A$ and $B$, of sizes respectively $m \times n$ and $n \times m$, such that $M = AB$ and $\tilde{M} = BA$. In the notation of \eqref{eq:notationMesons}, this is the necessary and sufficient condition to be able to write 
\begin{equation}
         \raisebox{-.5\height}{ \begin{tikzpicture}
            \node[gauge,label=below:{$n$}] (2) at (3,0) {};
            \node[gauge,label=below:{$m$}] (3) at (6,0) {};
            \node (3l) at (5,-1) {$\tilde{M}$}; 
            \node (2r) at (4,-1) {$M$};
            \draw[->,dotted] (3l)--(5.5,0);
            \draw[->,dotted] (2r)--(3.5,0);
            \draw (2)--(3);
        \end{tikzpicture} }
\end{equation}

Note that although the relation $\leftrightsquigarrow$ is reflexive and symmetric, it is not transitive, and therefore it does not define an equivalence relation. We begin with three elementary properties. The first one ensures that one can safely reduce $M$ and $\tilde{M}$ to their Jordan normal form, and the second and third one relates the characteristic polynomials and ranks of $M$ and $\tilde{M}$. 

\paragraph{Lemma 1. } \emph{Let $M \in \mathbb{C}^{m \times m}$, $\tilde{M} \in \mathbb{C}^{n \times n}$, $P \in \mathrm{GL}(m,\mathbb{C})$ and $Q \in \mathrm{GL}(n,\mathbb{C})$. We have 
\begin{equation}
M \leftrightsquigarrow \tilde{M} \qquad \Longleftrightarrow   \qquad P^{-1} M P \leftrightsquigarrow Q^{-1} \tilde{M} Q \, . 
\end{equation}}

\textit{Proof: } This follows from the fact that if $M = AB$ and $\tilde{M} = BA$, then $P^{-1} M P = (P^{-1} A Q)(Q^{-1} B P)$ and $Q^{-1} \tilde{M} Q = (Q^{-1} B P)(P^{-1} A Q)$.  

\paragraph{Lemma 2. } \emph{ Let $M \in \mathbb{C}^{m \times m}$, $\tilde{M} \in \mathbb{C}^{n \times n}$. If $M \leftrightsquigarrow \tilde{M}$ then $x^n \chi_M (x) = x^m \chi_{\tilde{M}} (x)$. }

\textit{Proof: } This is a consequence of $\det (AB) = \det (BA)$. 

\paragraph{Lemma 3. } \emph{ Let $A \in \mathbb{C}^{m \times n}$, $B \in \mathbb{C}^{n \times m}$. Then 
\begin{equation}
2 r (BA) - n \leq r(AB) \, . 
\end{equation}}

\textit{Proof: } This follows from Sylvester's rank inequality, which states that $r(A) + r(B) - n \leq r(AB)$. Since we also know that $r(BA) \leq r(A)$ and $r(BA) \leq r(B)$, the lemma follows. 

\vspace{1cm}

We now turn to the central point of the argument. The previous lemmas show how certain conjugacy invariant quantities (eigenvalues, rank) behave with respect to the relation \eqref{eq:relation}. It turns out one can entirely characterize this relation by inspecting the Jordan decompositions. 

\paragraph{Lemma 4. } \emph{Let $M \in \mathbb{C}^{m \times m}$, $\tilde{M} \in \mathbb{C}^{n \times n}$ with $m \leq n$. Assume that $M$ has Jordan decomposition $J_m (\lambda)$. 
\begin{itemize}
\item If $\lambda \neq 0$ then:  $M \leftrightsquigarrow \tilde{M}$ if and only if $\tilde{M}$ has Jordan decomposition $J_m (\lambda) \oplus (J_1(0))^{\oplus n-m}$
\item If $\lambda = 0$ then: :  $M \leftrightsquigarrow \tilde{M}$ if and only if $\tilde{M}$ has Jordan decomposition $J_m (0) \oplus (J_1(0))^{\oplus n-m}$ or $J_{m \pm 1} (0) \oplus (J_1(0))^{\oplus n-m \mp 1}$. 
\end{itemize}}

\textit{Proof: } Using Lemma 1, we can assume the matrices $\tilde{M}$ and $M$ are in the form of Jordan blocks. Let $A \in \mathbb{C}^{m \times n}$, $B \in \mathbb{C}^{n \times m}$ such that $M = AB$ and $\tilde{M} = BA$. We introduce the decomposition
\begin{equation}
A = (A_1 , A_2) \, , \qquad B = \left( \begin{array}{c} B_1 \\ B_2

\end{array} \right) 
\end{equation}
with $A_1 , B_1 \in \mathbb{C}^{m \times m}$. Then 
\begin{equation}
M = AB = (A_1 B_1 + A_2 B_2) \, , \qquad N = BA = \left( \begin{array}{cc}
B_1 A_1 & B_1 A_2 \\ B_2 A_1 & B_2 A_2
\end{array} \right) \, . 
\end{equation}
\begin{itemize}
\item We start with the case $\lambda \neq 0$. Using Lemma 2, we know that the eigenvalues of $\tilde{M}$ are $\lambda$ with multiplicity $m$ and $0$ with multiplicity $n-m$. Therefore we know that $\tilde{M}$, which is in Jordan form, is block diagonal with blocks of sizes $m$ and $n-m$. 
We get that $B_1 A_1 = J_{m} (\lambda)$ is a non-singular matrix since $\lambda \neq 0$, therefore $A_1$ is non-singular. Since $B_2 A_1 = 0$ because of the block diagonal form of $N$, this implies $B_2 = 0$, and therefore $B_2 A_2 = 0$. We conclude $\tilde{M} = J_m (\lambda) \oplus (J_1(0))^{\oplus n-m}$. The converse is deduced from an elementary computation. 
\item Let us now move on to $\lambda =0$. Now both $\tilde{M}$ and $M$ are nilpotent. Using Lemma 3, we have 
\begin{equation}
m-2 = 2 r(AB) - m \leq r(BA) \leq m \, , 
\end{equation}
so\footnote{One can also argue with the nilpotency index: if $k_M$ is the nilpotency index of $M$ and $k_{\tilde{M}}$ the nilpotency index of $\tilde{M}$, we have $\tilde{M}^{k_M +1} = B M^{k_M} A = 0$ so $k_{\tilde{M}} \leq k_M +1$. Conversely, $k_{M} \leq k_{\tilde{M}} +1$, so $|k_M - k_{\tilde{M}} |\leq 1$. } 
\begin{equation}
\label{eq:rankInequality}
|r(M) - r(\tilde{M})| \leq 1 \, . 
\end{equation}
Let's now have a closer look at the maps involved: 
\begin{equation}
\raisebox{-.5\height}{\begin{tikzpicture}
\node (1) at (0,0) {$\mathbb{C}^n$};
\node (2) at (2,0) {$\mathbb{C}^n$};
\node (3) at (4,0) {$\mathbb{C}^n$};
\node (4) at (1,2) {$\mathbb{C}^m$};
\node (5) at (3,2) {$\mathbb{C}^m$};
\draw[->] (1) to node[below,midway] {$\tilde{M}$} (2);
\draw[->] (2) to node[below,midway] {$\tilde{M}$} (3);
\draw[->] (4) to node[above,midway] {$M$} (5);
\draw[->] (1) to node[left,midway] {$A$} (4);
\draw[->] (2) to node[left,midway] {$A$} (5);
\draw[->] (4) to node[left,midway] {$B$} (2);
\draw[->] (5) to node[left,midway] {$B$} (3);
\end{tikzpicture}}
\end{equation}
Since the rank of $M$ is $m-1$, $r(ABA) \geq r(A) -1$. On the other hand $r(ABA) < r(A)$ (otherwise $BA=\tilde{M}$ would be invertible). Therefore $r(ABA) = r(A)-1$, so $\mathrm{Im}(ABA)$ is a vector subspace of $\mathrm{Im}(A)$ with codimension 1. Therefore $r(BA) - r(BABA)$ is either 0 or 1. It can't be 0 as $\tilde{M}=BA$ is nilpotent, so it is 1. This means that $r(\tilde{M}^2) = r(\tilde{M}) -1$, so there is at most one Jordan block in $\tilde{M}$ with size $>1$. From the inequality \eqref{eq:rankInequality}, this block has size $m-1$, $m$ or $m+1$, which proves the lemma (again, the converse is deduced from an elementary computation). 
\end{itemize}

From this, one obtains a complete characterization of the relation \eqref{eq:relation} as follows. 

\paragraph{Theorem 1. }\label{theorem1} \emph{Let $M \in \mathbb{C}^{m \times m}$, $\tilde{M} \in \mathbb{C}^{n \times n}$ with $m \leq n$. Assume that $M$ has a block decomposition $M_{\mathbf{p}} \oplus \bigoplus J_{m_i} (\lambda_i)$, with $M_{\mathbf{p}}$ nilpotent in the orbit $\mathcal{O}_{\mathbf{p}}$ and $\lambda_i \neq 0$ and $\tilde{M}$ has a block decomposition $\tilde{M}_{\mathbf{q}} \oplus \bigoplus J_{n_j} (\mu_j)$, with $\tilde{M}_{\mathbf{q}}$ nilpotent in the orbit $\mathcal{O}_{\mathbf{q}}$ and $\mu_j \neq 0$. \\  Then $M \leftrightsquigarrow \tilde{M}$ if and only if  up to reordering $m_i = n_j$ and $\lambda_i = \mu_j$ and the partitions $\mathbf{p}$ and $\mathbf{q}$ are related by 
\begin{equation}
| \mathbf{p}_k - \mathbf{q}_k | \leq 1
\end{equation}
for a certain ordering of the entries, and including zeros. } \\ 

In other words, the invertible Jordan blocks stay the same, and the nilpotent Jordan block can see their sizes change by at most one unit. 

\subsection*{Examples}

To illustrate let us give a few examples. The following transition is allowed:
\begin{equation}
      \begin{tikzpicture}
            \node[gauge,label=below:{$3$}] (3) at (6,0) {};
            \node[gauge,label=below:{$2$}] (4) at (9,0) {};
            \draw (3)--(4);
            \node (3r) at (6.3,-2) {\scalebox{.8}{$  \left(\begin{array}{ccc}
            -4/3&0&-1\\0&1&0\\4/9&0&1/3
            \end{array}\right)$}};
            \node (4l) at (8.7,-2) {\scalebox{.8}{$  \left(\begin{array}{cc}
            0&1\\1&0
            \end{array}\right)$}};
            \draw[->,dotted] (3r)--(6.5,0);
            \draw[->,dotted] (4l)--(8.5,0);
        \end{tikzpicture} 
\end{equation}
as the Jordan decompositions are respectively $J_1(1) \oplus J_1 (-1) \oplus J_1 (0)$ and $J_1(1) \oplus J_1 (-1)$. This particular example appears in equation \eqref{eq:exEa}. 

\section{Weyl orbits}\label{Weylsec}

In this section we consider framed quivers with a set of balanced gauge nodes. For framed quivers there is no condition \eqref{eq:FI_cond} on the FI parameters. If the quiver is viewed in it's unframed version, then the additional U$(1)$ gauge node will carry an FI parameter specified by solving \eqref{eq:FI_cond}, which could either be zero on non-zero.

For a balanced subset of nodes in the shape of the Dynkin diagram of an algebra $g$ we find, that several choices of FI parameters lead to the same deformation of the Higgs branch. The equivalent choices are related by an action of the Weyl group of $g$. This is visible particularly well from the linking number of NS5 branes in a Hanany-Witten system.

\paragraph{Example 1.} Let us consider a specific example, the quiver:
\begin{equation}
    \begin{tikzpicture}
            \node[gauge,label=below:{$1$}] (1) at (1,0) {};
            \node[gauge,label=below:{$2$}] (2) at (2,0) {};
            \node[gauge,label=below:{$3$}] (3) at (3,0) {};
            \node[flavour,label=above:{$4$}] (f3) at (3,1) {};
            \draw (1)--(2)--(3)--(f3);
            \draw[blue] \convexpath{1,3}{0.2cm};
    \end{tikzpicture}\;.
\end{equation}
The balanced nodes form an $A_3$ Dynkin diagram, circled in blue.

We can now turn on an FI term $\lambda$ on the U$(1)$ gauge node, which corresponds to the weight $[1,0,0]$ of the $A_3$ algebra.
\begin{equation}
    \begin{tikzpicture}
            \node[gaugeo,label=below:{$1$},label=above:{$\lambda$}] (1) at (1,0) {};
            \node[gauge,label=below:{$2$}] (2) at (2,0) {};
            \node[gauge,label=below:{$3$}] (3) at (3,0) {};
            \node[flavour,label=above:{$4$}] (f3) at (3,1) {};
            \draw (1)--(2)--(3)--(f3);
    \end{tikzpicture}\qquad\longleftrightarrow\qquad [1,0,0]
\end{equation}
The Weyl group orbit of this weight is
\begin{equation}
    \label{eq:orbitsA3}
    \{\;[1,0,0]\;,\;[-1,1,0]\;,\;[0,-1,1]\;,\;[0,0,-1]\;\}\;.
\end{equation}
This implies that the following FI deformations are equivalent:
\begin{equation}
    \begin{tikzpicture}
            \node[gaugeo,label=below:{$1$},label=above:{$\lambda$}] (1) at (1,0) {};
            \node[gauge,label=below:{$2$}] (2) at (2,0) {};
            \node[gauge,label=below:{$3$}] (3) at (3,0) {};
            \node[flavour,label=above:{$4$}] (f3) at (3,1) {};
            \draw (1)--(2)--(3)--(f3);
    \end{tikzpicture}\quad,\quad\begin{tikzpicture}
            \node[gaugeo,label=below:{$1$},label=above:{$-\lambda$}] (1) at (1,0) {};
            \node[gaugeo,label=below:{$2$},label=above:{$\lambda$}] (2) at (2,0) {};
            \node[gauge,label=below:{$3$}] (3) at (3,0) {};
            \node[flavour,label=above:{$4$}] (f3) at (3,1) {};
            \draw (1)--(2)--(3)--(f3);
    \end{tikzpicture}\quad,\quad\begin{tikzpicture}
            \node[gauge,label=below:{$1$}] (1) at (1,0) {};
            \node[gaugeo,label=below:{$2$},,label=above:{$-\lambda$}] (2) at (2,0) {};
            \node[gaugeo,label=below:{$3$},label=right:{$\lambda$}] (3) at (3,0) {};
            \node[flavour,label=above:{$4$}] (f3) at (3,1) {};
            \draw (1)--(2)--(3)--(f3);
    \end{tikzpicture}\quad,\quad\begin{tikzpicture}
            \node[gauge,label=below:{$1$}] (1) at (1,0) {};
            \node[gauge,label=below:{$2$}] (2) at (2,0) {};
            \node[gaugeo,label=below:{$3$},label=right:{$-\lambda$}] (3) at (3,0) {};
            \node[flavour,label=above:{$4$}] (f3) at (3,1) {};
            \draw (1)--(2)--(3)--(f3);
    \end{tikzpicture}
\end{equation}
We can use the brane system to check that all choices of FI parameters indeed lead to the same answer.
The brane system with FI parameters turned off, in the Coulomb phase, is
\begin{equation}
    \begin{tikzpicture}
        \draw[red] (-3,-1)--(-3,1) (-2,-1)--(-2,1) (-1,-1)--(-1,1) (2.5,-1)--(2.5,1);
        \draw[blue] (-0.5,-0.25)--(0.5,0.75) (0,-0.25)--(1,0.75) (0.5,-0.25)--(1.5,0.75) (1,-0.25)--(2,0.75);
        \draw (-3,-0.7)--(-2,-0.7);
        \draw (-2,-0.6)--(-1,-0.6) (-2,-0.8)--(-1,-0.8);
        \draw (-1,-0.5)--(2.5,-0.5) (-1,-0.7)--(2.5,-0.7) (-1,-0.9)--(2.5,-0.9);
    \end{tikzpicture}\;.
\end{equation}
In order to do the FI deformation, we move to the Higgs phase
\begin{equation}
    \begin{tikzpicture}
        \draw[red] (-3,-1)--(-3,1) (-2,-1)--(-2,1) (-1,-1)--(-1,1) (2.5,-1)--(2.5,1);
        \draw[blue] (-0.5,-0.25)--(0.5,0.75) (0,-0.25)--(1,0.75) (0.5,-0.25)--(1.5,0.75) (1,-0.25)--(2,0.75);
        \draw (-3,0.25)--(0,0.25) (-2,0.35)--(0.1,0.35) (-1,0.45)--(0.2,0.45);
        \draw (1.5,0.25)--(2.5,0.25) (1.1,0.35)--(2.5,0.35) (0.7,0.45)--(2.5,0.45);
        \draw[transform canvas={xshift=-0.1cm,yshift=-0.1cm}] (0,0.25)--(0.5,0.25);
        \draw[transform canvas={xshift=-0.25cm,yshift=-0.25cm}] (0,0.25)--(0.5,0.25);
        \draw[transform canvas={xshift=-0.4cm,yshift=-0.4cm}] (0,0.25)--(0.5,0.25);
        \draw[transform canvas={xshift=-0.175cm,yshift=-0.175cm}] (0.5,0.25)--(1,0.25);
        \draw[transform canvas={xshift=-0.325cm,yshift=-0.325cm}] (0.5,0.25)--(1,0.25);
        \draw[transform canvas={xshift=-0.25cm,yshift=-0.25cm}] (1,0.25)--(1.5,0.25);
    \end{tikzpicture}\;.
\end{equation}
For a more symmetric picture we perform a Hanany-Witten transition.
\begin{equation}
    \begin{tikzpicture}
        \draw[red] (-4,-1)--(-4,1) (-3,-1)--(-3,1) (-2,-1)--(-2,1) (-1,-1)--(-1,1);
        \draw[blue] (-0.5,-0.25)--(0.5,0.75) (0,-0.25)--(1,0.75) (0.5,-0.25)--(1.5,0.75) (1,-0.25)--(2,0.75);
        \draw (-3,0.25)--(0,0.25) (-2,0.35)--(0.1,0.35) (-1,0.45)--(0.2,0.45);
        \draw (-4,0.15)--(-0.1,0.15);
        \draw[transform canvas={xshift=-0.1cm,yshift=-0.1cm}] (0,0.25)--(0.5,0.25);
        \draw[transform canvas={xshift=-0.25cm,yshift=-0.25cm}] (0,0.25)--(0.5,0.25);
        \draw[transform canvas={xshift=-0.4cm,yshift=-0.4cm}] (0,0.25)--(0.5,0.25);
        \draw[transform canvas={xshift=-0.175cm,yshift=-0.175cm}] (0.5,0.25)--(1,0.25);
        \draw[transform canvas={xshift=-0.325cm,yshift=-0.325cm}] (0.5,0.25)--(1,0.25);
        \draw[transform canvas={xshift=-0.25cm,yshift=-0.25cm}] (1,0.25)--(1.5,0.25);
    \end{tikzpicture}\;.
\end{equation}
An FI deformation corresponds to moving an NS5 brane. The four choices in \eqref{eq:orbitsA3} correspond to moving one of the four NS5 branes (and the frozen D3 brane connected to it). No matter which NS5 brane is sent away to infinity, one obtains the same brane system:
\begin{equation}
    \begin{tikzpicture}
        \draw[red] (-3,-1)--(-3,1) (-2,-1)--(-2,1) (-1,-1)--(-1,1);
        \draw[blue] (-0.5,-0.25)--(0.5,0.75) (0,-0.25)--(1,0.75) (0.5,-0.25)--(1.5,0.75) (1,-0.25)--(2,0.75);
        \draw (-3,0.25)--(0,0.25) (-2,0.35)--(0.1,0.35) (-1,0.45)--(0.2,0.45);
        \draw[transform canvas={xshift=-0.1cm,yshift=-0.1cm}] (0,0.25)--(0.5,0.25);
        \draw[transform canvas={xshift=-0.25cm,yshift=-0.25cm}] (0,0.25)--(0.5,0.25);
        \draw[transform canvas={xshift=-0.4cm,yshift=-0.4cm}] (0,0.25)--(0.5,0.25);
        \draw[transform canvas={xshift=-0.175cm,yshift=-0.175cm}] (0.5,0.25)--(1,0.25);
        \draw[transform canvas={xshift=-0.325cm,yshift=-0.325cm}] (0.5,0.25)--(1,0.25);
        \draw[transform canvas={xshift=-0.25cm,yshift=-0.25cm}] (1,0.25)--(1.5,0.25);
    \end{tikzpicture}\;.
\end{equation}
Moving to the (partial) Coulomb phase, after performing some Hanany-Witten transitions
\begin{equation}
    \begin{tikzpicture}
        \draw[red] (-2,-1)--(-2,1) (-1,-1)--(-1,1) (2,-1)--(2,1);
        \draw[blue] (-0.5,-0.25)--(0.5,0.75) (0,-0.25)--(1,0.75) (0.5,-0.25)--(1.5,0.75) (2.5,-0.25)--(3.5,0.75);
        \draw (-2,-0.7)--(-1,-0.7);
        \draw (-1,-0.6)--(2,-0.6) (-1,-0.8)--(2,-0.8);
        \draw[transform canvas={xshift=-0.1cm,yshift=-0.1cm}] (0,0.25)--(0.5,0.25);
        \draw[transform canvas={xshift=-0.2cm,yshift=-0.2cm}] (0.5,0.25)--(1,0.25);
        \draw[transform canvas={xshift=-0.3cm,yshift=-0.3cm}] (1,0.25)--(3,0.25);
    \end{tikzpicture}\;,
\end{equation}
we can read off the resulting theory
\begin{equation}
    \begin{tikzpicture}
            \node[gauge,label=below:{$1$}] (2) at (2,0) {};
            \node[gauge,label=below:{$2$}] (3) at (3,0) {};
            \node[flavour,label=above:{$3$}] (f3) at (3,1) {};
            \draw (2)--(3)--(f3);
            \node at (5,0.5) {$+$ $3$ free hypers};
    \end{tikzpicture}\;.
\end{equation}

\paragraph{Example 2.} Based on the last example one could ask, whether it never matters which NS5 brane one moves in a brane system. This is obviously not the case, as we demonstrate with another simple example. Consider the theory
\begin{equation}
    \begin{tikzpicture}
            \node[gauge,label=below:{$1$}] (1) at (1,0) {};
            \node[gauge,label=below:{$2$}] (2) at (2,0) {};
            \node[gauge,label=below:{$3$}] (3) at (3,0) {};
            \node[flavour,label=above:{$5$}] (f3) at (3,1) {};
            \draw (1)--(2)--(3)--(f3);
            \draw[blue] \convexpath{1,2}{0.2cm};
    \end{tikzpicture}\;.
\end{equation}
The set of balanced nodes, circled in blue, make up an $A_2$ Dynkin diagram. Let us again turn on an FI term on the first node, corresponding to the weight $[1,0]$ of $A_2$. The Weyl group orbit of this weight is
\begin{equation}
    \{\;[1,0]\;,\;[-1,1]\;,\;[0,-1]\;\}\;.
\end{equation}
The brane system with FI parameters turned off, in the Coulomb phase, is
\begin{equation}
    \begin{tikzpicture}
        \draw[red] (-3,-1)--(-3,1) (-2,-1)--(-2,1) (-1,-1)--(-1,1) (3,-1)--(3,1);
        \draw[blue] (-0.5,-0.25)--(0.5,0.75) (0,-0.25)--(1,0.75) (0.5,-0.25)--(1.5,0.75) (1,-0.25)--(2,0.75) (1.5,-0.25)--(2.5,0.75);
        \draw (-3,-0.7)--(-2,-0.7);
        \draw (-2,-0.6)--(-1,-0.6) (-2,-0.8)--(-1,-0.8);
        \draw (-1,-0.5)--(3,-0.5) (-1,-0.7)--(3,-0.7) (-1,-0.9)--(3,-0.9);
    \end{tikzpicture}\;.
\end{equation}
We can again move to the Higgs phase
\begin{equation}
    \begin{tikzpicture}
        \draw[red] (-3,-1)--(-3,1) (-2,-1)--(-2,1) (-1,-1)--(-1,1) (3,-1)--(3,1);
        \draw[blue] (-0.5,-0.25)--(0.5,0.75) (0,-0.25)--(1,0.75) (0.5,-0.25)--(1.5,0.75) (1,-0.25)--(2,0.75) (1.5,-0.25)--(2.5,0.75);
        \draw (-3,0.25)--(0,0.25) (-2,0.35)--(0.1,0.35) (-1,0.45)--(0.2,0.45);
        \draw (2,0.25)--(3,0.25) (1.6,0.35)--(3,0.35) (1.2,0.45)--(3,0.45);
        \draw[transform canvas={xshift=-0.15cm,yshift=-0.15cm}] (0,0.25)--(0.5,0.25);
        \draw[transform canvas={xshift=-0.3cm,yshift=-0.3cm}] (0,0.25)--(0.5,0.25);
        \draw[transform canvas={xshift=-0.45cm,yshift=-0.45cm}] (0,0.25)--(0.5,0.25);
        \draw[transform canvas={xshift=-0.1cm,yshift=-0.1cm}] (0.5,0.25)--(1,0.25);
        \draw[transform canvas={xshift=-0.25cm,yshift=-0.25cm}] (0.5,0.25)--(1,0.25);
        \draw[transform canvas={xshift=-0.4cm,yshift=-0.4cm}] (0.5,0.25)--(1,0.25);
        \draw[transform canvas={xshift=-0.175cm,yshift=-0.175cm}] (1,0.25)--(1.5,0.25);
        \draw[transform canvas={xshift=-0.325cm,yshift=-0.325cm}] (1,0.25)--(1.5,0.25);
        \draw[transform canvas={xshift=-0.25cm,yshift=-0.25cm}] (1.5,0.25)--(2,0.25);
    \end{tikzpicture}\;,
\end{equation}
and perform a Hanany-Witten transition, bringing all NS5 branes to the left
\begin{equation}
    \begin{tikzpicture}
        \draw[red] (-4,-1)--(-4,1) (-3,-1)--(-3,1) (-2,-1)--(-2,1) (-1,-1)--(-1,1);
        \draw[blue] (-0.5,-0.25)--(0.5,0.75) (0,-0.25)--(1,0.75) (0.5,-0.25)--(1.5,0.75) (1,-0.25)--(2,0.75) (1.5,-0.25)--(2.5,0.75);
        \draw (-4,0.25)--(0,0.25) (-3,0.35)--(0.1,0.35) (-2,0.45)--(0.2,0.45);
        \draw (-1,0.55)--(0.3,0.55) (-1,0.65)--(0.9,0.65);
        \draw[transform canvas={xshift=-0.15cm,yshift=-0.15cm}] (0,0.25)--(0.5,0.25);
        \draw[transform canvas={xshift=-0.3cm,yshift=-0.3cm}] (0,0.25)--(0.5,0.25);
        \draw[transform canvas={xshift=-0.45cm,yshift=-0.45cm}] (0,0.25)--(0.5,0.25);
        \draw[transform canvas={xshift=-0.1cm,yshift=-0.1cm}] (0.5,0.25)--(1,0.25);
        \draw[transform canvas={xshift=-0.25cm,yshift=-0.25cm}] (0.5,0.25)--(1,0.25);
        \draw[transform canvas={xshift=-0.4cm,yshift=-0.4cm}] (0.5,0.25)--(1,0.25);
        \draw[transform canvas={xshift=-0.175cm,yshift=-0.175cm}] (1,0.25)--(1.5,0.25);
        \draw[transform canvas={xshift=-0.325cm,yshift=-0.325cm}] (1,0.25)--(1.5,0.25);
        \draw[transform canvas={xshift=-0.25cm,yshift=-0.25cm}] (1.5,0.25)--(2,0.25);
    \end{tikzpicture}\;.
\end{equation}
Here we see that moving the first three NS5 branes away leads to the same brane system. However, moving the last NS5 brane leads to a different brane system.\\

\paragraph{Applications:} The Weyl orbit can be used to simplify an FI deformation. Take for example the quiver and FI deformation
\begin{equation}
\label{eq:WeylOrbitD6Ex1}
    \begin{tikzpicture}
            \node[gauge,label=below:{$1$}] (1) at (1,0) {};
            \node[gaugeo,label=below:{$2$},label=above:{$\lambda$}] (2) at (2,0) {};
            \node[gaugeo,label=below:{$3$},label=above:{$-\lambda$}] (3) at (3,0) {};
            \node[gauge,label=below:{$4$}] (4) at (4,0) {};
            \node[gaugeo,label=left:{$1$},label=above:{$\lambda$}] (4u) at (4,1) {};
            \node[gauge,label=right:{$2$}] (5u) at (5,0.5) {};
            \node[gauge,label=right:{$2$}] (5d) at (5,-0.5) {};
            \draw (1)--(2)--(3)--(4)--(5u) (5d)--(4)--(4u);
    \end{tikzpicture}
\end{equation}
We can frame the quiver in the following way
\begin{equation}
    \begin{tikzpicture}
            \node[gauge,label=below:{$1$}] (1) at (1,0) {};
            \node[gaugeo,label=below:{$2$},label=above:{$\lambda$}] (2) at (2,0) {};
            \node[gaugeo,label=below:{$3$},label=above:{$-\lambda$}] (3) at (3,0) {};
            \node[gauge,label=below:{$4$}] (4) at (4,0) {};
            \node[flavour,label=left:{$1$}] (4u) at (4,1) {};
            \node[gauge,label=right:{$2$}] (5u) at (5,0.5) {};
            \node[gauge,label=right:{$2$}] (5d) at (5,-0.5) {};
            \draw (1)--(2)--(3)--(4)--(5u) (5d)--(4)--(4u);
    \end{tikzpicture}
\end{equation}
leaving intact the subset of balanced gauge nodes which make up a $D_6$ Dynkin diagram. The FI deformation corresponds to the weight $[0,1,-1,0,0,0]$. In the same Weyl orbit we have the weight $[1,0,0,0,0,0]$. Therefore we have an equivalent FI deformation
\begin{equation}
    \begin{tikzpicture}
            \node[gaugeo,label=below:{$1$},label=above:{$\lambda$}] (1) at (1,0) {};
            \node[gauge,label=below:{$2$}] (2) at (2,0) {};
            \node[gauge,label=below:{$3$}] (3) at (3,0) {};
            \node[gauge,label=below:{$4$}] (4) at (4,0) {};
            \node[flavour,label=left:{$1$}] (4u) at (4,1) {};
            \node[gauge,label=right:{$2$}] (5u) at (5,0.5) {};
            \node[gauge,label=right:{$2$}] (5d) at (5,-0.5) {};
            \draw (1)--(2)--(3)--(4)--(5u) (5d)--(4)--(4u);
    \end{tikzpicture}
\end{equation}
Which in its unframed form is
\begin{equation}
\label{eq:WeylOrbitD6Ex2}
    \begin{tikzpicture}
            \node[gaugeo,label=below:{$1$},label=above:{$\lambda$}] (1) at (1,0) {};
            \node[gauge,label=below:{$2$}] (2) at (2,0) {};
            \node[gauge,label=below:{$3$}] (3) at (3,0) {};
            \node[gauge,label=below:{$4$}] (4) at (4,0) {};
            \node[gaugeo,label=left:{$1$},label=above:{$-\lambda$}] (4u) at (4,1) {};
            \node[gauge,label=right:{$2$}] (5u) at (5,0.5) {};
            \node[gauge,label=right:{$2$}] (5d) at (5,-0.5) {};
            \draw (1)--(2)--(3)--(4)--(5u) (5d)--(4)--(4u);
    \end{tikzpicture}
\end{equation}
Indeed one can check that both FI deformations \eqref{eq:WeylOrbitD6Ex1} and \eqref{eq:WeylOrbitD6Ex2} are equivalent. Hence the action of the Weyl group can turn seemingly more complicated FI deformations into simpler ones, involving less nodes.

\subsection*{Linear Quivers}

We can do the general case of linear framed quivers. Let $n$ be a positive integer, we consider a linear framed quiver with gauge nodes $\mathrm{U}(k_i)$ and flavors $N_i$ for $i=1 , \dots , n$. This defines two partitions as follows: 
\begin{equation}
    \lambda_i = k_{i+1} - k_i + \sum\limits_{j=1}^i N_j
\end{equation}
\begin{equation}
    \mu^T_i = \sum\limits_{j=1}^{n+1-i} N_j \, . 
\end{equation}
These correspond to linking numbers of branes: the $\lambda_i$ ($i=1 , \dots , n$) are the linking numbers of the NS5 and the $\mu_j$ ($j=1, \dots , n^T$). Turning on FI corresponds to picking one NS5 and moving it out of the local brane system. Moving two branes with the same linking number gives equivalent FI deformations. In terms of partitions, we choose an integer $\ell$ in the partition $\lambda$. The resulting theory is defined by $\lambda '$, $\mu '$ where $\lambda '$ is obtained from $\lambda$ by removing one occurrence of $\ell$ and $\mu' = \mu - [1^{\ell} , 0 , \dots]$.

 \section{Hilbert Series Check of the Mirror Pair}
 \label{app:Mirror}
    
In this appendix we provide a computational check of the mirror pair  \eqref{mirrornew}-\eqref{newmirror} by checking the equality of the Hilbert series for the Higgs and Coulomb branches of both quivers. The first equality is 
\begin{equation}
\mathrm{HS} ( \mathcal{C} [ \ref{mirrornew} ] ; t ) = \mathrm{HS} (  \mathcal{H} [ \ref{newmirror} ] ; t ) \, , 
\end{equation} 
and it is straightforward to check, as the 1-form symmetry gauging in \eqref{newmirror} does not affect this Hilbert series, and one can use the standard hyper-K\"ahler quotient. We focus instead here on the other check, involving the Coulomb branch Hilbert series of \eqref{newmirror}, which is sensitive to the quotient. 
The computation of $\mathrm{HS} (  \mathcal{C} [ \ref{newmirror} ] ; t )$ uses the monopole formula. We spell out the details, as there are subtleties due to the SU gauge groups and the gauging of 1-form symmetry. 

\paragraph{Hilbert series before gauging the 1-form symmetry. }
Consider first the $\mathfrak{su}(N)$ algebra, with fundamental weights $(\varpi_1 , \dots , \varpi_{N-1})$. We parametrize its coweights, which are identified to the weights as the algebra is simply laced, by $\varpi = a_1 \varpi_1 + \dots + a_{N-1} \varpi_{N-1}$ with $a_i \in \mathbb{Z}$. The fundamental Weyl chamber is given by $a_i \in \mathbb{N}$. We parametrize coweights of the $\mathfrak{u}(1)$ algebra by $b \in \mathbb{Z}$, and the coweights of $\mathfrak{su}(N)$ by $c_i \in \mathbb{Z}$. The Hilbert series for the quiver before gauging the 1-form symmetry is
\begin{eqnarray}
& & \sum\limits_{a_1 , \dots , a_{N-1} \in \mathbb{N}} \sum\limits_{b \in \mathbb{Z}}  \sum\limits_{c_1 , \dots , c_{M-1} \in \mathbb{N}} t^{\Delta (a,b,c)}  \frac{P^{\mathrm{SU}} (a ; t) P^{\mathrm{SU}} (c ; t)}{1-t^2}\nonumber \\ & &  \hspace{4cm} \times   \delta \left( N| \sum\limits_{i=1}^{N-1} i a_i  \right)  \delta \left( M| \sum\limits_{i=1}^{M-1} i c_i \right)\, ,  
\end{eqnarray}
where $\Delta (a,b,c)$ is the monopole operator conformal dimension \cite{Cremonesi:2013lqa}, $P^{\mathrm{SU}} (a ; t)$ is a dressing factor defined below, and $\delta_{N| \sum i a_i}  \delta_{M| \sum i c_i}$ imposes that the sum is restricted to the weights of the Langlands dual groups $\mathrm{PSU}(N)$ and $\mathrm{PSU}(M)$, i.e. the root lattice. The dressing factor $P^{\mathrm{SU}} (a ; t)$ depends only on which of the $a_i$ are vanishing, encoding which facet of the Weyl chamber the weight belongs to. If there are $r$ strings of zeros of lengths $l_1 , \dots , l_r$ in the string $a_1 , \dots , a_{N-1}$ then 
\begin{equation}
P^{\mathrm{SU}} (a ; t) = \mathrm{PE} \left[ \left(N-1 - \sum\limits_{i=1}^r l_i \right) t^2 + \sum\limits_{i=1}^r \sum\limits_{j=1}^{l_i} t^{2+2j} \right] \, . 
\end{equation}

\paragraph{Gauging the 1-form symmetry. }
In order to implement gauging the 1-form symmetry, we use the embedding: 
\begin{equation}
\begin{array}{ccc}
\mathbb{Z} & \rightarrow & Z \left(  \mathrm{SU}(N) \times \mathrm{U}(1) \times   \mathrm{SU}(M)  \right)  \\ 
 I & \mapsto & (e^{-2 \pi i \frac{I}{N}} \mathbf{1}_{N} , e^{2 \pi i \frac{I}{NM}} ,  e^{-2 \pi i \frac{I}{M}} \mathbf{1}_{M}) 
\end{array}    
\end{equation}
specified by the charge of the hypers in \eqref{newmirror}. The kernel of this morphism is $NM \mathbb{Z}$ so the image is isomorphic to $\mathbb{Z}_{NM}$. We shift the magnetic lattices accordingly \cite{Bourget:2020xdz}, which leads us to introduce the sums 
\begin{eqnarray}
H[I] &=&  \sum\limits_{a_1 , \dots , a_{N-1} \in \mathbb{N}} \sum\limits_{b \in \mathbb{Z} + \frac{I}{MN}}  \sum\limits_{c_1 , \dots , c_{M-1} \in \mathbb{N}} t^{\Delta (a,b,c)}  \frac{P^{\mathrm{SU}} (a ; t) P^{\mathrm{SU}} (c ; t)}{1-t^2}  \nonumber \\ & & \times   \delta \left( N| \left(-I + \sum\limits_{i=1}^{N-1} i a_i \right)   \right) \delta \left( M| \left(-I + \sum\limits_{i=1}^{M-1} i c_i\right) \right) \, , 
\end{eqnarray}
for any integer $I \in \mathbb{Z}$. 
The final result is then given by 
\begin{equation}
\mathrm{HS} (  \mathcal{C} [ \ref{newmirror} ] ; t ) = \sum\limits_{I \in \mathbb{Z}_{NM}}   H[I] \, . 
\label{eq:hsFormula}
\end{equation}
One finds that this agrees with the Higgs branch Hilbert series for the mirror, thus providing a strong argument for the validity of the mirror pair.

 \section{Minimal Deformations of $ADE$ Singularities}
\label{app:E}

We perform the minimal deformations of ADE-type singularities using finite quiver subtraction. It is known that the minimal deformations of Kleinian singularities are in bijection with the fundamental weights of the associated algebra. Therefore we turn on FI parameters for the affine node as well as one of the other nodes for each such minimal deformation. The quivers to subtract are obtained by finding all minimal free sub-quivers which contain both nodes with non-zero FI.

\subsection{\texorpdfstring{Type $A_n$}{Type A}}

For type $A$, the only breaking pattern is $A_n \rightarrow A_k A_{n-k-1}$. The subtractions are shown in Section \ref{sec:Basic}. 

\subsection{\texorpdfstring{Type $D_n$}{Type D}}

\subsubsection{\texorpdfstring{$D_n \rightarrow A_{n-1}$}{}}

\begin{equation}
  \raisebox{-.5\height}{  \hspace*{-3cm}\begin{tikzpicture}
        \node (a) at (0,0) {
        $\begin{tikzpicture}
            \node[gauge,label=below:{$1$}] (1) at (1,0) {};
            \node[gauge,label=below:{$2$}] (2) at (2,0) {};
            \node (3) at (3,0) {$\cdots$};
            \node[gauge,label=below:{$2$}] (4) at (4,0) {};
            \node[gauge,label=below:{$1$}] (5) at (5,0) {};
            \node[gaugeo,label=right:{$1$}] (4u) at (4,1) {};
            \node[gaugeo,label=left:{$1$}] (2u) at (2,1) {};
            \draw (1)--(2)--(3)--(4)--(5) (4u)--(4) (2)--(2u);
        \end{tikzpicture}$
        };
        \node (d) at (0,-3.5) {
        $\begin{tikzpicture}
            \node at (1.5,.5) {$-$};
            \node[gauge,label=below:{$1$}] (2) at (2,0) {};
            \node (3) at (3,0) {$\cdots$};
            \node[gauge,label=below:{$1$}] (4) at (4,0) {};
            \node[gaugeo,label=right:{$1$}] (4u) at (4,1) {};
            \node[gaugeo,label=left:{$1$}] (2u) at (2,1) {};
            \draw (2)--(3)--(4)--(4u) (2)--(2u);
        \end{tikzpicture}$
        };        
        \node (e) at (0,-7) {
        $\begin{tikzpicture}
            \node[gauge,label=below:{$1$}] (7) at (7,0) {};
            \node[gauge,label=below:{$1$}] (8) at (8,0) {};
            \node (9) at (9,0) {$\cdots$};
            \node[gauge,label=below:{$1$}] (10) at (10,0) {};
            \node[gauge,label=below:{$1$}] (11) at (11,0) {};
            \node[gaugeo,label=right:{$1$}] (10u) at (9,1) {};
            \draw (7)--(8)--(9)--(10)--(11)--(10u)--(7);
        \end{tikzpicture}$
        };
        \draw[->] (a) .. controls (-3,-3.5) .. (e);
    \end{tikzpicture}}
\end{equation}  

\subsubsection{\texorpdfstring{$D_n \rightarrow A_{k-1} D_{n-k}$}{}}

See Figure \ref{fig:solutionsDtype}. 

\subsection{\texorpdfstring{Type $E_6$}{Type E6}}

\subsubsection{\texorpdfstring{$E_6\rightarrow D_5$}{}}

\begin{equation}
   \raisebox{-.5\height}{ \begin{tikzpicture}
        \node (a) at (0,0) {
        $\begin{tikzpicture}
            \node[gaugeo,label=below:{$1$}] (1) at (1,0) {};
            \node[gauge,label=below:{$2$}] (2) at (2,0) {};
            \node[gauge,label=below:{$3$}] (3) at (3,0) {};
            \node[gauge,label=below:{$2$}] (4) at (4,0) {};
            \node[gauge,label=below:{$1$}] (5) at (5,0) {};
            \node[gauge,label=left:{$2$}] (3u) at (3,1) {};
            \node[gaugeo,label=left:{$1$}] (3uu) at (3,2) {};
            \draw (1)--(2)--(3)--(4)--(5) (3)--(3u)--(3uu);
        \end{tikzpicture}$
        };
        \node (b) at (0,-3.5) {
        $\begin{tikzpicture}
            \node at (-0.5,1) {$-$};
            \node[gaugeo,label=below:{$1$}] (1) at (1,0) {};
            \node[gauge,label=below:{$1$}] (2) at (2,0) {};
            \node[gauge,label=below:{$1$}] (3) at (3,0) {};
            \node[gauge,label=left:{$1$}] (3u) at (3,1) {};
            \node[gaugeo,label=left:{$1$}] (3uu) at (3,2) {};
            \draw (1)--(2)--(3) (3)--(3u)--(3uu);
        \end{tikzpicture}$
        };
        \node (c) at (0,-7) {
        $\begin{tikzpicture}
            \node[gauge,label=below:{$1$}] (2) at (2,0) {};
            \node[gauge,label=below:{$2$}] (3) at (3,0) {};
            \node[gauge,label=below:{$2$}] (4) at (4,0) {};
            \node[gauge,label=below:{$1$}] (5) at (5,0) {};
            \node[gauge,label=left:{$1$}] (3u) at (3,1) {};
            \node[gaugeo,label=right:{$1$}] (o) at (4,1) {};
            \draw (2)--(3)--(4)--(5) (3)--(3u) (4)--(o);
        \end{tikzpicture}$
        };
        \draw[->] (a) .. controls (-3,-3.5) .. (c);
    \end{tikzpicture}}
\end{equation}

\subsubsection{\texorpdfstring{$E_6\rightarrow A_1A_4$}{}}

\begin{equation}
  \raisebox{-.5\height}{  \begin{tikzpicture}
        \node (a) at (0,0) {
        $\begin{tikzpicture}
            \node[gauge,label=below:{$1$}] (1) at (1,0) {};
            \node[gaugeo,label=below:{$2$}] (2) at (2,0) {};
            \node[gauge,label=below:{$3$}] (3) at (3,0) {};
            \node[gauge,label=below:{$2$}] (4) at (4,0) {};
            \node[gauge,label=below:{$1$}] (5) at (5,0) {};
            \node[gauge,label=left:{$2$}] (3u) at (3,1) {};
            \node[gaugeo,label=left:{$1$}] (3uu) at (3,2) {};
            \draw (1)--(2)--(3)--(4)--(5) (3)--(3u)--(3uu);
        \end{tikzpicture}$
        };
        \node (b1) at (-3,-3.5) {
        $\begin{tikzpicture}
            \node at (1.5,1) {$-$};
            \node[gaugeo,label=below:{$2$}] (2) at (2,0) {};
            \node[gauge,label=below:{$3$}] (3) at (3,0) {};
            \node[gauge,label=below:{$2$}] (4) at (4,0) {};
            \node[gauge,label=below:{$1$}] (5) at (5,0) {};
            \node[gauge,label=left:{$2$}] (3u) at (3,1) {};
            \node[gaugeo,label=left:{$1$}] (3uu) at (3,2) {};
            \draw (2)--(3)--(4)--(5) (3)--(3u)--(3uu);
        \end{tikzpicture}$
        };
        \node (b2) at (3,-3.5) {
        $\begin{tikzpicture}
            \node at (0.5,1) {$-$};
            \node[gauge,label=below:{$1$}] (1) at (1,0) {};
            \node[gaugeo,label=below:{$2$}] (2) at (2,0) {};
            \node[gauge,label=below:{$2$}] (3) at (3,0) {};
            \node[gauge,label=below:{$1$}] (4) at (4,0) {};
            \node[gauge,label=left:{$1$}] (3u) at (3,1) {};
            \node[gaugeo,label=left:{$1$}] (3uu) at (3,2) {};
            \draw (1)--(2)--(3)--(4) (3)--(3u)--(3uu);
        \end{tikzpicture}$
        };
        \node (c1) at (-3,-7) {
        $\begin{tikzpicture}
            \node[gaugeo,label=right:{$1$}] (o) at (1,1) {};
            \node[gauge,label=below:{$1$}] (1) at (1,0) {};
            \draw[transform canvas={xshift=-0.1cm}] (1)--(o);
            \draw[transform canvas={xshift=0.1cm}] (1)--(o);
        \end{tikzpicture}$
        };
        \node (c2) at (3,-7) {
        $\begin{tikzpicture}
            \node[gaugeo,label=right:{$1$}] (o) at (4,1) {};
            \node[gauge,label=below:{$1$}] (3) at (3,0) {};
            \node[gauge,label=below:{$1$}] (4) at (4,0) {};
            \node[gauge,label=below:{$1$}] (5) at (5,0) {};
            \node[gauge,label=left:{$1$}] (3u) at (3,1) {};
            \draw (3)--(4)--(5) (3)--(3u) (3u)--(o)--(5);
        \end{tikzpicture}$
        };
        \draw[->] (a) .. controls (-6,-2.5) and (-6,-4.5) .. (c1);
        \draw[->] (a) .. controls (0,-3.5) .. (c2);
    \end{tikzpicture}}
    \label{eq:subtractionsE6}
\end{equation}

\subsubsection{\texorpdfstring{$E_6\rightarrow A_2A_1A_2$}{}}

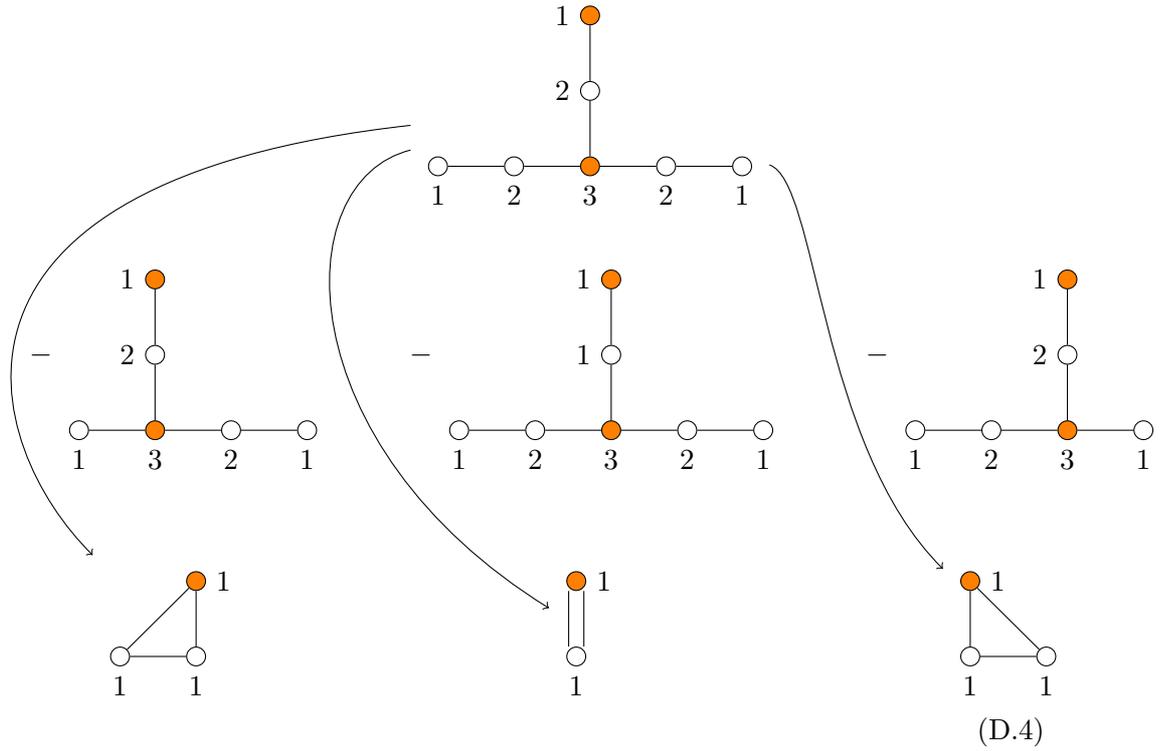
\begin{equation}
   \raisebox{-.5\height}{ \begin{tikzpicture}
        \node (a) at (0,0) {
        $\begin{tikzpicture}
            \node[gauge,label=below:{$1$}] (1) at (1,0) {};
            \node[gauge,label=below:{$2$}] (2) at (2,0) {};
            \node[gaugeo,label=below:{$3$}] (3) at (3,0) {};
            \node[gauge,label=below:{$2$}] (4) at (4,0) {};
            \node[gauge,label=below:{$1$}] (5) at (5,0) {};
            \node[gauge,label=left:{$2$}] (3u) at (3,1) {};
            \node[gaugeo,label=left:{$1$}] (3uu) at (3,2) {};
            \draw (1)--(2)--(3)--(4)--(5) (3)--(3u)--(3uu);
        \end{tikzpicture}$
        };
        \node (b1) at (-5.5,-3.5) {
        $\begin{tikzpicture}
            \node at (1.5,1) {$-$};
            \node[gauge,label=below:{$1$}] (2) at (2,0) {};
            \node[gaugeo,label=below:{$3$}] (3) at (3,0) {};
            \node[gauge,label=below:{$2$}] (4) at (4,0) {};
            \node[gauge,label=below:{$1$}] (5) at (5,0) {};
            \node[gauge,label=left:{$2$}] (3u) at (3,1) {};
            \node[gaugeo,label=left:{$1$}] (3uu) at (3,2) {};
            \draw (2)--(3)--(4)--(5) (3)--(3u)--(3uu);
        \end{tikzpicture}$
        };
        \node (b2) at (0,-3.5) {
        $\begin{tikzpicture}
            \node at (0.5,1) {$-$};
            \node[gauge,label=below:{$1$}] (1) at (1,0) {};
            \node[gauge,label=below:{$2$}] (2) at (2,0) {};
            \node[gaugeo,label=below:{$3$}] (3) at (3,0) {};
            \node[gauge,label=below:{$2$}] (4) at (4,0) {};
            \node[gauge,label=below:{$1$}] (5) at (5,0) {};
            \node[gauge,label=left:{$1$}] (3u) at (3,1) {};
            \node[gaugeo,label=left:{$1$}] (3uu) at (3,2) {};
            \draw (1)--(2)--(3)--(4)--(5) (3)--(3u)--(3uu);
        \end{tikzpicture}$
        };
        \node (b3) at (5.5,-3.5) {
        $\begin{tikzpicture}
            \node at (0.5,1) {$-$};
            \node[gauge,label=below:{$1$}] (1) at (1,0) {};
            \node[gauge,label=below:{$2$}] (2) at (2,0) {};
            \node[gaugeo,label=below:{$3$}] (3) at (3,0) {};
            \node[gauge,label=below:{$1$}] (4) at (4,0) {};
            \node[gauge,label=left:{$2$}] (3u) at (3,1) {};
            \node[gaugeo,label=left:{$1$}] (3uu) at (3,2) {};
            \draw (1)--(2)--(3)--(4) (3)--(3u)--(3uu);
        \end{tikzpicture}$
        };
        \node (c1) at (-5.5,-7) {
        $\begin{tikzpicture}
            \node[gaugeo,label=right:{$1$}] (o) at (2,1) {};
            \node[gauge,label=below:{$1$}] (1) at (1,0) {};
            \node[gauge,label=below:{$1$}] (2) at (2,0) {};
            \draw (1)--(2) (1)--(o)--(2);
        \end{tikzpicture}$
        };
        \node (c2) at (0,-7) {
        $\begin{tikzpicture}
            \node[gaugeo,label=right:{$1$}] (o) at (1,1) {};
            \node[gauge,label=below:{$1$}] (1) at (1,0) {};
            \draw[transform canvas={xshift=-0.1cm}] (1)--(o);
            \draw[transform canvas={xshift=0.1cm}] (1)--(o);
        \end{tikzpicture}$
        };
        \node (c3) at (5.5,-7) {
        $\begin{tikzpicture}
            \node[gaugeo,label=right:{$1$}] (o) at (1,1) {};
            \node[gauge,label=below:{$1$}] (1) at (1,0) {};
            \node[gauge,label=below:{$1$}] (2) at (2,0) {};
            \draw (1)--(2) (1)--(o)--(2);
        \end{tikzpicture}$
        };
        \draw[->] (a) .. controls (-9,-1) and (-8,-4.5) .. (c1);
        \draw[->] (a) .. controls (-4,-1) and (-4,-4.5) .. (c2);
        \draw[->] (a) .. controls (3,-1) and (3,-4.5) .. (c3);
    \end{tikzpicture}}
\end{equation}

An explicit solution for the first and third subtractions here is given as follows: 
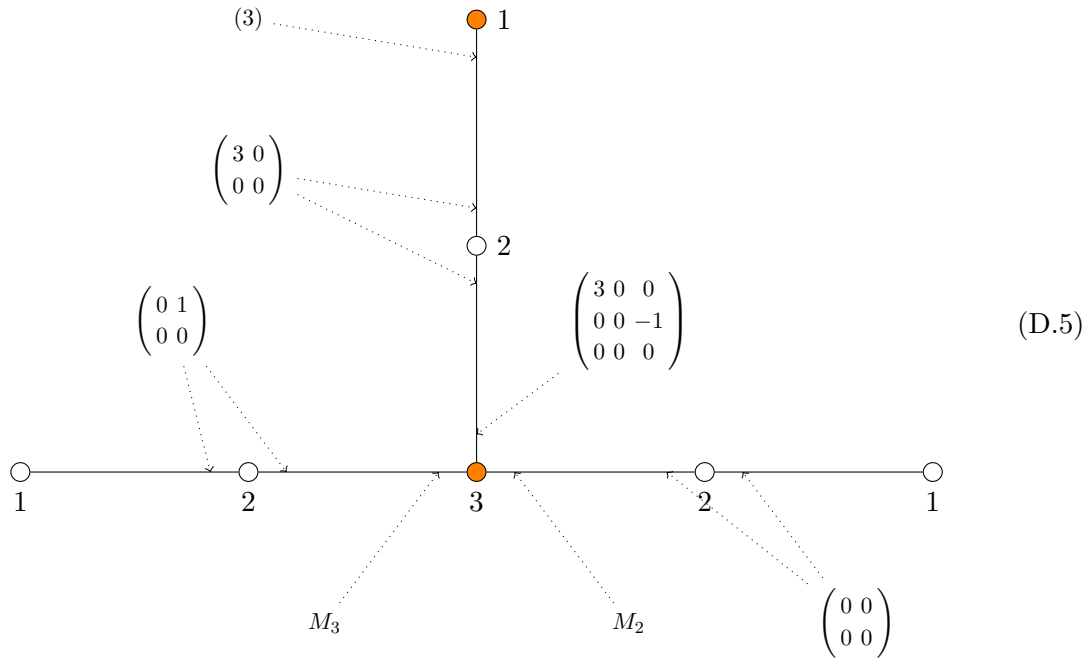
\begin{equation}
   \raisebox{-.5\height}{ \begin{tikzpicture}
            \node[gauge,label=below:{$1$}] (1) at (0,0) {};
            \node[gauge,label=below:{$2$}] (2) at (3,0) {};
            \node[gaugeo,label=below:{$3$}] (3) at (6,0) {};
            \node[gauge,label=below:{$2$}] (4) at (9,0) {};
            \node[gauge,label=below:{$1$}] (5) at (12,0) {};
            \node[gauge,label=right:{$2$}] (6) at (6,3) {};
            \node[gaugeo,label=right:{$1$}] (7) at (6,6) {};
            \draw (1)--(2)--(3)--(4)--(5) (3)--(6)--(7);
            \node (3l) at (4,-2) {\scalebox{.8}{$M_3$}}; 
            \node (3r) at (8,-2) {\scalebox{.8}{$M_2$}};
            \node (3u) at (8,+2) {\scalebox{.8}{$  \left(\begin{array}{ccc}
           3&0&0\\0&0&-1\\0&0&0
            \end{array}\right)$}};
            \node (2r) at (2,+2) {\scalebox{.8}{$  \left(\begin{array}{cc}
           0&1\\0&0
            \end{array}\right)$}};
            \node (6u) at (3,+4) {\scalebox{.8}{$  \left(\begin{array}{cc}
           3&0\\0&0
            \end{array}\right)$}};
            \node (7u) at (3,+6) {\scalebox{.8}{$ (3)$}};
            \node (4r) at (11,-2) {\scalebox{.8}{$  \left(\begin{array}{cc}
           0&0\\0&0
            \end{array}\right)$}};
            \draw[->,dotted] (3l)--(5.5,0);
            \draw[->,dotted] (3r)--(6.5,0);
            \draw[->,dotted] (3u)--(6,.5);
            \draw[->,dotted] (2r)--(3.5,0);
            \draw[->,dotted] (2r)--(2.5,0);
            \draw[->,dotted] (4r)--(9.5,0);
            \draw[->,dotted] (4r)--(8.5,0);
            \draw[->,dotted] (6u)--(6,3.5);
            \draw[->,dotted] (6u)--(6,2.5);
            \draw[->,dotted] (7u)--(6,5.5);
        \end{tikzpicture} }
        \label{eq:solutionE6-13}
\end{equation}
We note that the solution requires to find three matrices, say $M_1$, $M_2$ and $M_3$ such that $M_1$ has Jordan form $J_1 (3) \oplus J_2 (0)$, $M_2$ is nilpotent of order $2$ and $M_3$ is nilpotent of order 3, and $M_1 + M_2 + M_3 = \mathbf{1}_3$. 
Abstractly, we are looking for three endomorphisms of a 3-dimensional vector space satisfying with prescribed Jordan forms and constrained by an overall linear relation. Without loss of generality we can put one of the endomorphisms in Jordan form by picking an appropriate basis. The question becomes: write the following matrix as the sum of two nilpotent matrices $M_2$ and $M_3$ with nilpotency degrees 2 and 3: 
\begin{equation}
    \left(
\begin{array}{ccc}
           -2&0&0\\0&1&1\\0&0&1
            \end{array}\right) = M_2 + M_3 \, , \qquad M_2^2 = M_3^3 = 0 \, .  
            \label{eq:matrixToDecompose}
\end{equation}
One way to address this question is to write the most general nilpotent matrix (the orbit of $J_3(0)$) and impose that the difference with \eqref{eq:matrixToDecompose} squares to 0. This provides many solutions, none of which appear to be particularly simple. We pick one solution 
\begin{equation}
    M_2 = \left(
\begin{array}{ccc}
 \frac{8}{9} & -\frac{2}{3} & \frac{10}{9} \\
 \frac{4}{9} & -\frac{1}{3} & \frac{5}{9} \\
 -\frac{4}{9} & \frac{1}{3} & -\frac{5}{9} \\
\end{array}
\right) \qquad M_3 = \left(
\begin{array}{ccc}
 \frac{10}{9} & \frac{2}{3} & -\frac{10}{9} \\
 -\frac{4}{9} & -\frac{2}{3} & \frac{4}{9} \\
 \frac{4}{9} & -\frac{1}{3} & -\frac{4}{9} \\
\end{array}
\right) \, . 
\end{equation}

\subsubsection{\texorpdfstring{$E_6\rightarrow A_5$}{}}

\begin{equation}
  \raisebox{-.5\height}{  %
}
\end{equation}

\subsection{\texorpdfstring{Type $E_7$}{Type E7}}

\subsubsection{\texorpdfstring{$E_7\rightarrow E_6$}{}}

\begin{equation}
   \raisebox{-.5\height}{ %
}
\end{equation}

\subsubsection{\texorpdfstring{$E_7\rightarrow A_1D_5$}{}}

\begin{equation}
    %
\end{equation}

\subsubsection{\texorpdfstring{$E_7\rightarrow A_2A_4$}{}}

\begin{equation}
    %
\end{equation}

\begin{landscape}

\subsubsection{\texorpdfstring{$E_7\rightarrow A_3A_1A_2$}{}}

\begin{equation}
    \makebox[\textwidth][c]{%
}
\end{equation}

\end{landscape}

\subsubsection{\texorpdfstring{$E_7\rightarrow A_5A_1$}{}}

\begin{equation}
    %
\end{equation}

\subsubsection{\texorpdfstring{$E_7\rightarrow D_6$}{}}

\begin{equation}
  \raisebox{-.5\height}{  %
}
\end{equation}

\subsubsection{\texorpdfstring{$E_7\rightarrow A_6$}{}}

\begin{equation}
   \raisebox{-.5\height}{ %
}
\end{equation}

\subsection{\texorpdfstring{Type $E_8$}{Type E8}}

\subsubsection{\texorpdfstring{$E_8\rightarrow E_7$}{}}

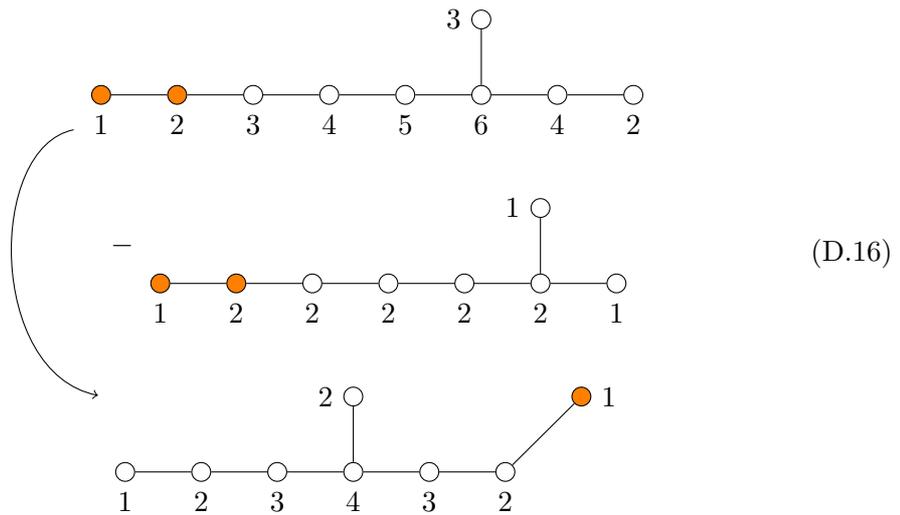
\begin{equation}
   \raisebox{-.5\height}{ %
}
\end{equation}

\subsubsection{\texorpdfstring{$E_8\rightarrow A_1E_6$}{}}

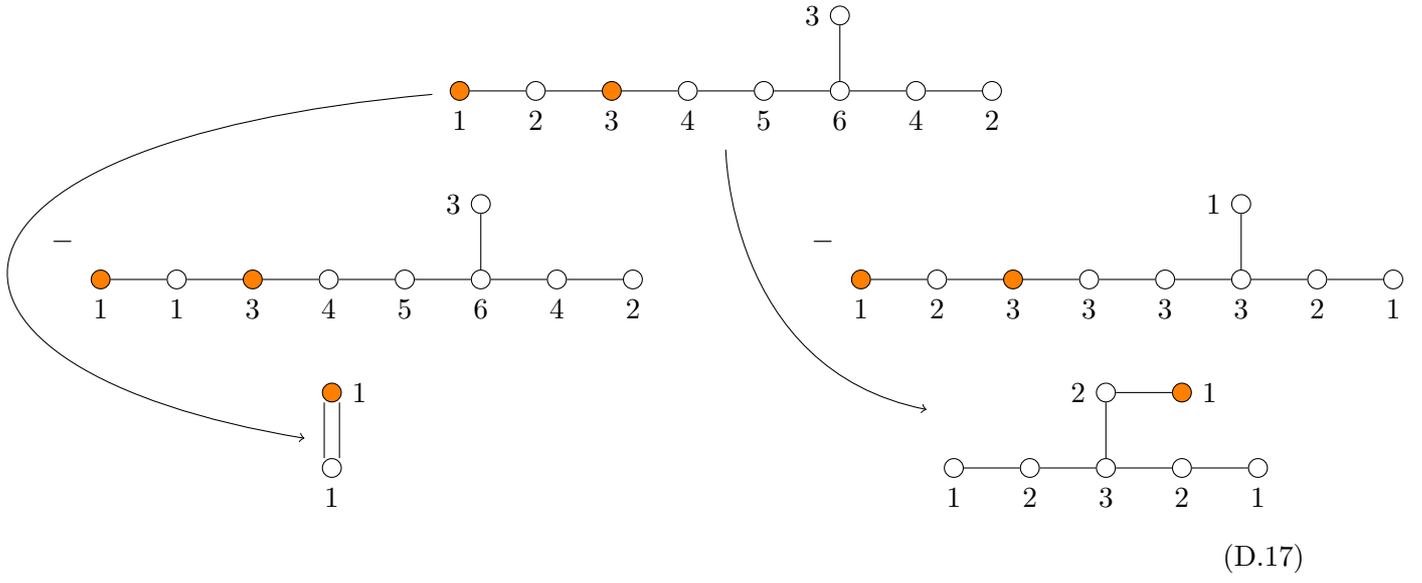
\begin{equation}
    \makebox[\textwidth][c]{%
}
\end{equation}

\subsubsection{\texorpdfstring{$E_8\rightarrow A_2D_5$}{}}

\begin{equation}
    \makebox[\textwidth][c]{%
}
\end{equation}

\subsubsection{\texorpdfstring{$E_8\rightarrow A_3A_4$}{}}

\begin{equation}
    \makebox[\textwidth][c]{%
}
\end{equation}

\begin{landscape}

\subsubsection{\texorpdfstring{$E_8\rightarrow A_4A_1A_2$}{}}

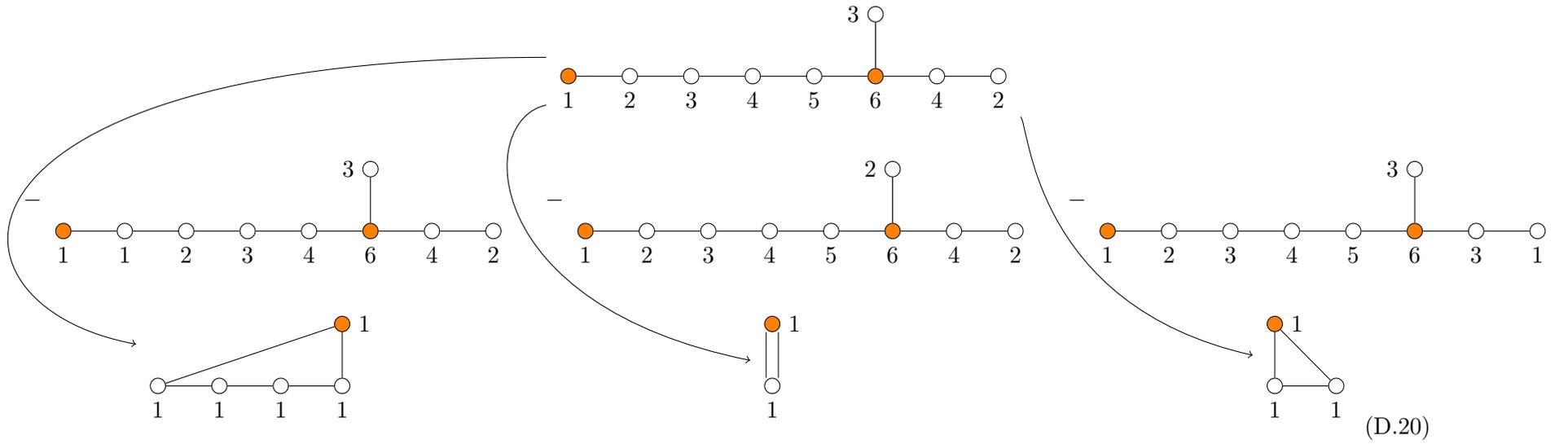
\begin{equation}
    \makebox[\textwidth][c]{%
}
\end{equation}

\end{landscape}

\subsubsection{\texorpdfstring{$E_8\rightarrow A_6A_1$}{}}

\begin{equation}
    \makebox[\textwidth][c]{%
}
\end{equation}

\subsubsection{\texorpdfstring{$E_8\rightarrow D_7$}{}}

\begin{equation}
  \raisebox{-.5\height}{  %
}
    \label{eq:E8D7}
\end{equation}

\subsubsection{\texorpdfstring{$E_8\rightarrow A_7$}{}}

\begin{equation}
   \raisebox{-.5\height}{ %
$
        };
        \draw[->] (a) .. controls (-5,-1) and (-5,-4) .. (c);
    \end{tikzpicture}}
\end{equation}

\providecommand{\href}[2]{#2}\begingroup\raggedright\endgroup

\end{document}